\documentclass[10pt]{book}

\usepackage[paperwidth=5in,paperheight=8in,includehead,
            inner=0.55in,outer=0.4in,top=0.56in,bottom=0.64in]{geometry}
\usepackage{amsmath}
\usepackage{amssymb}
\usepackage{graphicx}
\usepackage{adjustbox}
\usepackage[round,authoryear]{natbib}
\usepackage[hidelinks]{hyperref}
\usepackage{microtype}
\usepackage[most]{tcolorbox}
\usepackage{tikz}
\usetikzlibrary{arrows.meta,calc,bending,shapes.geometric}
\newcommand{\ngon}[1]{\tikz[baseline=-0.75ex]{\node[regular polygon,
  regular polygon sides=#1, draw, minimum size=2.5ex, inner sep=0pt,
  line width=0.5pt] {};}}
\newcommand{\sqdiag}{\tikz[baseline=0.5ex]{%
  \draw[line width=0.5pt] (0,0) rectangle (1.8ex,1.8ex);
  \draw[line width=0.5pt] (0,0) -- (1.8ex,1.8ex);}}
\newcommand{\sqstub}{\tikz[baseline=0.5ex]{%
  \draw[line width=0.5pt] (0,0) rectangle (1.8ex,1.8ex);
  \draw[line width=0.5pt] (1.8ex,1.8ex) -- ++(0.9ex,0.55ex);}}

\usepackage{fancyhdr}
\fancypagestyle{plain}{\fancyhf{}\fancyfoot[C]{\small\thepage}}

\makeatletter
\def\cleardoublepage{\clearpage\if@twoside\ifodd\c@page\else
  \hbox{}\thispagestyle{empty}\newpage\if@twocolumn\hbox{}\newpage\fi\fi\fi}
\makeatother
\colorlet{accent}{black!80}
\colorlet{accentB}{black!50}

\newtcolorbox{calculation}[1]{%
  breakable,
  enhanced,
  colback=black!4,
  colframe=black!45,
  boxrule=0.5pt,
  arc=2pt,
  left=6pt,right=6pt,top=6pt,bottom=6pt,
  fonttitle=\bfseries\small,
  fontupper=\small,
  coltitle=black,
  title={The mathematics behind it: #1 \hfill\textnormal{\itshape(this box can be skipped)}},
}

\begin{document}

\title{Local network growth:\\How simple rules drive network complexity}

\author{Alexei Vazquez}
\maketitle

\thispagestyle{empty}
\vspace*{\fill}
\begin{center}
{\itshape To the ICTP, where it all started\ldots\\[0.7em]
and Nodes \& Links, for a renaissance.}
\end{center}
\vspace*{\fill}
\clearpage

\chapter*{Preface}

In the year 2000 I was at the International Centre for Theoretical Physics 
(ICTP), located in Trieste, Italy. I had left my home country, Cuba, and gone 
there, with many dreams and much nostalgia. At the ICTP I indulged myself in 
the library, the scientific meetings, the computer cluster, and new friends. 
Then, on May 22--25, 2000, I attended the Research Workshop on Graph Theory 
and Statistical Physics. There was a presentation by an obscure Hungarian 
physicist by the name of Albert-L\'aszl\'o Barab\'asi: ``Emergence of scaling 
in complex networks.'' A small conference room and few attendees. Yet, it was 
eye-opening, a fascinating introduction to the properties of real networks 
and a hypothesis for their evolution. When I left that room I was charged, 
full of ideas. Two weeks later I had my first paper on network science: 
``Knowing a network by walking on it: emergence of scaling'' 
\citep{vazquez2000walk}. And right there, it all started.

Networks are everywhere, and so are the regularities they share. Whether one
looks at the Internet, the World Wide Web, a society, a cell, or the scientific
literature, the same statistical signatures recur: broad, often scale-free
degree distributions; short distances between nodes; high local clustering that
thins out in a regular hierarchy as degree grows; and systematic correlations
between the degrees of neighboring nodes. That these features appear across
systems with nothing else in common is one of the most striking findings of
network science. Explaining them has been the field's central occupation for
more than two decades.

The dominant explanation has rested on the preferential attachment mechanism, 
proposed by R\'eka Albert and Barab\'asi \citeyearpar{barabasi1999}: new nodes connect to existing ones
in proportion to their degree, and a power law follows. It is a powerful 
idea, but taken literally it is a \emph{global} rule, requiring each newcomer 
to weigh every node in the network against every other. Real web pages, 
proteins, and people do no such thing. They act on what is near them --- the 
page a surfer reaches, the friend a friend introduces, the gene that gets 
copied. Preferential attachment is a macroscopic perception. In the same way, our perception of temperature is a macroscopic consequence of air molecules heating our skin. This book is about the microscopic mechanism behind 
preferential attachment and beyond: the regularities of real networks 
\emph{emerge} from local rules. A growth rule that reaches just one step 
beyond a randomly chosen node, to one of its neighbors, already attaches 
preferentially by degree --- and the same locality that produces preferential 
attachment produces the clustering hierarchy, the degree correlations, the 
community structure, and the multiplicity of shortest paths, all as a 
package. Coincidentally, I consolidated these ideas while working as a postdoctoral research associate in Barab\'asi's group, then at the University of Notre Dame.

The chapters develop this thesis one mechanism at a time. After an introduction
that fixes the measures and states the argument, I treat growth by search
(citation and web graphs), by triadic closure (social networks), and by
duplication (protein interaction networks) --- the three local rules of my 2003
paper --- and then carry the same principle into more recent work: the
duplication--split structure of project schedules, the emergence of communities
from local rules, and the boosting of shortest path multiplicity. Each chapter
pairs a concrete local rule with a class of real networks it illuminates, and
each ends where it began, with the conclusion that locality is the common cause.

The book is intended for anyone curious about how complex systems organize 
themselves --- the interested general reader as much as the specialist. 
Readers drawn to the \emph{general} principle will find it running through 
every chapter. Mathematicians, computer scientists, and physicists, in 
particular, will recognize a single idea applied again and again: that a 
handful of simple local rules, with no global coordination, is enough to 
account for the statistics that recur across otherwise unrelated networks; 
the closing chapters on the emergence of communities and on the redundancy of 
shortest paths push that idea furthest. Readers drawn instead to a 
\emph{particular} system will find it given a chapter of its own. Social 
scientists will recognize their networks of acquaintance and collaboration in 
the friends-of-friends mechanism of Chapter~\ref{ch:friends}; biologists will 
find the evolution of protein interaction networks through gene duplication 
in Chapter~\ref{ch:duplication}; project managers and engineers will find the 
anatomy of project schedules --- and the surprising irrelevance of the 
critical path --- in the activity-specialization model of 
Chapter~\ref{ch:projects}; and those who study the web, citation, or the 
spread of information online will find the search-and-connect rule of 
Chapter~\ref{ch:search}. Graduate students entering the field will find the 
models developed in full. Whatever the entry point, the core ideas are told 
in plain language, and the more technical calculations are set in boxes that 
can be skipped without losing the thread. I have tried to keep the analysis 
self-contained, favoring mean-field arguments that make the mechanisms 
transparent over exhaustive rigor, and to let the empirical data set the 
agenda throughout. Readers will find the degree denoted $k$, the elements of 
a graph called nodes, and their relations called links, following the 
now-standard usage of network science.

Much of this material grew out of collaborations and conversations over many 
years, with special mentions to Alessandro Vespignani, Yamir Moreno, Romualdo 
Pastor-Satorras, Amos Maritan, Alessandro Flammini, Albert-L\'aszl\'o 
Barab\'asi, Sara Nadiv Soffer and Christos Ellinas. The original research 
papers on which the chapters draw are cited at the appropriate points and 
collected in the bibliography. Parts of the prose, and several of the 
figures, were prepared and edited with the help of Anthropic's Claude 
(Opus~4.8). I am grateful to the colleagues who shaped these ideas, and to 
the reader who, I hope, will come away persuaded that complex networks need 
not be designed --- that they can grow themselves, one local decision at a
time.

\emph{A note on the second edition.} The first edition argued its case largely
through models and mean-field reasoning; this one holds the models up to the
world. Real networks now run through the book --- drawn from public repositories
and measured with standard tools --- beginning in the introduction, where the
recurring regularities are shown in data rather than only described, and where a
closer look finds that one of them, the clustering hierarchy, is in good part an
artifact of how it is usually measured. The two closing chapters have grown the
most. The chapter on communities now makes exactly solvable, for several
idealized networks, the question of \emph{when} a network acquires communities,
and ends by measuring across close to a hundred real networks how their
\emph{number} grows with size; the chapter on shortest paths gains a mean-field
theory of redundancy and two constructions solved in closed form. Throughout, new
figures --- the first plots of real data in the book, alongside fresh schematics
--- aim to make the ideas visible as well as legible. The thesis is unchanged; the
evidence for it is stronger.

\emph{A note on the third edition.} The second edition ended its chapter on
communities with a number: the count of communities grows as a power of network
size, with an exponent above the self-similar one half. Three new chapters ask
what that number is a number \emph{of}, and the answer turns out to depend less
on the networks than on the instrument.

The first of them replaces the flat description of communities with a
\emph{nested} one, in which the groups are themselves grouped --- communities of
communities, the block structure described by a block model of its own rather
than spelled out entry by entry. That sounds like a technicality and is not. A
flat description pays a charge growing with the square of the number of groups,
and so cannot report more than about $\sqrt{n}$ of them however many are there;
and $\sqrt{n}$ is precisely the self-similar benchmark the measured exponent was
being held against, so the instrument could not have returned a value far above
it even in principle. Lift that ceiling and the exponent goes from two thirds to
one. The number of communities grows in proportion to size, their typical size
stops growing, and what a locally grown network converges to is not a shape at
all but a density --- so many communities per node, constant as the network grows
and set by the rule that grew it. That is the second edition's power law resolved
into something a network simply has, the way it has a clustering coefficient. It
comes with a caution the same chapter supplies: of the hierarchy such a fit
reports, only the bottom level is a measurement, and the tidy branching above it
is the prior's own scaffolding.

The other two chapters ask what the describer may \emph{name}. Giving the model a
vocabulary that can call a set of links a triangle or a square, rather than only
a pair of endpoints, can remove the communities altogether --- on networks whose
communities the second edition had proved exist. And on the two lattices whose
best description can be written down by hand rather than searched for, the
fitting misses it, by a tenth of the description length; there is no better cure
for confidence in one's own numbers than a case where the answer can be checked.

What emerges from following that is a single object rather than two. A network
motif and a network community are the same thing at two scales --- both say that
a set of nodes belongs together and how it is joined, the motif from a template
fixed in advance, the community bespoke and inferred --- and once they are
counted together, the quantity that survives every change of instrument is the
density of the two combined, constant in size and set by the growth rule. Local
rules do not converge to a shape that can be drawn. They converge to a rate at
which they keep writing structure, and the argument over how much of that
structure is ``motif'' and how much is ``community'' turns out to be an argument
about vocabulary. The three chapters are the case for that, and they are also, I
hope, an honest record of how often the instrument had to be rebuilt before the
measurement meant anything.

\vspace{1em}
\noindent Alexei Vazquez

\tableofcontents

\chapter{Introduction}
\label{ch:introduction}

\setcounter{page}{1}
\pagenumbering{arabic}

\section{Networks everywhere}

Look closely at almost any complex system and you will find, hiding inside 
it, a network. A network, in the sense used throughout this book, is nothing 
more exotic than a collection of things together with a record of which pairs 
of them are connected. The ``things'' are drawn as dots and called 
\emph{nodes}. A connection between two of them is drawn as a line and called 
a \emph{link}. That is the whole vocabulary needed to begin: a network is 
dots and lines, nodes and links.

What is remarkable is how many different systems fit this simple picture. The 
Internet is a network whose nodes are routers and whose links are the cables 
between them. The World Wide Web is a network of pages connected by the 
hyperlinks you click. A society is a network of people connected by 
acquaintance, friendship, collaboration, or contact. A living cell contains a 
network of proteins connected when two of them physically interact, and 
another of chemicals connected by the reactions that turn one into another. A 
project is decomposed into discrete activities and the logical dependencies 
between them. The scientific literature is a network of papers connected by 
citations, and of authors connected when they write a paper together 
\citep{newman2001,newman2003,albert2002,dorogovtsev2002}. These systems share 
almost nothing on the surface --- one is silicon and glass, another is flesh 
and acquaintance, another is ink and ideas --- and yet, viewed as networks, 
they turn out to look astonishingly alike.

That likeness is the subject of this book. Over the past few decades, as
researchers gathered maps of real networks ranging from a handful of nodes to
billions, the same handful of features kept reappearing, no matter the system.
Four of them organize everything that follows, so it is worth meeting them in
plain language before we make them precise.

The first is the \emph{small-world} property \citep{watts1998}. In a real
network you can usually get from any node to any other in only a few hops, even
when the network is enormous --- the famous ``six degrees of separation'' between
any two people on Earth is one example. At the same time these networks are
highly \emph{cliquish}: your friends tend to be friends of one another, far more
than chance would predict. Short hops everywhere, plus dense little
neighborhoods: that combination is the small world.

The second is the \emph{scale-free} property \citep{barabasi1999,faloutsos1999}.
If you count how many connections each node has, you find that most nodes have
very few, while a small number --- the \emph{hubs} --- have a huge number.
Hollywood has a few actors who have worked with almost everyone and a multitude
who have appeared in one film; the Web has a few pages everybody links to and
billions that are barely linked at all. There is no ``typical'' number of
connections in such a network, which is what the term \emph{scale-free} is
meant to capture.

The third, more subtle and the real connective tissue of this book, is that real
networks carry systematic \emph{correlations} and a \emph{hierarchy}
\citep{pastorsatorras2001,vazquez2002,ravasz2003,vazquez2025}. Whom a node
connects to is not random with respect to how connected those neighbors are, and
the cliquishness around a node depends in a regular way on how many connections
the node has. We will make both of these ideas concrete in a moment.

The fourth, more recent and pitched at a larger scale than the rest, is that real
networks divide into \emph{communities}: groups of nodes densely connected among
themselves but only loosely tied to the rest. People cluster into social circles,
web pages into topics, proteins into functional modules --- so that a network
often looks less like one seamless fabric than a patchwork of tight-knit groups
stitched loosely together. Detecting these communities, and explaining why they
arise, has grown into a research field of its own
\citep{girvan2002,fortunato2010,palla2005}.

The central question of the book is not \emph{whether} these patterns exist ---
the evidence is by now overwhelming --- but \emph{where they come from}. The
answer we will develop, over and over in different settings, is that they need
not be designed in. They appear all by themselves in networks that grow
according to \emph{local rules}: rules in which each newly arriving node decides
where to attach using only what is right in front of it --- a node it has already
reached, and that node's immediate neighbors. No node ever needs to survey the
whole network. That a global pattern can emerge from purely local decisions is,
in one sentence, the thesis of this book.

\section{Some basic measures}

To talk about networks precisely we need a few simple measures. None of them is
harder than counting; we introduce them gently here and will lean on them in
every chapter. Throughout, we picture a network with $n$ nodes. When it helps to
be formal we use the \emph{adjacency matrix} $a_{ij}$, a bookkeeping device that
is simply $1$ if nodes $i$ and $j$ are connected and $0$ if they are not --- a
giant table of yes/no answers to ``are these two linked?''

\paragraph{Degree and degree distribution.}
The \emph{degree} of a node is its number of links --- how many neighbors it has.
We write it $k$, so a node with $k=3$ has three connections. (In formulas, the
degree of node $i$ is $k_i = \sum_j a_{ij}$, which just says: add up the
yes-answers in row $i$ of the table.) The \emph{degree distribution} $p(k)$ is
the recipe for how degrees are spread across the network: it is the fraction of
nodes that have exactly $k$ links, equivalently the chance that a node picked at
random has degree $k$.

\begin{figure}[t]
\centering
\begin{adjustbox}{max width=\linewidth}\begin{tikzpicture}[
  sm/.style={circle,draw=black!60,fill=black!5,minimum size=4mm,inner sep=0pt},
  md/.style={circle,draw=black!60,fill=black!8,minimum size=6mm,inner sep=0pt},
  hub/.style={circle,draw=accent!80!black,fill=black!30,minimum size=10mm,inner sep=0pt},
  lnk/.style={draw=black!40,line width=0.7pt}]
  \node[hub] (H) at (1.7,0.6) {};
  \node[sm] (n1) at (0.1,1.5)  {};
  \node[sm] (n2) at (-0.2,0.3) {};
  \node[sm] (n3) at (0.7,-0.8) {};
  \node[sm] (n4) at (2.0,1.9)  {};
  \node[sm] (n5) at (3.1,1.2)  {};
  \node[md] (n6) at (3.3,-0.3) {};
  \node[sm] (n7) at (2.1,-0.9) {};
  \node[sm] (m)  at (4.6,0.4)  {};
  \draw[lnk] (H)--(n1) (H)--(n2) (H)--(n3) (H)--(n4) (H)--(n5) (H)--(n6) (H)--(n7);
  \draw[lnk] (n6)--(m) (n5)--(m);
  \node[accent,font=\scriptsize] at (1.7,-0.45) {hub, $k=7$};
  \node[font=\scriptsize] at (4.6,-0.25) {$k=2$};
\end{tikzpicture}\end{adjustbox}
\caption{\textbf{Degree.} A node's degree is simply its number of links. Drawn
here with node size growing with degree: a few hubs (such as the central one,
$k=7$) carry many links, while most nodes have only one or two. That imbalance ---
many small nodes, a few giant hubs --- is the hallmark of a scale-free
network.}
\label{fig:intro-degree}
\end{figure}
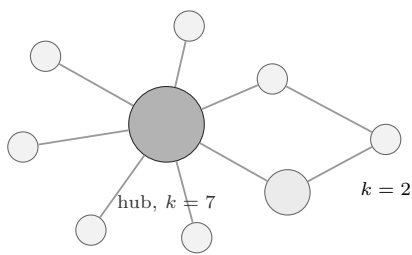

A network is called \emph{scale-free} when this distribution follows a
\emph{power law},
\begin{equation}
  p(k) \sim k^{-\gamma},
  \label{eq:intro-powerlaw}
\end{equation}
where the symbol $\sim$ means ``behaves like'' for large $k$, and $\gamma$
(gamma) is a number, the \emph{exponent}, typically between $2$ and $3$ for real
networks. A power law is just a rule that says ``ten times bigger is some fixed
number of times rarer.'' Unlike the bell-shaped curves of everyday statistics
--- human heights, say, which cluster tightly around an average --- a power law
has a long, heavy tail: extremely well-connected hubs are rare but nowhere near
as rare as a bell curve would allow. This is why scale-free networks are
hub-dominated, and why those hubs end up running much of the show
(Figure~\ref{fig:intro-degree}).

\begin{figure}[t]
\centering
\begin{adjustbox}{max width=\linewidth}\begin{tikzpicture}[
  nd/.style={circle,draw=black!60,fill=black!5,minimum size=4.5mm,inner sep=0pt},
  ce/.style={circle,draw=accent!80!black,fill=black!30,minimum size=5mm,inner sep=0pt},
  lnk/.style={draw=black!40,line width=0.7pt},
  cl/.style={draw=accent,line width=1.3pt}]
  \begin{scope}
    \node[ce] (C) at (0,0)      {};
    \node[nd] (u1) at (0,1.1)   {};
    \node[nd] (u2) at (1.05,0)  {};
    \node[nd] (u3) at (0,-1.1)  {};
    \node[nd] (u4) at (-1.05,0) {};
    \draw[lnk] (C)--(u1) (C)--(u2) (C)--(u3) (C)--(u4);
    \draw[cl] (u1)--(u2) (u2)--(u3) (u3)--(u4) (u4)--(u1);
    \node[font=\footnotesize] at (0,-1.7) {(a) high: $c\approx0.67$};
  \end{scope}
  \begin{scope}[xshift=4.6cm]
    \node[ce] (D) at (0,0)      {};
    \node[nd] (v1) at (0,1.1)   {};
    \node[nd] (v2) at (1.05,0)  {};
    \node[nd] (v3) at (0,-1.1)  {};
    \node[nd] (v4) at (-1.05,0) {};
    \draw[lnk] (D)--(v1) (D)--(v2) (D)--(v3) (D)--(v4);
    \node[font=\footnotesize] at (0,-1.7) {(b) low: $c=0$};
  \end{scope}
\end{tikzpicture}\end{adjustbox}
\caption{\textbf{Clustering coefficient.} The clustering of the central node
(gray) is the fraction of its neighbor pairs that are themselves linked. (a) Four
of the six neighbor pairs are linked (the bold links), so $c\approx0.67$ --- a
cliquish neighborhood. (b) No neighbors are linked to each other, so
$c=0$.}
\label{fig:intro-clustering}
\end{figure}
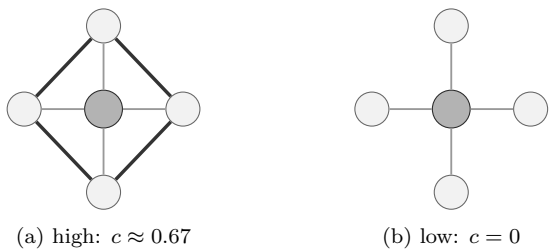

\paragraph{Clustering coefficient.}
The \emph{clustering coefficient} measures how cliquish a neighborhood is: of all
the pairs of your neighbors, what fraction are themselves connected? If you have four friends and
four of the possible six friendships among them actually exist, your clustering
coefficient is $4/6 \approx 0.67$. Formally, if node $i$ has $k_i$ neighbors and
$e_i$ of the possible links among those neighbors are present, then
\begin{equation}
  c_i = \frac{2 e_i}{k_i (k_i - 1)},
  \label{eq:intro-clustering}
\end{equation}
since $k_i(k_i-1)/2$ is the number of pairs that \emph{could} be linked. A
clustering coefficient near $1$ means a tight clique; near $0$ means your
contacts are strangers to each other. Averaged over all nodes we write $\langle
c\rangle$, where the angle brackets $\langle\cdot\rangle$ are standard shorthand
for ``average value.'' The striking empirical fact \citep{watts1998} is that
$\langle c\rangle$ in real networks is far larger than you would get by wiring
the same nodes up at random --- the quantitative fingerprint of those dense local
neighborhoods (Figure~\ref{fig:intro-clustering}).

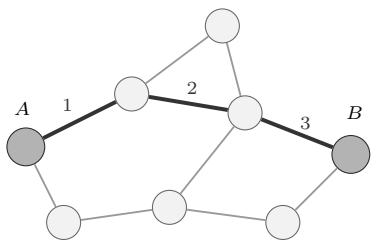
\begin{figure}[t]
\centering
\begin{adjustbox}{max width=\linewidth}\begin{tikzpicture}[
  nd/.style={circle,draw=black!60,fill=black!5,minimum size=4.5mm,inner sep=0pt},
  ep/.style={circle,draw=accent!80!black,fill=black!30,minimum size=5mm,inner sep=0pt},
  lnk/.style={draw=black!40,line width=0.7pt},
  pth/.style={draw=accent,line width=1.5pt}]
  \node[ep] (A)  at (0,0)      {};
  \node[nd] (p1) at (1.4,0.7)  {};
  \node[nd] (p2) at (2.9,0.45) {};
  \node[ep] (B)  at (4.3,-0.1) {};
  \node[nd] (x1) at (0.5,-1.0) {};
  \node[nd] (x2) at (1.9,-0.8) {};
  \node[nd] (x3) at (3.4,-1.0) {};
  \node[nd] (x4) at (2.6,1.6)  {};
  \draw[lnk] (A)--(x1) (x1)--(x2) (x2)--(x3) (x3)--(B) (p1)--(x4) (p2)--(x4) (x2)--(p2);
  \draw[pth] (A)--(p1) (p1)--(p2) (p2)--(B);
  \node[accent,font=\scriptsize] at (0.55,0.55) {1};
  \node[accent,font=\scriptsize] at (2.2,0.78)  {2};
  \node[accent,font=\scriptsize] at (3.7,0.32)  {3};
  \node[font=\scriptsize] at (-0.05,0.5) {$A$};
  \node[font=\scriptsize] at (4.35,0.45) {$B$};
\end{tikzpicture}\end{adjustbox}
\caption{\textbf{Distance.} The distance between two nodes is the number of links
on the shortest route between them. The shortest route from $A$ to $B$ here is
three hops (in bold); the lower route is longer. In real networks these shortest
distances stay small even as the network grows --- the small-world
effect.}
\label{fig:intro-distance}
\end{figure}

\paragraph{Average distance.}
The \emph{distance} between two nodes is the number of links on the shortest
route between them --- the fewest hops to get from one to the other. Averaged
over all pairs of nodes, this distance grows very slowly (logarithmically, or
even more slowly) as the network gets bigger. Slow growth of distance, together
with high clustering, is precisely the small-world effect named above
(Figure~\ref{fig:intro-distance}).

\paragraph{Degree correlations.}
Now to the more subtle pattern. Imagine following a link from a node out to one
of its neighbors and asking: how connected is the neighbor I land on? If the
network had no \emph{degree correlations}, the answer would not depend on where I
started. But real networks are correlated: in some, high-degree hubs tend to sit
next to low-degree nodes; in others, hubs cluster together with other hubs. The
cleanest way to see this is to record, for nodes of each degree $k$, the
\emph{average degree of their neighbors}, written $\langle k_{nn}\rangle(k)$ (the
subscript ``nn'' stands for nearest-neighbors) \citep{pastorsatorras2001}:
\begin{equation}
  \langle k_{nn} \rangle(k) = \sum_{k'} k'\, p(k' \mid k).
  \label{eq:intro-knn}
\end{equation}
Here $p(k' \mid k)$ is read ``the probability that a link from a degree-$k$ node
lands on a degree-$k'$ node,'' and the sum simply averages the landing degree
$k'$ weighted by how likely each is. Do not worry if the formula looks abstract;
the next box pins down the one fact we need from it, and you can take that fact
on faith and skip the box if you prefer.

It is worth knowing what this quantity would be if there were \emph{no}
correlations at all, because that is the yardstick against which real networks
are judged.

\begin{calculation}{the no-correlation baseline for neighbor degree}
If degrees are uncorrelated, the only reason a link is more likely to land on a
particular node is that the node has more links to land on. The chance of
arriving at a node of degree $k'$ is then proportional to $k'$ times how common
such nodes are, $k' p(k')$. Averaging the landing degree $k'$ with this weight
(and dividing by the total weight to normalize) gives
\begin{equation}
  \langle k_{nn} \rangle_{\text{unc}}
  = \frac{\sum_{k'} k'\cdot k' p(k')}{\sum_{k'} k' p(k')}
  = \frac{\langle k^2 \rangle}{\langle k \rangle},
  \label{eq:intro-knn-unc}
\end{equation}
where $\langle k\rangle$ is the average degree and $\langle k^2\rangle$ the
average of the squared degree.
\end{calculation}

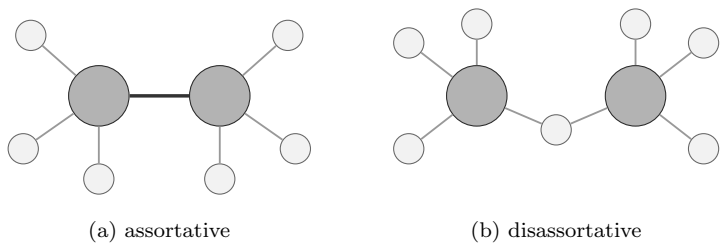
\begin{figure}[t]
\centering
\begin{adjustbox}{max width=\linewidth}\begin{tikzpicture}[
  nd/.style={circle,draw=black!60,fill=black!5,minimum size=4mm,inner sep=0pt},
  hub/.style={circle,draw=accent!80!black,fill=black!30,minimum size=8mm,inner sep=0pt},
  lnk/.style={draw=black!40,line width=0.7pt},
  hl/.style={draw=accent,line width=1.3pt}]
  \begin{scope}
    \node[hub] (Ha) at (0.7,0.3) {};
    \node[hub] (Hb) at (2.3,0.3) {};
    \draw[hl] (Ha)--(Hb);
    \node[nd] (a1) at (-0.2,1.1) {}; \node[nd] (a2) at (-0.3,-0.4) {}; \node[nd] (a3) at (0.7,-0.8) {};
    \node[nd] (b1) at (3.2,1.1)  {}; \node[nd] (b2) at (3.3,-0.4)  {}; \node[nd] (b3) at (2.3,-0.8) {};
    \draw[lnk] (Ha)--(a1) (Ha)--(a2) (Ha)--(a3) (Hb)--(b1) (Hb)--(b2) (Hb)--(b3);
    \node[font=\footnotesize] at (1.5,-1.5) {(a) assortative};
  \end{scope}
  \begin{scope}[xshift=5.2cm]
    \node[hub] (Hc) at (0.5,0.3)  {};
    \node[hub] (Hd) at (2.6,0.3)  {};
    \node[nd] (mid) at (1.55,-0.15) {};
    \draw[lnk] (Hc)--(mid) (Hd)--(mid);
    \node[nd] (c1) at (-0.4,1.0) {}; \node[nd] (c2) at (-0.4,-0.4) {}; \node[nd] (c3) at (0.5,1.25) {};
    \node[nd] (d1) at (3.5,1.0)  {}; \node[nd] (d2) at (3.5,-0.4)  {}; \node[nd] (d3) at (2.6,1.25) {};
    \draw[lnk] (Hc)--(c1) (Hc)--(c2) (Hc)--(c3) (Hd)--(d1) (Hd)--(d2) (Hd)--(d3);
    \node[font=\footnotesize] at (1.55,-1.5) {(b) disassortative};
  \end{scope}
\end{tikzpicture}\end{adjustbox}
\caption{\textbf{Degree correlations.} (a) In an \emph{assortative} network, hubs
tend to link to other hubs (``birds of a feather''). (b) In a
\emph{disassortative} one, hubs avoid each other and link instead to low-degree
nodes (``opposites attract''). Social networks lean assortative; technological and
biological ones, disassortative.}
\label{fig:intro-correlations}
\end{figure}

The result, Eq.~\eqref{eq:intro-knn-unc}, is the key takeaway: in a network with
no degree correlations, the average neighbor degree is the same flat number for
every node, regardless of that node's own degree. So when we plot $\langle
k_{nn}\rangle(k)$ against $k$ for a real network and the curve is \emph{not}
flat, the tilt tells us the kind of correlation present. If the curve slopes
\emph{down} --- well-connected nodes have poorly-connected neighbors --- the
network is called \emph{disassortative}; this is typical of technological and
biological networks. If it slopes \emph{up} --- the well-connected keep company
with the well-connected --- the network is \emph{assortative}, as social networks
tend to be \citep{newman2002}. Assortative and disassortative are simply the
``birds of a feather'' and ``opposites attract'' of network connectivity
(Figure~\ref{fig:intro-correlations}).

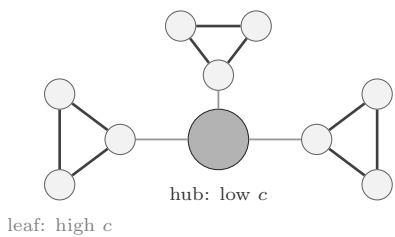
\begin{figure}[t]
\centering
\begin{adjustbox}{max width=\linewidth}\begin{tikzpicture}[
  nd/.style={circle,draw=black!60,fill=black!5,minimum size=4mm,inner sep=0pt},
  hub/.style={circle,draw=accent!80!black,fill=black!30,minimum size=8mm,inner sep=0pt},
  lnk/.style={draw=black!40,line width=0.7pt},
  tri/.style={draw=black!75,line width=1.0pt}]
  \node[hub] (H) at (2.0,0) {};
  \node[nd] (l1) at (-0.1,0.6) {}; \node[nd] (l2) at (-0.1,-0.6) {}; \node[nd] (l3) at (0.7,0) {};
  \draw[tri] (l1)--(l2) (l2)--(l3) (l3)--(l1);  \draw[lnk] (l3)--(H);
  \node[nd] (t1) at (1.5,1.5) {}; \node[nd] (t2) at (2.5,1.5) {}; \node[nd] (t3) at (2.0,0.85) {};
  \draw[tri] (t1)--(t2) (t2)--(t3) (t3)--(t1);  \draw[lnk] (t3)--(H);
  \node[nd] (r1) at (4.1,0.6) {}; \node[nd] (r2) at (4.1,-0.6) {}; \node[nd] (r3) at (3.3,0) {};
  \draw[tri] (r1)--(r2) (r2)--(r3) (r3)--(r1);  \draw[lnk] (r3)--(H);
  \node[accent,font=\scriptsize]  at (2.0,-0.72) {hub: low $c$};
  \node[accentB,font=\scriptsize] at (-0.1,-1.15) {leaf: high $c$};
\end{tikzpicture}\end{adjustbox}
\caption{\textbf{Clustering hierarchy.} Low-degree nodes sit inside tight clusters
where everyone is connected (high clustering, the dark triangles), while a hub
bridges many such clusters. The hub's own neighbors --- one drawn from each
cluster --- do not know each other, so the hub itself has low clustering.
Cliquishness therefore falls as degree rises.}
\label{fig:intro-hierarchy}
\end{figure}

\paragraph{Clustering hierarchy.}
Finally, we can combine the last two ideas by asking how cliquishness depends on
degree. Sorting nodes by their degree $k$ and averaging the clustering
coefficient within each group gives $\langle c\rangle(k)$. In a wide range of
real networks this quantity falls off as a power law,
\begin{equation}
  \langle c \rangle(k) \sim k^{-1},
  \label{eq:intro-hierarchy}
\end{equation}
meaning the more connections a node has, the \emph{less} cliquish its
neighborhood. This is the \emph{clustering hierarchy}, and it has an intuitive
reading \citep{ravasz2002,ravasz2003}: ordinary low-degree nodes sit inside
tight little communities where everyone knows everyone, while the big hubs act as
bridges between many such communities, so their own neighbors --- drawn from all
over --- mostly do not know each other. A small-town local has a dense circle of
mutual friends; a globe-trotting connector knows thousands of people who have
never met (Figure~\ref{fig:intro-hierarchy}).

That reading is appealing, and it is the one the literature has largely adopted.
It is also, in good part, wrong --- and the reason is worth a few pages, because
it is the first example in this book of a global pattern dissolving under
inspection \citep{soffer2005}.

Look again at the denominator of Eq.~\eqref{eq:intro-clustering}. We divided the
triangles $e_i$ around a node by $k_i(k_i-1)/2$, the number of \emph{pairs} of its
neighbors, on the grounds that this is how many links those neighbors could
possibly have among themselves. But could they? A neighbor $j$ has only $k_j$
links in total, and one of them is already spent reaching $i$. So $j$ can reach at
most $k_j-1$ of $i$'s other neighbors --- and if $k_j$ is small, that is nearly
none of them. The denominator counts pairs that the network's own degrees forbid.

The consequence is not subtle. Consider a hub of degree $k$ joined to $k$ nodes of
degree two, each of which spends its second link on one other neighbor of the hub
(Figure~\ref{fig:intro-doublestar}). Every neighbor of the hub is now connected to
another neighbor: they are as interconnected as their degrees permit, and no
rewiring could make them more so. There are $e_i=k/2$ links among them, so the
hub's clustering coefficient is
\begin{equation}
  c = \frac{2\,(k/2)}{k(k-1)} = \frac{1}{k-1},
  \label{eq:intro-hubc}
\end{equation}
which tends to zero as the hub grows. By the standard
measure this maximally-clustered neighborhood is reported as having no clustering
at all. A large-degree node attached to small-degree nodes will always be assigned
a small $c$, however its neighbors are wired --- and in a disassortative network,
that describes every hub. The measure and the correlations are entangled.

\begin{figure}[t]
\centering
\begin{adjustbox}{max width=\linewidth}\begin{tikzpicture}[
  nd/.style={circle,draw=black!60,fill=black!5,minimum size=3.6mm,inner sep=0pt},
  hub/.style={circle,draw=accent!80!black,fill=black!30,minimum size=7mm,inner sep=0pt},
  lnk/.style={draw=black!40,line width=0.7pt},
  tri/.style={draw=black!75,line width=1.0pt}]
  \node[hub] (H) at (0,0) {};
  \foreach \a/\i in {20/1, 55/2, 125/3, 160/4, 200/5, 235/6, 305/7, 340/8}
    \node[nd] (n\i) at ({2.35*cos(\a)},{1.45*sin(\a)}) {};
  \foreach \i in {1,...,8} \draw[lnk] (H)--(n\i);
  \draw[tri] (n1)--(n2) (n3)--(n4) (n5)--(n6) (n7)--(n8);
  \node[accent,font=\scriptsize] at (0,-0.62) {hub, $k=8$};
  \node[font=\scriptsize] at (0,-1.75) {every neighbor has degree $2$, and is paired with another};
\end{tikzpicture}\end{adjustbox}
\caption{\textbf{Why the standard clustering coefficient misleads.} The hub's
neighbors all have degree two, and each spends its spare link on another neighbor
of the hub. They are therefore \emph{as interconnected as their degrees allow} ---
no rewiring could add a single link among them. Yet the standard coefficient,
Eq.~\eqref{eq:intro-clustering}, divides by all $\binom{8}{2}=28$ neighbor pairs
and reports $c=4/28\approx0.14$, falling to zero as the hub grows. The pairs it
counts are pairs the degrees forbid.}
\label{fig:intro-doublestar}
\end{figure}

The repair is to divide not by the pairs one can \emph{draw} but by the links the
neighbors can \emph{afford}.

\begin{calculation}{clustering without the degree-correlation bias}
Let $\omega_i$ be the largest number of links that could exist among the $k_i$
neighbors of node $i$, given those neighbors' degrees. A first bound follows from
the argument above: neighbor $j$ can spend at most $\min(k_i,k_j)-1$ links on
$i$'s other neighbors --- at most $k_j-1$ because that is all it has left, and at
most $k_i-1$ because that is all the targets there are. Summing over neighbors and
halving, since each link is counted from both ends,
\begin{equation}
  \omega_i \;\le\; \Omega_i = \left\lfloor \frac{1}{2}
  \sum_{j \in \mathcal{N}(i)} \big[\min(k_i,k_j)-1\big] \right\rfloor
  \;\le\; \binom{k_i}{2},
  \label{eq:intro-omega}
\end{equation}
where $\mathcal{N}(i)$ denotes the neighbors of $i$. The last inequality is the
whole point: $\Omega_i$ can be far below $\binom{k_i}{2}$, and it is the standard
coefficient's insistence on the larger number that manufactures the hierarchy.
$\Omega_i$ is still only a bound, because it does not check that the links can
actually be arranged; $\omega_i$ itself is obtained by trying. Take the list
$\min(k_i,k_j)-1$ over all neighbors, sorted in decreasing order; spend the
largest entry's budget on as many of the next-largest entries as it can reach,
decrementing each; discard it and any entry now at zero; re-sort and repeat until
the list is empty. The number of links drawn is $\omega_i$. The corrected
clustering coefficient is then simply
\begin{equation}
  \tilde{c}_i = \frac{e_i}{\omega_i},
  \label{eq:intro-ctilde}
\end{equation}
the fraction of the links the neighbors could afford that they actually made.
Three properties follow. It is undefined when $\omega_i=0$ --- a hub of leaves has
no clustering to speak of, rather than zero clustering, which is the honest
verdict. It never falls below the old one, $\tilde{c}_i \ge c_i$. And when every
neighbor is at least as well connected as the node itself, the two coincide,
$\tilde{c}_i = c_i$, so nothing is changed where nothing was wrong.
\end{calculation}

\begin{figure}[t]
\centering
\includegraphics[width=\linewidth]{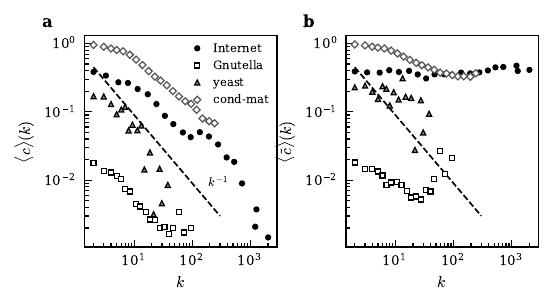}
\caption{\textbf{The clustering hierarchy, and what is left of it.} Four real
networks --- the Internet at the autonomous-system level, the Gnutella
file-sharing overlay, the yeast protein interaction network, and \texttt{cond-mat}
co-authorship --- drawn from the Netzschleuder repository
(Section~\ref{sec:intro-data}). (a) The standard clustering coefficient falls with
degree in every network, roughly as $k^{-1}$ (dashed guide): the hierarchy of
Eq.~\eqref{eq:intro-hierarchy}. (b) The same networks measured with
$\tilde{c}$, Eq.~\eqref{eq:intro-ctilde}, which divides by the links the neighbors
can afford rather than by the pairs one can draw. The vertical scale is the same
as in (a). The Internet's curve is now flat --- it varies by a factor of $1.5$
across four decades of degree --- and every other decay is much weakened. Most of
the hierarchy was in the measure, not in the network.}
\label{fig:data-clustering}
\end{figure}

Measure $\tilde{c}$ on real networks and the hierarchy largely evaporates
(Figure~\ref{fig:data-clustering}; the two panels share a vertical scale so that
the change can be read off directly). The four networks used here are introduced
properly in Section~\ref{sec:intro-data}, and are the ones this book keeps
returning to: the Internet at the level of its service providers, the Gnutella
file-sharing network, the protein interactions of yeast, and a physics
co-authorship network. For the Internet, $\langle c\rangle(k)$ falls by more than a
decade across four decades of degree; $\langle\tilde c\rangle(k)$ is \emph{flat}
--- it varies by a factor of $1.5$ over the whole range, and its fitted slope is
$+0.02$ where the uncorrected slope was $-0.76$. The same happens, less
completely, for the others: every slope moves sharply toward zero. What is left,
when the bias is stripped out, is a clustering coefficient that is either roughly
constant or decays only logarithmically with degree \citep{soffer2005} --- and a
logarithm, over the two or three decades of degree a real network offers, is very
nearly a constant.

So Eq.~\eqref{eq:intro-hierarchy} should be read with care. It is a robust
\emph{measurement}: plot $\langle c\rangle(k)$ for almost any real network and you
will see it. It is much weaker as \emph{evidence} for a designed hierarchy of
modules, because a large part of what it records is the degree correlations of the
network reflected back through a biased denominator. The clustering hierarchy and
the degree correlations are not two independent regularities; they are, in good
part, two views of one.

None of which makes the clustering itself an illusion. Two facts survive the
correction and matter for everything that follows. First, the corrected clustering
is \emph{high} --- $\tilde c \approx 0.39$ for the Internet, $0.80$ for
co-authorship --- so real networks really are cliquish; it is the \emph{trend} with
degree, not the cliquishness, that the bias manufactured. Second, and more
pointedly, that cliquishness is more than the degrees and their correlations can
account for: shuffle the links while holding every degree fixed and the clustering
drops, as Table~\ref{tab:intro-data} will show. The Internet is more clustered than
its own degree sequence and degree correlations can explain. Something built those
triangles, and the degrees do not know what.

The lesson is to read these measures together rather than in isolation --- a theme
that recurs throughout the book, where a single local rule will be seen to set the
clustering hierarchy and the degree correlations \emph{at once}, as facets of one
mechanism. When later chapters derive $\langle c\rangle(k)\sim k^{-1}$ from a local
rule, understand the claim precisely: the rule reproduces what the standard measure
reports of real networks, biases and all, which is the right target --- but the
deeper quantity it is matching is the pair of degree correlations and genuine
cliquishness that lie underneath.

\section{From global to local rules}

With the measures in hand we can state the puzzle this book exists to solve. The
modern story of network growth opens with the model of \citet{barabasi1999},
usually abbreviated BA after its authors. It rests on two ideas. The first is
\emph{growth}: networks are not built all at once but accumulate nodes over time,
the way the Web gains pages or the literature gains papers. The second is
\emph{preferential attachment}: a new node prefers to link to nodes that are
already well connected, with a probability proportional to their degree --- the
``rich get richer'' of networks. From these two ingredients a power-law degree
distribution follows reliably, and preferential attachment has rightly become the
textbook explanation for why so many networks are scale-free.

Two difficulties, though, drive everything that follows. The first is about
\emph{mechanism}. Preferential attachment, read literally, is a \emph{global}
rule: to attach in proportion to degree, a newcomer would have to know the degree
of every node in the entire network and weigh them all against each other.
That is implausible for the systems we care about. A scientist writing a new
paper has read a few dozen others, not the millions in existence; a new gene is a
copy of one existing gene, ignorant of the rest of the genome; a newcomer to a
town meets a handful of people, not the whole population. Real nodes act
\emph{locally}. So if preferential attachment is not literally true, why do so
many networks behave as if it were?

The second difficulty is that the BA model, for all its success with the degree
distribution, does not by itself reproduce the other patterns we just met. Its
clustering fades away as the network grows, it shows none of the clustering
hierarchy of Eq.~\eqref{eq:intro-hierarchy}, and its degree correlations are weak
\citep{albert2002,vazquez2003}. Something is missing.

The resolution pursued throughout this book is to throw out the global rule and
replace it with local ones, then watch what \emph{emerges}. By a \emph{local
rule} we mean a way of growing the network in which a newcomer reaches some node
--- chosen at random, or arrived at by surfing along links, or copied from an
existing node --- and then connects to that node and perhaps to a few of its
immediate neighbors, using nothing but that local information. The recurring and
rather beautiful discovery, established for several such rules
\citep{vazquez2003}, is that
\begin{quote}
\emph{an effective ``rich get richer'' behavior arises all on its own from
growth by local rules.}
\end{quote}
The reason, which we will meet many times, is almost a slogan: a well-connected
node is, by definition, a node that many local explorations will bump into,
because it has many links through which it can be reached. The preference for
hubs need not be legislated; it is a free by-product of poking around the network
locally. And because the very same local moves also create triangles (your
friend introduces you to their friend) and link nodes of unlike degree, the same
rules that produce the power law deliver, in one package, the clustering
hierarchy of Eq.~\eqref{eq:intro-hierarchy} and the degree correlations of
Eq.~\eqref{eq:intro-knn}. Three patterns that look independent turn out to share
a single root.

\section{What the data demand}
\label{sec:intro-data}

Any explanation has to answer to measurements, so it is worth recording up front
what those measurements actually look like. Rather than take this on trust, let us
measure four real networks and see. All four come from the Netzschleuder
repository \citep{netzschleuder}, a public catalogue of several hundred network
datasets, read here through the \texttt{graph-tool} library \citep{peixoto2014};
the code that fetches and measures them accompanies this book. They are: the
\emph{Internet} at the level of the service providers, the ``autonomous
systems,'' that own its routers; the \emph{Gnutella} peer-to-peer file-sharing
overlay; the protein interaction network of \emph{yeast}; and the
\emph{co-authorship} network of the \texttt{cond-mat} preprint archive, in which
two physicists are linked if they have written a paper together. Three
technological or biological, one social --- which turns out to be exactly the
division that matters.

Start with the small world, since it needs only two numbers per network
(Table~\ref{tab:intro-data}). Every one of these networks is small: the average
distance between two nodes is between four and seven hops, in networks of up to
sixty thousand nodes. And every one is far more cliquish than chance. Wiring the
same number of nodes at random with the same number of links --- the comparison
Watts and Strogatz made --- would give the clustering in the column headed
$\langle c\rangle_{\mathrm{ER}}$, which is between forty and two and a half
thousand times smaller than what is observed. The comparison is almost unfair,
though, because it also throws away the hubs, and hubs alone create some
clustering. The last of the clustering columns is therefore the honest one: it
shuffles the links at random while holding every node's degree fixed, and measures
what survives. Even against that stricter null the real networks are the more
cliquish, by factors of two to three hundred. The cliquishness is real, and it is
not merely a shadow of the degree distribution.

\begin{table}[t]
\centering
\footnotesize
\setlength{\tabcolsep}{3.2pt}
\caption{\textbf{Four real networks.} Measured on the largest connected component,
undirected. $\langle\tilde{c}\rangle$ is the clustering corrected for what the
neighbors' degrees allow, Eq.~\eqref{eq:intro-ctilde}; it never falls below
$\langle c\rangle$. $\langle c\rangle_{\mathrm{ER}}$ is the clustering of a random
network with the same $n$ and $\langle k\rangle$; $\langle c\rangle_{\mathrm{rew}}$
that of the same network with its links shuffled at random but every degree held
fixed. $\langle d\rangle$ is the average distance and $r$ the assortativity:
negative for the first three, positive for the last.}
\label{tab:intro-data}
\begin{tabular}{lrrrrrrrr}
\hline
Network & $n$ & $\langle k\rangle$ & $\langle c\rangle$ &
$\langle \tilde{c}\rangle$ &
$\langle c\rangle_{\mathrm{ER}}$ & $\langle c\rangle_{\mathrm{rew}}$ &
$\langle d\rangle$ & $r$ \\
\hline
Internet (AS)  & $22\,963$ & $4.2$ & $0.230$ & $0.386$ & $0.0002$ & $0.123$  & $3.9$ & $-0.052$ \\
Gnutella       & $62\,561$ & $4.7$ & $0.005$ & $0.012$ & $0.0001$ & $0.0002$ & $5.9$ & $-0.036$ \\
Yeast proteins & $1\,458$  & $2.7$ & $0.071$ & $0.215$ & $0.0018$ & $0.0050$ & $6.9$ & $-0.034$ \\
cond-mat       & $36\,458$ & $9.4$ & $0.657$ & $0.800$ & $0.0003$ & $0.0018$ & $5.5$ & $+0.055$ \\
\hline
\end{tabular}
\end{table}

\begin{figure}[t]
\centering
\includegraphics[width=\linewidth]{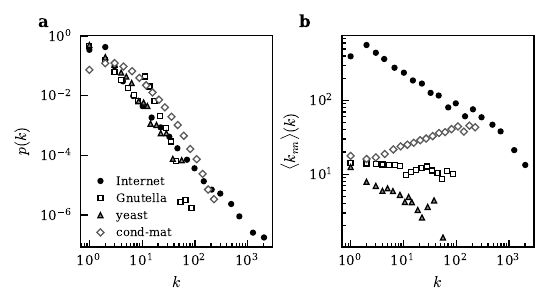}
\caption{\textbf{What the data demand.} The same four networks, from the
Netzschleuder repository through \texttt{graph-tool}. (a) The degree distributions
are broad and hub-dominated; the Internet spans four decades of degree. Gnutella's
protocol caps its peers, so its tail is the shortest. (b) The average neighbor
degree, where the two families part: falling for the Internet, Gnutella and the
yeast proteins --- \emph{disassortative} --- but rising for \texttt{cond-mat}
co-authorship, which is \emph{assortative}. The clustering of these same networks
is in Figure~\ref{fig:data-clustering}.}
\label{fig:data-degree}
\end{figure}

Now the measures that organize the rest of the book
(Figure~\ref{fig:data-degree}). The degree distributions are broad and
hub-dominated: the Internet runs from nodes with a single link to one with more
than two thousand, four decades of degree in a network four decades in size, and
the others do the same over a shorter range. Gnutella is the partial exception
worth naming --- its protocol caps how many peers a client will hold, so its tail
is cut short and it is the least scale-free of the four. Real data are like that;
the power law of Eq.~\eqref{eq:intro-powerlaw} is an idealization that some
systems wear better than others.

The clustering we have already been through: the hierarchy of
Eq.~\eqref{eq:intro-hierarchy} is there in all four
(Figure~\ref{fig:data-clustering}a), with measured slopes from $-0.6$ to $-1.3$,
and most of it dissolves once the degree-correlation bias is removed
(Figure~\ref{fig:data-clustering}b).

And then the two families part company, on the last measure and only on the last
measure. For the Internet, for Gnutella and for the yeast proteins, the average
neighbor degree \emph{falls} with degree: these networks are
\emph{disassortative}, their hubs presiding over peripheries of poorly-connected
nodes rather than huddling together. This is the signature of technological and
biological networks \citep{pastorsatorras2001,maslov2002,jeong2001}. For the
co-authorship network it \emph{rises}: prolific, well-connected authors tend to
write with one another, and the network is \emph{assortative}
\citep{newman2002}. The single number $r$ in Table~\ref{tab:intro-data} says the
same thing more briefly --- negative for the first three, positive for the fourth
--- and the sign of that number is the cleanest division in the data.

The contrast between the two families is instructive, because it shows that
whatever mechanism we propose must be able to produce \emph{either} sign of
correlation, depending on the microscopic move --- and, as the chapters will show,
local rules can do exactly that. It also tells us the two families are not two
kinds of network but one kind seen through one dial: everything else in
Figure~\ref{fig:data-degree} they share.

These regularities are robust, they hold across many orders of magnitude in
network size, and for years their shared origin was a mystery. Supplying that
origin, one mechanism and one class of networks at a time, is the work of this
book.

\section{Plan of the book}

The chapters proceed from mechanism to application, each built around one
concrete local rule and a family of real networks it illuminates. The reader
need not take any chapter on faith: in each, a simple local rule is stated, its
consequences are worked out (with the heavier algebra confined to boxes that can
be skipped), and the predictions are held up against real data.

\begin{itemize}
  \item \textbf{Search and connect.} A newcomer learns of existing nodes by
    exploring the network --- following references, surfing hyperlinks,
    performing a random walk --- and connects to what it finds. This is the
    natural model for searchable, directed networks such as the citation network
    and the World Wide Web, and it is where we show in detail how the simple act
    of surfing manufactures an effective ``rich get richer.''

  \item \textbf{Friends of friends.} A newcomer connects to a node and then to
    some of that node's neighbors --- you meet a person, then their friends. This
    rule for social network growth creates triangles directly, and it generates
    the assortative correlations and the cliquishness hierarchy of social
    networks.

  \item \textbf{Gene duplication.} Biological networks grow by copying: a node is
    duplicated together with its connections, after which links are gained and
    lost. Duplication is an intrinsically local move, and it explains the
    disassortative, hierarchical structure of protein interaction networks.

  \item \textbf{Activity specialization.} The schedule of a large project is a
    network of activities grown by a local refinement rule: each activity is
    either duplicated into parallel copies or split into specialized sequential
    parts. The same copying mechanism gives power-law degrees, while making the
    celebrated ``critical path'' vanish as projects grow and relocating delay
    risk in the network as a whole.

  \item \textbf{Emergence of communities.} Local rules do not only set degrees
    and correlations; they also organize nodes into communities --- tight groups
    with many links inside and few outside. We examine how this mesoscale
    structure arises from the same local dynamics, even when every node is
    otherwise identical.

  \item \textbf{Boosting shortest path multiplicity.} Finally we turn to the
    redundancy of routes through a network --- how many equally short paths join
    two nodes --- and show that local growth multiplies them, tying this
    redundancy to the very community structure of the previous chapter.
\end{itemize}

A single thread runs through all of them. In each case the apparently global
features of a network --- its scale-free degrees, its small-world distances, its
clustering hierarchy, its degree correlations, its communities, its redundant
paths --- turn out to be the collective, emergent consequence of simple rules that
no node applies with more than local knowledge. That is the sense in which the
title of this book should be read: not networks built by a designer, but networks
that grow themselves, one local decision at a time.


\chapter{Search and connect}
\label{ch:search}

\section{Finding what to link to}

The introduction left us with a puzzle. The ``rich get richer'' rule explains
why so many networks are scale-free, but as a literal recipe it asks the
impossible: a newcomer would have to know the degree of every node in the
network in order to favor the well-connected ones. In this chapter we solve the
puzzle for an important family of networks --- the \emph{searchable} ones --- by
replacing that impossible global rule with the most natural local act a newcomer
can perform. It \emph{searches} the network, moving from node to node along the
links, and connects to what it happens to find.

The systems to keep in mind are collections of documents that point at one
another. In the \emph{citation network}, each node is a published article and
each link is a reference from one article to another. In the World Wide Web,
each node is a page and each link is a hyperlink. These are \emph{directed}
networks: a reference, like a hyperlink, points one way. An article cites an
older one, not the reverse; a page links to another, which need not link back.
Because direction matters, each node now has two kinds of degree. Its
\emph{in-degree}, written $k^{in}$, counts the links pointing \emph{at} it ---
how many times a paper has been cited, how many pages link to a given page. Its
\emph{out-degree}, written $k^{out}$, counts the links pointing \emph{out} of it
--- how many references a paper makes, how many hyperlinks a page contains. It is
the in-degree that interests us, because in real document networks it is the
in-degree --- the count of incoming attention --- that is broadly distributed,
with a few hugely-cited classics and a long tail of papers cited once or never.

How does a new document come to point at the old ones? Not by consulting a
global popularity ranking. An author writing a new paper finds references by
reading a few papers and following \emph{their} reference lists, and then the
references of those, and so on \citep{price1965}. A web author discovers pages by
surfing from link to link and by typing queries into a search engine
\citep{brin1998}. In every case the information used is strictly local --- a
document one has reached, and the documents it points to. The claim of this
chapter is that this local searching, repeated by every newcomer, adds up to
exactly the ``rich get richer'' behavior that the introduction had to assume by
fiat, and that it brings the power-law degrees, the clustering hierarchy, and
the degree correlations along with it \citep{vazquez2001,vazquez2003}.

\section{Surfing as a random walk}

To turn ``surfing'' into something we can analyze, we model the surfer as a
\emph{random walker} on the network. A random walk is just what it sounds like:
start somewhere, and step to a randomly chosen neighbor, then to a neighbor of
\emph{that} node, and so on. Our surfer does this with a twist. From whatever
page it is on, with probability $q_e$ it follows one of that page's outgoing
hyperlinks, picked at random; with the remaining probability $1-q_e$ it gets
bored, stops following links, and jumps to a completely random page elsewhere.
The single dial $q_e$, a number between $0$ and $1$, sets how patiently the
surfer explores along links before restarting from scratch.

Now ask: in the long run, what fraction of its time does the walker spend at
each page? Call that fraction the page's \emph{visiting probability}, $v_i$ for
node $i$. A page is visited often if many links point to it and those linking
pages are themselves visited often --- a self-referential condition that turns
out to have a clean form. The remarkable thing, worth flagging before any
algebra, is that this visiting probability is \emph{exactly the quantity that
Google was built on}: it is the \emph{PageRank} of the page, the measure of
importance that ranks search results \citep{brin1998}. So the local rule we are
about to write is not a toy --- it is the very process real search engines compute
and then display to the authors who go on to create new links.

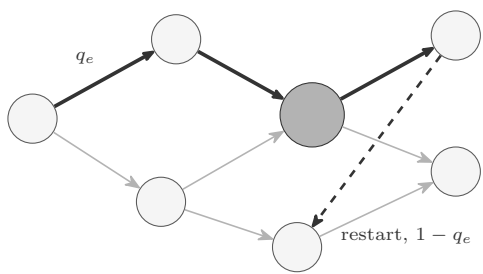
\begin{figure}[t]
\centering
\begin{adjustbox}{max width=\linewidth}\begin{tikzpicture}[>={Stealth[round,length=2.4mm]},
  nd/.style={circle,draw=black!65,fill=black!4,minimum size=6.5mm,inner sep=0pt},
  hub/.style={circle,draw=accent,fill=black!30,minimum size=8.5mm,inner sep=0pt},
  ex/.style={->,draw=black!30,line width=0.6pt},
  walk/.style={->,draw=accent,line width=1.4pt},
  jump/.style={->,draw=accent,line width=1.0pt,dashed}]
  \node[nd]  (n1) at (0,0)      {};
  \node[nd]  (n2) at (1.9,1.05) {};
  \node[nd]  (n3) at (1.7,-1.1) {};
  \node[hub] (n4) at (3.7,0.05) {};
  \node[nd]  (n5) at (3.5,-1.7) {};
  \node[nd]  (n6) at (5.6,1.1)  {};
  \node[nd]  (n7) at (5.6,-0.7) {};
  \draw[ex] (n1) -- (n3);
  \draw[ex] (n3) -- (n4);
  \draw[ex] (n3) -- (n5);
  \draw[ex] (n4) -- (n7);
  \draw[ex] (n5) -- (n7);
  \draw[walk] (n1) -- (n2);
  \draw[walk] (n2) -- (n4);
  \draw[walk] (n4) -- (n6);
  \draw[jump] (n6) -- (n5);
  \node[accent,font=\scriptsize] at (0.7,0.8)  {$q_e$};
  \node[accent,font=\scriptsize] at (4.95,-1.55) {restart, $1-q_e$};
\end{tikzpicture}\end{adjustbox}
\caption{\textbf{Surfing as a random walk.} From the page it is on, the searcher
follows one of the outgoing links with probability $q_e$ (solid arrows) or, with
probability $1-q_e$, jumps to a page chosen at random (dashed arrow). A page
reached along many routes --- like the highlighted hub --- is visited most often;
this visiting frequency is the page's PageRank, and it grows in proportion to the
number of links that point at the page.}
\label{fig:walk-surf}
\end{figure}

The picture to keep in mind is Figure~\ref{fig:walk-surf}. We would like to know
how a page's visiting probability depends on its
in-degree. Getting there exactly is hard, so we make the standard physicist's
move of replacing fluctuating quantities by their averages --- a so-called
\emph{mean-field approximation}, ``mean'' as in average. The next box carries
this out; the one sentence to carry away is stated right after it, so the box
can be skipped.

\begin{calculation}{from surfing to the visiting probability}
The visiting probability of node $i$ obeys a balance condition: the walker
arrives at $i$ either by a random jump (probability $1-q_e$, landing on any of
the $n$ nodes equally) or by following a link into $i$ from some node $j$ that
points to it,
\begin{equation}
  v_i = \frac{1-q_e}{n} + q_e \sum_j a_{ij}\, \frac{v_j}{k^{out}_j},
  \label{eq:walk-pagerank}
\end{equation}
where $a_{ij}=1$ when $j$ points to $i$, and the factor $1/k^{out}_j$ accounts
for $j$ spreading its attention over its $k^{out}_j$ outgoing links. This is the
PageRank equation. To simplify it, replace the sum over the specific nodes
pointing to $i$ by an average. A node is reached along an incoming link from a
node of out-degree $k^{out}$ with probability $k^{out} p_{k^{out}}/\langle
k^{out}\rangle$, and that link is then followed with probability $1/k^{out}$; the
two factors of $k^{out}$ cancel, so each incoming link contributes the same
average amount $\Theta = \langle v\rangle/\langle k^{out}\rangle$. Since node $i$
has $k^{in}_i$ incoming links,
\begin{equation}
  v_i = \frac{1-q_e}{n} + q_e\, \Theta\, k^{in}_i.
  \label{eq:walk-vi-mf}
\end{equation}
\end{calculation}

The payoff is Eq.~\eqref{eq:walk-vi-mf}, and it is simple to state in words: a
page's visiting probability grows in a straight line with its in-degree. Double
the number of links pointing at a page and you roughly double how often the
surfer lands on it. That straight-line relationship is the seed from which
everything in this chapter grows.

\section{An effective ``rich get richer''}

Being visited is not the same as being linked to --- you read many pages you do
not cite. So suppose that each time the surfer visits a page it creates a new
incoming link to it only with some probability $q_v$. We can then ask how fast a
page accumulates new links as a function of how many it already has. This rate
is the network-growth analog of preferential attachment, and computing it is the
crux of the chapter. The derivation is short; its conclusion is in the sentence
after the box.

\begin{calculation}{the rate at which a page gains links}
Let $\nu_s$ surfers be active and $\nu_a$ new pages be added per unit time. The
node count $n$ and link count $e$ grow as
\begin{equation}
  \frac{\partial n}{\partial t} = \nu_a, \qquad
  \frac{\partial e}{\partial t} = \nu_s\, q_v\, \langle v \rangle\, n,
  \label{eq:walk-growth}
\end{equation}
because each of the $\langle v\rangle n$ visits in a walk makes a link with
probability $q_v$. Integrating fixes the average degree and hence $\Theta =
\nu_a/(q_v\nu_s n)$. The rate at which a page of in-degree $k^{in}$ gains its
next link is $A(k^{in}) = q_v\, v(k^{in})$; substituting
Eq.~\eqref{eq:walk-vi-mf} and the value of $\Theta$,
\begin{equation}
  A(k^{in}) = \frac{1}{n}\left[\, q_v(1-q_e) + q_e\, \frac{\nu_a}{\nu_s}\, k^{in} \right].
  \label{eq:walk-attach}
\end{equation}
\end{calculation}

Equation~\eqref{eq:walk-attach} is the central result, and it deserves to be
read slowly. The rate at which a page attracts new links has two parts: a
constant piece, the same for every page, which gives even a brand-new page a
chance of being noticed; and a piece \emph{proportional to its current
in-degree}. That second piece is preferential attachment --- the rich getting
richer --- but notice where it came from. The surfer never consulted a global
ranking of pages; it only followed links, one at a time. The bias toward
well-connected pages appeared on its own, for the simple reason that a page with
many incoming links is a page the wandering surfer reaches by many routes. We
have \emph{derived} the rule the introduction had to assume. The same logic
underlies the ``redirection'' trick of \citet{krapivsky2001}, in which a newcomer
points either to a random node or to whatever that node points to --- a one-step
walk that yields the same effective preference.

\section{The degree distribution}

A linear attachment rate is exactly the ingredient needed to produce a power
law. Working out the resulting distribution of in-degrees requires solving a
balance equation for how many pages sit at each in-degree, which we do in the
box. Readers willing to take the outcome on trust can jump to the discussion
after it.

\begin{calculation}{the in-degree distribution and its exponent}
Let $n_{k^{in}}$ be the number of nodes with in-degree $k^{in}$. It changes as
nodes climb from one in-degree to the next by gaining links, with new nodes
injected at in-degree zero:
\begin{equation}
  \frac{\partial n_{k^{in}}}{\partial t}
  = \nu_s\, A_{k^{in}-1}\, n_{k^{in}-1}
  - \nu_s\, A_{k^{in}}\, n_{k^{in}}
  + \nu_a\, \delta_{k^{in},0}\,,
  \label{eq:walk-rate}
\end{equation}
the symbol $\delta_{k^{in},0}$ being $1$ only at $k^{in}=0$. For a network
growing at a constant rate the ratio $\nu_s/\nu_a=\alpha$ is fixed and the shape
of the distribution settles down, $n_{k^{in}}(t)=N p_{k^{in}}$. Solving for that
stationary shape gives (with $\Gamma$ the gamma function, a smooth
generalization of the factorial)
\begin{equation}
  p_{k^{in}} = \frac{1}{1+a}\,
  \frac{\Gamma[a(\gamma-1)+k^{in}]}{\Gamma[a(\gamma-1)]}\;
  \frac{\Gamma[(1+a)(\gamma-1)+1]}{\Gamma[(1+a)(\gamma-1)+k^{in}+1]},
  \label{eq:walk-pk}
\end{equation}
\begin{equation}
  \gamma = 1 + \frac{1}{q_e}, \qquad a = \alpha\, q_v\,(1-q_e),
  \label{eq:walk-gamma}
\end{equation}
which for large in-degree is the power law
\begin{equation}
  p_{k^{in}} \sim (k^{in})^{-\gamma}.
  \label{eq:walk-asym}
\end{equation}
\end{calculation}

Two features of the result, Eq.~\eqref{eq:walk-gamma}, are worth dwelling on.
First, the exponent $\gamma$ is at least $2$ --- exactly the range seen in real
citation and web networks. Second, and more surprising, $\gamma$ depends
\emph{only} on $q_e$, the surfer's patience for following links. It does not
depend on $q_v$, how often readers actually cite what they read, nor on the
growth rate of the network. Whether authors are generous or stingy with
citations changes how many links there are, but not the \emph{shape} of the
distribution; only the manner of \emph{exploring} the network does that. As the
surfer explores ever more deeply ($q_e\to 1$) the exponent drops to its minimum
of $2$, the most hub-dominated case; as the surfer restarts constantly ($q_e\to
0$) the exponent grows without bound and the heavy tail of hubs disappears.

\section{Cliquishness and correlations}

The same walk that sets the degrees also lays down triangles, so it predicts the
clustering hierarchy of the introduction with no extra assumptions. The reason
is simple to picture: whenever the surfer steps from a page to one of that
page's neighbors and links to both, it has closed a triangle. A short
calculation turns this picture into a formula.

\begin{calculation}{how clustering falls with degree}

Counting the links $e_i$ among the neighbors of node $i$ created by the walk 
gives $\partial e_i/\partial t \approx (1+q_e)\,\partial k^{in}_i/\partial 
t$. Integrating this differential equation, assuming $e = 0$ for $k = 1$, we 
obtain $e = (1 + q_e) (k - 1)$. Inserting this into the definition of the 
clustering coefficient,
$\langle c\rangle_k = 2e(k)/[k(k-1)]$, yields
\begin{equation}
  \langle c \rangle_k \approx \frac{2(1+q_e)}{k}.
  \label{eq:walk-ck}
\end{equation}
\end{calculation}

The message of Eq.~\eqref{eq:walk-ck} is that cliquishness is inversely
proportional to degree: a node with twice the degree has roughly half the
clustering. This is precisely the clustering hierarchy of
Eq.~\eqref{eq:intro-hierarchy}, seen in the Internet, the Web, and the protein
network --- and here it falls out for free from growth by searching. The
intuition is that a node gains a triangle every time the surfer steps from it to
a neighbor, so its triangles grow in step with its degree, while the number of
\emph{possible} triangles grows much faster, like the square of the degree;
their ratio, the clustering, therefore falls off as one over the degree.

The degree correlations are governed by the same single dial $q_e$. Numerical
simulation of the model shows that when $q_e$ is small the network looks
essentially uncorrelated --- the average neighbor degree is flat, as in the
baseline of Eq.~\eqref{eq:intro-knn-unc} --- while as $q_e$ grows the average
neighbor degree develops the downward slope of a \emph{disassortative} network,
hubs sitting amid poorly-connected pages. This matches the real Web, whose
measured exponent $\gamma\approx 2.1$ corresponds through
Eq.~\eqref{eq:walk-gamma} to a patient surfer with $q_e>0.5$, squarely in the
disassortative regime. One knob, $q_e$, thus fixes the degree-distribution
exponent, the cliquishness hierarchy, and the sign of the correlations all at
once --- tying together what the introduction presented as three separate facts.

\section{Two ways to search}

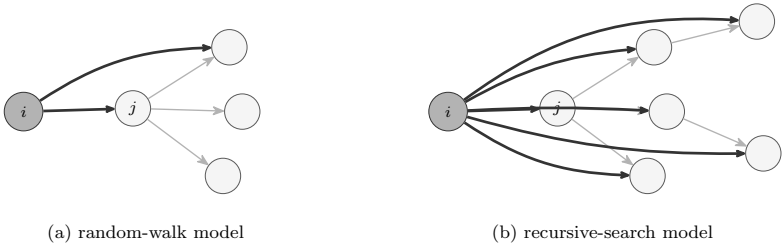
\begin{figure}[t]
\centering
\begin{adjustbox}{max width=\linewidth}\begin{tikzpicture}[>={Stealth[round,length=2.2mm]},
  nd/.style={circle,draw=black!65,fill=black!4,minimum size=5.5mm,inner sep=0pt,font=\scriptsize},
  new/.style={circle,draw=accent,fill=black!30,minimum size=6mm,inner sep=0pt,font=\scriptsize},
  ex/.style={->,draw=black!30,line width=0.6pt},
  add/.style={->,draw=accent,line width=1.2pt}]
  \begin{scope}
    \node[nd]  (j)  at (0,0)       {$j$};
    \node[nd]  (m1) at (1.5,0.95)  {};
    \node[nd]  (m2) at (1.7,-0.05) {};
    \node[nd]  (m3) at (1.4,-1.05) {};
    \draw[ex] (j)--(m1); \draw[ex] (j)--(m2); \draw[ex] (j)--(m3);
    \node[new] (i) at (-1.7,-0.05) {$i$};
    \draw[add] (i)--(j);
    \draw[add] (i) to[bend left=16] (m1);
    \node[font=\footnotesize] at (0.2,-1.95) {(a) random-walk model};
  \end{scope}
  \begin{scope}[xshift=6.6cm]
    \node[nd]  (jj) at (0,0)       {$j$};
    \node[nd]  (q1) at (1.5,0.95)  {};
    \node[nd]  (q2) at (1.7,-0.05) {};
    \node[nd]  (q3) at (1.4,-1.05) {};
    \node[nd]  (p1) at (3.1,1.35)  {};
    \node[nd]  (p2) at (3.2,-0.7)  {};
    \draw[ex] (jj)--(q1); \draw[ex] (jj)--(q2); \draw[ex] (jj)--(q3);
    \draw[ex] (q1)--(p1); \draw[ex] (q2)--(p2);
    \node[new] (ii) at (-1.7,-0.05) {$i$};
    \draw[add] (ii)--(jj);
    \draw[add] (ii) to[bend left=12]  (q1);
    \draw[add] (ii) to[bend left=3]   (q2);
    \draw[add] (ii) to[bend right=16] (q3);
    \draw[add] (ii) to[bend left=20]  (p1);
    \draw[add] (ii) to[bend right=8]  (p2);
    \node[font=\footnotesize] at (0.7,-1.95) {(b) recursive-search model};
  \end{scope}
\end{tikzpicture}\end{adjustbox}
\caption{\textbf{Two ways to search, when a new node $i$ joins.} (a) In the
random-walk model, $i$ links to a randomly chosen node $j$ and then takes a
single step to one of $j$'s neighbors, linking to it as well. (b) In the
recursive-search model, $i$ follows \emph{all} of $j$'s links, and then their
links in turn, linking to every node its widening search reaches. New links are
drawn in bold, pre-existing links thin and gray.}
\label{fig:walk-models}
\end{figure}

It helps to see two concrete versions of the searching rule, because they behave
differently and bracket the range of possibilities. Figure~\ref{fig:walk-models}
contrasts them.

\paragraph{The random-walk model.} The simplest version uses a single surfer
that always links what it visits and starts a fresh search from a new node each
time it stops. Stated as a recipe, beginning from a couple of connected nodes,
repeat:
\begin{itemize}
  \item \emph{Add.} Create a new node with one link pointing to a randomly
    chosen existing node.
  \item \emph{Walk.} Having just linked to a node, then with probability $q_e$
    also link to one of that node's neighbors; otherwise go back to the add
    step.
\end{itemize}
Here the surfer takes a single step along the network before possibly 
stopping. This model produces the power-law tail in Eq.~\eqref{eq:walk-asym} 
and the clustering law of Eq.~\eqref{eq:walk-ck} (the latter in very good 
agreement with simulation), while the predicted exponent is a slight 
overestimate --- the price of the mean-field approximation in the box above.

\paragraph{The recursive-search model.} A more thorough searcher does not 
stop after one step but follows \emph{all} the links of each node it reaches, 
then all the links of those, exploring an ever-widening territory 
\citep{vazquez2001}. Specifically,
\begin{itemize}
  \item \emph{Add.} Create a new node with one link pointing to a randomly
    chosen existing node.
  \item \emph{Walk.} If a link is created to a node in the network then
    with probability $q_e$ a link is also created to each of its nearest
    neighbors. When no link is created go to the add rule.
\end{itemize}
This version has a richer behavior --- in fact a genuine \emph{phase transition},
a sudden change of character as the dial $q_e$ is turned, of the kind familiar
from water freezing into ice. The next box records the two cases that can be
solved exactly and the location of the transition; the meaning is summarized
afterward.

\begin{calculation}{the recursive searcher and its phase transition}

For $q_e=0$ only the add step operates, every node is equally likely to be 
linked and $A(k^{in}) = 1/n$ independently of $k^{in}$. In this case the 
steady state solution $n_{k^{in}}(n) = n\, p_{k^{in}}$ of Eq.~\eqref{eq:walk-rate} is
\begin{equation}
  p_{k^{in}} = 2^{-(k^{in}+1)}.
  \label{eq:walk-rec0}
\end{equation}
For $q_e=1$ the recursion reaches every neighbor, the attachment rate becomes 
exactly linear, $A(k^{in})=(1+k^{in})/n$ --- genuine preferential attachment, 
generated locally --- and the steady state solution $n_{k^{in}}(n) = n\, p_{k^{in}}$ of Eq.~\eqref{eq:walk-rate} is
\begin{equation}
  p_{k^{in}} = \frac{1}{(k^{in}+1)(k^{in}+2)} \sim (k^{in})^{-2}.
  \label{eq:walk-rec1}
\end{equation}
In between, we don't have an analytical solution. Yet, the average out-degree 
in the steady state is
\begin{equation}
  \langle k^{out}\rangle = \frac{1 - \sqrt{1-4 q_e}}{2 q_e},
  \label{eq:walk-rec-dou}
\end{equation}
which is finite only for $q_e \le q_c = 1/4$; above this threshold it diverges.
\end{calculation}

The exact results bracket the behavior: a fast-decaying exponential when the
searcher barely explores, and a clean power law with exponent $2$ when it
explores exhaustively. The new feature is the threshold $q_c$ in
Eq.~\eqref{eq:walk-rec-dou}. Below it the network has a well-defined average
connectivity; above it the average degree runs away and the in-degree
distribution locks onto the power-law exponent $\gamma=2$, no matter the precise
value of $q_e$. Simulations confirm this picture, with the threshold near $q_c
\approx 0.5$.

The striking part is that scale-invariance --- the power law --- is not produced
by carefully tuning $q_e$ to one magic value, but holds across a whole range of
$q_e$ above the threshold. Physicists have a name for systems that sit
automatically at such a critical, scale-free point without fine-tuning:
\emph{self-organized criticality}, the textbook example being a sandpile that,
however you pour sand on it, keeps adjusting itself to the brink of avalanching.
Here the slow addition of new pages and the fast exploration of the existing
network play the roles of slow pouring and fast avalanching. Read this way, the
ubiquity of scale-free citation and web networks is not a delicate coincidence
but the generic, self-organized outcome of building a network by searching it.

\section{Summary}

Searchable networks grow by a rule with nothing global in it: a newcomer
explores the existing network along its links and connects to what it finds.
Modeling the explorer as a random walker --- the very PageRank process that ranks
web pages --- we found that a page's visiting probability rises in a straight line
with its in-degree, which converts the local act of walking into an effective
``rich get richer.'' From that single mechanism follow a power-law in-degree
distribution whose exponent $\gamma=1+1/q_e$ is set purely by how the network is
explored, a clustering hierarchy in which cliquishness falls as one over the
degree, and tunable, generically disassortative degree correlations. Letting the
searcher explore exhaustively turns the smooth picture into a self-organized
critical transition with a robust exponent of $2$. The puzzle of where
preferential attachment ``comes from,'' for this family of networks, is thereby
dissolved: it comes from walking.


\chapter{Friends of friends}
\label{ch:friends}

\section{Triadic closure as a growth rule}

The previous chapter built networks by exploration. This one builds them by
introduction. Think about how you came to know the people you know. Only rarely
did you befriend a complete stranger; far more often a friend introduced you to
one of \emph{their} friends. This everyday mechanism --- a new tie forming
between two people who already share a common acquaintance --- has a name in
sociology: \emph{triadic closure}, because it ``closes'' a triangle by adding
the third side to a path of two. Two people who both know a third are, in a
sense, a triangle waiting to happen, and triadic closure is the event that
completes it. Sociologists have long seen this as the engine of social structure
\citep{granovetter1973}. If most new links close triangles, then triangles pile
up, and the network acquires the high cliquishness --- the large clustering
coefficient of the introduction --- that sets social networks apart from random
ones.

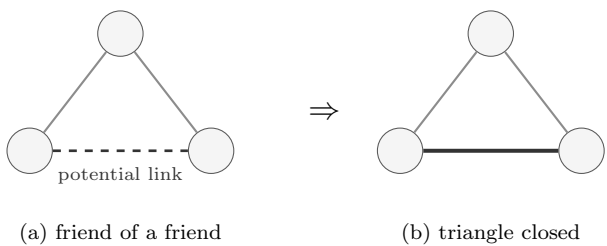
\begin{figure}[t]
\centering
\begin{adjustbox}{max width=\linewidth}\begin{tikzpicture}[
  nd/.style={circle,draw=black!65,fill=black!4,minimum size=6mm,inner sep=0pt},
  lnk/.style={draw=black!45,line width=0.8pt},
  pot/.style={draw=accent,line width=1.0pt,dashed},
  newl/.style={draw=accent,line width=1.5pt}]
  \begin{scope}
    \node[nd] (A) at (0,1.15)    {};
    \node[nd] (B) at (-1.2,-0.4) {};
    \node[nd] (C) at (1.2,-0.4)  {};
    \draw[lnk] (A)--(B);
    \draw[lnk] (A)--(C);
    \draw[pot] (B)--(C);
    \node[accent,font=\scriptsize] at (0,-0.72) {potential link};
    \node[font=\footnotesize] at (0,-1.5) {(a) friend of a friend};
  \end{scope}
  \node at (2.7,0.1) {\large $\Rightarrow$};
  \begin{scope}[xshift=4.9cm]
    \node[nd] (A2) at (0,1.15)    {};
    \node[nd] (B2) at (-1.2,-0.4) {};
    \node[nd] (C2) at (1.2,-0.4)  {};
    \draw[lnk] (A2)--(B2);
    \draw[lnk] (A2)--(C2);
    \draw[newl] (B2)--(C2);
    \node[font=\footnotesize] at (0,-1.5) {(b) triangle closed};
  \end{scope}
\end{tikzpicture}\end{adjustbox}
\caption{\textbf{Triadic closure.} Two people (bottom) who already share a common
friend (top) form a \emph{potential link} --- a friend-of-a-friend pair, drawn
dashed. Closing the triangle turns that potential link into a real one, the
elementary step by which the friends-of-friends rule builds social structure.}
\label{fig:fr-closure}
\end{figure}

Figure~\ref{fig:fr-closure} shows this elementary step. Notice that it is a
thoroughly \emph{local} rule. To introduce two friends you
need to know only one node and its neighbors: who you know, and who they know.
That is exactly the kind of local information this book is built on, and the
question, as always, is what such a rule produces in the aggregate. The answer,
worked out below following \citet{vazquez2003} and reviewed in
\citet{vazquez2025}, is that connecting friends-of-friends generates the whole
suite of social-network signatures at once: an effective ``rich get richer'' and
hence a power-law degree distribution, the clustering hierarchy in which
cliquishness falls with degree, and --- in pointed contrast to the technological
and biological networks of the other chapters --- \emph{assortative} degree
correlations, the tendency of well-connected people to know one another.

Models in this spirit have a history. \citet{davidsen2002} introduced an
acquaintance-network model (the DEB model, after its authors) in which people are
repeatedly introduced to friends of friends, producing a small world with high
clustering; \citet{holme2002} bolted an explicit triangle-forming step onto
preferential attachment to give it tunable clustering. What follows organizes
these ideas around a single bookkeeping device that makes the mechanism
transparent --- the \emph{potential link}.

\section{Potential links}

The key idea is that, under triadic closure, a pair of nodes is not just
``connected'' or ``not connected.'' A pair that already shares a common neighbor
is a triangle waiting to close --- a candidate in a way that an arbitrary pair of
strangers is not. So we sort every pair of nodes into one of three states:
\begin{itemize}
  \item \emph{disconnected} ($s$): no link and no common neighbor --- two
    strangers with no mutual acquaintance;
  \item connected by a \emph{potential link} ($p$): no link yet, but at least one
    common neighbor --- two friends-of-a-friend, poised to be introduced;
  \item connected by a \emph{link} ($e$): already friends.
\end{itemize}
The \emph{potential link} is the formal stand-in for ``friend of a friend.'' We
write $k_i$ for the ordinary degree of node $i$ (its number of friends) and
$k^*_i$ for its \emph{potential degree}, the number of friends-of-friends it has
--- the number of potential links waiting at it.

Triadic closure is then a story about a pair changing state: a potential link
($p$) becomes a real link ($e$) when the triangle closes. To track this we need
to know the rates at which pairs hop between the three states, and one
relationship among those rates is the mechanical heart of the model. It says
that potential links are born in proportion to a node's degree --- introduce
yourself to someone with many friends and you instantly acquire many
friends-of-friends. The next box sets up the bookkeeping and draws out its main
consequence; the conclusion is restated plainly afterward, so the box may be
skipped.

\begin{calculation}{from the three-state bookkeeping to ``rich get richer''}
Let $\nu_{x\to y}$ be the rate (per unit of network size $n$) at which a pair
hops from state $x$ to state $y$. Ignoring fluctuations, the average degree and
potential degree of node $i$ evolve as
\begin{align}
  n\,\frac{\partial k_i}{\partial n}
    &= \nu_{s\to e}\,\hat k_i + \nu_{p\to e}\,k^*_i
       - (\nu_{e\to s}+\nu_{e\to p})\,k_i, \nonumber\\
  n\,\frac{\partial k^*_i}{\partial n}
    &= \nu_{s\to p}\,\hat k_i + \nu_{e\to p}\,k_i
       - (\nu_{p\to s}+\nu_{p\to e})\,k^*_i,
  \label{eq:fr-master}
\end{align}
where $\hat k_i = n - k_i - k^*_i$ is the number of nodes that are neither
friends nor friends-of-friends of $i$. The link between the two kinds of
connection comes from the geometry of triangles: when a new node $i$ attaches to
an existing node $j$, it instantly shares the common neighbor $j$ with \emph{all}
of $j$'s neighbors, gaining a potential link to each. Hence potential links are
spawned in proportion to degree,
\begin{equation}
  \nu_{s\to p} = \nu_{s\to e}\, k_i, \qquad
  \nu_{p\to s} = \nu_{e\to s}\, k_i.
  \label{eq:fr-coupling}
\end{equation}
Now make two modeling choices. First, neglect link deletion, $\nu_{e\to s}=0$ ---
accurate for, say, a co-authorship network, where having written a paper together
is a fact that cannot later become false. Second, make closing a triangle far
likelier than linking two strangers,
\begin{equation}
  \nu_{s\to e} = \frac{\mu_0}{n^2}, \qquad
  \nu_{p\to e} = \frac{\mu_1}{n},
  \label{eq:fr-rates}
\end{equation}
the stranger-linking rate carrying an extra factor of $1/n$. Substituting these
into Eq.~\eqref{eq:fr-master} collapses the system to
\begin{equation}
  n\,\frac{\partial k_i}{\partial n} = \mu_0 + \mu_1\, k^*_i, \qquad
  n\,\frac{\partial k^*_i}{\partial n} = \mu_0\, k_i - \mu_1\, k^*_i.
  \label{eq:fr-reduced}
\end{equation}
\end{calculation}

The upshot of Eq.~\eqref{eq:fr-reduced} is that a node's degree grows at a rate
proportional to how many friends-of-friends it has --- and a node with many
friends has, through them, many friends-of-friends. So once again the rich get
richer, but this time the preference is manufactured by triadic closure rather
than by surfing: a well-connected person is simply involved in more potential
introductions. As in the previous chapter, no node ever consulted a global
ranking; the bias toward hubs is a free by-product of a purely local act.

\section{Degree distribution and clustering hierarchy}

Because the growth rate is linear in degree, the same reasoning that gave a power
law in the last chapter applies here, and it simultaneously fixes the
cliquishness. The box carries out both; the two results and their meaning are in
the paragraph after it.

\begin{calculation}{the power law and the clustering hierarchy}
Integrating the linear system~\eqref{eq:fr-reduced} gives power-law growth of
both degrees with the age of the node,
\begin{equation}
  k_i(n) = k_0 \left(\frac{n}{n_i}\right)^{\beta}, \qquad
  k^*_i(n) = k^*_0 \left(\frac{n}{n_i}\right)^{\beta},
  \label{eq:fr-growth}
\end{equation}
where $n_i$ is the network size when node $i$ arrived and
\begin{equation}
  \beta = \frac{\mu_1}{2}\left(-1 + \sqrt{1 + 4\,\frac{\mu_0}{\mu_1}}\,\right).
  \label{eq:fr-beta}
\end{equation}
With nodes added at a constant rate, $P(n_i{=}n)=1/n$, a node has degree above $k$
only when it is old enough for Eq.~\eqref{eq:fr-growth} to exceed $k$; averaging
over its birth time gives the cumulative distribution
\begin{equation}
\adjustbox{max width=\linewidth}{$\displaystyle
  P(k_i > k) = P\!\left[k_0\left(\frac{n}{n_i}\right)^{\beta} > k\right]
  = \int_0^n \frac{dn_i}{n}\,
    \Theta\!\left[k_0\left(\frac{n}{n_i}\right)^{\beta} - k\right]
$}
  \label{eq:fr-cumulative}
\end{equation}
with $\Theta[\cdot]$ the step function, $1$ when its bracket holds and $0$
otherwise. Differentiating it leaves a power-law degree distribution,
\begin{equation}
  p_k = \frac{\partial P(k_i > k)}{\partial k} \sim k^{-\gamma}, \qquad
  \gamma = 1 + \frac{1}{\beta}.
  \label{eq:fr-pk}
\end{equation}
For the clustering, the dominant way node $i$ gains a triangle is a potential
link closing: when the potential link from $i$ to $j$ (sharing neighbor $k$)
becomes real, $i$ gains both the neighbor $j$ and a link $j$--$k$ among its
neighbors. Thus $\partial e_i/\partial n = \nu_{p\to e}\,k^*_i = \mu_1 k^*_i/n$,
and integrating with Eq.~\eqref{eq:fr-growth},
\begin{equation}
  \langle c\rangle_k = \frac{2 e(k)}{k(k-1)} \approx \frac{2\mu_1}{k}.
  \label{eq:fr-ck}
\end{equation}
\end{calculation}

Two familiar results emerge. First, Eq.~\eqref{eq:fr-pk}: the degree
distribution is again a power law. It is worth knowing exactly where the power
law comes from --- it comes from the proportionality in
Eq.~\eqref{eq:fr-coupling}, the fact that potential links are spawned in
proportion to degree. Were they instead spawned at a flat rate, the mechanism
would still create triangles but the degree distribution would decay
exponentially, with no fat tail of hubs. Scale-freeness is the fingerprint of
that degree-proportional spawning. Second, Eq.~\eqref{eq:fr-ck}: cliquishness
again falls as one over the degree --- the clustering hierarchy of the
introduction, now arriving by a third route (after search, and before
duplication in the next chapter). Here its origin is especially vivid: every
triangle a node owns is just a friend-of-a-friend pair that has been introduced.

\section{Connecting nearest neighbors}

The continuum theory is borne out by a minimal simulation, a growing version of
the DEB rule \citep{davidsen2002}. Starting from a single node and no links,
repeat:
\begin{itemize}
  \item with probability $1-u$, \emph{add} a new node and link it to a randomly
    chosen node $j$ --- which by definition creates a potential link between the
    newcomer and every neighbor of $j$;
  \item with probability $u$, \emph{close} a triangle: pick a potential link at
    random and turn it into a real link.
\end{itemize}
The single dial $u$ is the rate of triangle-closing relative to adding newcomers.
Feeding the mapping $\mu_0=1$, $\mu_1=u/(1-u)$ into the formulas above gives the
exponent in closed form, as the short box records.

\begin{calculation}{the exponent as a function of the closure rate}
With $\mu_0=1$ and $\mu_1=u/(1-u)$, Eqs.~\eqref{eq:fr-beta} and~\eqref{eq:fr-pk}
give
\begin{equation}
  \gamma(u) = 1 + \frac{2(1-u)}{u}
    \left(-1 + \sqrt{1 + 4\,\frac{1-u}{u}}\,\right)^{-1},
  \label{eq:fr-gammau}
\end{equation}
with limiting values
\begin{equation}
  \gamma(0) = \infty, \qquad \gamma(1) = 2.
  \label{eq:fr-limits}
\end{equation}
\end{calculation}

In words, the exponent is tunable by how eagerly the network closes triangles.
When closure is frequent and newcomers rare ($u\to 1$) the exponent sinks to its
minimum of $2$, the most hub-dominated case; when newcomers arrive so fast that
triangles never get a chance to close ($u\to 0$) the exponent blows up and the
heavy tail of hubs is lost. Simulations up to a million nodes confirm the
power-law degree distribution (the mean-field formula slightly overestimating the
exponent, as in the previous chapter), and they confirm the clustering hierarchy,
with cliquishness falling as a power of degree.

\section{Why social networks are assortative}

The most telling prediction is about the degree correlations --- and here the
friends-of-friends rule parts company with every other rule in this book.
Measuring the average neighbor degree $\langle k_{nn}\rangle(k)$ in the simulated
network, one finds it \emph{rises} with degree: the well-connected are connected
to the well-connected. The network is \emph{assortative}
\citep{newman2002} --- ``birds of a feather'' --- exactly as measured in real
collaboration and co-authorship networks, and exactly opposite to the
disassortative Internet and protein networks of the introduction.

Why the opposite sign? The reason becomes obvious once seen. Every new link here
closes a triangle, joining two nodes that share a common neighbor. But a node
that serves as the common neighbor of many pairs is, by definition, a high-degree
node, and the two nodes it brings together are drawn from its own neighborhood,
which is better connected than the network at large. So closure tends to wire
together nodes of \emph{similar}, and similarly high, degree. The very mechanism
that the disassortative models of the other chapters lack --- linking within a
neighborhood rather than reaching out to a random or peripheral node --- is what
tips the correlations positive. This is the deeper lesson of the chapter, and of
the review that develops it \citep{vazquez2025}: the \emph{sign} of the degree
correlations is not an arbitrary fitting knob but a direct readout of the
microscopic rule. Search and duplication reach outward, to the periphery, and so
make hubs that shun one another; triadic closure works inward, within
neighborhoods, and so makes hubs that keep each other's company.

\section{Summary}

Modeling social growth by triadic closure --- a newcomer attaches to a node and
then, over time, to that node's friends --- and formalizing ``friend of a friend''
as a \emph{potential link}, we obtain a tractable theory in which potential links
are created in proportion to degree. From that one ingredient follow an effective
``rich get richer,'' a power-law degree distribution with a tunable exponent
$\gamma = 1 + 1/\beta$, a clustering hierarchy in which cliquishness falls as one
over the degree, and --- distinctively --- assortative degree correlations.
Connecting friends-of-friends thus reproduces, from purely local moves, the
characteristic shape of social networks, and explains why the sign of their
correlations differs from that of the technological and biological networks
treated elsewhere in this book.


\chapter{Gene duplication}
\label{ch:duplication}

\section{Networks that copy themselves}

The two rules of the preceding chapters each had a newcomer reach into the
existing network and \emph{choose} where to attach --- by surfing, or by closing
a triangle. Biology offers a third rule, and in some ways a more primitive one:
the network grows by \emph{copying} a node that is already there. The newcomer
does not pick its neighbors at all. It inherits them, wholesale, from a parent.

\begin{figure}[t]
\centering
\begin{adjustbox}{max width=\linewidth}\begin{tikzpicture}[
  nd/.style={circle,draw=black!65,fill=black!4,minimum size=5.5mm,inner sep=0pt,font=\scriptsize},
  cp/.style={circle,draw=accent,fill=black!30,minimum size=5.5mm,inner sep=0pt,font=\scriptsize},
  lnk/.style={draw=black!45,line width=0.8pt},
  newl/.style={draw=accent,line width=1.3pt},
  gone/.style={draw=black!22,line width=0.7pt,dotted}]
  \begin{scope}
    \node[nd] (i)  at (-0.2,0.6)  {$i$};
    \node[cp] (ip) at (-0.2,-0.6) {$i'$};
    \node[nd] (a)  at (1.5,1.05)  {};
    \node[nd] (b)  at (1.75,0.0)  {};
    \node[nd] (c)  at (1.5,-1.05) {};
    \draw[lnk]  (i)--(a); \draw[lnk] (i)--(b); \draw[lnk] (i)--(c);
    \draw[newl] (ip)--(a); \draw[newl] (ip)--(b); \draw[newl] (ip)--(c);
    \draw[newl] (i)--(ip);
    \node[font=\footnotesize] at (0.75,-1.85) {(a) duplication};
  \end{scope}
  \node at (3.25,0.0) {\large $\Rightarrow$};
  \begin{scope}[xshift=5.7cm]
    \node[nd] (i2)  at (-0.2,0.6)  {$i$};
    \node[cp] (ip2) at (-0.2,-0.6) {$i'$};
    \node[nd] (a2)  at (1.5,1.05)  {};
    \node[nd] (b2)  at (1.75,0.0)  {};
    \node[nd] (c2)  at (1.5,-1.05) {};
    \draw[lnk]  (i2)--(a2);
    \draw[newl] (ip2)--(b2);
    \draw[lnk]  (i2)--(c2);
    \draw[newl] (i2)--(ip2);
    \draw[gone] (ip2)--(a2);
    \draw[gone] (i2)--(b2);
    \draw[gone] (ip2)--(c2);
    \node[font=\footnotesize] at (0.75,-1.85) {(b) divergence};
  \end{scope}
\end{tikzpicture}\end{adjustbox}
\caption{\textbf{Duplication and divergence.} (a) A gene duplication adds a copy
$i'$ (gray) wired to every interaction partner of the ancestor $i$, plus a link
between the twins when the protein is self-interacting. (b) Divergence then prunes
the now-redundant interactions (dotted), each partner kept by only one twin, so
the copies drift apart.}
\label{fig:dup-diverge}
\end{figure}
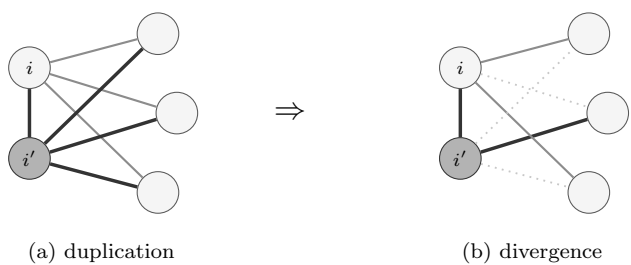

The setting is the inner machinery of the living cell. Genomes evolve in good
part through \emph{duplication} --- a gene, or sometimes an entire genome, is
accidentally copied --- followed by \emph{divergence}, the slow accumulation of
mutations that make the two copies drift apart. That duplication is a major
engine of evolution was argued classically by \citet{ohno1970}. To turn this into
a network statement we use the \emph{protein interaction network}: each node is a
protein (the product of a gene), and a link joins two proteins that physically
interact --- that bind to each other to do their job. Now watch what a gene
duplication does to this network. Right after the gene is copied, the cell makes
two identical proteins, so the new protein interacts with \emph{exactly the same
partners} as the old one: the duplicate node arrives already wired to all the
neighbors of its ancestor (Figure~\ref{fig:dup-diverge}). Mutations then add a few
interactions and, more often,
remove some, letting the twins diverge \citep{wagner2001,sole2002}. Both steps
are thoroughly local --- the copy needs to know only its parent and its parent's
neighbors, never the rest of the genome.

The same picture reaches beyond biology. A new web page is often made by copying
another page's hyperlinks and then editing the list --- duplication as copying,
divergence as editing. Following \citet{vazquez2003} and \citet{vazquez2003pin},
this chapter shows that copy-and-diverge, like the rules before it, produces an
effective ``rich get richer,'' a power-law degree distribution, the clustering
hierarchy, and --- as actually measured in protein networks \citep{maslov2002} ---
\emph{disassortative} correlations. It also reveals something new: a
\emph{multifractal} degree distribution, which we will unpack when we reach it, a
fingerprint left by the act of inheritance.

\section{Duplication and divergence, in their simplest form}

Start with the simplest version, treating duplication and divergence as
independent random events. The single idea that drives everything is this: a node
gains a new link whenever one of its \emph{neighbors} is duplicated, because the
neighbor's copy inherits a link back to it. A well-connected node has many
neighbors, so it has many chances to be on the receiving end of a duplication ---
and there, already, is the rich getting richer. The box makes this precise and
reads off the degree distribution; the conclusion follows it.

\begin{calculation}{degree growth and the power law from duplication}
In a mean-field (average-quantities) description, with increasing the number of
nodes $n$ the degree of node $i$ evolves as
\begin{equation}
  \frac{\partial k_i}{\partial n}
    = \nu_D\, k_i + \nu_C\,(n - k_i) - \nu_L\, k_i,
  \label{eq:dup-master}
\end{equation}
with $\nu_D$, $\nu_C$, $\nu_L$ the per-node rates of duplication, random link
creation, and link loss. The first term is the key one: it is proportional to
$k_i$ because each of $i$'s $k_i$ neighbors, if duplicated, hands $i$ a new link.
Each duplication adds one node, so $\nu_D = 1/n$; writing $\nu_C=\mu_0/n$ and
$\nu_L=\mu_1/n$ reduces the equation to
\begin{equation}
  n\,\frac{\partial k_i}{\partial n} = \mu_0 + (1-\mu_1)\, k_i,
  \label{eq:dup-reduced}
\end{equation}
whose growth rate is linear in the degree. Integrating it gives the degree of
node $i$ as the network grows,
\begin{equation}
  k_i(n) = \left(k_i(n_i) + \frac{\mu_0}{1-\mu_1}\right)
           \left(\frac{n}{n_i}\right)^{\beta} - \frac{\mu_0}{1-\mu_1},
  \label{eq:dup-growth}
\end{equation}
where $n_i$ and $k_i(n_i)$ are the network size and the node's degree at the
moment it was added, and the exponent is
\begin{equation}
  \beta = 1 - \mu_1 \qquad (\mu_1 < 1),
  \label{eq:dup-beta}
\end{equation}
the condition $\mu_1<1$ keeping the network from emptying out. To turn this growth
law into a degree distribution, note that node $i$ has degree above $k$ exactly
when its age and birth degree make Eq.~\eqref{eq:dup-growth} exceed $k$,
\begin{equation}
  P(k_i > k) = \Theta\!\left[\left(k_i(n_i) + \frac{\mu_0}{1-\mu_1}\right)
           \left(\frac{n}{n_i}\right)^{\beta} - \frac{\mu_0}{1-\mu_1} > k\right],
  \label{eq:dup-cumulative}
\end{equation}
with $\Theta[\cdot]$ the step function, $1$ when its bracket holds and $0$
otherwise.
Both its birth size $n_i$ and its birth degree $k_i(n_i)$ are random: for nodes
added at a constant rate $n_i$ is uniform, $P(n_i{=}n)=1/n$, while a node inherits
its ancestor's degree, so $P[k_i(n_i){=}k']=p_{k'}$ in the stationary state.
Averaging over both,
\begin{equation}
\adjustbox{max width=\linewidth}{$\displaystyle
  P(k_i > k) = \sum_{k'} p_{k'} \int_1^n \frac{dn_i}{n}\,
    \Theta\!\left[\left(k' + \frac{\mu_0}{1-\mu_1}\right)
           \left(\frac{n}{n_i}\right)^{\beta} - \frac{\mu_0}{1-\mu_1} > k\right]
$}
  \label{eq:dup-average}
\end{equation}
Differentiating this cumulative distribution and taking $n\gg1$ leaves,
for nodes added at a constant rate, a power-law degree distribution
\begin{equation}
  p_k = \frac{\partial P(k_i > k)}{\partial k}
      \sim \left(\frac{\mu_0}{1-\mu_1} + k\right)^{-\gamma}, \qquad
  \gamma = 1 + \frac{1}{1-\mu_1}.
  \label{eq:dup-pk}
\end{equation}
\end{calculation}

The result is once again a power law, Eq.~\eqref{eq:dup-pk}, and its exponent is
controlled entirely by $\mu_1$, the rate at which divergence removes links: the
more aggressively the duplicates shed interactions, the steeper the tail and the
fewer the giant hubs. As in the previous two chapters, the power law can be
traced to a single term that grows in proportion to degree --- here the
duplication term of Eq.~\eqref{eq:dup-reduced} --- and without it the distribution
would decay as a featureless exponential. Copying, all by itself, is a
rich-get-richer engine.

\section{Cliquishness from self-interaction}

If divergence added and removed links purely at random, this model would make no
triangles, and its cliquishness would fade to nothing in a large network. But
biology supplies one specific local source of triangles. Some proteins are
\emph{self-interacting} --- they bind to copies of themselves. When the gene for
such a protein is duplicated, the ancestor and its new twin can bind to
\emph{each other}, because each still carries the self-binding site
\citep{wagner2001}. That single interaction, laid down at the moment of
duplication, is the seed of a triangle. The short box turns this into the
clustering law.

\begin{calculation}{the clustering hierarchy from self-interaction}
Let a duplicated self-interacting protein bond to its ancestor with probability
$q_v$. Then whenever a neighbor of node $i$ is duplicated, $i$ gains not only the
copy as a new neighbor but, with probability $q_v$, a link between two of its
neighbors (the ancestor and the copy). Hence the number of links $e_i$ between the
neighbors of node $i$ change as
\begin{equation}
\frac{\partial e_i}{\partial n} \approx q_v\,\frac{\partial k_i}{\partial n}.
\end{equation}
Integrating, assuming $e_i=0$ when $k_i=1$, we get $e_i = q_v (k_i - 1)$.
Inserting this result into the clustering definition,
\begin{equation}
  \langle c\rangle_k = \frac{2 e(k)}{k(k-1)} \approx \frac{2 q_v}{k}.
  \label{eq:dup-ck}
\end{equation}
\end{calculation}

So cliquishness again falls as one over the degree, Eq.~\eqref{eq:dup-ck} --- the
clustering hierarchy of the introduction, reached now by a third independent
route. There is one telling refinement: because removing an interaction also
destroys triangles, and the biggest hubs have the most triangles to lose,
divergence makes the clustering of high-degree nodes decay a little \emph{faster}
than one over the degree. That slightly-too-steep falloff is a small signature
that distinguishes duplication from the search and closure mechanisms.

\section{When duplication and divergence are coupled}

In a real protein network the two processes cannot truly be independent. The
network does a job --- it keeps the cell alive --- so losing the wrong interaction
can be lethal, and which links a duplicate keeps is under natural selection. The
classical view of \citet{ohno1970} lets one redundant copy decay freely; the more
modern \emph{subfunctionalization} picture of \citet{force1999} has the two
copies divide the ancestor's duties between them, each keeping part, together
keeping the whole. Either way the fates of the two copies' links are
\emph{coupled}: a shared interaction is far more likely to be retained by one
copy than dropped by both. To capture this, \citet{vazquez2003pin} use the
following rule. Starting from two connected nodes, repeat:
\begin{itemize}
  \item \emph{Duplication.} Pick a node $i$ at random and add a copy $i'$ joined
    to every neighbor of $i$; with probability $q_v$ also add the link $i$--$i'$
    (the self-interaction case).
  \item \emph{Divergence.} For each node $j$ now joined to both $i$ and $i'$,
    delete one of the two redundant links $(i,j)$ or $(i',j)$, at random, with
    probability $1-q_e$.
\end{itemize}
The divergence step touches only the \emph{redundant} pairs just created, and
that is what couples it to duplication. Genomic data showing that duplicates
diverge soon after they are born \citep{wagner2001} justify treating divergence
as finishing before the next duplication, so the node count $n$ doubles as a
clock. A related model with lopsided divergence is studied in \citet{vazquez2003}.

Coupling makes the large-scale behavior considerably richer, in two ways worth
previewing before the box. First, a single number --- the retention parameter
$q_e$ --- decides whether the network stays sparse or grows ever denser, with a
sharp switch between the two regimes. Second, the degree distribution is no
longer a plain power law but a \emph{multifractal} one. A quick word on that
term: a plain power law is described by one exponent, the same at every scale; a
multifractal is described by a whole continuum of exponents, because different
``moments'' of the distribution (the average degree, the average of the squared
degree, and so on) scale with different powers of the network size. The box
records the transition and the moment exponents; the interpretation follows.

\begin{calculation}{the density transition and the multifractal spectrum}
Tracking the mean degree through one duplication event gives
\begin{equation}
\adjustbox{max width=\linewidth}{$\displaystyle
  \langle k\rangle(n+1)
   = \frac{n\langle k\rangle(n) + 2q_v + (2q_e - 1)\langle k\rangle(n)}{n+1}
$}
  \label{eq:dup-meandeg}
\end{equation}
which in the large-network limit switches behavior at $q_e=1/2$: for $q_e<1/2$
the mean degree saturates at the finite value $\langle k\rangle = 2q_v/(1-2q_e)$,
while for $q_e>1/2$ it grows without bound as $N^{2q_e-1}$. A rate-equation
treatment in the manner of \citet{kim2002} gives an attachment rate
$A(k)=(q_v+q_e k)/n$ --- linear in degree again --- and predicts that the $l$-th
moment of the degree distribution scales as $M_l\sim n^{\sigma_l(q_e)}$ with
\begin{equation}
  \sigma_l(q_e) = l\, q_e + 2\left[\left(\frac{1+q_e}{2}\right)^{l} - 1\right].
  \label{eq:dup-moments}
\end{equation}
For each $l$ there is a value $q_l$ at which $\sigma_l$ changes sign, with
$q_1=1/2$ and $q_2 = 2\sqrt{3}-3 \approx 0.46$.
\end{calculation}

The two messages of the box are these. First, Eq.~\eqref{eq:dup-meandeg} shows a
genuine switch at $q_e=1/2$: keep too few links and the network settles into a
sparse steady state; keep enough and it densifies forever. Second, and more
subtle, the moment exponents $\sigma_l$ in Eq.~\eqref{eq:dup-moments} depend on
$l$ in a curved, \emph{nonlinear} way --- the hallmark of a multifractal. Where
the earlier chapters gave clean single power laws, duplication gives a spectrum
of exponents, and the reason is exactly the inheritance at its core: each
newcomer copies the degree environment of its parent, injecting a kind of
heritable heterogeneity that a single exponent cannot capture. The deletion of
links from the copies is what converts that inheritance into multifractality.

\section{Disassortative correlations}

Simulations spanning network sizes from a thousand to a million nodes confirm the
analysis \citep{vazquez2003pin, vazquez2003}: the moment exponents match the multifractal prediction; the clustering
falls with degree as a power law with exponent at or above one (steepening as
link loss grows, just as Eq.~\eqref{eq:dup-ck} and its caveat foretold); and the
average neighbor degree \emph{decreases} with degree, $\langle k_{nn}\rangle(k)
\sim k^{-0.1}$. The duplication network is therefore \emph{disassortative} --- its
hubs are surrounded by poorly-connected partners rather than by each other.

This is the correct sign. Disassortative mixing is precisely what
\citet{maslov2002} measured in the protein interaction network of yeast, and it
files duplication alongside the technological networks of
Chapter~\ref{ch:search} and against the social networks of
Chapter~\ref{ch:friends}. With this third mechanism in hand, the three rules of
the book so far bracket the empirical landscape, as the table summarizes (here
$\beta$ is the rate at which cliquishness falls with degree, and $\alpha$ the
slope of the neighbor-degree curve --- positive for assortative, negative for
disassortative):
\begin{center}
\begin{tabular}{lcc}
\hline
Mechanism & $\langle c\rangle_k \sim k^{-\beta}$ & $\langle k_{nn}\rangle_k \sim k^{\alpha}$ \\
\hline
Connecting neighbors    & $0 < \beta < 1$ & $\alpha > 0$ \\
Random walk             & $\beta = 1$     & $\alpha \le 0$ \\
Duplication--divergence & $\beta \ge 1$   & $\alpha < 0$ \\
\hline
\end{tabular}
\end{center}
Every local rule produces a clustering hierarchy; what differs is the \emph{sign}
of the correlations, which sorts the rules into the assortative (social,
closure-driven) and disassortative (technological and biological, reach-driven)
classes seen in real data.

\section{The unifying principle}

Three different stories --- surfing a network, closing a triangle, copying a node
--- have each delivered the same effective ``rich get richer.'' That is no
accident, and the duplication picture makes the reason simple enough to state in
a single sentence.

Picking a node \emph{at random} plays no favorites: every node is equally likely,
hub or not. But none of the local rules in this book act on a randomly chosen
node. They act on a \emph{neighbor} of one --- the page a surfer steps to, the
friend a friend introduces, the partner a duplicated gene inherits. And the
chance that a particular node $i$ is the neighbor of a randomly chosen node is
\begin{equation}
  \frac{k_i}{\sum_j k_j},
  \label{eq:dup-unifying}
\end{equation}
which is exactly the linear preferential-attachment probability of the
Barab\'asi--Albert model \citep{barabasi1999}. High-degree nodes are
over-represented among the neighbors of a random node for the most elementary of
reasons: they belong to more neighborhoods, so more random nodes have them as a
neighbor. Preferential attachment never has to be imposed. It is induced
automatically the instant a growth rule takes one step out from a random node to
one of its neighbors. The clustering hierarchy has the same local root --- a rule
that makes one link to a node tends to make another among that node's neighbors
--- and the sign of the degree correlations depends on \emph{which} neighbor gets
chosen. Locality is the common cause, and the three mechanisms are three faces of
it.

\section{Summary}

Biological networks grow by duplication: a node is copied along with its
connections, and the copies then diverge by gaining and losing links. Because a
node gains a link whenever one of its neighbors is duplicated, the rule builds in
an effective ``rich get richer'' and hence a power-law degree distribution, while
duplications of self-interacting proteins generate a clustering hierarchy that
link loss steepens. Coupling divergence to duplication adds a density transition
at $q_e=1/2$ and a multifractal degree distribution born of inheritance, and the
model reproduces the disassortative correlations measured in real protein
networks. Together with the two previous chapters, duplication completes a simple
and general principle: a growth rule that attaches to a \emph{neighbor} of a
randomly chosen node attaches preferentially by degree --- and so preferential
attachment, the clustering hierarchy, and degree correlations are all the generic
harvest of locality.


\chapter{Activity specialization}
\label{ch:projects}

\section{Projects as growing networks}

The duplication rule of the previous chapter was born in biology, but copying is
not a biological privilege. This chapter applies the same idea to a very
different system --- the schedule of a large engineering project --- and gets, in
return, a surprising practical payoff: an explanation of why big projects so
reliably run late.

A project schedule can be drawn as a network, called an \emph{activity network}.
Its nodes are the individual activities, or tasks, that make up the project, and
its links record \emph{precedence}: a link from activity $a$ to activity $b$
means $b$ cannot begin until $a$ is finished --- you cannot paint a wall before it
is built. Because work only flows forward in time, the network has no loops; in
the jargon it is a \emph{directed acyclic graph}, with starting activities (no
predecessors) and finishing activities (no successors). For such a network there
is one object that has ruled project management since the 1950s: the
\emph{critical path}. This is the longest chain of dependent activities running
from start to finish --- the route through the project with no slack anywhere
along it, so that delaying \emph{any} activity on it pushes back the whole
project's end date by the same amount \citep{kelley1959}. The manager's
time-honored prescription is therefore simple: watch the critical path, and
guard it at all costs.

And yet projects keep finishing late. A survey of more than ten thousand of them
found only a few percent delivered on time and on budget, and fifteen years of
improvements in management practice barely moved that figure
\citep{flyvbjerg2003}. This stubborn record hints that the critical path --- a
single one-dimensional thread --- may be the wrong thing to watch in schedules
that have grown into genuinely complex networks. Following
\citet{vazquezprojects2023}, this chapter grows such schedules from a local rule,
shows that the critical path \emph{shrinks toward irrelevance} as projects get
bigger, and relocates the real source of delay risk in the network as a whole.

\section{The duplication--split rule}

Where does a schedule come from? Not in one piece. A plan is \emph{refined}: a
coarse activity is broken into finer ones, again and again, until the plan is
detailed enough to execute. There are two natural ways to break an activity in
two, and they correspond to two kinds of work. This is our local rule, with a
single dial, the duplication probability $q$. Beginning from two activities in
series, repeat: pick an existing activity at random and, with probability $q$,
\emph{duplicate} it, otherwise \emph{split} it.
\begin{itemize}
  \item \emph{Duplication} (probability $q$). A \emph{generic} activity is split
    into two smaller activities that run \emph{in parallel} --- think of doubling
    a crew so two teams do the same kind of work at once. Both copies inherit
    \emph{all} the predecessors and \emph{all} the successors of the parent.
    Duplication \emph{widens} the schedule, adding parallel strands of work.
  \item \emph{Split} (probability $1-q$). A \emph{specialized} activity is cut
    into two activities run \emph{in sequence} --- one specialist does the first
    part and hands off to another who does the second. The parent keeps its
    predecessors, the new activity takes over its successors, and a link joins
    the two. Splitting \emph{lengthens} the schedule, extending a chain.
\end{itemize}
\begin{figure}[t]
\centering
\begin{adjustbox}{max width=\linewidth}\begin{tikzpicture}[>={Stealth[round,length=2.2mm]},
  nd/.style={circle,draw=black!65,fill=black!4,minimum size=5.5mm,inner sep=0pt,font=\scriptsize},
  act/.style={circle,draw=accent,fill=black!30,minimum size=5.5mm,inner sep=0pt,font=\scriptsize},
  newl/.style={->,draw=accent,line width=1.2pt}]
  \begin{scope}
    \node[nd]  (p)  at (0,0)      {};
    \node[act] (A1) at (1.4,0.7)  {};
    \node[act] (A2) at (1.4,-0.7) {};
    \node[nd]  (s)  at (2.8,0)     {};
    \draw[newl] (p)--(A1); \draw[newl] (p)--(A2);
    \draw[newl] (A1)--(s); \draw[newl] (A2)--(s);
    \node[font=\footnotesize] at (1.4,-1.7) {(a) duplication: in parallel};
  \end{scope}
  \begin{scope}[xshift=5.6cm]
    \node[nd]  (p2) at (0,0)    {};
    \node[act] (B1) at (1.1,0)  {};
    \node[act] (B2) at (2.3,0)  {};
    \node[nd]  (s2) at (3.4,0)   {};
    \draw[newl] (p2)--(B1); \draw[newl] (B1)--(B2); \draw[newl] (B2)--(s2);
    \node[font=\footnotesize] at (1.7,-1.7) {(b) split: in sequence};
  \end{scope}
\end{tikzpicture}\end{adjustbox}
\caption{\textbf{Refining an activity.} A generic activity is replaced either by
(a) two copies that run \emph{in parallel}, both inheriting the predecessor (left,
gray) and successor (right, gray) of the parent --- duplication, which widens the
schedule --- or by (b) two specialized activities that run \emph{in sequence} --- a
split, which lengthens it.}
\label{fig:proj-dupsplit}
\end{figure}

So the dial $q$ measures how a project is organized (Figure~\ref{fig:proj-dupsplit}). Small $q$ means almost every
refinement is a split: work is broken into ever-longer chains of specialists, and
the schedule is nearly a single long line. Large $q$ means almost every
refinement is a duplication: generic work is parallelized into many simultaneous
strands, and the schedule fans out into a broad web. The same copying that
duplicated a gene in the previous chapter here duplicates an activity; the new
ingredient is the \emph{split} --- the act of \emph{specialization} that gives the
chapter its name --- and the forward-flowing, start-to-finish structure of a
schedule.

\section{Predecessors and successors: identical power laws}

Because the rule copies an activity together with its predecessors and
successors, it is a duplication rule of the kind that produced power-law degrees
in the previous chapter --- only now there are two degrees to track, the number of
predecessors (the in-degree) and the number of successors (the out-degree). The
box works out the distribution of predecessors by following how many activities
have each number of them; the same argument applies verbatim to successors. The
result is in the paragraph after.

\begin{calculation}{the distribution of predecessors}
Let $n_k(n)$ be the number of activities with $k$ predecessors when the schedule
has $n$ activities in all. Adding one activity changes this count by
\begin{equation}
\adjustbox{max width=\linewidth}{$\displaystyle
  n_k(n+1) = n_k(n)
   + q\left[\frac{k-1}{n}\,n_{k-1}(n) - \frac{k}{n}\,n_k(n)
            + \frac{1}{n}\,n_k(n)\right]
   + (1-q)\,\delta_{k,1}
$}
  \label{eq:proj-rate}
\end{equation}
the bracket collecting the duplication events (an activity climbs from $k-1$ to
$k$ predecessors when one of its predecessors is duplicated, and so on) and the
last term adding a fresh single-predecessor activity at each split. Looking for a
stable shape $n_k(n)=n\,p_k$ turns this into a recurrence,
\begin{equation}
  p_k = q\big[(k-1)\,p_{k-1} - k\,p_k + p_k\big] + (1-q)\,\delta_{k,1},
  \label{eq:proj-recurrence}
\end{equation}
which solves in closed form (with $\Gamma$ the gamma function) to
\begin{equation}
  p_k = \prod_{s=1}^{k-1}\frac{s}{\tfrac{1}{q}+s}\; p_1
      = \Gamma\!\left(\tfrac{1}{q}+1\right)
        \frac{\Gamma(k)}{\Gamma\!\left(\tfrac{1}{q}+k\right)}\, p_1,
  \label{eq:proj-pk}
\end{equation}
with the power-law tail
\begin{equation}
  p_k \sim k^{-1/q}.
  \label{eq:proj-asym}
\end{equation}
\end{calculation}

The result, Eq.~\eqref{eq:proj-asym}, is a power law with exponent $\gamma=1/q$.
A wide, duplication-heavy schedule (large $q$) has the heaviest tail of
highly-connected activities; at $q=1/2$ we recover the familiar value $\gamma=2$.
The very same reasoning, run on successors instead of predecessors, gives the
\emph{identical} distribution for the out-degree. That coincidence is the model's
signature prediction --- the spread of predecessor counts and the spread of
successor counts should look the same --- and it follows directly from the
even-handed way duplication treats the two ends of an activity.

\section{The vanishing critical path}

The degrees are a local property; the critical path is a global one. Remarkably,
the same rule fixes how the critical path grows with the size of the project, and
the answer overturns the manager's intuition. The box does the short
calculation; the consequences follow.

\begin{calculation}{how the critical path scales}
Let $c$ be the number of activities on the critical path of an $n$-activity
schedule. The path can only lengthen when a \emph{split} lands on it (a split
inserts an activity in series), and the chance that the activity picked for
refinement lies on the critical path is $c/n$. Hence
\begin{equation}
  \frac{dc}{dn} = (1-q)\,\frac{c}{n},
  \label{eq:proj-cp}
\end{equation}
which integrates to $c\sim n^{1-q}$, so the \emph{fraction} of activities on the
critical path shrinks as
\begin{equation}
  \frac{c}{n} \sim n^{-\alpha(q)}, \qquad \alpha(q) \le q.
  \label{eq:proj-cpfrac}
\end{equation}
Because distances here grow as a power of size, $c\sim n^{1/D}$, the network is a
\emph{fractal} with dimension
\begin{equation}
  D = \frac{1}{1-\alpha(q)}.
  \label{eq:proj-fractal}
\end{equation}
\end{calculation}

The headline is Eq.~\eqref{eq:proj-cpfrac}: for any amount of duplication
($q>0$), the critical path is a \emph{vanishing fraction} of the project as the
project grows. The bigger and more parallel the schedule, the smaller the slice
of the work that the manager's cherished path actually covers --- so watching the
critical path means watching an ever-shrinking sliver of the whole. (Simulations
confirm the power-law shrinkage, with the measured exponent $\alpha$ a little
below the simple bound $q$; for instance $\alpha\approx 0.11,\,0.28,\,0.43$ at
$q=0.20,\,0.50,\,0.70$.)

The scaling also tells us what \emph{kind} of network a schedule is. In a
small-world network, distances grow only logarithmically with size --- doubling
the network barely lengthens the longest path. Here distances grow as a
\emph{power} of size, the mark of a \emph{fractal} network: one whose ``size''
and ``length'' are tied together by a fixed dimension $D$, much as a physical
fractal looks the same at every magnification. Duplication--split schedules are
thus a class apart --- power-law degrees like a scale-free network, but power-law
(not logarithmic) distances, unlike a small world. Specialization builds the long
fractal chains; duplication stacks them up in parallel.

\section{Real project schedules}

This is not just a story about a model. \citet{vazquezprojects2023} examined
$77$ real construction projects, together worth more than ten billion dollars,
which yielded $323$ activity networks --- snapshots of projects at different
stages --- ranging from about a hundred to sixteen thousand activities. Three
predictions survive the encounter with data. First, the predecessor and successor
counts are each power-law distributed, and fitting them \emph{separately} yields
duplication indices $q$ that agree, just as the model demands they should.
Second, those indices fall between roughly $0.1$ and $0.5$, most above $0.2$:
real schedules sit on the split-dominated, nearly-linear side --- but with enough
duplication to matter. Third, the share of activities on the critical path falls
with project size right across the collection, confirming that the critical path
really does dwindle as projects grow.

\section{Delay risk as percolation}

If the critical path fades to insignificance and yet projects are still late,
where does the delay risk live? The answer is that it is spread across the whole
network, and it is governed by \emph{percolation} --- the same mathematics that
describes how water seeps through porous rock or how a forest fire spreads from
tree to tree.

Picture an outside shock --- bad weather, a strike, a pandemic --- that pushes
some activities past their planned finish. A delay only passes to a downstream
activity if it is large enough to use up the spare time, the \emph{free float},
built into that dependency; otherwise the buffer absorbs it and the delay stops
there. Now treat each dependency as a channel that passes a delay along with some
probability $p$, and ask what fraction of activities end up delayed beyond their
own slack --- the \emph{delay incidence}. This is exactly a percolation problem:
delays spread through the network of dependencies like fire through a forest, and
percolation theory says such spreading has a sharp \emph{threshold}
\citep{albert2000,ellinas2019,schwartz2002},
\begin{equation}
  p_c = \frac{1}{\langle k\rangle},
  \label{eq:proj-pc}
\end{equation}
where $\langle k\rangle$ is the average number of downstream dependencies per
activity. The meaning of a threshold is stark: below $p_c$ delays stay local and
project-wide overruns essentially never happen; above $p_c$ a finite fraction of
projects suffer a system-wide cascade --- the fire jumps the firebreaks.

The data choose between two stories cleanly. If the critical path governed
delay, the incidence ought to track $p\,c$ (transmission times critical-path
size); in the real projects that correlation is negligible. If percolation
governs delay, the incidence ought to switch on exactly at $p-p_c=0$; and that is
what the construction projects show --- below threshold, delay incidence stays
under one percent; above it, the incidence climbs, reaching fifteen percent of
all activities in the worst cases. The true control knob of project delay is
$p-p_c$, a property of the \emph{whole} network, not $p\,c$, a property of one
path through it.

Here is the chapter's practical lesson, and its link to the rest of the book. The
very same local rule that sets the degree distributions also sets the average
degree $\langle k\rangle$, and through Eq.~\eqref{eq:proj-pc} the threshold for
catastrophic delay. A project's tendency to run late is written into the network
its activities have grown into --- not into any single path through it.

\section{Summary}

Project schedules are forward-flowing networks of activities, grown by a local
refinement rule: pick an activity and either duplicate it into parallel copies
(probability $q$) or split it into specialized sequential parts. This
duplication--split rule predicts identical power-law distributions for the
numbers of predecessors and successors, $p_k\sim k^{-1/q}$; a critical path that
grows only as $c\sim n^{1-q}$ and so vanishes as a fraction of large projects;
and a fractal, rather than small-world, geometry. All three hold in $77$ real
construction projects. And because the critical path becomes irrelevant at scale,
delay risk is better read through percolation, with a threshold $p_c=1/\langle
k\rangle$ fixed by the grown network's average degree --- so that project
performance, in the end, is a property of the activity network as a whole.


\chapter{Emergence of communities}
\label{ch:communities}

\section{Communities nobody designed}

So far the local rules of this book have explained the \emph{degree} structure of
networks --- their power laws, their clustering hierarchy, their correlations.
This chapter turns to a feature of an altogether different kind: \emph{community
structure}. A community is a group of nodes that are densely connected among
themselves but only sparsely connected to the rest of the network --- a clump
that hangs together. They are everywhere in real data: people fall into social
circles, web pages into topics, proteins into functional modules, each group
talking mostly to itself. Splitting a network into such groups is what
researchers mean by \emph{community detection}, and a network that splits cleanly
is said to have strong community structure.

Where do these communities come from? The usual answer points to differences
between the nodes themselves --- \emph{node heterogeneity}, in the jargon. People
of the same profession or politics, web pages on the same subject, proteins with
the same biological function: nodes of a kind, the story goes, are simply more
likely to link to one another, and that built-in preference is what carves the
network into groups. On this view, communities are stamped onto the network by
properties the nodes carried in from the start.

The thesis of this chapter, following \citet{vazquezcommunities2025}, is that no
such stamp is needed. \emph{Communities emerge from local growth rules even when
every node is identical.} The first hint is decades old: \citet{jin2001} showed
that a social network grown purely by triadic closure --- the friends-of-friends
rule of Chapter~\ref{ch:friends} --- develops communities although its nodes are
all of one type, and the same was later seen in the gene-duplication models of
Chapter~\ref{ch:duplication} \citep{sole2002}. If a featureless local rule
segregates a network into groups all by itself, then community structure joins
preferential attachment and the clustering hierarchy as one more thing that
\emph{emerges} from locality rather than being designed in.

To make that claim sharp we should ask it as a quantitative question, and the
right way to frame it comes from a corner of mathematics called \emph{Ramsey
theory}. Ramsey theory studies how large a random structure must become before
some orderly pattern is forced to appear. Its best-known nugget is a party fact:
in any gathering of six people, there are always either three who all know one
another or three who are all mutual strangers --- order of that kind is
unavoidable once the group is big enough. We will ask the analogous question of
growing networks: given a local growth rule, how large must the network become
before it is virtually \emph{certain} to contain communities? This chapter is, in
effect, a Ramsey theory of community structure in growing networks.

\section{The Ramsey community number}

Before we can measure when communities appear, we need a trustworthy way to
\emph{detect} them, and here some care is required, because the most popular
method can be fooled. That method, \emph{modularity optimization}, looks for the
division of a network that packs the most links inside groups; the trouble is
that it will happily report groups even in a completely random network that has
no real structure at all \citep{newman2006,fortunato2010}. To avoid being
deceived, the work relies instead on two more discerning tools: the
\emph{stochastic block model}, which fits the network to a statistical model of
blocks and keeps only the structure that genuinely earns its keep
\citep{karrer2011}, cross-checked against \emph{Infomap}, which finds groups by
watching where a random walker gets trapped \citep{rosvall2008}. Both, crucially,
report \emph{no} communities in a truly random network.

That last point also supplies the essential control. To be sure a community is
the work of the growth rule and not merely of the degrees, one compares the grown
network against a \emph{rewired} version of itself --- the links shuffled at
random while every node keeps its number of connections (this shuffling is called
the configuration model). If the communities survive the shuffle, they were never
more than an accident of the degrees; if they vanish, they were the doing of the
local rule. For every local model in this chapter, the shuffled version has no
communities at all --- so whatever structure we find is genuinely built by the
dynamics.

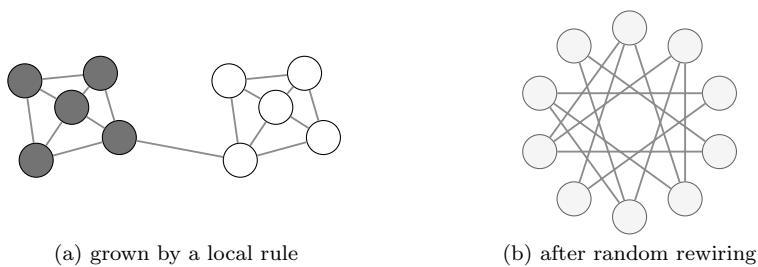
\begin{figure}[t]
\centering
\begin{adjustbox}{max width=\linewidth}\begin{tikzpicture}[
  nd/.style={circle,draw=black!60,fill=black!4,minimum size=4.5mm,inner sep=0pt},
  cL/.style={circle,draw=black,fill=black!55,minimum size=4.5mm,inner sep=0pt},
  cR/.style={circle,draw=black,fill=white,minimum size=4.5mm,inner sep=0pt},
  lnk/.style={draw=black!45,line width=0.7pt}]
  \begin{scope}
    \node[cL] (La) at (0.0,0.9)  {};
    \node[cL] (Lb) at (1.0,1.0)  {};
    \node[cL] (Lc) at (1.25,0.15){};
    \node[cL] (Ld) at (0.15,-0.15){};
    \node[cL] (Le) at (0.62,0.55){};
    \draw[lnk] (La)--(Lb) (La)--(Le) (La)--(Ld) (Lb)--(Lc)
               (Lb)--(Le) (Lc)--(Ld) (Lc)--(Le) (Ld)--(Le);
    \node[cR] (Ra) at (2.70,0.9)  {};
    \node[cR] (Rb) at (3.70,1.0)  {};
    \node[cR] (Rc) at (3.95,0.15) {};
    \node[cR] (Rd) at (2.85,-0.15){};
    \node[cR] (Re) at (3.32,0.55) {};
    \draw[lnk] (Ra)--(Rb) (Ra)--(Re) (Ra)--(Rd) (Rb)--(Rc)
               (Rb)--(Re) (Rc)--(Rd) (Rc)--(Re) (Rd)--(Re);
    \draw[lnk] (Lc)--(Rd);
    \node[font=\footnotesize] at (2.0,-1.4) {(a) grown by a local rule};
  \end{scope}
  \begin{scope}[xshift=6.4cm]
    \foreach \i in {0,...,9}
      \node[nd] (d\i) at ({1.6+1.25*cos(90+36*\i)},{0.35+1.25*sin(90+36*\i)}) {};
    \draw[lnk] (d0)--(d4) (d0)--(d6) (d0)--(d3) (d1)--(d5) (d1)--(d7)
               (d2)--(d6) (d2)--(d8) (d2)--(d5) (d3)--(d7) (d3)--(d9)
               (d4)--(d8) (d5)--(d9) (d6)--(d9);
    \node[font=\footnotesize] at (1.6,-1.4) {(b) after random rewiring};
  \end{scope}
\end{tikzpicture}\end{adjustbox}
\caption{\textbf{Communities emerge from local rules.} (a) A network grown by a
local rule settles into dense communities (two here, one dark gray, one white) joined by
only a few links. (b) Shuffling the links at random while keeping every node's
number of connections destroys the communities --- so the structure was the work
of the rule, not of the degrees alone.}
\label{fig:com-rewire}
\end{figure}

With detection settled, we can pose the question precisely
(Figure~\ref{fig:com-rewire}). The box below states
the two definitions that do the work; both are short, and their meaning is given
in words right after, so the box can be skipped.

\begin{calculation}{the communities likelihood and the Ramsey community number}
For a growth rule $f_G$ that produces a network of $n$ nodes, let a detection
method $f_\kappa$ return the number $\kappa$ of communities it finds. Define the
\emph{communities likelihood} as the probability --- estimated by generating many
networks --- that an $n$-node instance has more than one community,
\begin{equation}
  P_\kappa(n) = \mathrm{Prob}\big\{\, \kappa \ge 2 \,\big\}.
  \label{eq:com-pk}
\end{equation}
The \emph{Ramsey community number} is then the smallest size at which communities
are all but guaranteed, at a confidence $1-\epsilon$ (we use $\epsilon=0.05$,
i.e.\ $95\%$):
\begin{equation}
  r_\kappa = \min\{\, n : P_\kappa(n) \ge 1-\epsilon \,\}.
  \label{eq:com-rk}
\end{equation}
A rule is said to have the \emph{emergent communities property} when such an
$r_\kappa$ exists.
\end{calculation}

In plain terms: $P_\kappa(n)$ is the chance that a network grown to size $n$ has
communities, and the Ramsey community number $r_\kappa$ is the size beyond which
that chance is essentially $(1-\epsilon)100\%$. A \emph{finite} $r_\kappa$ is the fingerprint
of emergence --- grow the network past it and communities are a near-certainty. A
rule for which no such size exists, no matter how large the network, simply does
not make communities on its own. The whole question ``do communities emerge from
this rule?'' thus reduces to ``is $r_\kappa$ finite?''

One caution should be lodged now and collected later. The definition above is
stated \emph{relative to a detection method} $f_\kappa$: $r_\kappa$ is a property
of the network \emph{and} of the rule used to interrogate it, and the two cannot
be separated. For most of this chapter that dependence is a technicality, the
choice of method shifting the numbers without touching the conclusions. In the
last part of the chapter, where we can compute $r_\kappa$ exactly, it will turn
out to be rather more than a technicality.

\section{Local rules grow communities}

We now line up two local rules against two controls. The local rules --- local
search and duplication--split, both met earlier in this book --- have the emergent
communities property; the controls --- reshuffling a network's links, and growing
one with no locality at all --- do not. Holding the four side by side is the whole
argument of the chapter in miniature.

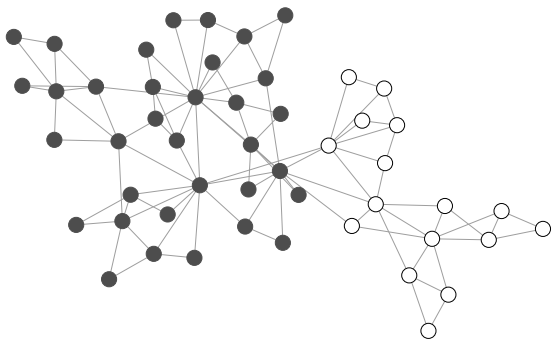
\begin{figure}[t]
\centering
\begin{adjustbox}{max width=\linewidth}\begin{tikzpicture}[
  netlink/.style={draw=black!38,line width=0.35pt},
  nA/.style={circle,draw=black!70,fill=black!70,line width=0.3pt,inner sep=0pt,minimum size=0.20cm},
  nB/.style={circle,draw=black,fill=white,line width=0.3pt,inner sep=0pt,minimum size=0.20cm}]
  \draw[netlink] (3.520,2.115) -- (2.461,1.926);
  \draw[netlink] (3.520,2.115) -- (4.165,2.451);
  \draw[netlink] (3.520,2.115) -- (4.788,1.675);
  \draw[netlink] (3.520,2.115) -- (4.472,1.388);
  \draw[netlink] (3.520,2.115) -- (2.404,3.090);
  \draw[netlink] (3.520,2.115) -- (3.136,2.464);
  \draw[netlink] (3.520,2.115) -- (3.332,3.341);
  \draw[netlink] (3.520,2.115) -- (3.767,1.799);
  \draw[netlink] (3.520,2.115) -- (3.060,1.380);
  \draw[netlink] (3.520,2.115) -- (3.559,1.165);
  \draw[netlink] (3.520,2.115) -- (3.104,1.871);
  \draw[netlink] (2.461,1.926) -- (4.165,2.451);
  \draw[netlink] (2.461,1.926) -- (2.404,3.090);
  \draw[netlink] (2.461,1.926) -- (1.385,2.508);
  \draw[netlink] (2.461,1.926) -- (2.156,2.518);
  \draw[netlink] (2.461,1.926) -- (1.436,1.452);
  \draw[netlink] (2.461,1.926) -- (3.060,1.380);
  \draw[netlink] (2.461,1.926) -- (1.545,1.801);
  \draw[netlink] (2.461,1.926) -- (1.851,1.019);
  \draw[netlink] (2.461,1.926) -- (2.035,1.540);
  \draw[netlink] (2.461,1.926) -- (2.389,0.966);
  \draw[netlink] (4.165,2.451) -- (4.788,1.675);
  \draw[netlink] (4.165,2.451) -- (4.911,2.219);
  \draw[netlink] (4.165,2.451) -- (5.070,2.721);
  \draw[netlink] (4.165,2.451) -- (4.899,3.206);
  \draw[netlink] (4.165,2.451) -- (4.432,3.356);
  \draw[netlink] (4.165,2.451) -- (4.609,2.781);
  \draw[netlink] (4.788,1.675) -- (4.472,1.388);
  \draw[netlink] (4.788,1.675) -- (5.533,1.218);
  \draw[netlink] (4.788,1.675) -- (4.911,2.219);
  \draw[netlink] (4.788,1.675) -- (5.231,0.734);
  \draw[netlink] (4.788,1.675) -- (5.703,1.652);
  \draw[netlink] (4.472,1.388) -- (5.533,1.218);
  \draw[netlink] (2.404,3.090) -- (1.385,2.508);
  \draw[netlink] (2.404,3.090) -- (1.087,3.231);
  \draw[netlink] (2.404,3.090) -- (3.136,2.464);
  \draw[netlink] (2.404,3.090) -- (2.156,2.518);
  \draw[netlink] (2.404,3.090) -- (3.332,3.341);
  \draw[netlink] (2.404,3.090) -- (2.943,3.019);
  \draw[netlink] (2.404,3.090) -- (3.050,3.895);
  \draw[netlink] (2.404,3.090) -- (2.629,3.551);
  \draw[netlink] (2.404,3.090) -- (2.568,4.112);
  \draw[netlink] (2.404,3.090) -- (1.839,3.227);
  \draw[netlink] (2.404,3.090) -- (2.111,4.109);
  \draw[netlink] (2.404,3.090) -- (1.752,3.720);
  \draw[netlink] (1.385,2.508) -- (1.087,3.231);
  \draw[netlink] (1.385,2.508) -- (0.562,3.169);
  \draw[netlink] (1.385,2.508) -- (1.436,1.452);
  \draw[netlink] (1.385,2.508) -- (0.536,2.527);
  \draw[netlink] (5.533,1.218) -- (5.231,0.734);
  \draw[netlink] (5.533,1.218) -- (5.703,1.652);
  \draw[netlink] (5.533,1.218) -- (5.749,0.479);
  \draw[netlink] (5.533,1.218) -- (6.283,1.204);
  \draw[netlink] (5.533,1.218) -- (6.452,1.585);
  \draw[netlink] (4.911,2.219) -- (5.070,2.721);
  \draw[netlink] (1.087,3.231) -- (0.562,3.169);
  \draw[netlink] (1.087,3.231) -- (0.539,3.797);
  \draw[netlink] (1.087,3.231) -- (0.112,3.242);
  \draw[netlink] (3.136,2.464) -- (3.767,1.799);
  \draw[netlink] (3.136,2.464) -- (2.943,3.019);
  \draw[netlink] (3.136,2.464) -- (3.530,2.871);
  \draw[netlink] (3.136,2.464) -- (3.104,1.871);
  \draw[netlink] (5.231,0.734) -- (5.749,0.479);
  \draw[netlink] (5.231,0.734) -- (5.485,0.000);
  \draw[netlink] (5.070,2.721) -- (4.899,3.206);
  \draw[netlink] (5.070,2.721) -- (4.609,2.781);
  \draw[netlink] (2.156,2.518) -- (1.839,3.227);
  \draw[netlink] (2.156,2.518) -- (1.874,2.807);
  \draw[netlink] (0.562,3.169) -- (0.539,3.797);
  \draw[netlink] (0.562,3.169) -- (0.536,2.527);
  \draw[netlink] (0.562,3.169) -- (0.112,3.242);
  \draw[netlink] (0.562,3.169) -- (0.000,3.889);
  \draw[netlink] (4.899,3.206) -- (4.432,3.356);
  \draw[netlink] (1.436,1.452) -- (1.545,1.801);
  \draw[netlink] (1.436,1.452) -- (1.851,1.019);
  \draw[netlink] (1.436,1.452) -- (0.825,1.402);
  \draw[netlink] (1.436,1.452) -- (1.260,0.683);
  \draw[netlink] (3.332,3.341) -- (3.050,3.895);
  \draw[netlink] (3.332,3.341) -- (3.591,4.174);
  \draw[netlink] (0.539,3.797) -- (0.000,3.889);
  \draw[netlink] (5.703,1.652) -- (6.283,1.204);
  \draw[netlink] (5.749,0.479) -- (5.485,0.000);
  \draw[netlink] (2.943,3.019) -- (2.629,3.551);
  \draw[netlink] (2.943,3.019) -- (3.530,2.871);
  \draw[netlink] (3.050,3.895) -- (2.568,4.112);
  \draw[netlink] (3.050,3.895) -- (3.591,4.174);
  \draw[netlink] (3.060,1.380) -- (3.559,1.165);
  \draw[netlink] (6.283,1.204) -- (6.452,1.585);
  \draw[netlink] (6.283,1.204) -- (7.000,1.349);
  \draw[netlink] (2.568,4.112) -- (2.111,4.109);
  \draw[netlink] (6.452,1.585) -- (7.000,1.349);
  \draw[netlink] (1.545,1.801) -- (0.825,1.402);
  \draw[netlink] (1.545,1.801) -- (2.035,1.540);
  \draw[netlink] (1.851,1.019) -- (2.389,0.966);
  \draw[netlink] (1.851,1.019) -- (1.260,0.683);
  \draw[netlink] (1.839,3.227) -- (1.752,3.720);
  \draw[netlink] (1.839,3.227) -- (1.874,2.807);
  \node[nA] at (3.520,2.115) {};
  \node[nA] at (2.461,1.926) {};
  \node[nB] at (4.165,2.451) {};
  \node[nB] at (4.788,1.675) {};
  \node[nB] at (4.472,1.388) {};
  \node[nA] at (2.404,3.090) {};
  \node[nA] at (1.385,2.508) {};
  \node[nB] at (5.533,1.218) {};
  \node[nB] at (4.911,2.219) {};
  \node[nA] at (1.087,3.231) {};
  \node[nA] at (3.136,2.464) {};
  \node[nB] at (5.231,0.734) {};
  \node[nB] at (5.070,2.721) {};
  \node[nA] at (2.156,2.518) {};
  \node[nA] at (0.562,3.169) {};
  \node[nB] at (4.899,3.206) {};
  \node[nA] at (1.436,1.452) {};
  \node[nA] at (3.332,3.341) {};
  \node[nA] at (0.539,3.797) {};
  \node[nA] at (3.767,1.799) {};
  \node[nB] at (5.703,1.652) {};
  \node[nB] at (5.749,0.479) {};
  \node[nB] at (4.432,3.356) {};
  \node[nA] at (2.943,3.019) {};
  \node[nA] at (3.050,3.895) {};
  \node[nA] at (0.536,2.527) {};
  \node[nA] at (2.629,3.551) {};
  \node[nA] at (3.060,1.380) {};
  \node[nA] at (0.112,3.242) {};
  \node[nB] at (6.283,1.204) {};
  \node[nA] at (2.568,4.112) {};
  \node[nA] at (3.591,4.174) {};
  \node[nB] at (6.452,1.585) {};
  \node[nA] at (1.545,1.801) {};
  \node[nA] at (3.530,2.871) {};
  \node[nA] at (3.559,1.165) {};
  \node[nA] at (1.851,1.019) {};
  \node[nA] at (0.825,1.402) {};
  \node[nA] at (0.000,3.889) {};
  \node[nB] at (7.000,1.349) {};
  \node[nA] at (1.839,3.227) {};
  \node[nA] at (2.111,4.109) {};
  \node[nA] at (2.035,1.540) {};
  \node[nB] at (4.609,2.781) {};
  \node[nA] at (2.389,0.966) {};
  \node[nA] at (1.752,3.720) {};
  \node[nA] at (3.104,1.871) {};
  \node[nA] at (1.260,0.683) {};
  \node[nB] at (5.485,0.000) {};
  \node[nA] at (1.874,2.807) {};
\end{tikzpicture}\end{adjustbox}
\caption{\textbf{Communities in a local-search network.} An instance of the
local-search model $LS(n{=}50,\ell{=}1)$. The stochastic block model
resolves two communities, shown dark gray and white; they emerge even though the
rule treats every node alike and never mentions groups.}
\label{fig:com-ls}
\end{figure}

\paragraph{Local search.} The model $LS(n,\ell)$ is the search rule of
Chapter~\ref{ch:search} in undirected form \citep{vazquez2001}: each new node
takes a short walk of $\ell$ steps from a random starting node and links to every
node it visits. It carries preferential attachment and high clustering, and ---
though nothing in the rule ever mentions groups --- the block model resolves clear
communities, with $P_\kappa$ climbing from near zero around $n=10$ to a
near-certain one beyond about $n=100$, and staying there out to a hundred thousand
nodes \citep{vazquezcommunities2025}. Figure~\ref{fig:com-ls} shows one such network.

\begin{figure}[t]
\centering
\begin{adjustbox}{max width=\linewidth}\begin{tikzpicture}[
  netlink/.style={draw=black!38,line width=0.35pt},
  nA/.style={circle,draw=black!70,fill=black!70,line width=0.3pt,inner sep=0pt,minimum size=0.20cm},
  nB/.style={circle,draw=black,fill=white,line width=0.3pt,inner sep=0pt,minimum size=0.20cm}]
  \draw[netlink] (1.363,3.902) -- (1.829,3.777);
  \draw[netlink] (1.829,3.777) -- (2.156,3.406);
  \draw[netlink] (1.829,3.777) -- (2.129,4.241);
  \draw[netlink] (1.829,3.777) -- (1.104,3.216);
  \draw[netlink] (1.829,3.777) -- (1.121,4.647);
  \draw[netlink] (1.829,3.777) -- (1.660,3.275);
  \draw[netlink] (1.829,3.777) -- (2.999,3.419);
  \draw[netlink] (4.531,1.266) -- (3.820,1.711);
  \draw[netlink] (4.531,1.266) -- (4.470,0.559);
  \draw[netlink] (4.531,1.266) -- (5.367,1.624);
  \draw[netlink] (5.159,0.034) -- (5.245,0.612);
  \draw[netlink] (5.159,0.034) -- (4.652,0.000);
  \draw[netlink] (4.200,2.925) -- (5.039,2.711);
  \draw[netlink] (4.200,2.925) -- (4.186,2.247);
  \draw[netlink] (4.200,2.925) -- (4.685,2.096);
  \draw[netlink] (4.200,2.925) -- (3.849,2.360);
  \draw[netlink] (4.200,2.925) -- (3.736,2.957);
  \draw[netlink] (4.200,2.925) -- (4.672,3.645);
  \draw[netlink] (4.200,2.925) -- (3.737,3.342);
  \draw[netlink] (4.200,2.925) -- (2.999,3.419);
  \draw[netlink] (4.200,2.925) -- (4.843,3.182);
  \draw[netlink] (0.892,2.657) -- (1.104,3.216);
  \draw[netlink] (5.039,2.711) -- (5.811,2.581);
  \draw[netlink] (0.792,6.149) -- (0.602,5.534);
  \draw[netlink] (1.104,3.216) -- (0.509,3.076);
  \draw[netlink] (1.121,4.647) -- (0.636,4.313);
  \draw[netlink] (1.121,4.647) -- (0.602,5.534);
  \draw[netlink] (1.121,4.647) -- (1.492,5.026);
  \draw[netlink] (1.121,4.647) -- (0.638,4.815);
  \draw[netlink] (4.186,2.247) -- (3.820,1.711);
  \draw[netlink] (3.820,1.711) -- (3.038,1.708);
  \draw[netlink] (3.820,1.711) -- (3.849,2.360);
  \draw[netlink] (3.820,1.711) -- (3.432,1.888);
  \draw[netlink] (3.820,1.711) -- (3.329,1.446);
  \draw[netlink] (4.685,2.096) -- (5.026,1.288);
  \draw[netlink] (3.320,2.849) -- (3.105,2.133);
  \draw[netlink] (3.320,2.849) -- (3.736,2.957);
  \draw[netlink] (3.320,2.849) -- (3.737,3.342);
  \draw[netlink] (0.602,5.534) -- (0.313,6.121);
  \draw[netlink] (0.602,5.534) -- (1.104,5.765);
  \draw[netlink] (0.602,5.534) -- (0.000,5.775);
  \draw[netlink] (0.602,5.534) -- (0.069,5.291);
  \draw[netlink] (5.245,0.612) -- (5.026,1.288);
  \draw[netlink] (4.470,0.559) -- (4.652,0.000);
  \draw[netlink] (3.105,2.133) -- (3.038,1.708);
  \draw[netlink] (3.105,2.133) -- (3.432,1.888);
  \draw[netlink] (3.105,2.133) -- (3.329,1.446);
  \draw[netlink] (6.008,2.122) -- (5.367,1.624);
  \draw[netlink] (6.008,2.122) -- (5.995,3.007);
  \draw[netlink] (6.008,2.122) -- (6.606,1.896);
  \draw[netlink] (5.478,3.238) -- (5.995,3.007);
  \draw[netlink] (5.478,3.238) -- (4.843,3.182);
  \draw[netlink] (5.995,3.007) -- (5.903,3.683);
  \draw[netlink] (5.811,2.581) -- (6.538,2.527);
  \draw[netlink] (7.000,2.213) -- (6.606,1.896);
  \draw[netlink] (7.000,2.213) -- (6.538,2.527);
  \draw[netlink] (4.672,3.645) -- (5.316,3.935);
  \draw[netlink] (5.316,3.935) -- (5.903,3.683);
  \node[nB] at (1.363,3.902) {};
  \node[nB] at (1.829,3.777) {};
  \node[nB] at (2.156,3.406) {};
  \node[nA] at (4.531,1.266) {};
  \node[nA] at (5.159,0.034) {};
  \node[nA] at (4.200,2.925) {};
  \node[nB] at (0.892,2.657) {};
  \node[nB] at (2.129,4.241) {};
  \node[nA] at (5.039,2.711) {};
  \node[nB] at (0.792,6.149) {};
  \node[nB] at (1.104,3.216) {};
  \node[nB] at (1.121,4.647) {};
  \node[nA] at (4.186,2.247) {};
  \node[nB] at (1.660,3.275) {};
  \node[nA] at (3.820,1.711) {};
  \node[nB] at (0.636,4.313) {};
  \node[nA] at (4.685,2.096) {};
  \node[nB] at (0.509,3.076) {};
  \node[nA] at (3.320,2.849) {};
  \node[nB] at (0.602,5.534) {};
  \node[nA] at (5.245,0.612) {};
  \node[nA] at (4.470,0.559) {};
  \node[nA] at (3.105,2.133) {};
  \node[nA] at (3.038,1.708) {};
  \node[nA] at (6.008,2.122) {};
  \node[nA] at (5.367,1.624) {};
  \node[nA] at (5.478,3.238) {};
  \node[nB] at (0.313,6.121) {};
  \node[nB] at (1.104,5.765) {};
  \node[nA] at (3.849,2.360) {};
  \node[nB] at (1.492,5.026) {};
  \node[nA] at (3.736,2.957) {};
  \node[nA] at (4.652,0.000) {};
  \node[nA] at (5.995,3.007) {};
  \node[nA] at (5.811,2.581) {};
  \node[nA] at (7.000,2.213) {};
  \node[nB] at (0.000,5.775) {};
  \node[nA] at (6.606,1.896) {};
  \node[nA] at (4.672,3.645) {};
  \node[nA] at (3.737,3.342) {};
  \node[nA] at (5.316,3.935) {};
  \node[nA] at (5.026,1.288) {};
  \node[nA] at (2.999,3.419) {};
  \node[nA] at (6.538,2.527) {};
  \node[nA] at (4.843,3.182) {};
  \node[nA] at (3.432,1.888) {};
  \node[nA] at (5.903,3.683) {};
  \node[nA] at (3.329,1.446) {};
  \node[nB] at (0.638,4.815) {};
  \node[nB] at (0.069,5.291) {};
\end{tikzpicture}\end{adjustbox}
\caption{\textbf{Communities in a duplication--split network.} An instance of the
duplication--split model $DS(n{=}50,q{=}0.3)$. Two communities (dark gray
and white) again emerge --- even though this rule builds no triangles at all, so the
network is far less cliquish than the local-search one of
Figure~\ref{fig:com-ls}.}
\label{fig:com-ds}
\end{figure}
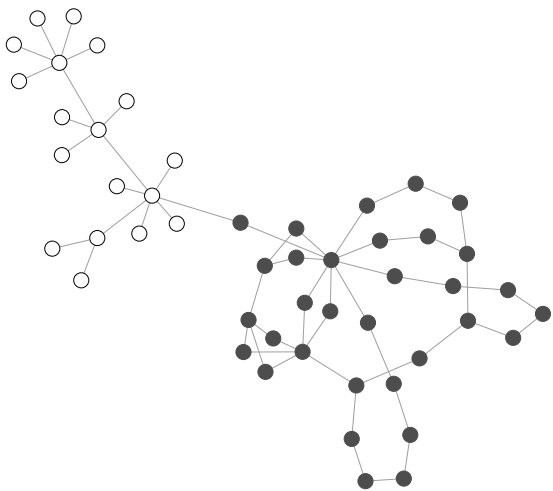

\paragraph{Duplication--split.} The model $DS(n,q)$ is the undirected
cousin of the project-schedule rule of Chapter~\ref{ch:projects}: a new node
either copies all of a random node's links (probability $q$) or splits one of its
links in two. This model is the decisive case, because it builds \emph{no
triangles whatsoever} --- duplication makes none, and splitting destroys any that
might form --- and yet communities emerge just as reliably. The lesson is sharp: a
network does not need to be cliquish to have communities. What it needs is
locality, not triangles. Figure~\ref{fig:com-ds} shows an instance: two
communities, and not a triangle in sight.

\paragraph{Degree-preserving randomization.} To be sure the communities are the
work of the local rule and not of the degrees alone, we need a way to wipe out the
local structure while leaving every degree untouched. The \emph{configuration
model}, met when we defined the Ramsey number, does exactly that: it takes a
finished network and reshuffles its links at random, keeping each node's number of
connections fixed but scrambling \emph{which} nodes are joined. Applied to a
local-search or a duplication--split network, this reshuffling destroys the
communities outright --- the randomized versions have no finite Ramsey number at
all (the two ``randomized'' rows in the table below). Whatever the local rules
built, the bare list of degrees cannot sustain.

\paragraph{Barab\'asi--Albert.} Finally, $BA(n,m)$, the canonical
preferential-attachment model, is our \emph{non-local} control: a growing,
rich-get-richer network in which each newcomer attaches its $m$ links to nodes
drawn from the \emph{whole} graph, never once consulting a neighborhood. Despite
its growth and its rich-get-richer rule it has no emergent communities (the last
row of the table), and the next section is devoted to teasing apart exactly why.

A few representative Ramsey numbers, based on numerical estimates for $\epsilon=0.05$
\citep{vazquezcommunities2025}, make the pattern concrete:

\begin{center}
\begin{tabular}{lcc}
\hline
Model & $r_\kappa$ (block model) & $r_\kappa$ (Infomap) \\
\hline
$LS(\ell=1)$               & $81$   & $222$  \\
$DS(q=0.5)$             & $98$   & $622$  \\
$LS(\ell=1)$, randomized   & ---    & ---    \\
$DS(q=0.5)$, randomized & ---    & ---    \\
$BA(m=2)$               & ---    & ---    \\
\hline
\end{tabular}
\end{center}
The two columns already show that $r_\kappa$ depends on how one looks: Infomap
needs two to six times the size the block model does. The dependence goes deeper
than a choice of algorithm. Chapter~\ref{ch:motifs} fits these same rules with a
block model that can also describe its links as planted triangles and squares,
and finds $r_\kappa=200$ for $LS(\ell=1)$ against the $81$ above, while
$DS(q=0.5)$ does not move at all --- because local search's structure \emph{is}
triangles, and a method that can name a triangle stops counting it as evidence
for a community. Read the numbers below as properties of a rule seen through an
instrument, which is what the definition says they are.

Every genuinely local rule has a finite Ramsey number --- grow it past a few
hundred nodes and communities are certain. The rules without local structure ---
the bottom three rows, the two reshuffled networks and the Barab\'asi--Albert
model --- have none: the dashes mean communities never become certain, however
large the network. The exact figures differ between the
two detection methods --- there is no single ``true'' definition of a community,
and where the line falls depends a little on how one draws it --- but the
\emph{split} between local rules (finite) and non-local ones (infinite) is the
same either way. Local structure is a necessary condition for communities to
emerge.

\section{When globality fails: the Barab\'asi--Albert model}

The natural control is the Barab\'asi--Albert model \citep{barabasi1999}, the
textbook growing network with \emph{no} locality at all: each newcomer attaches
its $m$ links to nodes chosen from the \emph{whole} network in proportion to
their degree. It has growth and preferential attachment, but no step that ever
looks at a node's neighborhood --- and it does \emph{not} have the emergent
communities property. Its community likelihood rises only to about one-half near
a thousand nodes and then \emph{falls} as the network grows further, never
reaching certainty. The little structure it has comes from the mere age-ordering
of its nodes, not from any local rule, and it washes out at scale.

The boundary case is illuminating. Give the model a baseline ``attractiveness''
$a$, so that a node is chosen with a probability proportional to $a +
k_i/\sum_j k_j$ --- a fixed head-start $a$ on top of the usual degree preference.
This tunes the degree exponent to $\gamma=2+a/m$. Tracking the Ramsey number
across these variants reveals a clean rule of thumb: communities emerge exactly
when the network is \emph{clustered} enough. The Ramsey number falls as the
average clustering coefficient rises, and \emph{diverges} --- communities never
become certain --- as the clustering goes to zero \citep{vazquezcommunities2025}. The standard Barab\'asi--Albert
model sits right at that vanishing point, because its clustering decays toward
zero as it grows (like $(\ln n)^2/n$): the bigger it gets, the fewer short cycles
it has, and short cycles are what local structure is made of. So even here the
principle holds --- where the dynamics happen to generate appreciable clustering
(an \emph{implicit} bit of local structure), communities appear; where clustering
fades, they do not.

\section{How local must ``local'' be?}

If locality is the cause, exactly how local must a connection be to count? Two
dials settle it: the length of the cycles a rule creates, and the distance it
reaches across. Triadic closure closes triangles (cycles of length three); the
duplication--split rule works with squares (length four). To sweep the cycle
length freely, use the \emph{bubble model}: at each step a chain of $L$ new nodes
is added and its two ends are attached to the two ends of a randomly chosen
existing link, creating a fresh cycle of length $L+2$. As $L$ grows --- as the
rule weaves ever longer cycles --- the Ramsey number climbs with it, and steeply:
from $r_\kappa=108$ for the shortest cycles to more than fifty thousand by $L=9$
\citep{vazquezcommunities2025}, as the table below records. The climb is not quite
smooth --- riding on top of it is a regular even--odd alternation, the odd-length
cycles bringing communities on sooner than the even cycles around them, an effect
plainest for the shorter cycles --- but the overall message is unmistakable: short
cycles, tight local feedback, bring communities on soonest.

The alternation is left unexplained here, and it does not stay that way. An odd
cycle is what makes a network non-bipartite, so the rules that bring communities
on sooner are exactly the ones that cannot be two-coloured;
Chapter~\ref{ch:bubbles} finds the same parity deciding a quite different
question, whether those communities survive being renamed as motifs, and takes
the coincidence of the two as the reason to look at this family closely.

\begin{center}
\begin{tabular}{ccc}
\hline
$L$ & cycle length $L+2$ & $r_\kappa$ \\
\hline
1 & 3  & 108 \\
2 & 4  & 326 \\
3 & 5  & 246 \\
4 & 6  & 2201 \\
5 & 7  & 752 \\
6 & 8  & 6595 \\
7 & 9  & 2740 \\
8 & 10 & 15863 \\
9 & 11 & 56160 \\
\hline
\end{tabular}
\end{center}

Distance matters just as much. In the bubble model the chain's two ends attach to
nodes that are a single step apart (the ends of one link). Loosen this so the
ends may attach to nodes up to $W$ steps apart, with the chain length fixed at
$L=1$. For $W=1$ communities still emerge with certainty, at $r_\kappa=108$;
stretching to $W=2$ pushes the threshold up roughly tenfold, to $r_\kappa=1090$,
but the communities still come. At $W=3$ --- and likewise if the ends are joined
to nodes picked anywhere at random --- they stop appearing altogether, the Ramsey
number ceasing to exist (the dash in the table) \citep{vazquezcommunities2025}.

\begin{center}
\begin{tabular}{cc}
\hline
$W$ & $r_\kappa$ \\
\hline
1 & 108 \\
2 & 1090 \\
3 & --- \\
\hline
\end{tabular}
\end{center}

For this model, ``local'' means roughly two steps: communities form
when new links reinforce ties among nearby nodes, and dissolve once links are
allowed to leap across the network. This is the whole book's premise made
quantitative. It is not growth, nor preferential attachment, nor even clustering
in the abstract that carves a network into communities --- it is the
\emph{locality} of the rule by which it grew.

\section{Networks we can solve exactly}

Everything so far has been measured, not proved. The Ramsey numbers of the last
three sections came from growing many networks and running a detection algorithm
on each --- honest work, but it leaves the central question in the dark. We know
\emph{that} communities appear beyond a certain size. We do not know \emph{why},
and we cannot say what the size depends on.

To answer that we need cases simple enough to solve with pencil and paper. Three
have recently been found, and the rest of this chapter is their story. Each strips
the problem down to a different bone. The \emph{ring} \citep{vazquezring2026}
removes growth, randomness, and heterogeneity all at once, leaving nothing but
local wiring. The \emph{diamond lattice} \citep{vazquezrg2026} --- the same
lattice whose shortest paths we shall count in Chapter~\ref{ch:distance} --- lets
us watch, exactly, why the evidence for communities accumulates at all. The
\emph{pseudofractal web} \citep{vazquezpseudofractal2026} asks whether the
communities we find are real or an illusion cast by the hubs.

A warning before we start: these three networks are not \emph{grown} at all. They
are fixed, deterministic constructions, and no detection algorithm is ever run on
them. What they isolate is locality itself, stripped of every other ingredient,
and what they answer is not ``can an algorithm find the communities?'' but ``at
what size does the split become the better description?'' That is the question the
Ramsey number was always asking, and here we can answer it exactly.

The trick that makes all three solvable is to use the block model not as an
algorithm but as a \emph{formula}. Given a network and a candidate split, the
probability of the data under each hypothesis can be integrated in closed form,
and the two compared.

\begin{calculation}{the block model as a formula}
Suppose the nodes are divided into blocks, and that the chance of a link between
two nodes depends only on which blocks they are in: $\theta_1$ within block~1,
$\theta_2$ within block~2, $\theta_{12}$ between. We do not know these
probabilities, so we average over them, giving each a $\mathrm{Beta}(\alpha,\alpha)$
prior, and likewise average over the label prior $\pi$ that decides how nodes are
apportioned:
\begin{equation}
\begin{split}
  P = \frac{1}{Z}\int
  &\prod_{i\in\{1,2\}}\theta_i^{\,E_i}(1-\theta_i)^{\binom{n_i}{2}-E_i}\\
  &\times\theta_{12}^{\,E_{12}}(1-\theta_{12})^{n_1n_2-E_{12}}\\
  &\times\pi^{n_1}(1-\pi)^{n-n_1}\,P(\theta)P(\pi)\,d\theta\,d\pi .
\end{split}
\label{eq:com-model}
\end{equation}
Here $n_i$ is the size of block $i$, $E_i$ the number of links inside it, and
$E_{12}$ the number crossing between. Every integral is of the same standard shape,
\begin{equation}
  \int_0^1 x^{E+\alpha-1}(1-x)^{N-E+\alpha-1}\,dx
  = \mathrm{B}(E{+}\alpha,\,N{-}E{+}\alpha),
  \label{eq:com-betaint}
\end{equation}
where $\mathrm{B}$ is the Beta function. So each hypothesis collapses to a product
of Beta functions, and their ratio $R$ --- the \emph{evidence ratio} --- can be
evaluated exactly. All we ever need is a handful of link counts.
\end{calculation}

Writing $R$ for the ratio of the evidence for the split to that for no split, the
posterior weight of the partition is
\begin{equation}
  P(\mathrm{split}) = \frac{R}{1+R} = \frac{1}{1+e^{-\log R}},
  \label{eq:com-psplit}
\end{equation}
so $\log R>0$ means the split is the better description, and the Ramsey number is
the smallest size at which $P(\mathrm{split})$ reaches the certainty $q$ we demand.

\section{The ring wants to be broken}

Begin with the least promising network imaginable. Place $n$ nodes on a circle and
join each to its $c$ nearest neighbors on either side --- the \emph{circulant
ring} $C_n(1,\dots,c)$, the unrewired lattice that Watts and Strogatz started
from. Every node is identical. Every node has the same degree $2c$, and the ring
has $cn$ links. Nothing grows, nothing is random, and there is no group of nodes
that belongs together more than any other. If any network has no communities, this
one does not.

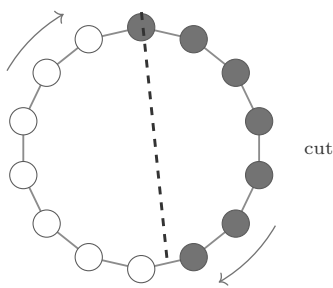
\begin{figure}[t]
\centering
\begin{adjustbox}{max width=\linewidth}\begin{tikzpicture}[
  nd/.style={circle,draw=black!65,minimum size=3.6mm,inner sep=0pt},
  cA/.style={nd,fill=black!55},
  cB/.style={nd,fill=white},
  lnk/.style={draw=black!45,line width=0.7pt},
  cut/.style={draw=accent,line width=1.1pt,dashed}]
  \def\N{14}
  \foreach \i in {1,...,\N}{
    \pgfmathsetmacro\ang{90-(\i-1)*360/\N}
    \coordinate (p\i) at (\ang:1.6);
  }
  \foreach \i in {1,...,\N}{
    \pgfmathtruncatemacro\nextj{mod(\i,\N)+1}
    \draw[lnk] (p\i)--(p\nextj);
  }
  \foreach \i in {1,...,7}{\node[cA] at (p\i){};}
  \foreach \i in {8,...,\N}{\node[cB] at (p\i){};}
  \draw[cut] ($(p14)!0.5!(p1)$)++(0.35,0.28) -- ($(p7)!0.5!(p8)$)++(-0.35,-0.28);
  \node[accent,font=\scriptsize] at (2.35,0) {cut};
  \draw[->,black!55,line width=0.5pt] (150:2.05) arc (150:120:2.05);
  \draw[->,black!55,line width=0.5pt] (-30:2.05) arc (-30:-60:2.05);
\end{tikzpicture}\end{adjustbox}
\caption{\textbf{Cutting the ring.} The block model splits the ring into two
contiguous arcs (dark and white), preferring this to no split at all once the ring
is large enough. But every node looks the same, so the cut (dashed) could sit
between any pair of beads and score exactly the same: the model says the ring
\emph{must} be cut, without saying where. The two arcs are real communities in the
way a magnet has a real direction --- fixed in any one outcome, undetermined by the
symmetry.}
\label{fig:com-cutring}
\end{figure}

Ask the block model anyway. The candidate split is into two contiguous arcs
(Figure~\ref{fig:com-cutring}), and the counts it needs are elementary.

\begin{calculation}{the ring's link counts and its evidence ratio}
Take an arc of $n_i$ consecutive nodes. A link of span $k$ joins nodes $k$ apart,
and $n_i-k$ such links fit inside the arc, so summing over the spans
$k=1,\dots,c$,
\begin{equation}
  E_i = \sum_{k=1}^{c}(n_i-k) = c\,n_i - \frac{c(c+1)}{2},
  \label{eq:com-ringEi}
\end{equation}
and the pairs inside the arc that are \emph{not} linked number
$\binom{n_i}{2}-E_i=\binom{n_i-c}{2}$. Only links near the two cut points cross
between the arcs, and counting them by span gives
\begin{equation}
  E_{12} = 2\sum_{k=1}^{c} k = c(c+1),
  \label{eq:com-ringE12}
\end{equation}
independent of $n$. As a check, $E_1+E_2+E_{12}=cn$. For the balanced split
$n_1=n_2=m\equiv n/2$; the unpartitioned ring has all $cn$ links among
$\binom{n}{2}$ pairs. Feeding these into Eq.~\eqref{eq:com-model} and using
Eq.~\eqref{eq:com-betaint} on every integral,
\begin{equation}
\begin{split}
  R = \Big[ &\mathrm{B}\big(cm{-}\tfrac{c(c+1)}{2}{+}\alpha,\ \tbinom{m-c}{2}{+}\alpha\big)^{2}\\
  \times{} &\mathrm{B}\big(c(c{+}1){+}\alpha,\ m^{2}{-}c(c{+}1){+}\alpha\big)\\
  \times{} &\mathrm{B}(m{+}\alpha,\ m{+}\alpha) \Big]\\
  \Big/ \Big[ &\mathrm{B}(\alpha,\alpha)^{2}\,
  \mathrm{B}\big(cn{+}\alpha,\ \tbinom{n}{2}{-}cn{+}\alpha\big)\\
  \times{} &\mathrm{B}(\alpha,\ n{+}\alpha) \Big].
\end{split}
\label{eq:com-ringR}
\end{equation}
No approximation has been made: this is exact for every $n$, $c$ and $\alpha$.
\end{calculation}

\begin{figure}[t]
\centering
\includegraphics[width=\linewidth]{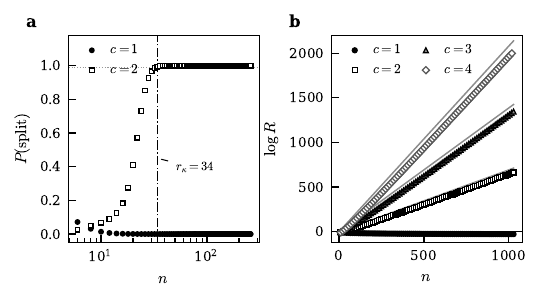}
\caption{\textbf{The ring wants to be broken.} (a) The probability that a block
model prefers two arcs to no split. The plain cycle ($c=1$) is never split; the
next-nearest-neighbor ring ($c=2$) is split beyond $r_\kappa\simeq34$ nodes
(dash-dotted line; dotted line marks $99\%$ certainty). (b) The evidence for the
split against the prediction $\log R\simeq(c-1)\,n\ln 2$ (lines),
Eq.~\eqref{eq:com-ringlaw}, for $c=1,\dots,4$. The cycle lies exactly on zero: it
is balanced, not reluctant.}
\label{fig:com-ring}
\end{figure}

Evaluating Eq.~\eqref{eq:com-ringR} gives an answer that depends, with startling
sharpness, on the range $c$ (Figure~\ref{fig:com-ring}a). The plain cycle, $c=1$,
is \emph{never} split: its evidence ratio decays as a power of the size,
\begin{equation}
  \log R \sim A - (2\alpha+3)\log n \longrightarrow -\infty,
  \label{eq:com-ringc1}
\end{equation}
so that $P(\mathrm{split})\sim n^{-(2\alpha+3)}$ vanishes and $r_\kappa=\infty$.
But the next-nearest-neighbor ring, $c=2$, is split as soon as it exceeds about
thirty-five nodes --- $r_\kappa\simeq34$ at $95\%$ certainty --- and beyond that
the evidence piles up in proportion to the size, $\log R\sim(\ln 2)\,n$. A
featureless ring of beads, given one extra reach, acquires communities.

Why should one extra neighbor make such a difference? The published answer is that
a plain cycle offers no density contrast for the block model to exploit. That
cannot be right: an arc of a cycle is twice as dense as the whole cycle, and so is
an arc of any other ring --- the contrast is a factor of two for \emph{every} $c$.
The real reason is a cancellation, and it is worth seeing because it generalizes.

\begin{calculation}{why the ring breaks, and when}
Two quantities compete. Cutting the ring into two arcs of $n/2$ makes each arc
look twice as dense as the whole ring did: the arcs hold nearly all $cn$ links but
only about half the node pairs. Each link therefore gains $\ln 2$ of evidence, for
a total \emph{gain} of $cn\ln 2$. Against this stands the price of naming the
split --- of saying \emph{which} nodes go in which block. That price is the label
term of Eq.~\eqref{eq:com-ringR}, and at $\alpha=1$ it is not merely of order
$n\ln 2$ but exactly a binomial coefficient. With $n=2m$,
\begin{equation}
  \frac{\mathrm{B}(m{+}1,m{+}1)}{\mathrm{B}(1,n{+}1)}
  = \frac{(m!)^{2}}{(2m{+}1)!}\,(n{+}1)
  = \frac{(m!)^{2}}{n!}
  = \binom{n}{n/2}^{-1},
  \label{eq:com-labelexact}
\end{equation}
so the label term contributes
\begin{equation}
  -\log\binom{n}{n/2} \simeq -\,n\ln 2,
  \label{eq:com-label}
\end{equation}
the entropy of labeling $n$ nodes with two labels. Subtracting the price from the
gain,
\begin{equation}
  \log R \;\simeq\; \big(cn - n\big)\ln 2 \;=\; (c-1)\,n\ln 2 .
  \label{eq:com-ringlaw}
\end{equation}
Since $cn$ is the number of links $m$, this reads
$\log R\simeq(m-n)\ln 2$: the ring is split exactly when its links
outnumber its nodes.
\end{calculation}

Equation~\eqref{eq:com-ringlaw} goes beyond what the ring paper states --- it
reports only the $c=2$ case --- but it is easily checked against the exact formula
for $c=1,\dots,4$, and it passes (Figure~\ref{fig:com-ring}b). It also dissolves
the mystery of the cycle. The cycle is not \emph{reluctant} to break; it is
exactly \emph{balanced}. A cycle has precisely as many links as nodes, so gain and
price cancel to the last term, and the fate of the split is left to the smaller
corrections of Eq.~\eqref{eq:com-ringc1}, which happen to be negative. Move to
$c=2$ and the cancellation is broken by a whole factor of $n$. The celebrated
contrast between the cycle and the ring is thus not a transition at all but a
knife edge: $c=1$ sits at the zero of a perfectly smooth family, and the evidence
grows linearly in $c-1$ thereafter.

And now the connection this book has been building toward. The quantity
$m-n$ is, up to one, the number of \emph{independent cycles} a
connected network carries. A tree has none, and is never split. A single ring has
exactly one, and sits exactly on the fence. A network with an extensive number of
independent cycles is split with an evidence proportional to how many it has. The
same short loops that Chapter~\ref{ch:distance} will show multiplying shortest
paths are what carve a network into communities here --- two faces, again, of the
one local coin.

There remains something troubling about calling the two arcs communities, and it
should be said plainly. The ring looks the same from
every node; rotate it by one bead and it is unchanged. So if cutting it at some
point is favored, then cutting it at any \emph{other} point is favored exactly as
much: all $n$ rotations of the balanced split have identical evidence. The block
model is not telling us where the communities are. It is telling us that the ring
\emph{must be cut somewhere}, while remaining perfectly silent about where --- and
it is the ring's own symmetry that keeps it silent. Physicists know this situation
well. A magnet below its critical temperature must point somewhere, though nothing
in its physics prefers north to south; the symmetry of the laws survives in the
multiplicity of the outcomes, not in any one of them. The ring's communities are
of exactly this kind: real, in that the split genuinely describes the network
better; degenerate, in that no particular split is preferred. The ring does not
have communities so much as it wants to be broken.

\section{Why communities emerge at all}

The ring shows that locality suffices. It does not explain why size should be the
deciding variable --- why every one of these networks has a threshold at all. For
that we turn to the diamond lattice, which we shall meet again in
Chapter~\ref{ch:distance}: begin with a single link between two poles $A$ and $B$,
and at each generation replace \emph{every} link by $b$ parallel paths of $s$
links each. Take $b=s=2$, the classic Migdal--Kadanoff cell
\citep{migdal1975,kadanoff1976}. Generation $t$ has
\begin{equation}
  m = (bs)^t = 4^t \ \text{links}, \qquad
  n_t = \tfrac{1}{3}\big(2\cdot 4^t + 4\big) \ \text{nodes},
  \label{eq:com-diasizes}
\end{equation}
and a node born at generation $k$ ends with degree $2\cdot2^{t-k}$, while the two
poles are permanent hubs of degree $b^t=2^t$. The lattice is strongly
degree-heterogeneous, which is what will make the next section's question sharp.

\begin{figure}[t]
\centering
\includegraphics[width=0.5\linewidth]{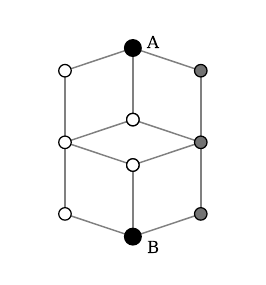}
\caption{\textbf{The diamond's two communities.} The $b=s=2$ diamond lattice at
generation $t=2$. Its two bundles --- the left half (open circles) and the right
half (gray) --- are the two communities; they touch only at the poles $A$ and $B$
(black), so the only links crossing between them are the few at top and bottom.
As the lattice grows, that seam becomes an ever fainter share of each bundle's
interior, which is why the split sharpens without limit.}
\label{fig:com-diamond}
\end{figure}

The $b$ bundles hanging between the poles are the obvious candidate communities
(Figure~\ref{fig:com-diamond}). Take block~1 to be one bundle together with the two
poles, block~2 the other bundle. Because the bundles meet \emph{only} at the poles,
every crossing link joins a pole to a bundle-2 node, and the counts close in
themselves.

\begin{calculation}{the counts, and the map that carries them}
For the bundle cut of the $b=s=2$ diamond,
\begin{equation}
\begin{aligned}
  e_{11} &= \tfrac{4^t}{2}, &\qquad e_{22} &= \tfrac{4^t}{2}-2^t, \\
  e_{12} &= 2^t, &\qquad
  \kappa_1 &= 4^t+2^t, \\
  \kappa_2 &= 4^t-2^t, & &
\end{aligned}
\label{eq:com-diacounts}
\end{equation}
where $e_{rs}$ counts links between blocks $r$ and $s$ and $\kappa_r$ is the sum
of degrees in block $r$. The seam is \emph{subextensive}: $e_{12}=2^t=\sqrt{m}$,
so the fraction of links crossing the cut decays geometrically,
$e_{12}/m=(1/s)^t\to0$.

The point of the lattice is that these tallies renormalize \emph{exactly}. Collect
them in $\mathbf{v}=(e_{11},e_{22},e_{12},\kappa_1,\kappa_2)^{\top}$; then
$\mathbf{v}(t{+}1)=M\mathbf{v}(t)$ with
\begin{equation}
  M=\begin{pmatrix}
     4&0&0&0&0\\ 0&4&2&0&0\\ 0&0&2&0&0\\ 0&0&-2&4&0\\ 0&0&2&0&4
    \end{pmatrix}.
  \label{eq:com-M}
\end{equation}
The entries read off the construction. A link inside either block is replaced by
four links whose new nodes inherit its bundle label, so within-block links
quadruple. A seam link joins a pole to a bundle-2 node; of its four children, the
two at the pole stay seam links and the two at the far end are absorbed into
block~2 --- whence $e_{12}'=2e_{12}$, with the displaced pair reappearing in
$e_{22}'$. The spectrum is what matters:
\begin{equation}
  \lambda_{\mathrm{vol}} = bs, \qquad \lambda_{\mathrm{seam}} = b .
  \label{eq:com-eigen}
\end{equation}
Read backwards, $M^{-1}$ is precisely the Migdal--Kadanoff decimation that strips
the youngest generation; read forwards, $M$ is the growth generator of the
detection problem.
\end{calculation}

Two eigenvalues, and their ratio is the whole story:
\begin{equation}
  \frac{\lambda_{\mathrm{seam}}}{\lambda_{\mathrm{vol}}} = \frac{1}{s} .
  \label{eq:com-gap}
\end{equation}
For $s\ge2$ the seam is left behind by the bulk, so the boundary between the
communities grows ever fainter relative to their interiors, and the split gets
better and better with size. For $s=1$ the two rates are equal, the seam keeps
pace, and the lattice never separates --- the diamond's exact counterpart of the
plain cycle, and a pleasing echo of it. In a sentence: $b$ sets how many
communities there are, and $s$ sets how sharply they are separated.

To weigh the evidence we need a null model that already knows the degrees ---
otherwise, as the next section will show, we would only be measuring the hubs. The
degree-corrected block model \citep{karrer2011} supplies one.

\begin{calculation}{the degree-corrected evidence}
Let each node carry a propensity $\theta_i=k_i/\sqrt{2m}$ and let the expected
number of links between $i$ and $j$ be $\theta_i\theta_j\,\omega_{g_ig_j}$, where
$g_i$ is the block of $i$. Setting every $\omega_{rs}=1$ reproduces the
configuration model exactly --- a network with the same degrees and nothing else
--- so $\omega_{rs}$ measures departure from chance. The expected block-pair
weights are
\begin{equation}
  \Omega_{rr} = \frac{\kappa_r^2}{4m}, \qquad
  \Omega_{rs} = \frac{\kappa_r\kappa_s}{2m}\quad(r\neq s).
  \label{eq:com-Omega}
\end{equation}
Giving each $\omega_{rs}$ a $\mathrm{Gamma}(\alpha,\alpha)$ prior, of mean one and
so centered on the configuration model, and integrating it out,
\begin{equation}
  Z(e,\Omega) = \frac{\alpha^\alpha}{\Gamma(\alpha)}
  \int_0^\infty \omega^{e+\alpha-1}e^{-(\Omega+\alpha)\omega}\,d\omega
  = \frac{\alpha^\alpha}{\Gamma(\alpha)}\frac{\Gamma(e{+}\alpha)}{(\Omega{+}\alpha)^{e+\alpha}} .
  \label{eq:com-Zdc}
\end{equation}
The evidence ratio for the split against no split is then
\begin{equation}
  R_{\mathrm{dc}} = \frac{Z(e_{11},\Omega_{11})\,Z(e_{22},\Omega_{22})\,Z(e_{12},\Omega_{12})}
  {Z(m,\,m)} .
  \label{eq:com-Rdc}
\end{equation}
For large counts, Stirling turns this into something recognizable,
\begin{equation}
  \log R_{\mathrm{dc}} \simeq \sum_{r\le s} e_{rs}\,\ln\frac{e_{rs}}{\Omega_{rs}},
  \label{eq:com-KL}
\end{equation}
the surprise --- in the information-theorist's sense --- of finding the links where
they are rather than where the degrees alone would have put them. This is the
Bayesian, degree-corrected counterpart of modularity.
\end{calculation}

Everything now follows from two ratios. Write $u_{\mathrm{in}}=e_{rr}/\Omega_{rr}$
for the density inside a block relative to chance, and
$u_{\times}=e_{12}/\Omega_{12}$ for the seam. Inserting
Eq.~\eqref{eq:com-diacounts} into Eq.~\eqref{eq:com-Omega} gives them exactly, as
rational functions of $x=2^t$:
\begin{equation}
  u_{\mathrm{in}} = \frac{2x^2}{(x\pm1)^2},
  \qquad
  u_{\times} = \frac{2x}{x^2-1}.
  \label{eq:com-densities}
\end{equation}

\begin{figure}[t]
\centering
\includegraphics[width=\linewidth]{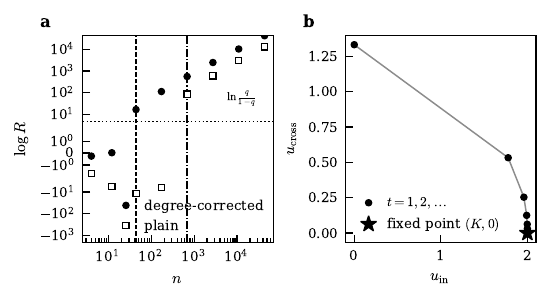}
\caption{\textbf{The Ramsey number as a crossing.} (a) Evidence for the split on
the diamond lattice against size. $r_\kappa$ is simply where the curve clears the
threshold $\ln[q/(1-q)]$ (dotted, $q=0.99$): at $n=44$ for a degree-corrected
block model (dashed line) but only at $n=684$ for a plain one (dash-dotted).
(b) The densities of Eq.~\eqref{eq:com-densities} flow to the community fixed
point $(K,0)$: the interiors settle at $K=2$ times chance, the seam fades to
nothing.}
\label{fig:com-rg}
\end{figure}

\begin{calculation}{the evidence per link}
Under the flow, Eq.~\eqref{eq:com-densities} converges to the \emph{community
fixed point} (Figure~\ref{fig:com-rg}b),
\begin{equation}
  (u_{\mathrm{in}}, u_{\times}) \longrightarrow (K, 0),
  \label{eq:com-fixed}
\end{equation}
with $K=b$ the number of communities, the seam density decaying as
$(\lambda_{\mathrm{seam}}/\lambda_{\mathrm{vol}})^t=(1/s)^t$. At the fixed point
each within-block term of Eq.~\eqref{eq:com-KL} contributes $\tfrac{m}{2}\ln K$
and the cross term is subleading, so the evidence density converges to a pure
number:
\begin{equation}
  \frac{\log R_{\mathrm{dc}}}{m} \longrightarrow \ln K .
  \label{eq:com-lnK}
\end{equation}
The reason is pure counting, and owes nothing to the diamond. If $K$ equal,
well-separated communities hold all the links, each holds $e_{rr}=m/K$ of them,
while the configuration model --- knowing only the degrees --- expects
$\Omega_{rr}=m/K^2$. The ratio is $K$ identically, and every link contributes
exactly $\ln K$: the entropy of resolving which of $K$ communities it belongs to.
This is the same $\ln 2$ per link that broke the ring.
\end{calculation}

Equation~\eqref{eq:com-lnK} explains the emergence of communities in one line, and
it is the answer this chapter has been missing. The evidence for a partition is
\emph{extensive}: it grows in proportion to the number of links, each one
contributing its $\ln K$. The price of positing the partition --- the extra block
affinities a model must fit --- grows only like the \emph{logarithm} of the size. A
quantity growing like $m\ln K$ must eventually overtake one growing like $\ln m$,
and once it does it never looks back. Emergence is not a delicate phenomenon
requiring a special mechanism; it is arithmetic. All that remains to ask is
\emph{where} the crossing happens:
\begin{equation}
  r_\kappa(q) = n_t\big|_{t_\star}, \qquad
  t_\star = \min\Big\{t : \log R_{\mathrm{dc}}(t) \ge \ln\tfrac{q}{1-q}\Big\},
  \label{eq:com-cross}
\end{equation}
the threshold $\ln[q/(1-q)]$ being set by the certainty $q$ we demand
(Figure~\ref{fig:com-rg}a). With $\log R_{\mathrm{dc}}\simeq(\ln K)(bs)^t$ from
Eq.~\eqref{eq:com-lnK}, the crossing generation is
\begin{equation}
  t_\star \simeq \Big\lceil \log_{bs}\frac{\ln[q/(1-q)]}{\ln K} \Big\rceil .
  \label{eq:com-tstar}
\end{equation}
This is the precise sense in which the Ramsey number is a
re\-nor\-mal\-iza\-tion-group
crossing: it is the generation at which a running, extensive evidence --- driven by
the volume eigenvalue, once the flow has turned onto the community fixed point ---
first clears a fixed detection threshold.

The whole $(b,s)$ family can be done at once, and the result is the closest thing
this book has to a formula for emergence.

\begin{calculation}{the Ramsey number in closed form}
The key structural fact is a lemma: for $s\ge2$, at every generation each link
either lies inside a single bundle or joins a pole to a bundle node; no link ever
joins two distinct bundles. (Induction on the replacement: a link inside a bundle
spawns links inside that bundle, and a pole-to-bundle link spawns pole-to-bundle
links and bundle-interior links.) So write $c_t$ for the number of pole-to-bundle
links of one bundle and $i_t$ for its interior links. Replacement gives the
triangular recursion
\begin{equation}
  c_{t+1} = b\,c_t, \qquad i_{t+1} = bs\,i_t + b(s-1)\,c_t,
  \label{eq:com-recbs}
\end{equation}
whose spectrum can be read straight off the diagonal --- $\lambda_{\mathrm{vol}}=bs$
and $\lambda_{\mathrm{seam}}=b$, now proved for all $(b,s)$, not just the symmetric
cell where the two roles cannot be told apart. With $c_1=2$, $i_1=s-2$,
\begin{equation}
  c_t = 2\,b^{\,t-1}, \qquad i_t = b^{\,t-1}\big(s^t-2\big).
  \label{eq:com-icbs}
\end{equation}
The block statistics follow: $e_{00}=i_t+c_t=b^{\,t-1}s^t$ for the pole block,
$e_{rr}=i_t$ and $e_{0r}=c_t$ for each bundle $r\ge1$, and $e_{rs}=0$ for distinct
bundles. The degree sums are $\kappa_r=2b^{\,t-1}(s^t{-}1)$ and
$\kappa_0=2b^{\,t-1}(s^t{-}1{+}b)$, and since each replaced link adds $b(s{-}1)$
nodes,
\begin{equation}
  n_t = 2 + \frac{b(s{-}1)\big[(bs)^t-1\big]}{bs-1}.
  \label{eq:com-nbs}
\end{equation}
The densities, with $x=s^t$, are again rational and again flow to $(b,0)$:
\begin{equation}
  \frac{e_{00}}{\Omega_{00}} = \frac{b\,x^2}{(x{+}b{-}1)^2},
  \quad
  \frac{e_{rr}}{\Omega_{rr}} = \frac{b\,x(x{-}2)}{(x{-}1)^2},
  \quad
  \frac{e_{0r}}{\Omega_{0r}} \to \frac{b}{s^t}.
  \label{eq:com-ratiosbs}
\end{equation}
The $\ln b$ contributions sum to $m\ln b$ \emph{exactly}, and the only deficit is
the seam's --- the price of the $b^t$ crossing links still present:
\begin{equation}
  \log R_{\mathrm{dc}} = m\ln b\,\big[1-\delta_t\big],
  \qquad
  \delta_t = \frac{2(b{-}1)}{b\ln b}\,\frac{t\ln s+1}{s^{t}} .
  \label{eq:com-delta}
\end{equation}
Setting this equal to the threshold and inverting Eq.~\eqref{eq:com-nbs},
\begin{equation}
\begin{aligned}
  r_\kappa(b,s;q) &= 2 + \frac{b(s{-}1)}{bs-1}\Big[(bs)^{t_\star}-1\Big], \\
  t_\star &= \Big\lceil \log_{bs}\Big(\ln\tfrac{q}{1-q}\Big/\ln b\Big) \Big\rceil .
\end{aligned}
\label{eq:com-rkbs}
\end{equation}
\end{calculation}

Equation~\eqref{eq:com-rkbs} deserves a moment. The Ramsey community number, which
three sections ago was a number we could only obtain by simulating thousands of
networks, is here a formula. And look at what survives in it: $b$, through the
entropy accumulated per link, and $s$, through the volume growth. The degrees drop
out. The priors drop out. The details of the lattice drop out. Emergence, it turns
out, depends on almost nothing about the network except how fast its communities
grow and how fast their boundary does.

\section{Communities, or just degrees?}

One doubt remains, and it is the sharpest of all. Every network in this book is
hub-dominated. A block model handed a scale-free network can improve its fit
simply by putting the hubs in one group and the rest in another --- not because
there is any community structure, but because there is a degree distribution to
absorb. How do we know that what we have been calling communities is not just the
hubs in disguise?

\begin{figure}[t]
\centering
\includegraphics[width=\linewidth]{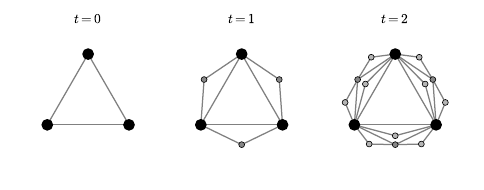}
\caption{\textbf{Building the pseudofractal web.} The deterministic rule, shown for
the first three generations. Start from a triangle ($t=0$); at each step, every
existing link gives birth to a new node joined to both its ends, and the old links
are kept. The three corners of the original triangle (black) never stop gaining
neighbors and become the towering hubs; each later node is fainter the later it was
born.}
\label{fig:com-pseudofractal}
\end{figure}

The pseudofractal web of \citet{dorogovtsevgoltsev2002} settles it. Its rule is as
local as they come: start from a triangle, and at every generation let each
existing link spawn a new node joined to both of its ends, keeping all the old
links (Figure~\ref{fig:com-pseudofractal}). What grows is deterministic, and
elementary bookkeeping gives
\begin{equation}
  n_t = \tfrac{3}{2}\big(3^{t}+1\big), \qquad E_t = 3^{\,t+1},
  \label{eq:com-pfNE}
\end{equation}
so the mean degree tends to four. A node born at generation $\tau$ ends with
degree $2^{\,t-\tau+1}$, and counting nodes by birth generation gives a power-law
degree distribution, $p(k)\sim k^{-\gamma}$ with
\begin{equation}
  \gamma = 1 + \frac{\ln 3}{\ln 2} \simeq 2.585,
  \label{eq:com-pfgamma}
\end{equation}
squarely in the scale-free range of real networks. Three seed hubs of degree
$2^{t+1}$ tower over everything. This web has exactly the degree heterogeneity
that makes the doubt bite.

It also has two rival splits with completely different characters. The
\emph{branch} cut follows the recursive construction, separating one branch of the
web from the others. The \emph{hub--leaf} cut ignores the construction entirely and
simply puts the high-degree nodes on one side.

\begin{calculation}{the branch cut, and what a degree-blind model makes of it}
Label the three seed links $0,1,2$; every new node and both its links inherit the
label of the link that spawned it, so each non-seed node sits in exactly one
branch. Take block~1 to be one branch plus the three seed hubs, block~2 the other
two branches. The counts are exact:
\begin{equation}
\begin{aligned}
  n_1 &= \tfrac{1}{2}\big(3^{t}+5\big), &\qquad n_2 &= 3^{t}-1, \\
  E_1 &= 3^{t}+2, &\qquad E_2 &= 2\cdot 3^{t}+2-2^{\,t+2}, \\
  E_{12} &= 2^{\,t+2}-4, & &
\end{aligned}
\label{eq:com-pfcounts}
\end{equation}
with $n_1+n_2=n_t$ and $E_1+E_2+E_{12}=E_t$. The seam is thin: $E_{12}$ grows only
as $2^{t}=n^{\ln2/\ln3}\simeq n^{0.631}$, so the fraction of crossing links
$E_{12}/E_t\sim(2/3)^{t}\to0$ --- the structural fact that drives everything.
Feeding Eq.~\eqref{eq:com-pfcounts} into the plain evidence
Eq.~\eqref{eq:com-model} and applying Stirling gives a linear growth,
\begin{equation}
  \log R \sim \Big(\ln 3 - \tfrac{2}{3}\ln 2\Big)\,n \approx 0.6365\,n,
  \label{eq:com-pfslope}
\end{equation}
independent of the prior.
\end{calculation}

Now correct for the degrees. The block degree sums of the branch cut are
\begin{equation}
  \kappa_1 = 2\cdot 3^{t}+2^{\,t+2}, \qquad
  \kappa_2 = 4\cdot 3^{t}-2^{\,t+2},
  \label{eq:com-pfkappas}
\end{equation}
with $\kappa_1+\kappa_2=2E_t$, and inserting these into
Eq.~\eqref{eq:com-Omega} gives densities that \emph{exceed} the configuration
model's expectation inside the blocks while the crossing density vanishes:
\begin{equation}
  \frac{E_1}{\Omega_{11}}\to 3, \qquad
  \frac{E_2}{\Omega_{22}}\to \tfrac{3}{2}, \qquad
  \frac{E_{12}}{\Omega_{12}}\sim\Big(\tfrac{2}{3}\Big)^{t}\to 0 .
  \label{eq:com-pfratios}
\end{equation}
The branch cut therefore survives the correction, and Eq.~\eqref{eq:com-KL} gives
its slope:
\begin{equation}
  \log R_{\mathrm{dc}} \sim \Big(2\ln 3 - \tfrac{4}{3}\ln 2\Big)\,n \approx 1.2730\,n,
  \label{eq:com-pfdcslope}
\end{equation}
exactly \emph{twice} the plain slope of Eq.~\eqref{eq:com-pfslope}. Correcting for
the degrees does not weaken the case for these communities. It doubles it.

\begin{figure}[t]
\centering
\includegraphics[width=\linewidth]{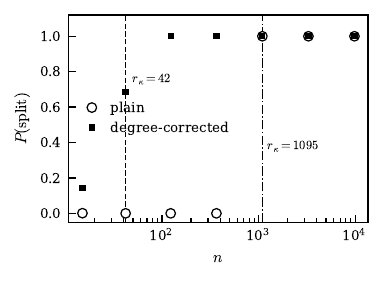}
\caption{\textbf{Communities, or just degrees?} The pseudofractal web under a
plain block model (open circles) and one corrected for the degree sequence (filled
squares). The corrected model is certain of the recursive communities from $n=42$;
the plain one denies them until $n=1095$. The degree gradient, far from
manufacturing the split, \emph{delays} the detector that cannot see past it.}
\label{fig:com-degree}
\end{figure}

The result is a reversal, and it is damning for the naive method. The plain block
model prefers the hub--leaf cut --- it is chasing the degree gradient, exactly as
feared. Correct for degrees and the hub--leaf cut collapses while the branch cut
strengthens: the recursive communities are real, and they are what a degree-blind
detector was missing while it stared at the hubs. The two verdicts on \emph{when}
disagree just as violently (Figure~\ref{fig:com-degree}). The plain model reports
no communities until $n=1095$; the corrected model has been sure of them since
$n=42$. Same network, same partition, same question --- and a twenty-five-fold
disagreement about the answer. The direction of that disagreement is the
counterintuitive part: the degree gradient, far from manufacturing the split,
\emph{delays} the detector that cannot see past it.

The moral is not that one number is right and the other wrong. It is that
$r_\kappa$ was never a property of the network alone. It is a property of the
network together with the question we ask of it, and the warning lodged when we
defined it now has to be paid. The exactly solvable cases are valuable precisely
because they let us see this: a quantity that looked like a fact about the web
turns out to depend, by a factor of twenty-five, on which null model we hold it
against. That does not dissolve the chapter's thesis --- under \emph{either} rule
the local construction has a finite $r_\kappa$, while a network with no local
structure has none --- but it does mean that a bare Ramsey number, quoted without
its detection rule, says less than it appears to.

How much less is worth knowing in advance, because the answer is larger than a
factor of twenty-five. Both rules compared here describe a network as blocks and
links, and differ only in what they expect the degrees to do. Give the detector a
richer vocabulary --- let it call a set of links a triangle or a square, and not
merely a pair of endpoints --- and Chapter~\ref{ch:motifs} finds that these two
lattices have no communities at all: each is tiled exactly by copies of the motif
its own rule lays down, and the cheapest description of it uses one group and no
partition whatever. The communities established in this section are real, and
they are real \emph{relative to a vocabulary that cannot name a triangle}. That
is the same warning as the one above, pushed as far as it goes.

\section{How many communities?}
\label{sec:com-howmany}

We have been asking whether a network splits in two. That was a convenience, and
it is time to drop it. A self-similar network has structure at every scale, so
why stop at the first cut?

Take the pseudofractal web again, and cut it three ways instead of two, one block
per branch. The symmetry now works for us rather than against us, and the counts
are cleaner than the branch cut's:
\begin{equation}
  n_r = \tfrac{1}{2}\big(3^{t}+1\big), \quad
  \kappa_r = 2\cdot 3^{t}, \quad
  e_{rr} = 3^{t}-2^{t}, \quad
  e_{rs} = 2^{t}.
  \label{eq:com-three}
\end{equation}
The densities behave exactly as the diamond's fixed point promised, with $K=3$:
\begin{equation}
  \frac{e_{rr}}{\Omega_{rr}} = 3\big(1-(2/3)^{t}\big)\to 3,
  \qquad
  \frac{e_{rs}}{\Omega_{rs}} = \tfrac{3}{2}(2/3)^{t}\to 0,
  \label{eq:com-threeratios}
\end{equation}
and the evidence density is $\ln 3$ per link, just as Eq.~\eqref{eq:com-lnK}
requires of three well-separated communities:
\begin{equation}
  \log R_{\mathrm{dc}}^{(3)} \sim 2\ln 3\,n \approx 2.197\,n .
  \label{eq:com-slope3}
\end{equation}
This beats the two-block cut of Eq.~\eqref{eq:com-pfdcslope} by exactly
$\tfrac{4}{3}\ln 2\,n$ --- the evidence the branch cut forfeited by lumping two
communities together. Three blocks describe the web better than two. And having
seen that, there is no reason to stop: removing the links of generation $d$ leaves
$3^{d+1}$ statistically identical sub-webs, and each level of refinement adds
another $\simeq2\ln3\,n$ to the evidence.

What stops it? The seams do. Every cut has to pay for the links it severs, and the
deeper the cut the more of them there are. On the diamond the competition can be
written down exactly. A depth-$d$ cut severs $c(d)=2^{t+d}$ links, leaving a
fraction $e_{\mathrm{in}}/m=1-2^{d-t}$ inside blocks --- each level doubling the
interface, which is the seam eigenvalue $\lambda_{\mathrm{seam}}=b=2$ acting once
per level. Balancing that against the reward for splitting the degree sum evenly
gives a modularity
\begin{equation}
  Q(d) = \frac{e_{\mathrm{in}}}{m} - \gamma\frac{\sum_c \kappa_c^2}{(2m)^2}
  = 1 - 2^{\,d-t} - \gamma\,2^{-d} + O(2^{-t}),
  \label{eq:com-Qd}
\end{equation}
where $\gamma$ is the resolution at which we choose to look. Maximizing over the
depth gives $2^{2d}=\gamma 2^{t}$, that is
\begin{equation}
  d_\star = \tfrac{1}{2}\big(t+\log_2\gamma\big),
  \qquad
  q_{\mathrm{opt}} = 2^{d_\star} = \sqrt{\gamma}\;2^{t/2} \sim \sqrt{\gamma\,n}.
  \label{eq:com-qopt}
\end{equation}
The number of communities that best describes the network is not two, and is not
any fixed number: it \emph{grows with the network}, as its square root. The web
does the same, with $K_{\mathrm{opt}}=3^{\lfloor t/2\rfloor}\sim\sqrt{n}$. The
optimum sits at the geometric midpoint of the network's own hierarchy --- about
$\sqrt{n}$ communities of about $\sqrt{n}$ nodes each --- and the two-block cut we
labored over is merely the top of that tree. It is worth noting that $\sqrt{n}$ is
the same scale that Chapter~\ref{ch:distance} will find in the exponent of the
shortest path multiplicity on this very lattice; both are the seam scale $b^t$ in
disguise.

\section{Communities as a state of matter}

There is one more thing the diamond will give us, and it is the strangest. So far
communities have been an inference: the partition a statistician prefers. But the
quantity being optimized is a Hamiltonian, and a Hamiltonian is physics. Written
in the form of \citet{reichardt2004},
\begin{equation}
  \mathcal{H} = -\sum_{i\neq j}\big(A_{ij}-\gamma\,p_{ij}\big)\,\delta(\sigma_i,\sigma_j),
  \qquad p_{ij} = \frac{\kappa_i\kappa_j}{2m},
  \label{eq:com-rbH}
\end{equation}
where $\sigma_i$ labels the community of node $i$ and $A_{ij}$ is the adjacency
matrix, the ground state is the community partition, and at $\gamma=1$ this is
modularity itself. For two communities the labels are Ising spins, and the
Hamiltonian rearranges into something a physicist recognizes at sight:
\begin{equation}
  \mathcal{H} = -\sum_{\langle ij\rangle}\sigma_i\sigma_j
  + \frac{\gamma}{4m}\Big[M_\kappa^2 - \sum_i \kappa_i^2\Big],
  \qquad
  M_\kappa = \sum_i \kappa_i\sigma_i .
  \label{eq:com-rbIsing}
\end{equation}
A ferromagnet --- neighbors want to agree, which is cohesion --- plus an
infinite-range, degree-weighted penalty that punishes the whole network for
agreeing too much, which is what forces a split.

Because the penalty couples the spins only through the single collective variable
$M_\kappa$, the model can be solved exactly on the diamond by decimation, and the
answer is a genuine phase. Take the correlation between the two poles,
$-\langle\sigma_A\sigma_B\rangle$, as the order parameter. Below the ferromagnetic
critical temperature
\begin{equation}
  \frac{k_B T_c}{J} = 1.641,
  \label{eq:com-Tc}
\end{equation}
the two poles lock into \emph{opposite} communities for any resolution $\gamma>0$,
however small --- the degree-weighted penalty bites hardest on the two
highest-degree nodes --- and the order survives as $n\to\infty$. The resolution
selects \emph{which} partition orders, not \emph{whether} it does. The contrast
with a plain ferromagnet is the whole point: an ordinary Ising model on the same
lattice has no such phase, its ground state being the trivial single community. It
is the community penalty that changes the ground state into a partition.

Push to the growing hierarchy of the last section and the picture completes.
With $q=2^d$ colors the model becomes a Potts model, whose series and parallel
decimations are
\begin{equation}
\begin{aligned}
  A_{\mathrm{ser}} &= A_1A_2 + (q{-}1)B_1B_2, \\
  B_{\mathrm{ser}} &= A_1B_2 + B_1A_2 + (q{-}2)B_1B_2, \\
  A_{\mathrm{par}} &= A_1A_2, \qquad B_{\mathrm{par}} = B_1B_2,
\end{aligned}
\label{eq:com-pottsRG}
\end{equation}
with $A$ the weight for two ends in the same community and $B$ for different ones.
Solving these, every $q>2$ orders through a \emph{first-order} transition --- the
partition snaps into place rather than fading in, as is known for Potts models on
hierarchical lattices \citep{griffiths1982} --- at a temperature that falls only
logarithmically with the
number of communities,
\begin{equation}
  \frac{k_B T_c(q)}{J} \sim \frac{1}{c\,\ln q}, \qquad c\approx 0.70 .
  \label{eq:com-Tcq}
\end{equation}
So the hierarchy orders level by level, coarsest first, each finer level freezing
at a lower temperature; and since $q_{\mathrm{opt}}\sim\sqrt{n}$, the finest level
still orders at $k_BT_c/J\sim1/\ln n$. Every thermodynamically stable level of the
hierarchy survives as the network grows.

The conclusion is worth stating baldly. On this lattice the community structure is
not merely detectable, in the sense that a statistician would infer it. It is
\emph{ordered}, in the sense that a physicist would recognize: a phase of matter,
with a critical temperature, an order parameter, and a cascade of transitions
building the hierarchy from the top down. Communities, born of nothing but local
wiring, turn out to be a state that a network can be in.

\section{How many communities do real networks have?}
\label{sec:com-count}

Everything so far has been theory: models grown by hand, and lattices solved with
pencil and paper. It is time to ask the real world. The
question of Section~\ref{sec:com-howmany} --- \emph{how many} communities a network
has, and how that number grows --- was answered there only for self-similar
constructions, where the best
description settled on about $\sqrt{n}$ communities of $\sqrt{n}$ nodes each. Does
anything so clean survive contact with the messy networks of the world? Following
\citet{vazquezcount2026}, we can now say that it does, and better than the models
led us to expect.

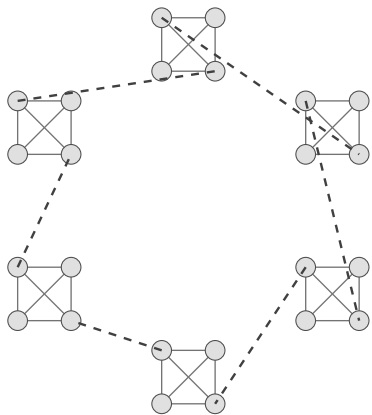
\begin{figure}[t]
\centering
\begin{adjustbox}{max width=\linewidth}\begin{tikzpicture}[
  nd/.style={circle,draw=black!70,fill=black!12,minimum size=2.6mm,inner sep=0pt},
  iedge/.style={draw=black!55,line width=0.5pt},
  br/.style={draw=black!75,line width=0.9pt,dashed}]
  \def\L{6}
  \foreach \c in {1,...,\L}{
    \pgfmathsetmacro\ca{90-(\c-1)*360/\L}
    \coordinate (C\c) at (\ca:2.2);
    \foreach \j in {1,...,4}{
      \pgfmathsetmacro\ja{(\j-1)*90+45}
      \coordinate (v\c\j) at ($(C\c)+(\ja:0.5)$);
    }
    \foreach \a/\b in {1/2,1/3,1/4,2/3,2/4,3/4}{\draw[iedge] (v\c\a)--(v\c\b);}
    \foreach \j in {1,...,4}{\node[nd] at (v\c\j){};}
  }
  \foreach \c in {1,...,\L}{
    \pgfmathtruncatemacro\d{mod(\c,\L)+1}
    \draw[br] (v\c2)--(v\d4);
  }
\end{tikzpicture}\end{adjustbox}
\caption{\textbf{The modular limit: the caveman graph.} A ring of tight cliques
(here six cliques of four nodes), each joined to its neighbors by a single bridge
(dashed). Each clique is so much denser inside than out that it is unmistakably one
community, so the number of communities simply counts the cliques and grows in
proportion to size, $B\propto n$ ($\beta=1$). The self-similar networks of
Section~\ref{sec:com-howmany} sit at the opposite limit, $\beta=\tfrac12$; real
networks fall between.}
\label{fig:com-caveman}
\end{figure}

Two idealized limits fix the range of possibilities
(Figure~\ref{fig:com-caveman}). At one end is a network built of dense modules,
like Watts's \emph{caveman} graph --- a ring of tight cliques, each clique so much
denser inside than out that it is unmistakably its own community
\citep{watts1999caveman}. There the number of communities $B$ simply counts the
cliques, so it grows in direct proportion to size, $B\propto n$: the \emph{modular}
limit, exponent one. At the other end are the self-similar networks of
Section~\ref{sec:com-howmany}, whose community count grows only as the square root,
$B\propto\sqrt{n}$: the \emph{fractal} limit, exponent one half. Writing
$B\propto n^{\beta}$, every network must fall somewhere in $\tfrac12\le\beta\le1$.
Where is an empirical question, and until it is measured we do not know whether
real networks resemble stacks of modules, self-similar fractals, or neither.

\begin{figure}[t]
\centering
\includegraphics[width=\linewidth]{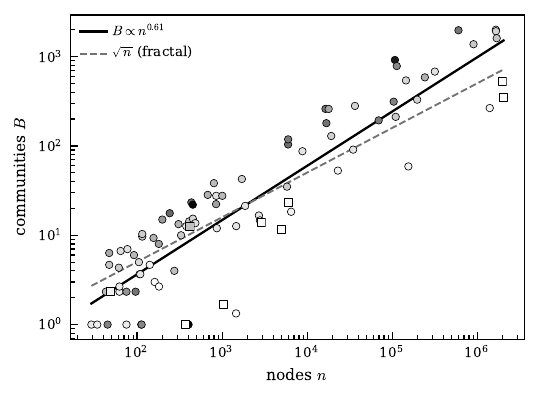}
\caption{\textbf{The number of communities grows as a power law of size.} Number
of communities $B$ against number of nodes $n$ for close to a hundred real
networks (log--log), each counted with the degree-corrected block model. The solid
line is the fitted power law $B\propto n^{0.61}$; the dashed line is the fractal
$\sqrt{n}$. Denser networks (darker fill) sit above the line, sparse ones below;
geographic, spatially embedded networks (squares) are the sparsest and fall
furthest below. The exponent lies well above the fractal one half.}
\label{fig:com-count}
\end{figure}

So measure it. Take close to a hundred undirected networks from the Netzschleuder
repository \citep{netzschleuder}, chosen only to span the widest possible range of
sizes --- from Zachary's thirty-four-member karate club to networks of more than
two million nodes --- and drawn from social, biological, infrastructural and
collaboration settings with nothing else in common. Count the communities of each
with the degree-corrected block model, letting the method infer the number of
groups from the data rather than fixing it in advance \citep{peixoto2013,karrer2011}.
The result is a clean straight line on logarithmic axes, across more than four
decades of size (Figure~\ref{fig:com-count}):
\begin{equation}
  B \propto n^{\beta}, \qquad \beta \approx 0.61,
  \label{eq:com-beta}
\end{equation}
with a bootstrap $95\%$ confidence interval, $[0.55,0.67]$, that lies wholly above
one half and excludes it. The karate club resolves into a single group, a football
schedule into ten, the Internet into several dozen, the largest social and
collaboration networks into hundreds --- and one exponent threads them all,
although they share no common origin, rule, or scale. Real networks are
\emph{super-fractal}: they carry more resolvable community structure than
self-similar models predict, yet they fall far short of the modular extreme.

The scatter about the line is not noise but signal: it tracks how dense the
network is. At a fixed number of nodes, a denser web carries more information and
supports a finer division, so dense networks (the darker points in
Figure~\ref{fig:com-count}) sit above the line and sparse ones below, a trend that
is statistically firm across the sample. The extreme cases are the geographic
networks --- roads, power grids, transport --- which are the sparsest of all and
lie furthest below; their position is explained by their low density, not by their
spatial embedding as such.

\begin{figure}[t]
\centering
\includegraphics[width=\linewidth]{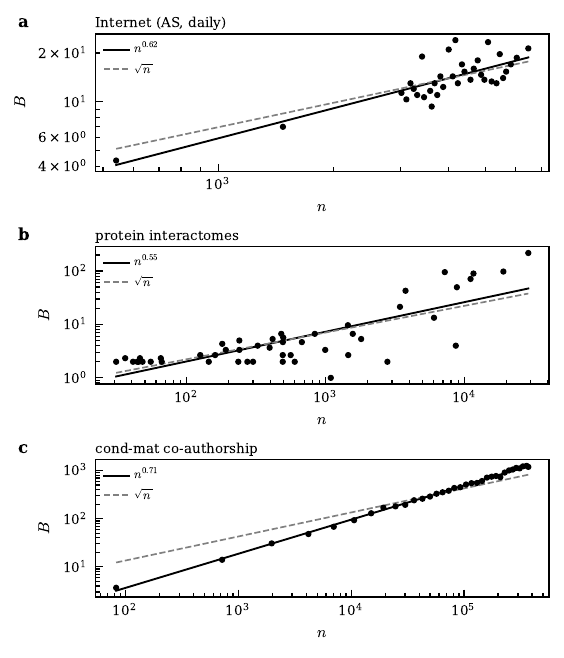}
\caption{\textbf{The power law holds within a single growing system.} Number of
communities $B$ against size $n$ (log--log) as three systems grow: (a) the Internet
at the autonomous-system level, from daily snapshots; (b) protein interactomes
across organisms; (c) the \texttt{cond-mat} co-authorship network, accumulated year
by year to nearly $400{,}000$ authors. Each is a power law (solid line, fitted
exponent) above the fractal $\sqrt{n}$ (dashed) --- the same super-fractal range as
the cross-section of Figure~\ref{fig:com-count}, but now with the same rule laying
down every point.}
\label{fig:com-count-systems}
\end{figure}

A cross-section of unrelated networks is one kind of evidence; following a
\emph{single} system as it grows, so that the same mechanism lays down every
point, is a sharper one. Do this for the three systems of the previous chapter ---
the very networks whose shortest paths we counted there --- and each again traces a
power law (Figure~\ref{fig:com-count-systems}): the Internet at the
autonomous-system level, $\beta\approx0.62$; the protein interactomes across
organisms, $\approx0.55$; and the condensed-matter co-authorship network,
accumulated year by year to nearly four hundred thousand authors, $\approx0.71$.
Grown by its own rules, each system multiplies its communities as a power law of
size, with an exponent again between one half and two thirds.

\begin{figure}[t]
\centering
\includegraphics[width=\linewidth]{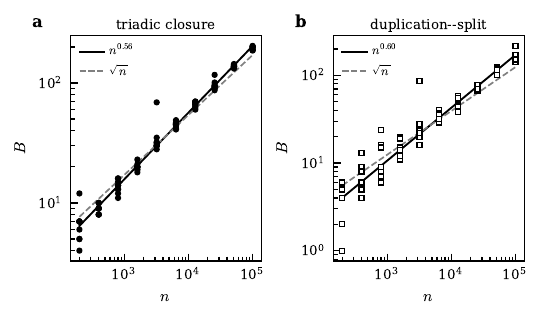}
\caption{\textbf{Local rules reproduce the super-fractal exponent.} Number of
communities against size for the two local-growth models grown to $10^5$ nodes:
(a) triadic closure and (b) duplication--split. Each follows a power law (solid
line) with an exponent above the fractal $\sqrt{n}$ (dashed), in the same
super-fractal range as the real networks of Figure~\ref{fig:com-count}.}
\label{fig:com-count-models}
\end{figure}

And the book's own local rules do the same. Grow the triadic-closure and
duplication--split models --- the social and biological rules of
Chapters~\ref{ch:friends} and~\ref{ch:duplication} --- to a hundred thousand nodes
and count their communities exactly as for the data, and both yield clean power
laws with super-fractal exponents, $\beta\approx0.56$ and $0.60$, their confidence
intervals again above one half (Figure~\ref{fig:com-count-models}). The
super-fractal exponent is thus not a peculiarity of the data but a consequence of
local growth: the same rules that make communities emerge at all, earlier in this
chapter, also fix how their number multiplies with size, and fix it in the range
the real networks occupy.

One qualification has to be entered here, and it is taken up properly in
Chapter~\ref{ch:graphons}. The counts above come from a flat block model, whose
own resolution cannot exceed $\sqrt{n}$ groups however many are present. That
ceiling has the same functional form as the fractal limit the exponent is being
compared against, so the instrument cannot report a value far above one half even
in principle. Refitting these two rules with a model whose resolution is not so
bounded raises $\beta$ from $0.56$ and $0.60$ to about one on the very same
networks, and refitting the real networks of Figure~\ref{fig:com-count} the same
way raises the exponent of Eq.~\eqref{eq:com-beta} from $0.63$ to $0.80$, with an
interval lying wholly above two thirds. What survives that is the inequality
$\beta>\tfrac12$, which only strengthens; what does not survive unqualified is
the particular value and the picture of a balance struck midway between fractal
and modular.

The reading of Eq.~\eqref{eq:com-beta} is simple and worth stating plainly. The
natural number of parts into which a network divides is not fixed but grows with
the network, as $n^{\beta}$, while the typical community swells as $n^{1-\beta}$.
With $\tfrac12<\beta<\tfrac23$ both grow without bound, yet the communities remain
a vanishing fraction of the whole: one can neither summarize an arbitrarily large
network with a fixed handful of groups, nor shatter it into ever more without
limit. The exponent sets the balance --- and that it sits reliably above one half,
across systems that share nothing else, is the finding. The number of communities
is a stable, measurable property of how a network is organized, and network growth
by local rules is what herds a system's components into that many clusters.

\section{Summary}

Community structure, long blamed on built-in differences between nodes, is
instead something that \emph{emerges} from growing a network by local rules, even
when every node is identical. Made precise through the Ramsey community number
$r_\kappa$ --- the size beyond which communities are all but certain --- the
statement is clean: the local rules of earlier chapters (search and
duplication--split) have a finite $r_\kappa$, while their
degree-preserving shuffles, and the non-local Barab\'asi--Albert model, have none.
Triangles are not required (duplication--split makes none and still segregates),
but locality is: communities come soonest for rules that close short cycles
between nearby nodes, and disappear once links may span the whole network.

Three networks simple enough to solve exactly turn all of that from a measurement
into a theory. A featureless ring of beads, with no growth and no randomness
whatever, is already split in two once each bead reaches beyond its nearest
neighbor. The evidence for the split is $\log R\simeq(m-n)\ln 2$: a
gain of $\ln 2$ for every link, against a price of $n\ln 2$ for saying which node
goes where. So a network is broken into communities precisely when it carries more
links than nodes --- which is to say, when it carries an extensive number of
independent cycles. A tree never breaks; a single cycle sits exactly on the fence,
which is why the plain cycle is not reluctant to split but exactly balanced. The
same short loops that Chapter~\ref{ch:distance} shows multiplying shortest paths
are what carve a network into communities.

The diamond lattice says why any threshold exists at all. Its block-model tallies
renormalize exactly, under a map with just two eigenvalues --- $bs$ for the bulk of
a community, $b$ for the seam between them --- whose ratio $1/s$ decides
everything: for $s\ge2$ the seam is left behind and the communities sharpen without
limit, while for $s=1$ it keeps pace and the lattice never separates, the exact
counterpart of the plain cycle. The densities flow to a fixed point where each
community is $K$ times denser than chance and the seam has vanished, so every link
contributes $\ln K$ --- the entropy of resolving which community it belongs to ---
and the evidence is \emph{extensive} while the price of positing a partition grows
only \emph{logarithmically}. A quantity growing like $m\ln K$ must overtake one
growing like $\ln m$. Emergence is not delicate; it is arithmetic, and $r_\kappa$
merely records where the crossing falls --- now a closed formula in which the
degrees, the priors, and the lattice all drop out, leaving only how fast the
communities grow and how fast their boundary does.

The pseudofractal web supplies the caution and the reassurance together. Correct
for the degree sequence --- score the split against a null that already knows every
node's degree --- and the same web's Ramsey number moves from $1095$ to $42$, while
the preferred cut flips from the hubs to the branches. The communities are real:
they not only survive the correction but double their evidence under it, and the
degree gradient turns out to \emph{delay} a degree-blind detector rather than fool
it. But $r_\kappa$ belongs to the network and the question together, never to the
network alone.

Two closing surprises from the solvable models. Asking how many communities,
rather than whether there are two, gives an answer that grows with the network:
about $\sqrt{n}$ communities of about $\sqrt{n}$ nodes, the geometric midpoint of
the network's own hierarchy, and the same $\sqrt{n}$ that governs shortest path
multiplicity on the same lattice. And the partition is not merely something a
statistician infers: written as a Hamiltonian it has a critical temperature, an
order parameter, and a cascade of first-order transitions that build the hierarchy
level by level. Communities are a state the network is in.

Held up to the real world, the count grows faster than the fractal models predict.
Across close to a hundred real networks, and within single systems followed as
they grow, the number of communities is a power law of size, $B\propto n^{\beta}$
with $\beta$ above the self-similar $\sqrt{n}$ --- measured here as between one
half and two thirds, though Chapter~\ref{ch:graphons} shows that the upper end of
that range is set by the resolving power of the method as much as by the
networks, and that lifting it pushes the exponent towards the modular $n$. The
book's own local rules reproduce whichever value the method reports, so the
inequality, at least, is one more regularity that locality accounts for.

Having already accounted for degree distributions, the clustering hierarchy, and
degree correlations, this chapter adds community structure --- both its emergence
and the power law that counts it --- to the list of global patterns that locality
alone can explain \citep{vazquez2025}.


\chapter{Nested communities and network limits}
\label{ch:graphons}

\section{A photograph of the whole}
\label{sec:gr-regimes}

The last chapter counted communities. It found that the number grows as a power
of the size, $B\propto n^{\beta}$ with $\beta\approx0.61$, and compared that
exponent with the $\sqrt{n}$ of the self-similar models. This chapter asks what
that comparison means, and the answer turns out to reach further than the
question.

\begin{figure}[t]
\centering
\includegraphics[width=\linewidth]{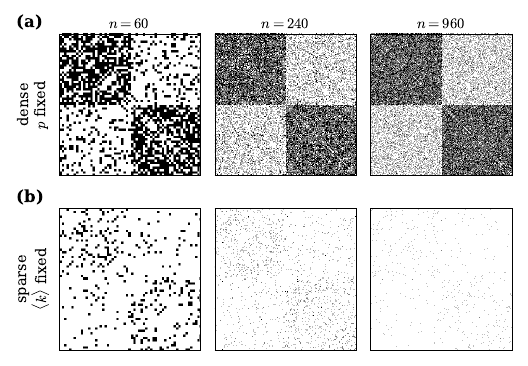}
\caption{\textbf{The limit picture, under two scalings.} Adjacency matrices of the
same two-block structure at $n=60$, $240$, $960$, nodes ordered by block.
(a)~Dense: the link probabilities are held fixed, and the picture sharpens onto a
step function --- the graphon. (b)~Sparse: the mean degree is held fixed instead,
so the probabilities fall as $1/n$. The structure is identical, and just as
detectable, but the picture fades to white. Real networks are of the second kind.}
\label{fig:gr-limit}
\end{figure}

Start with a way of looking at a network that we have not used yet. Draw its
adjacency matrix as a picture: one pixel for each pair of nodes, black if they
are linked, white if they are not. Order the nodes so that similar ones sit
together. Now grow the network and watch the picture. If the network is
\emph{dense} --- if each node links to a fixed \emph{fraction} of the others ---
the picture settles down. The individual pixels become too small to see, and what
survives is a pattern of greys: a smooth function $W(x,y)$ on the unit square
giving, for each pair of positions, the probability of a link
(Figure~\ref{fig:gr-limit}a). That limiting function is called a \emph{graphon},
and it is the object that a growing dense network converges to
\citep{lovasz2006,borgs2008}. Two networks with the same graphon are the same
network as far as any large-scale statistic can tell.

This is a powerful idea, and it has a large and beautiful theory attached to it.
It is also, applied naively to the networks in this book, useless --- and the
precise way in which it fails is instructive. The reason is the second row of
Figure~\ref{fig:gr-limit}. Real networks are not dense. The Internet, the
interactome, a co-authorship graph, all keep a roughly fixed mean degree as they
grow; so does every model in this book. When the mean degree is fixed the link
probability falls like $1/n$, the pixels thin out, and the picture converges ---
to white. The limit exists, and it has thrown away everything we care about.

The obvious repair is to turn up the contrast: divide the whole picture by its
own average darkness before taking the limit, so that what survives is the
\emph{pattern} of the greys rather than their absolute level. This works, and it
is the substance of a theory of sparse graph limits developed by
\citet{borgs2019}. Rescaling costs something. Dividing by a shrinking average
makes the values grow without bound, so the limit is no longer a probability
between zero and one; it is what analysts call an \emph{$L^p$ function}, meaning
one that may spike as high as it likes provided the average of its $p$-th power
stays finite. A spike can be tall or it can be broad, but not both. That is
exactly the licence a hub needs --- a few nodes joined to a large fraction of the
rest --- and it is why the theory has to be rebuilt around a class of functions
wider than the bounded ones. The gain is real. It is the first limit theory that copes with hubs, and with the
power-law degree distributions that come with them.

There are then three regimes, not two, and it is worth knowing which one a
network is in. Dense networks have ordinary bounded graphons. Networks that are
sparse but \emph{densifying} --- the link probability falling to zero while the
mean degree still creeps up --- have $L^p$ graphons. And networks of genuinely
fixed mean degree have neither. The rescaling that saves the middle case needs the
mean degree to grow without bound, and when it does not, the rescaled picture
concentrates instead of settling: for the ring it piles up along the diagonal,
approaching not a function but a line of infinite density.

Which regime a real system is in is a question with an answer, and the systems
followed through this book do not give the same one. Track the mean degree against
size and fit $\langle k\rangle\propto n^{\delta}$: the Internet at the
autonomous-system level, over $733$ snapshots, gives $\delta=-0.04$ --- flat, or
very slightly thinning, a mean degree that stays near four while the network grows
sixty-fold. That is the third regime, and no limit theory yet reaches it. The
cond-mat co-authorship network gives $\delta=+0.36$, and its mean degree climbs
from $2.2$ to $19$ as it grows from four hundred to four hundred thousand authors.
That is the middle regime: co-authorship densifies, and it is the one system here
that the $L^p$ theory actually covers.

The interactomes appear to densify too, more gently, but the appearance should not
be trusted: that series runs across \emph{organisms} rather than across time, and
a better-studied organism has both more proteins catalogued and more interactions
recorded per protein. The trend there measures the attention of biologists at
least as much as the architecture of cells.

Even the middle regime is tighter than it looks, in a way the book's own models
can measure. Since the exponent $p$ says how tall a spike the class will
tolerate, membership in $L^p$ is a condition on how extreme the hubs are allowed
to be: for a degree distribution $p_k\sim k^{-\gamma}$ it holds when
$p<\gamma-1$, while the theory itself needs $p>1$, so there is a usable window
only for $\gamma>2$. The
pseudofractal web of the last chapter has $\gamma\simeq2.585$, leaving the window
$1<p<1.585$; real networks, whose exponents cluster between two and three, are all
squeezed into the same narrow slot, and it closes entirely as $\gamma\to2$. The
limit theory reaches the scale-free world, but only just.

Being in the third regime is worth stating plainly, because it explains the shape
of the previous chapter. The evidence for a partition there grew like $m\ln K$, in
proportion to the number of \emph{links}; the graphon limit normalizes by the
number of \emph{pairs}, which is $n^2$. The two scales differ by the mean degree, and in
the sparse world that ratio is the whole story. It is why the last chapter had to
compute at finite $n$ and could not simply take a limit. The finite-size
calculation was not a convenience. It was the only instrument that sees anything.

\section{The block model is a graphon}
\label{sec:gr-blockmodel}

Set the difficulty aside for a moment, because the language is useful even where
the limit theorem is not.

A stochastic block model \emph{is} a graphon --- the simplest interesting kind. If
the nodes fall into blocks and the link probability depends only on which blocks
the endpoints are in, then $W$ is a step function: constant on each rectangle,
jumping at the block boundaries. Fitting a block model to a network is fitting a
step function to a picture, and asking how many blocks it needs is asking how
finely the picture must be pixelated before the description stops improving.

The degree-corrected model of the last chapter is a graphon too, and its shape
says exactly what degree correction \emph{does}.

\begin{calculation}{degree correction, as a factorization}
Let the graphon factorize as
\begin{equation}
  W(x,y) = f(x)\,f(y)\;w\big(g(x),g(y)\big),
  \label{eq:gr-factor}
\end{equation}
where $g(x)$ names the block at position $x$, $w$ is constant on block pairs, and
$f$ is an arbitrary non-negative function. The first two factors form a
\emph{rank-one} graphon: a product $f(x)f(y)$, in which a node's propensity to
link is a single number and nothing about \emph{whom} it links to. That is the
configuration model, and $f$ is the degree profile. Comparing with the discrete
version of the last chapter, $\theta_i=k_i/\sqrt{2m}$ is the empirical estimate of
$f$, and $\omega_{rs}$ is the block-constant residual $w$. Setting $w\equiv1$
recovers the pure degree model, which is exactly the statement made there that
$\omega_{rs}=1$ reproduces the configuration model.
\end{calculation}

So the two block models of the last chapter differ in one respect only: the
baseline they measure structure against. The plain model tests the network
against a \emph{constant} graphon; the degree-corrected model tests it against the
best \emph{rank-one} graphon. Degree correction is the removal of the rank-one
part, nothing more and nothing less.

Seen this way the pseudofractal reversal loses its strangeness. A degree-blind
model has no rank-one factor to spend, so the only way it can express a steep
degree gradient is through the block structure itself --- and it duly spends its
blocks on separating hubs from leaves, which is the hub--leaf cut it was found
preferring. Give it a rank-one factor and the gradient is absorbed there, leaving
the blocks free to describe what actually remains: the branch geometry. The
correction does not weaken the case for the branch cut because the branch cut was
never what the degrees were explaining.

The evidence itself has a graphon reading as well. The large-count form of the
degree-corrected evidence --- the sum of $e_{rs}\ln(e_{rs}/\Omega_{rs})$ over
block pairs, met in the last chapter --- is $m$ times a relative entropy between
the fitted residual $w$ and the flat one --- the same functional that governs large deviations for
random graphs \citep{chatterjee2011}. This sharpens a claim made in passing there.
The result $\log R_{\mathrm{dc}}/m\to\ln K$ was said to owe nothing to the diamond
lattice; now we can say why. Relative entropy is a property of the limiting shape,
not of the construction that produced it, and any arrangement of $K$ equal
well-separated blocks has the same one.

\section{How much can be seen}

Now to the question the last chapter raised without quite asking: what sets the
number of blocks?

Fitting a step function to a picture is a balancing act familiar from any
histogram. Too few bins and the description is crude --- real variation is
averaged away. Too many and each bin holds too little data to estimate, so the
description is noise. The optimum sits where the two costs meet, and it moves
with the amount of data. This is the classical bias--variance trade-off, and for
networks it has been worked out precisely.

\begin{calculation}{the cost of a $k$-block description}
Three terms compete. Smoothness here is measured by the \emph{H\"older
exponent}, which grades how violently a surface is allowed to change when one
moves a short distance across it: a graphon has smoothness $\alpha$ if, for some
constant $M$,
\begin{equation}
  |W(x,y)-W(x',y')| \;\le\; M\big(|x-x'|^{\alpha}+|y-y'|^{\alpha}\big),
  \label{eq:gr-holder}
\end{equation}
so that $\alpha=1$ is the Lipschitz case --- a surface of bounded slope --- and
smaller $\alpha$ lets it swing more violently over a short distance. The
\emph{approximation} error of describing such a surface by $k$ steps then follows
in a line: cover the square with $k\times k$ cells of side $1/k$ and fit a constant
on each; two points in a cell are at most $1/k$ apart, so
Eq.~\eqref{eq:gr-holder} caps the deviation inside it at $Mk^{-\alpha}$, and since
the risk is a mean \emph{squared} error, squaring gives $k^{-2\alpha}$. Against it
run two estimation costs, established by
\citet{gao2015}: a term $k^2/n^2$ for fitting the $k^2$ block parameters
themselves, and a term $n^{-1}\ln k$ for the harder problem of not knowing which
node belongs to which block. The total is
\begin{equation}
  \mathrm{risk}(k) \;\asymp\; k^{-2\alpha} \;+\; \frac{\ln k}{n} \;+\; \frac{k^2}{n^2}.
  \label{eq:gr-risk}
\end{equation}
The two estimation terms both \emph{grow} with $k$, so on their own they would
push towards a single block; it is the approximation error, falling with $k$, that
creates an interior optimum. Balancing $k^{-2\alpha}$ against $k^2/n^2$ gives
\begin{equation}
  k^* \;\propto\; n^{\,1/(1+\alpha)} .
  \label{eq:gr-kstar}
\end{equation}
\end{calculation}

\begin{figure}[t]
\centering
\includegraphics[width=\linewidth]{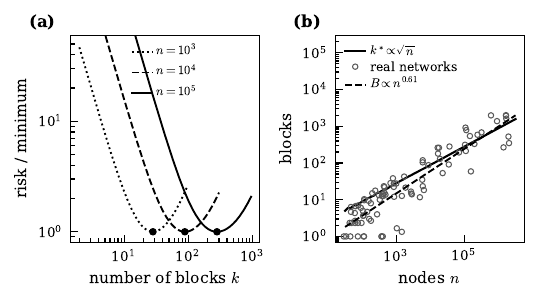}
\caption{\textbf{How many blocks can be seen.} (a)~The risk of a $k$-block
description, Eq.~\eqref{eq:gr-risk}, for three network sizes, each normalized to
its own minimum (dots). The optimum moves right as $\sqrt{n}$. (b)~That optimum
against size, with the community counts of the real networks of the last chapter
and their fitted power law. The vertical placement of the $k^*$ line depends on
the graphon's smoothness constant and carries no meaning; its slope of one half
does.}
\label{fig:gr-resolution}
\end{figure}

Equation~\eqref{eq:gr-kstar} is the key. For a Lipschitz graphon, $\alpha=1$ and
the optimal description uses
\begin{equation}
  k^* \propto \sqrt{n}
  \label{eq:gr-sqrtn}
\end{equation}
blocks --- which is precisely the $\sqrt{n}$ that \citet{olhede2014} propose for
the network histogram, and which Figure~\ref{fig:gr-resolution}a recovers by
direct minimization of Eq.~\eqref{eq:gr-risk}.

And here the chapter's opening question returns, transformed. The $\sqrt{n}$ that
the last chapter used as its fractal yardstick --- the self-similar prediction of
$\sqrt{n}$ communities of $\sqrt{n}$ nodes --- turns out to be the same $\sqrt{n}$
that statistics arrives at from an entirely different direction, as the resolution
at which a smooth shape is best described. The local rule models and the statisticians'
histograms agree on a number for reasons that have nothing to do with each other.

\section{Rough, not smooth}

Take the agreement seriously and the measured exponent stops being a curiosity and
becomes a measurement of something.

The real networks gave $\beta\approx0.61$, with a bootstrap interval
$[0.55,0.67]$ that excludes one half. Read through Eq.~\eqref{eq:gr-kstar}, an
exponent is no longer just a number above or below a fractal benchmark: it is a
statement about the smoothness of the shape being described. Inverting
$\beta=1/(1+\alpha)$,
\begin{equation}
  \alpha \;=\; \frac{1}{\beta}-1 \;\approx\; 0.65,
  \qquad \alpha\in[0.49,\,0.82],
  \label{eq:gr-alpha}
\end{equation}
and the interval lies wholly below $\alpha=1$. Direct minimization of
Eq.~\eqref{eq:gr-risk} at $\alpha=0.65$ returns an exponent of $0.611$, against
the measured $0.608$.

So the excess over $\sqrt{n}$ is not merely ``more communities than a fractal.''
It says that if real networks are converging to any limiting shape at all, that
shape is \emph{rough}: not Lipschitz, not smooth in the sense the estimation
theory assumes, but H\"older-continuous with an exponent around two thirds. A
smooth graphon --- any smooth graphon, of any shape, at any scale, since the
smoothness constant sets only the height of the line in
Figure~\ref{fig:gr-resolution}b and not its slope --- supports a description that
refines as $\sqrt{n}$ and no faster. Real networks refine faster than that. They
keep resolving new structure at every scale, at a rate a smooth limit cannot
supply.

This is the same conclusion the last chapter reached in its own language, and the
same one this book has been reaching since the introduction, arrived at from the
outside. A network built by a local rule has no smooth limiting shape to converge
to, because the rule keeps writing structure at whatever scale the network has
currently reached.

It is worth knowing that this is not a quarrel with the limit theory but an answer
to a question its own authors left open. Introducing a network generator built from
a singular measure rather than a function, \citet{palla2010} --- Lov\'asz among
them --- put the alternatives as plainly as they can be put:

\begin{quote}\small
Should we expect that large real graphs converge to some limiting network in a
strict sense of the convergence? Or, alternatively, their structure cannot be
mapped onto a fixed function, and only an ever-changing (with the size of the
network) measure (in the infinite network size limit becoming singular) can be used
to reflect the underlying structural complexity?
\end{quote}

\noindent That was posed in 2010 and left as a question. Equation~\eqref{eq:gr-alpha}
is a measurement bearing on it, taken across a hundred networks, and it comes down
on the second side: no fixed smooth function will do, because the description keeps
refining faster than any such function allows. The rest of this chapter is about how
far that answer can be trusted.

\section{Five meanings of one number}
\label{sec:gr-five}

A caution belongs here, and it is a sharp one. Five different arguments in this
book and its neighbouring literature all deliver $\sqrt{n}$, and they are not the
same statement:

\begin{itemize}
\item the \emph{fractal} $\sqrt{n}$ of the self-similar models --- a prediction
  about how many communities the networks themselves hold;
\item the \emph{geometric} $\sqrt{n}$ of Chapter~\ref{ch:distance} --- the
  multiplicity of shortest paths on the same lattice, a prediction about how many
  ways there are to cross it;
\item the \emph{constructive} $\sqrt{n}$ --- the resolution any cascade of equal
  cells is forced to, once its mean degree is held fixed;
\item the \emph{statistical} $\sqrt{n}$ of Eq.~\eqref{eq:gr-sqrtn} --- the best
  resolution obtainable for a smooth shape, a statement about descriptions;
\item the \emph{algorithmic} $\sqrt{n}$ --- the resolution limit of the
  non-nested block model, which cannot report more than of order $\sqrt{n}$
  groups no matter how many are present \citep{peixoto2014hier}, and which is
  derived, along with the ceiling that replaces it, in
  Section~\ref{sec:gr-twopriors}.
\end{itemize}

\noindent The last of these has an older and better-known cousin. Long before
anyone examined the block model's ceiling, \citet{fortunato2007} showed that
modularity --- then the standard objective for community detection --- cannot
resolve communities smaller than about $\sqrt{m}$ links in a network of $m$
links, and so merges small ones no matter how clearly separated they are. The
mechanism is different: modularity's limit comes from comparing against a global
null model, the block model's from the cost of transmitting a matrix. But the
lesson is the same, and it is one this chapter will keep returning to. A number
of communities reported by a method is a statement about the method as much as
about the network, and the only way to tell the two apart is to change the
method and watch what moves.

The third of these is the one just met, and it is worth making explicit, because it
shows how little the number has to assume. Build a network from a singular measure
in the manner of \citet{palla2010}: divide the square into $m^2$ cells, give each a
probability, and multiply the whole pattern into itself $k$ times, so that the
picture has $m^k$ rows and grows sharper with every iteration. If the cells are all
the same size, the mean degree of the resulting network is $N m^{-2k}$, so holding
it fixed while refining the measure forces
\begin{equation}
  m^{k} \;=\; \sqrt{N/\langle k\rangle} \;\propto\; \sqrt{N} .
  \label{eq:gr-cascade}
\end{equation}
The construction resolves $\sqrt{N}$ rows because sparsity leaves it no choice. No
network, no estimator and no smoothness assumption enters that argument --- only
the bookkeeping of keeping a self-similar object sparse.

Three of the five now have reasons, and they are the same reason. The fractal and
geometric counts are a lattice's blocks and its routes tallied at the halfway point
of its own construction, and the constructive count is what any equal-celled cascade
must do to stay sparse; all three are self-similarity keeping its books. That is why
$\beta=1/2$ is the natural exponent for a self-similar object, and why measuring
$\beta>1/2$ says the object is not one --- the multifractal case, where the cells
are allowed to differ in size, has no such constraint and its exponent is free.
The remaining two agree with them for no reason at all.

The algorithmic $\sqrt{n}$ is a property of neither the network nor the ideal
description but of the instrument, and it deserves attention because the counts of
the last chapter were produced by exactly that instrument: a flat degree-corrected
block model selected by description length. Its ceiling and the null hypothesis under test
have the same functional form, which is an uncomfortable coincidence.

The bias runs in the reassuring direction --- a ceiling can only suppress the
counts of the largest networks and so pull the fitted exponent \emph{down},
meaning $\beta>1/2$ is measured in spite of the instrument rather than because of
it. But the conclusion that survives is correspondingly narrower: these data
establish an exponent above one half over four decades of size, and cannot by
themselves certify that it persists asymptotically, because the estimator that
produced them is not able to report an asymptotic exponent above one half. Only
the nested block model, which lifts the ceiling to order $n/\ln n$, can settle
that, and re-running the census with it is the obvious next measurement. Should
the exponent survive, Eq.~\eqref{eq:gr-alpha} converts it directly into a
statement about the roughness of the limit.

\section{What the ring cannot tell us}

One more thing the limit view clarifies, and it is a caveat about a result rather
than a strengthening of one.

The ring of the last chapter was found to prefer being split in two, for every
range $c\ge2$. But every node of the ring is identical to every other, and the
model was explicit that there is no preferred place to cut: rotate the proposed
partition by one bead and the evidence is unchanged. In the language of limits
that is a symmetry, and a continuous one. The ring's local limit --- what an
observer sitting on a node sees, out to any fixed distance
\citep{benjamini2001} --- is an infinite line, the same in every direction, with
no communities anywhere in it.

So the ring exhibits, in the cleanest possible form, a distinction the previous
chapter's machinery does not itself draw. Evidence for splitting is not the same
as the existence of a partition to be found. The ring supplies the first without
the second: a whole continuum of cuts, all equally good, none of them recoverable
from the data, and a limit object with no structure at all. What the evidence
ratio reports is that one block is a poor description --- a true and interesting
fact --- and not that two particular blocks are the right one. For the diamond
lattice and the pseudofractal web, whose partitions are pinned by their
construction, no such degeneracy arises and the identification is sound. It is the
homogeneous case that needs the care.

The ring is also the only one of the three entitled to that local limit, and the
reason is worth recording, because it marks out exactly where this book's networks
fall. A local limit of the kind just used requires an absolute ceiling on the
degree: every node must look like every other from close up, out to a bounded
number of neighbours. The ring obliges --- every node has degree $2c$ and always
will. The diamond lattice and the pseudofractal web do not: their poles grow
without bound, reaching degree $2^t$, so no ceiling exists and the local theory
does not apply to them. Nor does the theory at the other end, which needs the
number of links to outgrow the number of nodes, while theirs stays proportional
to it.

That is a genuine gap, and the book's central examples sit in it: too heavy-tailed
for the local theory, too sparse for the global one. It is also a gap the subject
knows about and is working to close --- a recent line of work \citep{backhausz2018}
proposes treating a network not as a picture at all but as an \emph{operator}, and
recovers dense graphons and local limits as two special cases of one notion of
convergence. Whether it reaches the case that matters here --- bounded mean degree
with unbounded hubs --- is precisely the question to put to it.

\section{What clustering rules out}
\label{sec:gr-clustering}

Everything so far has been about \emph{shape} --- whether a limiting picture
exists and how finely it can be resolved. There is a blunter test, and the
introduction ran it before any of this machinery was in place.

A graphon model, of any kind, draws its links independently once the positions of
the nodes are fixed. That is what makes it tractable, and it has a consequence
that cannot be argued away: a triangle needs three links, each drawn separately,
so triangles are as rare as three coincidences. In a sparse network the clustering
coefficient of such a model collapses to the link density itself, a number of
order $\langle k\rangle/n$ --- vanishing. Real networks are not like that, and the
introduction already measured by how much. Its column $C_{\mathrm{ER}}$, computed
there as the clustering of a random network of the same size and density, is
precisely the graphon prediction:

\begin{center}
\begin{tabular}{lccc}
\hline
Network & $C$ & $C_{\mathrm{ER}}$ & $C/C_{\mathrm{ER}}$ \\
\hline
Internet (AS)          & $0.230$  & $0.00018$ & $1254$ \\
Gnutella               & $0.0055$ & $0.000076$ & $72$ \\
Yeast proteins         & $0.071$  & $0.0018$  & $39$ \\
cond-mat co-authorship & $0.657$  & $0.00026$ & $2541$ \\
\hline
\end{tabular}
\end{center}

Between forty and two and a half thousand times too many triangles. No adjustment
of the shape $W$ repairs a discrepancy of that size, because the shape is not what
is wrong: the independence is.

But the argument has to be made carefully, and the careful version is more
interesting than the crude one. The comparison above is against a \emph{constant}
graphon, and we have already seen in
Section~\ref{sec:gr-blockmodel} that the constant is the weaker of the two
baselines. Against the \emph{rank-one} graphon --- the network reshuffled to
preserve every degree --- the verdict divides. The Internet's clustering falls only
from $0.230$ to $0.123$ under that reshuffling, a factor of $1.9$: its degrees
alone, hubs and all, manufacture most of its triangles, because a node of enormous
degree is shared by many pairs and closes their triangles by accident. The
co-authorship network, whose degrees are far less extreme, still stands $356$ times
above its own reshuffling.

So the two baselines separate the systems exactly as they separated the
pseudofractal's two cuts. Where the degrees are wild, degree heterogeneity accounts
for much of the clustering and little is left to explain. Where they are tame, the
whole of it survives the strongest degree-preserving null available --- and that
residue is what Chapter~\ref{ch:friends} was written to explain. Triadic closure is
not an embellishment on a graphon; it is the thing the graphon cannot do.

\section{Locality, or exchangeability}

There is a reason all of this bears on the book's thesis rather than merely
decorating it, and it is worth ending on.

A graphon is not only a limit; it is a complete description of a certain
\emph{kind} of random network --- those whose nodes are \emph{exchangeable},
meaning the labels carry no information and any two nodes are, before we look,
the same. Shuffle the names of the nodes and the probability of getting that
network back is unchanged; nothing about a node except its hidden coordinate
$x$ can matter \citep{aldous1981,diaconis2008}. Every graphon model is exchangeable and every
exchangeable model is a graphon. Exchangeability is a strong assumption, and what
it excludes is precisely geometry: a node's position, its neighbourhood, its
history. It excludes locality.

The chapters before this one have argued that networks are built by nodes acting
on what is near them --- the page a surfer reached, the friend who made the
introduction, the gene that was copied. A rule of that kind produces a network in
which position matters, in which where a node attached is written into the
structure around it. Such a network is not exchangeable, and so it has no graphon
in the ordinary sense. The failure of the limit theory to say anything useful
about real networks, which began this chapter as an inconvenience, is at bottom
the same fact this book has been documenting all along, in a different vocabulary.
Sparsity and roughness and the absence of a smooth limit are what locality looks
like when a statistician goes looking for a shape.

What the sparse limits should be instead is an open and active question, and the
theory of sparse exchangeable graphs \citep{caron2017} is one serious attempt at
it. But it would be wrong to end by filing all of this under work not yet begun,
because on one point the previous chapter has already done a piece of it without
saying so.

The companion to the $L^p$ theory \citep{borgs2018} asks what it should
\emph{mean} for a sparse network to converge, and proves that several answers that
look quite different are the same answer. One can ask that the rescaled picture
settle down. One can ask that the \emph{quotients} settle --- the tables of
block-to-block densities produced by chopping the network into parts. One can ask
that the best partition at fixed block sizes settle, or that the sum over all such
partitions, weighted, settle: a ground state energy and a free energy, in the
language of physics. These are all equivalent.

That list is the last chapter's apparatus, item by item. A quotient is exactly the
table of counts $e_{rs}$ against expectations $\Omega_{rs}$. Fixed block sizes are
exactly the candidate split. The weighted sum over labelings, with the label prior
integrated out, is exactly the evidence. And the renormalization flow
$(u_{\mathrm{in}},u_\times)\to(K,0)$ is a quotient converging --- proved by hand,
for one lattice, in precisely the sense the theorem intends. That is why the
resulting $\ln K$ per link is so indifferent to the lattice that produced it:
convergence of quotients is a statement about the limiting shape, and any
arrangement with the same limiting quotient must give the same free energy.

The hypotheses of that theorem do not cover a bounded-degree lattice, so what the
last chapter proved is not a corollary of it; the flow had to be computed. But the
agreement of the two routes is worth more than either alone, and it marks out what
a limit theory for locally grown networks would have to look like. There is also,
on the other side of Eq.~\eqref{eq:gr-alpha}, a number measured across a hundred
real networks, still waiting to be explained. The rest of this chapter explains
it, and the explanation turns out to be less about the networks than about the
two instruments used to count them.

\section{Two priors for the same partition}
\label{sec:gr-twopriors}

The last chapter counted communities with a block model, and this one has spent
its length on what such a model can and cannot represent. It is time to ask a
narrower question, and one that turns out to matter more than it looks: given
that a block model is being fitted at all, \emph{which} block model, and does the
choice change the answer?

The choice at issue is between the flat model of Chapter~\ref{ch:communities} and
the nested one --- both in the microcanonical, degree-corrected form developed by
\citet{peixoto2014mcmc,peixoto2019bayesian} on the degree-corrected model of
\citet{karrer2011}, and both fitted here with the same software --- and it is
worth being precise about the difference, because it is
not a difference in what the two can \emph{express}. Both describe a network as
$B$ groups together with the matrix $e_{rs}$ of link counts running between them,
both infer $B$ from the data rather than being told it, and any partition one can
write down the other can write down too. What differs is the price of naming it.

The flat model transmits the matrix under a uniform prior, as though its
$B(B+1)/2$ entries were arbitrary numbers with no relation to one another, which
costs about $\tfrac12B^2\log_2 m$ bits for a network of $m$ links. That charge
grows with the square of the number of groups while the saving from splitting a
group grows more slowly, and where the two meet is a ceiling: no flat fit
recovers more than about $\sqrt n$ groups, however many are present
\citep{peixoto2014hier}.

The nested model spends less on the very same object, by declining to treat it as
arbitrary. The counts $e_{rs}$ are themselves a network --- one node per group,
one weighted link per pair --- and it is described by a block model of its own,
which is described by another, and so on up. A group-level structure that is
sparse or modular, as it usually is, then costs a fraction of what spelling out
the matrix would; the quadratic charge disappears, and the ceiling rises to around
$n/\log n$: some $8700$ groups at $n=10^5$ against $316$.

\begin{calculation}{the price of a block structure, and the ceiling it sets}
Both models describe a network by a partition $b$ and the matrix $e_{rs}$ of link
counts, and score it by the description length
$\Sigma=-\log_2P(A\mid e,b)-\log_2P(e,b)$. The first term is the likelihood, which
they share. The second is the price of the description, which they do not.

The flat model takes $e_{rs}$ to be an arbitrary symmetric matrix, so it must
transmit $B(B+1)/2$ independent entries, each of size up to $m$. That costs about
$\tfrac12B^{2}\log_2 m$ bits. Splitting one group into two saves at most of order
$m/B$ bits of likelihood, so the two terms balance at
\begin{equation}
  B_{\max} \;\propto\; \sqrt{m} \;\propto\; \sqrt{n}
  \label{eq:gr-flatceiling}
\end{equation}
for a network of fixed mean degree \citep{peixoto2014hier} --- the
\emph{algorithmic} $\sqrt n$ of Section~\ref{sec:gr-five}, and the only one of the
five that is a fact about a method rather than about a network. Past that point every
further split costs more to announce than it saves, whatever the network looks
like. The ceiling is a property of the code, not of the data.

The nested model refuses the premise that $e_{rs}$ is arbitrary. The counts are
themselves a network --- one node per group, one weighted link per pair --- so
describe them with a block model, whose own count matrix is described by another,
and so on until one group remains. Each level costs a fraction of what spelling
out its matrix would, provided the group-level structure is itself sparse or
modular, which it generally is. The quadratic charge never appears, and the
ceiling rises to
\begin{equation}
  B_{\max} \;\propto\; n/\log n,
  \label{eq:gr-nestedceiling}
\end{equation}
some $8700$ groups at $n=10^{5}$ against $316$. Setting the hierarchy to a single
level recovers the flat model exactly, so the comparison between them is strictly
nested and an extra level can only pay for itself or not.

The same arithmetic answers a question the next chapter has to face, so it is
worth doing here. Suppose one wanted a block model over $j$-sets of nodes rather
than pairs --- a natural thought if the object generating the network acts on
triples or quadruples. Its count tensor has about $B^{j}/j!$ entries, and
repeating the balance above gives
\begin{equation}
  B_{\max} \;\approx\; \bigl[m\,(j-1)!\,/\ln m\bigr]^{1/j},
  \label{eq:gr-bmax}
\end{equation}
which is $\sqrt n$ at $j=2$ and $n^{1/4}$ at $j=4$ once $m$ is of order $n$:
eighteen groups against three hundred and sixteen at $n=10^{5}$. Raising the arity
costs resolution, and costs it fast. A model that can no longer see communities is
a poor instrument for asking what communities are made of, which is why
Chapter~\ref{ch:motifs} reaches for a richer \emph{vocabulary} inside the ordinary
pair model rather than for a tensor over $j$-sets.
\end{calculation}

\begin{figure}[t]
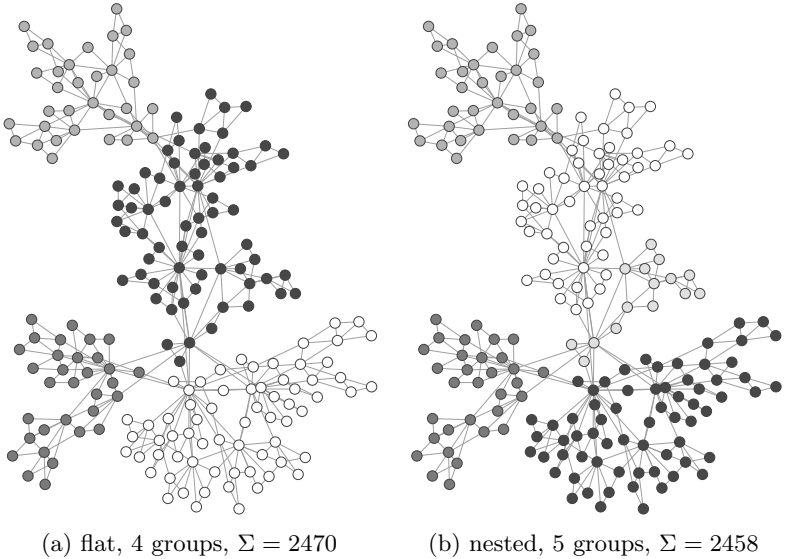

\centering
\begin{minipage}{0.48\linewidth}\centering
\begin{adjustbox}{max width=\linewidth}\input{netfig-fn-flat.tex}\end{adjustbox}\\[2pt]
{\small (a) flat, $4$ groups, $\Sigma=2470$}
\end{minipage}\hfill
\begin{minipage}{0.48\linewidth}\centering
\begin{adjustbox}{max width=\linewidth}\input{netfig-fn-nested.tex}\end{adjustbox}\\[2pt]
{\small (b) nested, $5$ groups, $\Sigma=2458$}
\end{minipage}
\caption{\textbf{The same network, divided by the two priors.} An instance of
$LS(n{=}200,\ell{=}1)$, laid out once and coloured twice. The flat model resolves
four groups; the nested model splits the larger of them and resolves five, and it
does so while spending twelve bits \emph{fewer} on the description as a whole ---
which is the sense in which the finer division is not merely permitted but
preferred. At $n=100$ the two return the same three groups: the ceiling has to be
approached before it can be felt.}
\label{fig:gr-flatnested}
\end{figure}

So the two differ in a prior and not in a likelihood. Given the same partition and
the same counts they assign the network the same probability; they disagree only
about how surprising the counts were. Figure~\ref{fig:gr-flatnested} shows what
that disagreement looks like on a single network.

\section{Which prior the data prefers}
\label{sec:gr-whichprior}

A difference of prior invites an obvious objection. If the two models disagree
only in what they expected before seeing anything, on what grounds is either to
be believed?

The answer is that a prior is a code, and a code can be measured. The whole point
of the description length is that it needs no external arbiter: the better prior
is the one that assigns a shorter description to the data that actually occurred.
Both codes are complete, both are fixed before the answer is known, and comparing
their lengths is a test rather than a preference.

The comparison is also structurally safe, because the nested model with a single
level \emph{is} the flat model. The hierarchy is a strict generalisation, so it
can only help, up to the small cost of announcing how many levels there are. One
is not choosing between rival beliefs but asking whether an extra layer of
description pays for itself.

On real networks it pays, and handsomely. Fitting both models to fourteen of the
networks of Chapter~\ref{ch:communities}, and reading off the description length
each achieves:

\begin{center}
\small
\begin{tabular}{lrrrr}
\hline
network & $n$ & $\Sigma$ flat & $\Sigma$ nested & difference\\
\hline
netscience              & $379$    & $5\,775$     & $5\,583$     & $-192$\\
kegg\_metabolic         & $1\,865$ & $40\,142$    & $39\,709$    & $-433$\\
urban\_streets          & $2\,870$ & $43\,262$    & $42\,793$    & $-470$\\
celegans\_interactome   & $5\,966$ & $619\,605$   & $606\,016$   & $-13\,589$\\
ppi\_interolog\_human   & $16\,820$& $3\,041\,650$& $3\,037\,343$& $-4\,307$\\
internet\_as            & $22\,963$& $404\,104$   & $399\,172$   & $-4\,932$\\
\hline
\end{tabular}
\end{center}

\noindent Across the fourteen the nested model is shorter on thirteen, by $1772$
bits on average. The one exception is a tie, on a network the model puts in a
single group, where there is no group-level graph to compress and the hierarchy
has nothing to do.

The flat model is therefore not wrong. It is a code that pays for arbitrariness it
does not need, and the data says so, in bits. Its $\sqrt n$ ceiling is a property
of that code and not of networks --- which matters, because $\sqrt n$ is exactly
the fractal limit the last chapter's exponent was being compared against.

\section{How many communities?}
\label{sec:gr-howmanyagain}

Chapter~\ref{ch:communities} measured $B\propto n^{\beta}$ with
$\beta\approx0.61$ and called real networks \emph{super-fractal}: above the
self-similar $\tfrac12$, below the modular $1$. Everything above says that number
should be re-examined with the other prior, and it does not survive intact.

\begin{figure}[t]
  \centering
  \includegraphics[width=\linewidth]{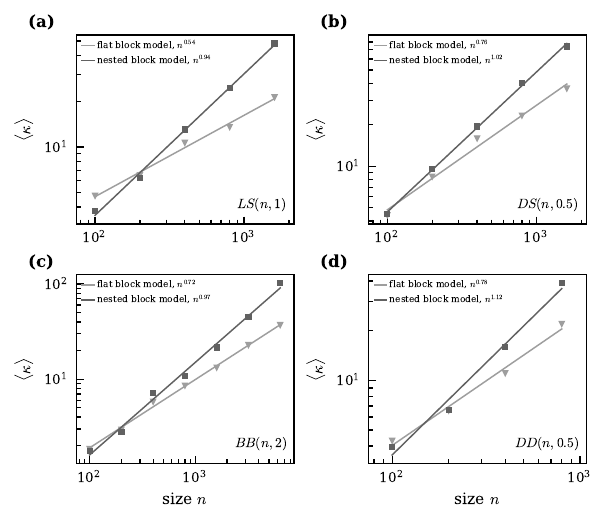}
  \caption{\textbf{How many communities, under each prior.} The mean number of
  communities against size for four growth rules, averaged over twenty networks
  at each size, with the best power law fitted to each method and its exponent
  in the legend. The flat model comes out near its own $\sqrt n$ ceiling
  throughout; the nested model grows almost in proportion to $n$.}
  \label{fig:gr-scaling}
\end{figure}

Take the book's own growth rules first, where the networks can be made to order
and grown as far as patience allows. Fitting both models to the same realisations
gives the table this chapter and the next both work from
(Figure~\ref{fig:gr-scaling}):

\begin{center}
\small
\begin{tabular}{lrrrrr}
\hline
 & & \multicolumn{2}{c}{$\beta$, growth of $B$} & & \\
rule & $n$ & flat & nested & $b$ & $\langle n_B\rangle$\\
\hline
$LS(n,1)$ & $100$--$1600$ & $0.54$ & $0.94$ & $0.032\pm0.003$ & $31$\\
$BB(n,1)$ & $400$--$3200$ & $0.48$ & $0.89$ & $0.026\pm0.003$ & $38$\\
$DS(n,1/3)$ & $100$--$1600$ & $0.78$ & $0.92$ & $0.017\pm0.002$ & $57$\\
$DS(n,0.5)$ & $100$--$1600$ & $0.76$ & $1.02$ & $0.048\pm0.002$ & $21$\\
$DD(n,0.5)$ & $100$--$800$ & $0.78$ & $1.12$ & $0.040\pm0.006$ & $25$\\
$BB(n,2)$ & $100$--$6400$ & $0.72$ & $0.97$ & $0.015\pm0.002$ & $66$\\
\hline
pseudofractal & $366$--$9843$ & $0.61$ & $1.02$ & $0.015\pm0.002$ & $65$\\
diamond & $684$--$10924$ & $0.71$ & $1.05$ & $0.009\pm0.002$ & $113$\\
\hline
\end{tabular}
\end{center}

\noindent The first six rows are the random rules of
Chapters~\ref{ch:friends}--\ref{ch:duplication}, the last two the deterministic
lattices, whose smallest generations are dropped because a lattice the fit puts
in one or two groups is below the resolution of either method and would
contribute a slope fitted to nothing. The last two columns are read off the
nested fit and are the subject of Section~\ref{sec:gr-density}; the exponents
come first.

\noindent The flat model reproduces Chapter~\ref{ch:communities} --- it was that
chapter's method --- and it reproduces it while sitting at its own ceiling. Its
column is remarkably uniform in doing so: eight rules of quite different
character, deterministic lattices among them, and every one of them reads between
$0.48$ and $0.78$. A ceiling produces exactly that kind of agreement, and a
network property has no reason to.

Lift the ceiling and every rule moves to the modular limit: the exponents cluster
on one, from $0.89$ to $1.12$, the number of communities grows in proportion to
the size, and their typical size stops growing. At $n=1600$ local search divides
into groups of about thirty-four nodes and duplication--split into groups of
about twenty-one, and those figures barely change as the networks are grown
further.

The same question on the real networks gives a smaller shift in the same
direction. Refitting close to ninety of them, from twenty-nine nodes to three
hundred thousand, both ways:

\begin{center}
\begin{tabular}{lrl}
\hline
 & $\beta$ & $95\%$ interval\\
\hline
flat, as Chapter~\ref{ch:communities} published it & $0.633$ & $[0.57,\,0.69]$\\
flat, refitted here                                & $0.630$ & $[0.57,\,0.69]$\\
nested                                             & $\mathbf{0.804}$ & $[0.72,\,0.88]$\\
\hline
\end{tabular}
\end{center}

\noindent The refit reproduces the published column to three decimals, which is
the check that this is the same measurement and not a different one. The nested
interval lies wholly above two thirds.

What survives of Chapter~\ref{ch:communities}, then, is the inequality. Every
method puts $\beta$ above one half, and lifting the resolution only raises it
further, so real networks are certainly not self-similar in their community
structure. What does not survive is the particular value, and with it the picture
of a balance struck midway between two limits. On the growth rules the answer is
the modular limit outright. On the real networks the census gives $0.80$ rather
than $0.61$ --- still short of $1$, but no longer comfortably inside the range the
last chapter named. The exponent was, in part, a property of the instrument.

That $0.80$ will not survive Section~\ref{sec:gr-within} either, and for a reason
that has nothing to do with priors. But the resolution question has to be settled
before the measurement question can be asked, so it is settled here first.

\section{The density of communities}
\label{sec:gr-density}

An exponent of one is a special number, and it is worth pausing on what it would
mean if it were exact. If $B$ grows in proportion to $n$, then the ratio
\begin{equation}
  b \;=\; B/n, \qquad \langle n_B\rangle \;=\; 1/b \;=\; n/B,
  \label{eq:gr-density}
\end{equation}
stops depending on size. Call $b$ the \emph{density of communities} --- how many
of them there are per node --- and $\langle n_B\rangle$ its reciprocal, the mean
number of nodes per community. Neither is then a rate of growth any longer; each
is a \emph{quantity}, a number a network of any size simply has, and one may
tabulate it the way one tabulates a clustering coefficient or a degree exponent.
On the growth rules this is very nearly what happens, and it is the last two
columns of the table above.

The exponents straddle one, and the density is flat --- to within a few per cent
for local search and duplication--split across sixteen-fold in size, for the
bubble rule across sixty-four-fold, and for the two lattices across a range of
sixteen and twenty-seven. The rules are not saying the same thing: their
densities differ by a factor of five, and the typical community runs from
twenty-one nodes under duplication--split to a hundred and thirteen on the
diamond lattice. Those numbers are as much a signature of the rule as its degree
exponent is.
Nothing in the rules sets a community size directly --- there is no parameter with
units of nodes anywhere in them --- and yet each settles on one and holds it while
the network grows by orders of magnitude around it.

That is a genuine finding, and it changes the reading of the whole exercise. It is
tempting to describe it as the limit these networks converge to, and it is nearly
the opposite. A graphon is a picture in which the group structure coarsens: as
$n$ grows the number of groups stays put and each group swells in proportion, so
that the labels shrink onto a fixed kernel on the unit square. Constant density is
that picture refused. Communities of twenty or a hundred nodes never swell, and no
matter how far the network is grown, the resolution needed to see its structure
keeps pace with the network itself. This is a second obstruction, distinct from
the sparsity of Section~\ref{sec:gr-regimes}: a sparse network could still have a
fixed community structure, and these do not. What the growth rules converge to
under a nested description is not a shape one could draw on the square. It is a
density.

On real networks the picture looked, at first, close but not the same.

\begin{center}
\small
\begin{tabular}{lrrrrr}
\hline
networks & no. & $n$ & $\beta$ & $\langle n_B\rangle$ & $b$ slope\\
\hline
social & $47$ & $34$--$317080$ & $0.80$ & $27$ & $-0.20$\\
biological & $15$ & $47$--$16820$ & $0.78$ & $38$ & $-0.22$\\
infrastructure & $10$ & $29$--$34761$ & $0.73$ & $165$ & $-0.27$\\
information & $9$ & $112$--$145145$ & $0.95$ & $27$ & $-0.05$\\
\hline
\emph{all} & $81$ & $29$--$317080$ & $0.80$ & $31$ & $-0.20$\\
\hline
\end{tabular}
\end{center}

\noindent The community sizes are of the same order the rules produce --- around
thirty nodes, with the infrastructure networks the exception at a hundred and
sixty. But the exponent came out below one almost everywhere, and the last column
says what that costs: $b$ falling as roughly $n^{-0.2}$, $\langle n_B\rangle$ growing
about fourfold across the four decades the census spans. Taken at face value, that
is a shortfall the growth rules would have to explain.

It is not to be taken at face value, and seeing why is worth more than the number
was.

\section{What a census exponent measures}
\label{sec:gr-within}

The two exponents being compared are not measurements of the same thing. The
rules' $\beta$ comes from a \emph{growth series}: one mechanism, one object,
followed through a sequence of sizes. The census $\beta$ comes from a
\emph{cross-section}: eighty-one different networks, built by different processes,
recorded by different projects, with size as the only thing ordering them. The
first says how the number of communities changes as a network grows. The second
says how it varies between systems that happen to differ in size. Nothing
guarantees these are the same number, and a census that mixes kinds of network
will report the second while looking like the first.

The census behaves the way a cross-section behaves. Restrict it to the sizes the
rules were actually run at, $n$ between $100$ and $1600$, and it returns
$\beta=0.58$; take all of it and it returns $0.80$. A genuine scaling law does not
move by two tenths depending on which of its instances one includes. Two further
symptoms point the same way. Networks sitting above the fitted line are markedly
denser than those below --- median mean degree $13.4$ against $5.8$, with the
residual correlating with mean degree at Spearman $\rho=+0.43$ --- and the
spatially embedded networks sit systematically below it. Neither of those moves
the exponent much when removed, which is the point: the scatter the line is drawn
through is not error but structure, so the interval quoted around $0.80$ is
narrower than the honest uncertainty.

The way to settle it is to measure the thing the rules measure. A few real systems
were recorded repeatedly \emph{while they grew}, and can be fitted along their own
histories rather than against each other: the Internet at the autonomous-system
level, from daily route-collector snapshots; the co-authorship network of the
history of science, whose links carry the year they were made and can be
accumulated; and the physical interactome of yeast, accumulated by the year each
interaction was published.

\begin{center}
\footnotesize\setlength{\tabcolsep}{3pt}
\begin{tabular}{llrrr}
\hline
system & $n$ & pts & $\beta$ & $\langle n_B\rangle$\\
\hline
Internet, autonomous sys. & $2086$--$6474$ & $38$ & $1.20\,[0.94,1.55]$ & $124$\\
co-authorship, hist. of sci. & $2886$--$182964$ & $18$ & $1.01\,[0.96,1.06]$ & $23$\\
yeast interactome, BioGRID & $47$--$6976$ & $13$ & $0.90\,[0.84,1.26]$ & $48$\\
\hline
\emph{census}, all kinds & $29$--$317\,080$ & $81$ & $0.80\,[0.72,0.88]$ & $31$\\
\hline
\end{tabular}
\end{center}

\noindent Every one of them exceeds the census figure, and the two that are
genuinely growing networks --- the Internet and the collaboration graph --- come
out at one, with the co-authorship interval excluding $0.80$ outright. Their
community sizes are the ones the rules produce: twenty-three nodes for
co-authorship, against twenty-one for duplication--split.

So the shortfall was never a property of real networks. It was a property of
averaging over kinds of them. The density of communities is constant, in the
systems where the question has been asked properly, and the growth rules were
right about it.

The third row is the exception that shows the mechanism. Yeast's interactome is
not a growing network at all --- the cell's protein interactions are not being
created as we watch, they are being \emph{found} --- so its series measures how
communities scale as a fixed object is discovered. It is the lowest of the three
and the only one whose central value falls short of one. That is the same confound
this chapter already flagged for the tree-of-life series, where the trend runs
across organisms and a better-studied organism has both more proteins catalogued
and more interactions recorded per protein. Here it runs across time instead, and
it can be quantified: incomplete observation of a network does depress the
apparent number of communities, and a census that mixes catalogued networks with
grown ones will inherit some of that depression.

Three cautions, so the table is not asked to carry more than it can. The
Internet series covers only a threefold range in size, which is why its interval
is so much wider than the others; co-authorship, at twenty-onefold, is the one
that carries the weight. Three route-collector days were dropped because the
collector was partly down; they are snapshots of less of the Internet, not of a
smaller one, and keeping them lowers the exponent to $0.96$, which changes nothing
about the conclusion. Four co-authorship snapshots were dropped for a reason worth
naming, since it is the chapter's own lesson turned on itself: they had been fitted
with fewer restarts of the minimiser than the rest, to save time, and a single
restart lands in a worse optimum more often on a large network than on a small one.
That puts a downward trend into the slope which belongs to the schedule and not to
the data. The tell was that the $2005$ snapshot came back with fewer communities
than the $1995$ one, which strictly contains it --- a cumulative series cannot lose
groups as it grows. And three systems are three, not a survey: what has been
shown is that the census exponent is not the growth exponent, and that where the
growth exponent can be measured it is one, not that every network in the world
obeys it.

The general lesson outlasts the number. An exponent fitted across a heterogeneous
collection of networks is a statement about the collection. It answers a question
about how systems differ, and it will be read --- it was read, in
Chapter~\ref{ch:communities} and in the first half of this section --- as a
statement about how a network grows. The two coincide only if the collection is
homogeneous in everything except size, which no census of real networks is.

One caution belongs with any use of these numbers. A density of communities is
not a property of a network alone; it is a property of a network together with the
code used to describe it, and this chapter has spent its length showing how much
that matters. Under the flat prior the same census gives $\beta=0.63$. The tables
above mean what they say only because the model is fixed --- nested,
degree-corrected, minimum description length --- and a density quoted without
naming the model is not a measurement. That is a familiar situation rather than a
disqualifying one. The clustering coefficient has two standard definitions that
disagree, and the introduction had to say which one it meant before it could say
anything about hierarchy.

One thing the change of prior does \emph{not} touch is the Ramsey community
number. Fitted both ways, local search crosses the ninety-five per cent threshold
at $70$ against $80$, duplication--split at $90$ against $100$,
duplication--divergence at $40$ either way. Resolution decides how finely a
network can be cut; it says almost nothing about whether there is anything to cut
at all. That second question depends on something else entirely, and is taken up
in the next chapter.

\section{The limit we found}
\label{sec:gr-conclusion}

This chapter set out to ask what the limit theories can say about networks grown
by local rules, and the answer arrived from an unexpected direction.

The expected answer was a shape. That is what a limit theory offers: let the
network grow, let the pixels shrink, and read off the function on the unit square
that survives. The chapter's first half is a catalogue of why that does not
happen here. Real networks and the local rule models sit in the third regime, of fixed
mean degree, where the rescaling that saves the sparse case has nothing to
rescale by. Their adjacency pictures are rough rather than smooth, H\"older with
an exponent near two thirds, so the histogram theorems that promise $\sqrt n$
cells are promising more than can be delivered. And their clustering does not
vanish, which any bounded graphon says it must.

There was a natural guess about what to reach for instead, and the shape of the
guess was: more arguments. A local rule does not act on pairs. Triadic closure
acts on \emph{triples} --- a node, its neighbour, and that neighbour's neighbour
--- and duplication acts on a node, its copy, and their shared neighbours, a
configuration of four. If the statistics a rule generates are to be carried by a
limiting object at all, the object may need arguments enough to hold the rule's
own footprint: a kernel $[0,1]^3\to[0,1]$ where the rule closes triangles,
$[0,1]^4\to[0,1]$ where it duplicates, with the fitted block model as the
two-variable shadow of it. Chapter~\ref{ch:motifs} takes that guess apart, and
finds it half right: a two-variable object provably loses the triangles, but a
higher-arity kernel indexed by \emph{nodes} does not obviously get them back, and
the natural object turns out to be indexed by pairs instead.

Meanwhile this chapter found something the guess does not mention, by asking a
narrower question and following it further than intended. Fit the nested model
instead of the flat one and the number of communities grows in proportion to the
size. The density $b=B/n$ of Eq.~\eqref{eq:gr-density} stops depending on $n$; it
becomes a number a network has, the way it has a clustering coefficient. The growth rules hold it flat over one to
two decades and disagree about its value by a factor of five, twenty-one nodes
per community under duplication--split against a hundred and thirteen on the
diamond lattice. On
the real systems where the question can be asked properly --- those recorded
repeatedly while they grew --- it is flat too, and the collaboration graph settles
at twenty-three nodes per community, next to duplication--split's twenty-one.

It is tempting to go one step further and say that the limit is a \emph{nested}
structure: communities of communities, all the way up, with the same constant
governing every level. The fitted hierarchies appear to say so. Read off the level
sizes of a nested fit and the branching ratio is large at the bottom --- that is
the community size --- and then settles at two to four for every level above it,
on every rule, which looks like self-similarity.

It is not, and the check that shows it is the one this chapter has been preaching.
Fit the same model to a network built with no hierarchy above its bottom level: a
planted partition of sixty-four equal groups and nothing else. The model recovers
the sixty-four exactly, and then builds a four-four-four tree on top of them. The
upper branching is the prior's own scaffolding. (On a network with no communities
at all the model correctly returns a single level, so it is not inventing
structure from nothing --- it is organising real groups into a tree that the data
never asked for.) The bottom level is a measurement. The levels above it are a
description device, and no claim about a self-similar hierarchy survives them.

What survives is not one number but a small set of possibilities, and that is the
finding.

\begin{quote}
\emph{Conjecture.} For large $n$ a locally grown network does not converge to a
kernel on the unit square --- not of two arguments, and not of any fixed number.
It converges \emph{locally}: the neighbourhood of a typical node settles down,
and with it a community of bounded size. What the limit is characterised by is
intensive, chief among these quantities the density of communities $b=B/n$,
equivalently the mean community size $\langle n_B\rangle$, constant in $n$ and set
by the growth rule.
\end{quote}

\noindent The word \emph{locally} is doing the work the word \emph{almost} used
to do, and it points at a theory that already exists: local weak convergence
\citep{benjamini2001}, in which one keeps a bounded neighbourhood around a
randomly chosen root and asks what it looks like, rather than coarsening the
whole picture onto a fixed shape.
That is the natural home for a network whose communities never grow.

And it explains why a graphon was never going to work, in a way that is worth
stating in one sentence. A graphon is the picture in which group structure
\emph{coarsens}: as $n$ grows the groups stay fixed in number and swell in
proportion, until the labels shrink onto a kernel one could draw. Constant density
refuses that twice over. The communities never swell, so there is nothing to shrink
onto; and the resolution needed to see the structure keeps pace with the network
forever, so no finite description ever suffices. This is an obstruction distinct
from sparsity --- a sparse network could perfectly well have a fixed community
structure, and these do not.

Whether some higher-order exchangeable object could hold a bounded community
around a typical node is exactly what remains open, and nothing here settles it.
What is settled is that the two-variable picture cannot, and that looking for the
answer as a picture at all may have been the wrong instinct.

A constant density is not, however, the only thing a locally grown network can
have, and the conjecture is better read as naming one of three outcomes than as
naming the only one. There are three, and the next chapter measures which rule
falls into which:

\begin{itemize}
\item \emph{No communities.} The fit returns a single group at any size, and
  $r_\kappa$ is infinite. Chapter~\ref{ch:communities}'s non-local control
  $BA(n,2)$ is the standing example, and locality is what the other rules have
  that it does not.
\item \emph{Constant density.} $\beta=1$, and $b=B/n$ is a number the rule has.
  This is where every rule in the table above sits, and it is what the conjecture
  describes.
\item \emph{Fractal.} Communities emerge and their number grows, but more slowly
  than the network: $0<\beta<1$, so the density falls away to nothing. It is a
  coherent possibility and no rule in this book has been found in it.
\end{itemize}

\noindent The first class is the one that turns out to matter, and a qualification
belongs with the conjecture rather than after it, because the next chapter finds
the counterexample and it is better met head-on. Every density quoted above was
measured with a block model, whose vocabulary cannot name a triangle.
Chapter~\ref{ch:motifs} fits the same networks with one that can, and on the
diamond lattice --- a rule as local as any in this book --- the communities do not
merely thin out, they go: the best available description uses a single group at
every size, and it is the description minimum description length prefers. The
pseudofractal web goes the same way. So a network can have a density of
communities under one description and none at all under a better one.

That is not a defeat for the conjecture so much as a statement of what it is
conjecturing about. A limit class, like an exponent, is a property of a network
together with the code used to describe it; this chapter has said as much twice
already, once for the prior and once for the sample. What the lattices add is that
improving the code can move a network between classes and not merely shift its
exponent --- and, since the improvement in their case was a decomposition written
down by hand rather than found by fitting, that ``the best available
description'' is a moving target. The conjecture should therefore be read as a claim about the best
available description, and it stands or falls on whether the networks that matter
keep their density when the description improves --- which is measured, for what
it can reach, in Section~\ref{sec:gr-decides}. Chapter~\ref{ch:bubbles} then
finds that the answer turns on whether a rule's motifs overlap, and that a
deterministic rule, whose motifs do not, is also the case where the fitting is
least to be trusted.

Two things are worth carrying forward. The first is a warning about instruments,
learned twice in this chapter and at some cost. The exponent $\beta$ moved from
$0.63$ to $0.80$ when the prior changed, and from $0.80$ to $1$ when the sample
stopped mixing kinds of network; neither move had anything to do with the
networks. A number measured through a method is a statement about the pair, and
the only way to tell which half is speaking is to change the method and watch.
The second is the density itself, which the next chapter will try to move by
changing the vocabulary rather than the resolution. On most of the rules it will
fail, because their motifs pile onto each other and a decomposition that gives
each link to one of them can take up only half the network. On the three whose
motifs lie separately --- exactly so on the two lattices, nearly so on
$BB(n,2)$ --- it will succeed completely, taking the communities away rather than
merely thinning them, and those three are the reason the classes have to be named
at all.

\chapter{Communities with network motifs}
\label{ch:motifs}

The last chapter asked how finely a network can be divided, and answered it with
two priors over the same object. This one asks a different question: what the
divider is allowed to \emph{know} about the thing it is dividing.

Start with cycles, because they are where a network's local structure shows
itself. Every network has cycles of every length, and their numbers are not
remotely comparable: the yeast interactome holds a couple of hundred triangles
and ten thousand cycles of length eight, and a duplication--divergence network of
two hundred nodes holds none of length three, five or seven and seven million of
length eight. The counts climb steeply with length, and the steepness is not a
free parameter --- it is set by the short cycles.

That is the fact this chapter is built on, and it is worth stating at the outset.
The number of $h$-cycles through a node of degree $k$ is fixed by the degree
distribution together with one three-point quantity, the clustering coefficient,
and the length enters only as an exponent: the same closing probability applied
$h-2$ times around the loop (Section~\ref{sec:gr-conjecture}). Long cycles are not an
independent feature of a network. They are what short cycles compound into. A
description that has the triangles right has the whole spectrum right, wherever
triangles exist at all.

Wherever they do not, the compounding starts from something else, and the
question of \emph{which} something else turns out to be answerable by
measurement. A rule that closes squares rather than triangles leaves a spectrum
with a different shape, and one that closes pentagons a different shape again.

All of which bears on communities in a way that is not obvious until it is
stated. A block model divides a network on the evidence of where its links fall.
Clustering is evidence: a knot of mutually connected nodes looks like a
community. But if the network was grown by a rule that closes triangles
everywhere, the knots are not communities at all, and a method that could
\emph{name} a triangle would have an innocent explanation available where a
method without one has only the partition. So the question of this chapter is
whether giving a community model a vocabulary of motifs changes what it reports.
The answer, worked out over the sections that follow, is that it usually changes
when communities appear and not how many there are --- but that on two rules out
of eight it does something larger, and moves the network into a different limit
class altogether.

The concern is not new, and neither is the vocabulary. Recurring small subgraphs
were named \emph{network motifs} by \citet{milo2002}, who found them at
frequencies no random graph with the same degrees would produce; random graph
ensembles built to contain prescribed subgraphs rather than prescribed links were
constructed by \citet{karrer2010subgraphs}, and the model of
Section~\ref{sec:gr-motifsbm} is a block-structured version of theirs. Looking for
communities with motifs rather than with links is older still:
\citet{arenas2008motif} rewrote Newman--Girvan modularity so that what it counts
inside a group is copies of a chosen motif rather than edges, and showed that the
partition one gets depends on which motif is chosen --- the same observation this
chapter arrives at from the side of description length, and reached there by
changing an objective rather than a code. Closest of
all, \citet{peixoto2022triadic} put exactly the suspicion above to work: a block
model in which each link is generated either by community membership or by
triadic closure over an existing path, with the assignment inferred alongside the
partition. On real networks it removes a good deal of what a plain block model
had been calling community structure. What follows differs in two ways. The
alphabet is not fixed to the triangle --- squares and bicliques are in it too,
because Chapter~\ref{ch:duplication}'s rules do not close triangles --- and the
target is not any one network but the \emph{scaling}, how the number of
communities grows with size and whether that exponent is a property of the
network or of the vocabulary used to describe it.

\section{Why a few motifs might be enough}
\label{sec:gr-conjecture}

Chapter~\ref{ch:graphons} ended by conjecturing that a locally grown network
converges to no kernel on the unit square at all, of two arguments or of any
number, and that what characterises its limit is instead intensive: a density of
communities, constant in size and set by the rule. It reached that conclusion by
abandoning an earlier guess --- that the limit is a kernel of \emph{higher
arity}, three arguments where the rule closes triangles and four where it
duplicates, with the fitted block model as the two-variable shadow of it.

This section takes the abandoned guess seriously for as long as it is useful,
which is not all the way. Its purpose is not to settle what the limit object is;
that question is left where Chapter~\ref{ch:graphons} left it. Its purpose is to
license an instrument. If a two-variable description can be shown to lose
something with a definite shape, and the shape can be shown to be fixed by the
shortest cycle the rule closes, then the way to build a better community model is
not to raise the arity of everything --- which is unaffordable, as we shall see
--- but to hand the ordinary pair model a small vocabulary of short motifs. That
is the conjecture this section exists to state. Everything after it is a test of
the instrument, not of the conjecture.

\subsection*{What a two-variable fit loses}

Take the two lattices whose construction is known exactly, so that every quantity
can be computed rather than estimated.

\begin{calculation}{the higher-order structure a two-variable model can hold}
The pseudofractal web of \citet{dorogovtsevgoltsev2002} has $n=(3^{t+1}+3)/2$
nodes and $3^{t+1}$ links, and every node added closes exactly one triangle, so
it holds exactly $n-2$ of them. Fit Chapter~\ref{ch:communities}'s branch cut and
ask each fitted model how many it expects, via $\mathrm{tr}(P^3)/6$ for its
expected adjacency matrix $P$.

The plain block model assigns one density per block pair. In a sparse network
those densities fall like $1/n$, and a triple of blocks contributes
$n^3\times n^{-3}$: the expected count is $O(1)$, and numerically it
\emph{saturates} at $21.3$ triangles however large the web grows.

The degree-corrected model does better, and still fails. Its expected count
weights each block by the degree sum $Q_r=\sum_{i\in r}k_i(k_i-1)$ rather than by
its size --- $k(k-1)$ and not $k^2$, because a triangle leaves a node by a
different link than it entered by, and the fit holds the degrees fixed --- and
for this web $Q\sim4^{t}$ while the degree sum is $S\sim3^{t}$, so the count
grows as $(Q/S)^3\sim(64/27)^t$ against a truth growing as $3^t$. The deficit
diverges, as $(81/64)^{t}=n^{0.214}$. Allowing the number of blocks to grow as
the data demand, $B\propto n^{\beta}$, helps but does not close it: $B$ blocks of
$n/B$ nodes contribute $O(1)$ triangles each, hence $n^{\beta}$ in all, still
short of $\Theta(n)$ by $n^{1-\beta}$.

The diamond lattice moves the arity by one and makes the failure worse. It is
bipartite and holds no triangle at all; its rule replaces each link with two
parallel paths, so its motif is the four-cycle, and after $t$ generations it has
$m=4^t$ links, $n=(2\cdot4^t+4)/3$ nodes and exactly $4^{t-1}$ four-cycles ---
only those closed at the \emph{last} generation are still four-cycles, earlier
ones having had their links subdivided into longer loops. Evaluating
$\mathrm{tr}(P^4)/8$ on the same cut, the plain model again saturates, at
$81/4=20.25$ four-cycles for a lattice of any size. The degree-corrected model is
where the two lattices part company: its count is governed by the ratio of the
excess-degree sum to the degree sum, and for this lattice that ratio is not a
power of $n$ at all but
\begin{equation}
  Q/S \;=\; \sum_i k_i(k_i-1) \Big/ \sum_i k_i \;=\; t
  \label{eq:gr-QS}
\end{equation}
exactly, so the expected four-cycle count converges to $t^4/4$:
\emph{poly\-log\-arithmic} in $n$, against a truth growing as $n/6$. At $t=26$ the
lattice holds $1.1\times10^{15}$ four-cycles and the fitted degree-corrected
model expects $1.2\times10^{5}$ of them.
\end{calculation}

So a two-variable description preserves what the \emph{cut metric} sees --- the
yardstick the limit theory uses, which judges two networks close when no way of
splitting the nodes into two sets reveals a difference in the density of links
across the split, and so registers block densities, degrees, the evidence for a
partition, everything Chapter~\ref{ch:graphons} computed --- and loses the
higher-order structure entirely, by a margin that widens without bound.

The gap between the two lattices --- $n^{0.214}$ against $n/(\ln n)^4$ --- is the
one thing they genuinely differ in, and it is worth naming because it recurs. The
pseudofractal's degrees are wild, $\gamma\simeq2.585$, and a rank-one model with
hubs that extreme manufactures a great many triangles by accident: three links
that all happen to touch the same few enormous nodes close a triangle whether or
not any rule intended it. The diamond's degrees are milder, $\gamma=3$ exactly,
which is the marginal case where $\sum k_i^2$ outgrows $\sum k_i$ only
logarithmically --- and a rank-one model then has nothing to fake with. The
two-variable model's apparent grasp of higher-order structure is not a grasp at
all; it is borrowed from degree heterogeneity, and it fails as the degrees become
tame. That is the mechanism behind something already measured in
Section~\ref{sec:gr-clustering}, where the degree-preserving null came within a
factor of $1.9$ of the Internet's clustering, whose degrees are extreme, and fell
$356$ times short for co-authorship, whose degrees are not. Degree-corrected
models do not capture clustering. They capture it exactly where the degrees are
heavy enough to counterfeit it.

One warning for anyone testing this at accessible sizes: the failure is not
visible early, and it changes sign. At $n\simeq3000$ the degree-corrected model
\emph{over}counts triangles fivefold, and only past $n\sim10^8$ does it cross
over to undercounting. A study stopping at ten thousand nodes would conclude the
opposite of the truth.

\subsection*{What is lost has a shape}

Triangles and four-cycles are two points. Asking the question at every cycle
length turns the finding from a shortfall into something with more content ---
and it is the content that motivates the alphabet.

\begin{calculation}{why a two-variable model has a one-parameter spectrum}
For a model with independent links and expected adjacency $P$, the expected number
of $L$-cycles is $\mathrm{tr}(P^L)/2L$. When $P$ is a $K$-block fit this collapses
onto a $K\times K$ \emph{transfer matrix}
\begin{equation}
  T_{rs} \;=\; \omega_{rs}\sqrt{Q_rQ_s}
  \qquad\text{(degree-corrected)},
  \label{eq:gr-transfer}
\end{equation}
with $Q_r=\sum_{i\in r}k_i(k_i-1)$, and $T_{rs}=p_{rs}\sqrt{n_rn_s}$ for the plain
model, so that the count at every length is $\mathrm{tr}(T^L)/2L$. The entire motif
spectrum is generated by one small matrix, and for large $L$ it is
$\lambda^L/2L$ for the leading eigenvalue $\lambda$ --- a smooth geometric profile.
A two-variable model has no freedom to put a feature at one particular length,
whatever its blocks are doing.

The weight is $k(k-1)$ rather than $k^2$ because a cycle leaves a node by a
different link than the one it arrived on, and the degree-corrected fit holds the
degree sequence fixed: a node offers $k(k-1)$ ordered pairs of links, not $k^2$.
The one-block case is then $\langle k^2-k\rangle/\langle k\rangle$, which is the
leading eigenvalue of the \emph{non-backtracking} operator, which counts walks
forbidden to turn round and retrace the link they arrived on --- the natural
object for cycles, since a cycle never does. The distinction is invisible in the
limits taken above, where the ratio diverges and the $-1$ is lost, but it is
worth a factor of $25$ at $L=8$ on a real network, where the ratio is nearer $3$.
\end{calculation}

The lattices do nothing but put features at particular lengths. The diamond is
built from four-cycles and carries cycles of length $4$, $8$, $16$, and so on ---
and \emph{exactly none} at any other length. Counted directly at $t=5$: zero
$3$-cycles, $256$ four-cycles, zero at five, six and seven, $1024$ eight-cycles.
The fitted models, meanwhile, expect $8.3$ triangles, $42$ five-cycles and $104$
six-cycles under the plain fit, and $40$, $668$ and $2994$ under the
degree-corrected one, at lengths where the lattice contains nothing whatever.

So the failure is not that the numbers are too small. It is that a two-variable
model places mass at every length while a locally grown lattice places it at a
few, and no adjustment of blocks or degrees can move a smooth profile onto a
sparse set of lengths. What a richer description buys is not a bigger number but
the ability to have a shape --- and a \emph{shape} supported on a few short
lengths is exactly what a small alphabet of motifs is.

That the shape is generated by one eigenvalue also fixes how the two lattices
differ. For a degree distribution $p(k)\sim k^{-\gamma}$,
\begin{equation}
  \lambda \;\sim\; n^{(3-\gamma)/(\gamma-1)} \quad (\gamma<3), \qquad
  \lambda \sim \ln n \quad (\gamma=3),
  \label{eq:gr-lambda}
\end{equation}
which gives the pseudofractal's exponent $0.2619$ and the diamond's $Q/S=t$. One
quantity governs both earlier results: the deficit at length $L$ is of order
$n/\lambda^{L}$, undercounting while $\lambda^L<n$ and overcounting after. The
rank-one part does not describe higher-order structure at any length. It
manufactures a geometric series and happens to cross the truth once.

\begin{figure}[t]
  \centering
  \includegraphics[width=\linewidth]{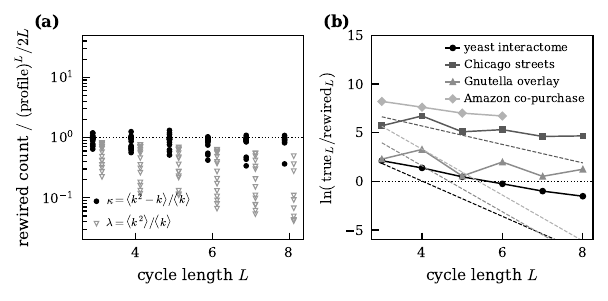}
  \caption{The two halves of the prediction, on real networks. \textbf{(a)} A
  degree-preserving rewiring's own cycle counts, divided by the geometric profile
  its transfer matrix predicts. With the excess-degree branching factor
  $\langle k^2-k\rangle/\langle k\rangle$ the ratio sits on $1$ at every length;
  with $\langle k^2\rangle/\langle k\rangle$ it drifts away by an order of
  magnitude. \textbf{(b)} The deficit $\ln(\text{true}_L/\text{rewired}_L)$ for
  four systems, against the line of slope $-\ln\lambda$ through
  $L^\ast=\ln n/\ln\lambda$ that Eq.~\eqref{eq:gr-lambda} predicts (dashed). The
  measured curves are large and nearly flat.}
  \label{fig:gr-crossover}
\end{figure}

That crossing is a falsifiable prediction, and it was tested on $44$ real
networks, with cycles counted exactly at $L=3\ldots8$ and each fitted model
represented by a sample from it rather than by the formula. Half of it came back
confirmed, and it is the half the argument stands on: a fitted model's own motif
spectrum really is generated by its transfer matrix, the sampled counts sitting
on $\lambda^L/2L$ to within one per cent at every length
(Figure~\ref{fig:gr-crossover}a). Eq.~\eqref{eq:gr-transfer} is not an
approximation that degrades. The crossing itself does not survive: real networks
carry many more cycles than their rewirings at \emph{every} length alike --- five
times more at the median, four thousand times for a large co-purchase network ---
and the excess barely decays, the measured slope being about a tenth of
$-\ln\lambda$, with thirty-nine of forty-two networks departing from the
predicted slope by more than two standard errors
(Figure~\ref{fig:gr-crossover}b). The deficit is a level, not a slope. The
lattices' motif counts are extensive and flat in $L$, so dividing by a geometric
profile leaves a geometric deficit and a crossing; real networks' cycle counts
grow with $L$ at very nearly the rate their rewirings do. The crossing is a
property of the deterministic webs, and the $L^\ast$ formula should be read as a
statement about them.

None of that touches the support argument, which the same measurement instead
strengthened by turning up a real network with the diamond's pathology.
Fullerenes are molecules, and their cycle spectra are exactly $12$ five-cycles
and $n/2-10$ six-cycles and \emph{nothing at any other length} --- twelve
pentagons at every size, which is Euler's theorem. Both fitted models put
positive mass everywhere: for $\mathrm{C}_{6000}$, five triangles, forty-three
four-cycles, twenty-seven seven-cycles. There the deficit is not fivefold or four
thousandfold, it is infinite in both directions, and no choice of blocks or
degrees can repair it.

\subsection*{How short a vocabulary has to be}

The shape is supported on a few lengths. How few, and which, is not something to
be postulated, because the cycle spectrum of a real network is already known to
be generated by a local quantity of arity three. Write $p_k$ for the degree
distribution and $c_k$ for the clustering coefficient of a node of degree $k$.
Then the number of $h$-cycles through a node of that degree
is~\citep{vazquez2005cycles}
\begin{equation}
  \frac{N_h}{n} \;=\; g_h\sum_{k}p_k\,\frac{(h-1)!}{2}\binom{k}{h-1}\,c_k^{\,h-2},
  \label{eq:gr-cycgen}
\end{equation}
with $g_h$ a correction for multiple counting: choose $h-1$ neighbours of a
central node, order them, and close the cycle by requiring $h-2$ links among
consecutive ones, each present with probability $c_k$. The degree distribution
and one three-point quantity fix the count at \emph{every} length, and the
measurement holds at $h=3,4,5$ across co-authorship, the Internet and a semantic
network~\citep{vazquez2005cycles}.

This is the licence the alphabet needs. The tower of longer cycles requires no
mechanism of its own: it is the product $c_k^{\,h-2}$, the same closing
probability applied $h-2$ times around the loop. Arity three buys not just the
triangle but the whole spectrum built from triangles, and for a great many real
networks triangles do the job.

They cannot do it for a network with no triangles. There $c_k=0$ identically,
Eq.~\eqref{eq:gr-cycgen} vanishes at every $h\ge3$, and the failure is not an
underestimate but an empty prediction --- against a network full of cycles. The
book's own duplication--split rule is exactly that case, built so that
duplication makes no triangles and splitting destroys any that might form.
Counting $DS(n,1/3)$ directly at $n=2000$: no triangles at all, then
$1.1\times10^4$ squares, $7.9\times10^3$ five-cycles and $1.3\times10^7$
eight-cycles. Nothing about the network is short of cycles; the arity-three
generator simply cannot see any of them.

So the order required is set by the shortest cycle the rule closes, and it is
measurable rather than postulated. The bubble family makes it a dial, since
$BB(n,L)$ closes a fresh cycle of length $L+2$ at every step:

\begin{center}
\begin{tabular}{llll}
\hline
model & motif & $c$ & cycle lengths present, $L=3\ldots10$\\
\hline
$BB(n,1)$ & triangle & $0.74$ & every length\\
$BB(n,2)$ & square & $0$ & $4,6,8,10$\\
$BB(n,3)$ & pentagon & $0$ & $5,8$\\
$BB(n,4)$ & hexagon & $0$ & $6,10$\\
\hline
\end{tabular}
\end{center}

\noindent Only the first is within reach of Eq.~\eqref{eq:gr-cycgen}. The others
place their mass exactly where the motif and its gluings put it: two faces
sharing a link close a cycle of length $a+b-2$, so pentagons give $5$ and $8$,
hexagons $6$ and $10$, and everything else is empty. That is the same arithmetic
the fullerenes obey --- twelve pentagons and $n/2-10$ hexagons at any size, then
$60$ nine-cycles and $3n/2-60$ ten-cycles, with seven and eight empty because no
two pentagons touch. A molecule and a growth rule, agreeing on a rule about which
templates may be adjacent.

\subsection*{The conjecture, and why not a tensor}

Three things have now been established, and together they say what to build. A
two-variable fit loses the higher-order structure by a diverging margin. What it
loses is not a number but a shape, supported on a few short lengths. And which
lengths are populated is fixed by the shortest cycle the rule closes, everything
longer following from it by compounding. Hence:

\begin{quote}
\emph{Conjecture.} A small alphabet --- the link, the triangle, the square, and
whatever else a rule's own footprint requires --- is sufficient vocabulary for
describing a locally grown network. Naming the shortest cycle a rule closes
accounts for the higher-order structure a two-variable model loses, and no
higher-arity object is needed to hold it.
\end{quote}

\noindent The alternative is to raise the arity of the model itself, and it is
worth saying why that is not done here. In the dense case the question is already
settled against it: an exchangeable array admits the representation
$f(\xi_\emptyset,\xi_i,\xi_j,\xi_{ij})$, and for a simple dense graph the
pair-level variable $\xi_{ij}$ can always be absorbed into a threshold, leaving a
two-variable kernel. Exchangeability \emph{forces} the graphon form; no
higher-order object can add anything. Whatever raising the arity buys, it cannot
be that, and its content lies entirely in the sparse regime --- which is the
regime this whole book is about, and the reason the clustering of
Section~\ref{sec:gr-clustering} does not vanish when a graphon says it must.
There the pair-level randomness does not wash out, and the natural object keeps
arguments indexed by \emph{pairs} rather than by nodes: not $W(x_i,x_j,x_k)$ but
something closer to $W(x_{ij},x_{ik},x_{jk})$, which is the shape the limit
theory of hypergraphs already takes. The inferential side of that idea has been
built: \citet{young2021hypergraph} ask, of an ordinary pairwise network, which of
its cliques are better explained as single higher-order interactions than as the
links they appear to be, and answer by description length. It is the same
accounting used below, with hyperlinks in place of motifs.

In practice, though, the tensor is unaffordable, and Chapter~\ref{ch:graphons}
priced it exactly: a block model over $j$-sets has a count tensor with about
$B^j/j!$ entries, and balancing that against what a split saves gives
$B_{\max}\approx[m(j-1)!/\ln m]^{1/j}$, which is $\sqrt n$ at $j=2$ and $n^{1/4}$
at $j=4$ --- eighteen groups where the pair model recovers three hundred and
sixteen (Eq.~\eqref{eq:gr-bmax}). A model that can no longer see communities is a
poor instrument for asking what communities are made of. The conjecture's virtue
is that it buys the shape without paying that price, and the next section builds
the instrument that follows from it.

\section{Fitting the order, not assuming it}
\label{sec:gr-motifsbm}

All of that still asks the reader to accept the arity of each rule as given. It
can instead be \emph{inferred}, by the same criterion that fits a block model in
the first place: minimum description length. The posterior of
Section~\ref{sec:gr-blockmodel} is \emph{microcanonical} --- it fixes the link
counts exactly rather than only on average, so that a model is a set of
constraints and the description length counts the networks satisfying them --- and
$\Sigma=-\log_2P(A,e,b)$ makes no use of the fact that links join two nodes. Nothing stops us handing it a
model built from something else.

\begin{calculation}{an alphabet of motifs, priced}
The obvious something else --- a block model over $4$-sets, its count matrix
replaced by a four-index tensor --- is the wrong choice, and
Section~\ref{sec:gr-twopriors} has already priced it: a $j$-index tensor drops the
resolution to $B_{\max}\approx[m(j-1)!/\ln m]^{1/j}$, which at $j=4$ is $n^{1/4}$,
eighteen groups where the pair model recovers three hundred and sixteen. A model
that can no longer see communities is a poor instrument for asking what
communities are made of.

The affordable object keeps the ordinary block structure and changes what the
blocks are built from. Every link is assigned to exactly one \emph{motif
instance} drawn from a small alphabet --- a link, a triangle, a square --- with
motifs of arity three and above planted inside a single block, and instances
wired by \emph{stub matching}: each node is given a number of loose link-ends,
or stubs, and the ends are then paired off at random, so that a node's share of
the motifs is fixed by how many stubs it was given. That generative recipe is
\citeauthor{karrer2010subgraphs}'s subgraph configuration model
\citeyearpar{karrer2010subgraphs}, run inside blocks. Planting $c$
copies of a motif that spans $s$ nodes inside one block can be done in
\begin{equation}
  \Omega \;=\; \frac{(sc)!}{\prod_i d_i!\;c!\;a^{c}}
  \label{eq:gr-omega}
\end{equation}
ways, where $d_i$ is how many of them node $i$ takes part in and $a$ is the size
of the motif's automorphism group --- the number of relabellings of its own nodes
that leave it unchanged, which is $2$ for a link, $6$ for a triangle and $8$ for
a square. Dividing by $a^{c}$ is what stops the same motif being counted once for
each way of writing it down. $\Sigma$ is then the sum of $\log_2\Omega$ over the
blocks together with the cost of the partition and of the counts. The parameters
number one count per motif type per block plus the block matrix, $3B+B^2$ for the
alphabet used here, rather than the $B^4/24$ of the tensor, so the
resolution ceiling stays at Eq.~\eqref{eq:gr-flatceiling}'s. The price is that a triangle straddling
two blocks is not representable and is described as three links instead.
\end{calculation}

Two features make this a test rather than a fitting exercise. With the alphabet
reduced to the link alone, $\Sigma$ is exactly the description length of the
ordinary degree-corrected model, so enlarging the alphabet is a strictly nested
comparison and a negative $\Delta\Sigma$ means the motifs paid for themselves. And
the criterion can tell a planted triangle from three incidental links, because
$c$ planted triangles are more constrained --- hence cheaper to specify --- than
the $3c$ independent links they account for, while the extra counts cost only a
little. Where triangles are unavoidable rather than planted the arithmetic goes
the other way, and on a nine-node graph carrying $26$ of its $36$ possible links
the criterion correctly plants none at all.

Run against the networks whose construction is known exactly, it recovers the
rule:

\begin{center}
\begin{tabular}{lrrl}
\hline
network & $n$ & $\Delta\Sigma$ (bits) & selected\\
\hline
pseudofractal, $t=5$      & 366 & $-398$ & triangle\\
$BB(n,1)$                 & 500 & $-259$ & triangle\\
diamond, $t=5$            & 684 & $-1032$ & square\\
$DS(n,1/3)$               & 500 & $-39$  & square\\
$BB(n,2)$                 & 500 & $-204$ & square\\
$BB(n,3)$, pentagons      & 500 & $+21$  & none\\
$BB(n,4)$, hexagons       & 502 & $+19$  & none\\
fullerene $\mathrm{C}_{240}$ & 240 & $+17$ & none\\
\hline
\end{tabular}
\end{center}

\noindent The last three rows are the ones that matter. Their motifs lie outside
the alphabet, and there the enlarged model costs bits and buys nothing --- the
$+17$ is precisely the price of announcing two totals that turn out to be zero.
It does not offer a square when the network was built from pentagons. A planted
network with triangles in one community and squares in another is recovered with
both, each in its own block; and the diamond at $t=4$ is decomposed into exactly
$64=4^{t-1}$ squares, the number it actually contains.

Two cautions. The decomposition assigns each link to one instance, so motifs that
overlap cannot all be represented: the pseudofractal holds $n-2$ triangles but at
most a third of its links' worth are disjoint, and $138$ were found. The
selection is meaningful; the counts are not estimates. And the search is greedy
in two coupled directions at once, which can trap it --- on the diamond, the
branch cut of Chapter~\ref{ch:communities} places \emph{every} four-cycle across
the cut, so from that partition the model finds no motif at all, and only a
different starting partition reveals them.

So the conjecture of Section~\ref{sec:gr-conjecture} holds up as far as the
instrument can be checked against networks whose construction is known. A symbol
of arity three pays for itself where the rule closes triangles, one of arity four
where it closes squares, and neither pays where the rule closes something longer
--- which is the conjecture's own statement, arrived at by measurement rather
than by assumption, and the alphabet is doing its job when it declines as well as
when it accepts. That is as much as the vocabulary question needs; what the
vocabulary then \emph{does} to a community count is a separate matter, and the
rest of the chapter is about that.

\section{Two limit classes}
\label{sec:gr-classes}

The instrument is built. It can be run on every rule in this book, and against
the two block models it contains as special cases, so that the same networks are
counted three ways: by a flat block model, by a nested one, and by the nested one
with an alphabet of motifs. What separates the three is worth pinning down before
the numbers arrive, because the whole chapter turns on it.

\begin{calculation}{the three methods, and what separates them}
All three minimize the same description length,
\begin{equation}
  \Sigma \;=\; -\log_2 P(A\,|\,\text{model}) \;-\; \log_2 P(\text{model}),
  \label{eq:gr-sigma}
\end{equation}
the bits needed to transmit the network together with the model used to compress
it, and all three infer the number of communities rather than being told it. All
three are also \emph{degree-corrected} in the sense of
Eq.~\eqref{eq:gr-factor}: each node carries its own propensity to link, so that a
hub and a leaf can sit in the same community without the fit having to separate
them on degree alone. Chapter~\ref{ch:communities} showed why that matters --- the
degree-blind fit cuts the pseudofractal into hubs and leaves rather than into
branches. With that held fixed, the three differ only in what the model is
permitted to say.

\emph{Flat block model.} Blocks, and a $B\times B$ matrix of link counts between
them under an uninformative prior. Describing that matrix costs about
$\tfrac12B^2\log_2 m$ bits for a network of $m$ links, which is what caps the
recoverable groups at $B_{\max}\propto\sqrt n$. This is the method of
Chapter~\ref{ch:communities}.

\emph{Nested block model.} The same, except that the matrix of link counts is
itself described by a block model, and that one by another, in a hierarchy. The
cost of the top-level description falls away and the ceiling rises to
$B_{\max}\propto n/\log n$ \citep{peixoto2014hier}.

\emph{Nested $+$ motifs.} The nested model, with every link additionally assigned
to one motif instance --- a link, a triangle or a square --- as in
Eq.~\eqref{eq:gr-omega}. With the alphabet reduced to the link alone this is
exactly the nested model, so the comparison is strictly nested and
$\Delta\Sigma<0$ means the motifs paid for themselves.

The first two differ in resolving power and nothing else; the second and third
differ in vocabulary and nothing else. So a change between the first two is about
how finely a method can see, and a change between the last two is about what it
is willing to call a community.
\end{calculation}

\noindent Two quantities are read off each fit. The first is
Chapter~\ref{ch:communities}'s Ramsey community number $r_\kappa$, the size at
which communities become certain --- $P_\kappa\ge0.95$ over a hundred
realisations up to $n=200$ and twenty above it --- which asks \emph{whether} a
network has communities. The second is Chapter~\ref{ch:graphons}'s exponent
$\beta$ in $\kappa\propto n^{\beta}$, which asks \emph{how many}. Here is
everything, for every rule, all three ways.

\begin{center}
\footnotesize\setlength{\tabcolsep}{3pt}
\begin{tabular}{llrrrrrrr}
\hline
 & & \multicolumn{3}{c}{$r_\kappa$} & \multicolumn{3}{c}{$\beta$} & links\\
rule & motif & flat & nested & $+$mot. & flat & nested & $+$mot. & named\\
\hline
\multicolumn{9}{l}{\emph{constant density}}\\
\quad $LS(n,1)$ & $\triangle$ & $70$ & $80$ & $200$ & $0.54$ & $0.94$ & $1.04$ & $62\%$\\
\quad $BB(n,1)$ & $\triangle$ & $90$ & $100$ & $400$ & $0.48$ & $0.89$ & $0.89$ & $54\%$\\
\quad $DS(n,1/3)$ & $\square$ & $300$ & $250$ & $250$ & $0.78$ & $0.92$ & $1.04$ & $26\%$\\
\quad $DS(n,0.5)$ & $\square$ & $90$ & $100$ & $100$ & $0.76$ & $1.02$ & $1.01$ & $12\%$\\
\quad $DD(n,0.5)$ & $K_{2,d}$ & $40$ & $40$ & $40$ & $0.78$ & $1.12$ & $1.11$ & $0\%$\\
\hline
\multicolumn{9}{l}{\emph{no communities}}\\
\quad $BA(n,2)^{\dagger}$ & --- & never & --- & --- & --- & --- & --- & ---\\
\quad $BB(n,2)$ & $\square$ & $400$ & $400$ & never & $0.72$ & $0.97$ & --- & $57\%$\\
\quad pseudofractal & $\triangle$ & $366$ & $366$ & $366$ & $0.61$ & $1.02$ & --- & $100\%$\\
\quad diamond & $\square$ & $44$ & $44$ & $684$ & $0.71$ & $1.05$ & --- & $100\%$\\
\hline
\end{tabular}
\end{center}

\noindent The motif column is the shortest cycle the rule closes --- $\triangle$
for a triangle, $\square$ for a square, $K_{2,d}$ for duplication's biclique,
which has no symbol and is the one motif in the book of no fixed size. The last
column is the fraction of a network's links that the alphabet actually succeeds
in naming at the largest size fitted, and it is there because it turns out to be
what decides the rest. The dagger marks the one row taken from
Chapter~\ref{ch:communities} rather than measured here.

The lattices need one remark, since they are deterministic --- one network per
generation, not a hundred realisations per size. For them $P_\kappa$ is a step
function rather than an estimate, so $r_\kappa$ is exactly the smallest
generation whose fit returns more than one group, and the only imprecision is the
grid: a lattice quadruples at every step, so the figures quoted are the first
generation past the onset and the true crossing lies somewhere in the factor of
four below. Their exponents are fitted from $n=366$ and $n=684$ upward, the
smaller generations being below the resolution of any of the three methods.

Read down the table and the rules fall into two classes, which are the organising
fact of this chapter and the sections that follow. A third is definable and turns
out to be empty, which is worth setting out alongside them.

\emph{No communities.} $P_\kappa$ never reaches certainty; the best description
uses a single group at arbitrarily large size. Chapter~\ref{ch:communities}'s
non-local control $BA(n,2)$ is the canonical member --- preferential attachment
with no step that ever looks at a neighbourhood, and no communities at any size.
It is the one row here not measured with this chapter's instrument, and is marked
accordingly: the ``never'' is that chapter's flat fit, and neither the nested
model nor the alphabet has been run on it, so it stands as the class's definition
rather than as a comparison. What is new is that a network with a perfectly good
local rule can be moved into the same class by nothing more than a change of
vocabulary. $BB(n,2)$ has a Ramsey number of $400$ under both block models and
none at all once its squares can be named, and both deterministic lattices go the
same way for a reason Section~\ref{sec:gr-square} makes exact.

\emph{Constant density.} $\beta=1$, so $b=\kappa/n$ stops depending on size and
becomes a number the rule has, in the sense of Chapter~\ref{ch:graphons}'s
Eq.~\eqref{eq:gr-density}. Five of the measured rules are here, the nested model
and the alphabet agreeing on every one of them to within about a tenth.

\emph{Fractal}, the one that is empty. Communities would emerge, their number
would grow, and the density would not survive: $0<\beta<1$, so $b\to0$ as a power
of the size. Nothing in this book does that. The diamond lattice appeared to for
a while, at $\beta=0.31$, and Section~\ref{sec:gr-square} is the account of why
it does not --- the exponent was a partition the search was reporting and did not
need. Nothing rules the case out, and it is defined here rather than dropped
because it is exactly what a real network might turn out to do; the note at the
end of Section~\ref{sec:gr-decides} says what would have to be measured.

Three things about that classification need saying immediately, because they
qualify everything after it.

The first is that a class is a property of a network \emph{together with the code
used to describe it}, and the table shows this in the plainest possible way:
three of the eight rules change class when the vocabulary changes --- $BB(n,2)$
and both lattices, all of them out of the constant-density class and into the
first --- and nothing about the networks changes at all. This is the same lesson Chapter~\ref{ch:graphons}
learned twice --- once for the prior and once for the sample --- and it is
sharper here, because a prior moved an exponent while a vocabulary moves a rule
from one class to another.

The second is what the flat column says. Under the flat block model \emph{every}
rule in the table reads $\beta$ between $0.48$ and $0.78$: on that instrument the
whole book is in the fractal class. It is not. The flat model is sitting on its
own $\sqrt n$ ceiling, as Chapter~\ref{ch:graphons} established, and the apparent
fractality is the ceiling talking. Any claim that a network is fractal in this
sense has to be made with a model that could have reported the other two --- and,
as the diamond turns out to show, with a fit one has some reason to think is the
best available.

The third is that the classification is a statement about the best description
anyone has found, and not about the best there is. That distinction is usually
academic and on the two lattices it is not: both were placed in other classes by
a fit that turned out to be beaten, at every size, by a decomposition one can
write down by hand. Section~\ref{sec:gr-square} does that, and the fractal class
is empty as a result. Where a better description is found, a row moves.

\section{The rules that keep their density}
\label{sec:gr-chapter6}

Five of the random rules are in the constant-density class, and they get there by
two quite different routes: either the alphabet names their motif and the density
survives anyway, or the alphabet cannot name it at all. Both routes are worth
following, because between them they say what the vocabulary is and is not able
to do.

Start with what naming a motif does to the \emph{onset}. The two rules
Chapter~\ref{ch:communities} published Ramsey numbers for are local search, at
$81$, and duplication--split, at $98$. The master table puts the first at $70$
under the flat model and $80$ under the nested one, and the second at $90$ and
$100$.

The two block-model columns bracket the published numbers without
quite reproducing them, and the gap is worth a word, since the flat column is
supposed to be Chapter~\ref{ch:communities}'s own method. It is the same method
in the sense that matters --- a degree-corrected block model, its number of groups
inferred by description length --- but that names a family of procedures rather
than one. The published figures come from a separate
implementation~\citep{vazquezcommunities2025}; the fit is a stochastic search
whose answer depends on how many restarts it is given and on the version of the
library doing it; $P_\kappa$ is estimated from a hundred realisations; and
$r_\kappa$ can only land on one of the sizes actually grown, here $\ldots70$,
$80$, $90$, $100\ldots$, so both published values fall one step above the flat
column. Agreement to a grid step is the check that this is the same measurement.
Equality was never on offer, and the columns to compare against each other are
the three measured here, which do share an implementation.

\begin{figure}[t]
  \centering
  \includegraphics[width=\linewidth]{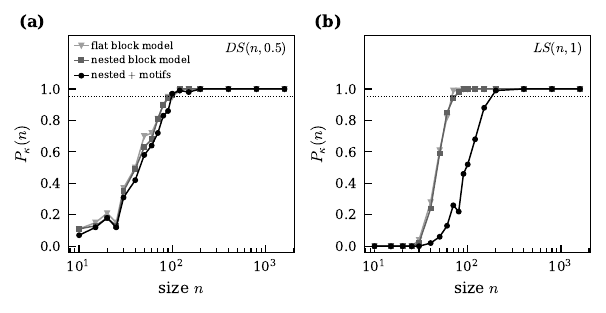}
  \caption{\textbf{The detection method is part of the answer.} Every panel
  compares three methods on identical networks: a flat block model, a nested one,
  and the nested one with triangles and squares in its alphabet.
  \textbf{(a)}~duplication--split, \textbf{(b)}~local search. The communities
  likelihood $P_\kappa(n)$, from a hundred networks at each size up to $n=200$
  and twenty above it, with the dotted line at the $95\%$ threshold whose
  crossing defines the Ramsey community number. Duplication--split makes no
  triangles and the three curves lie on top of one another; local search
  \emph{is} the triadic-closure rule, and naming its triangles moves the crossing
  from $80$ to $200$.}
  \label{fig:gr-ramsey}
\end{figure}

The motif alphabet then moves one of them and not the other. Local search
\emph{is} the triadic-closure rule, so its local structure is triangles; an
alphabet that contains the triangle books that structure as planted motifs
instead of reading it as evidence for communities, and the size at which
communities become certain is pushed from $80$ to $200$. Duplication--split makes
no triangles at all, its squares overlap heavily and compress little, and its
Ramsey number does not move (Figure~\ref{fig:gr-ramsey}).

$BB(n,1)$, which closes triangles by a quite different mechanism --- laying a
single new node across a random link rather than closing a path --- behaves like
local search and not like its own square-closing sibling. Its onset moves from
$100$ to $400$ and nothing else happens: past the threshold the motif model's
community count converges on the nested model's rather than collapsing away from
it, $5.6$ against $5.8$ at $n=200$ and $11.1$ against $11.9$ at $n=400$, and its
exponent is $0.892$ with the alphabet against $0.886$ without it.

The pseudofractal web looks at first like the same story told deterministically
--- $1.02$ against $0.99$, a lattice built entirely of triangles whose density
appears untouched by being allowed to name them --- and it is not. That reading
comes apart in Section~\ref{sec:gr-square}, and it comes apart in a way that says
something about the instrument rather than about the lattice. It is left out of
this section for that reason, and so is the diamond.

So where the alphabet names the rule's own motif, it postpones the onset by a
factor of two to four and leaves the exponent alone. That is the first route into
this class, and the more informative one, because the alphabet was working and
the density survived anyway.

\subsection*{Where the alphabet names nothing}

The second route is duplication with divergence, and its entry in the table is
the $0\%$ in the last column: at $n=800$ the alphabet succeeds in naming
essentially none of the network's links. That is not for want of squares.
$DD(n,q)$ copies a random node and deletes each copied link independently with
probability $q$; it never subdivides anything, so unlike duplication--split it
stays bipartite, and counted directly at $q=\tfrac12$ it has mass at the even
lengths and exactly nothing at the odd ones:

\begin{center}
\begin{tabular}{lrrrrrr}
\hline
 & \multicolumn{6}{c}{number of cycles of length $L$}\\
$DD(n,\tfrac12)$ & $L=3$ & $4$ & $5$ & $6$ & $7$ & $8$ \\
\hline
$n=100$ & $0$ & $694$ & $0$ & $9\,099$ & $0$ & $123\,001$\\
$n=200$ & $0$ & $3\,658$ & $0$ & $143\,822$ & $0$ & $7\,002\,135$\\
\hline
\end{tabular}
\end{center}

\noindent By the reading of Section~\ref{sec:gr-conjecture} this is a square-built
network if ever there was one, and the motif model declines to plant a single
square in it. Fitted at $q$ from $0.2$ to $0.6$ the enlarged alphabet buys
nothing: it costs a few tens of bits on average, and the scatter between repeated
fits of the same network is larger than the cost, so the only thing the number
says is that nothing was gained.

The reason is that duplication does not make a square. It makes a node with $d$
shared neighbours --- a complete bipartite $K_{2,d}$ --- holding $\binom{d}{2}$
four-cycles laid over one another. At $n=200$ and $q=\tfrac12$ the sharing runs
to fourteen common neighbours between one pair of nodes, so that pair alone
accounts for $91$ four-cycles, and across the network each link sits in some
twenty-four of them. A decomposition that gives every link to exactly one motif
can name a handful and must leave the rest unspoken.

That diagnosis blames the alphabet for lacking a symbol, which is a testable
complaint, and the test is to supply the symbol. Add the biclique $K_{2,d}$: two
hubs joined to the same $d$ leaves, $d+2$ nodes and $2d$ links. Its automorphism
group has size $|\mathrm{Aut}|=2\,d!$ --- the two hubs may swap, and the $d$
leaves permute freely --- which is what Eq.~\eqref{eq:gr-omega} divides by so
that one planted copy is charged once rather than once per way of writing it
down. It is exactly what duplication builds, and $d=2$ is the square, so the
family starts at $d=3$. On a network of planted bicliques it does what it should.
Twenty disjoint $K_{2,6}$, which is $240$ links:

\begin{center}
\begin{tabular}{lrl}
\hline
alphabet & $\Sigma$ (bits) & planted\\
\hline
link only            & $1820$ & ---\\
$+$ triangle, square & $1415$ & $60$ squares\\
$+$ biclique         & $\mathbf{1028}$ & $20$ $K_{2,6}$\\
\hline
\end{tabular}
\end{center}

\noindent The square alphabet carves each biclique into three disjoint squares
and pays $387$ bits more than naming it once. (Bicliques have to be offered
before squares, or the squares consume the links first and the larger object can
never be seen --- a reminder that a greedy search over a nested vocabulary must
work from the largest symbol down.)

Now duplication--divergence, with everything in one block. The model finds $760$
biclique candidates in $DD(n{=}200,q{=}\tfrac12)$ and plants twenty-four of them,
two as large as $K_{2,14}$, taking $\Sigma$ from $3711$ to $3458$: a saving of
$252$ bits where the square alphabet saved nothing at all. The vocabulary was not
the problem. Given the symbol, the model names the structure and profits by it.

What defeats it is the partition. Refit the same network with the block model's
own division into eleven groups, and the number of biclique candidates is
\emph{zero}. Not few --- none. Every duplicated pair has been separated from its
shared neighbours by a block boundary, and a motif that straddles a boundary is
not representable at all.

That is a more damaging limitation than the missing symbol, and it is
self-inflicted. The restriction of motifs to the inside of a single block is what
kept the parameter count at $3B+B^2$ rather than the $B^4/24$ of a tensor, and so
kept the resolution at $\sqrt{n}$ instead of $n^{1/4}$
(Section~\ref{sec:gr-motifsbm}). The saving is real. It also happens to break the
scheme on the one rule that most needs it, because duplication puts a copy and
its original in different groups --- they have the same neighbours but need not
be joined to each other, which is exactly the configuration a block model splits.
Worse, the two descriptions cannot be combined. One block with the bicliques
named costs $3458$ bits; eleven groups with nothing named costs $3163$; and there
is no fit that has both, because choosing the groups destroys the motifs. The
block structure wins by some three hundred bits, so the model reports communities
and no motifs --- but it reports them by default rather than by preference,
having been given no option that keeps the duplication visible.

Adding self-interaction does not rescue it. The duplication rule of
Chapter~\ref{ch:duplication} allows the copy to bind its own original with some
probability $p$, and since the pair already shares neighbours, every such link
closes triangles. At $p=0.1$ and $q=0.45$ the network is no longer bipartite, the
odd cycle lengths fill in, and the clustering coefficient rises from zero to
about $0.18$ and stays there as the network grows --- the interactome-like
behaviour that motivated the rule. The alphabet still plants nothing: at $n=100$,
$200$ and $400$ the enlarged model costs bits rather than saving them. The
triangles stack exactly as the squares did, all of those at a duplicated pair
sharing the one link between the pair, so that a single link can carry eleven of
them and an edge-disjoint decomposition can name at most one per pair.
Duplication defeats the alphabet whether or not it makes triangles, because
everything it builds is piled onto the same shared neighbourhood.

So duplication reaches the constant-density class by the uninteresting route: its
$\beta$ is $1.11$ with the alphabet and $1.12$ without it because the alphabet
never got started. A rule whose footprint is a biclique of variable size, laid
over a shared neighbourhood the partition wants to cut, is out of reach of a
scheme that names fixed objects inside single blocks --- and that is the rule
most often proposed for how protein interaction networks grow.

A last caution, which applies to every $\Delta\Sigma$ quoted in this chapter. The
fit is a stochastic search, and repeated fits of the \emph{same} network return
description lengths that differ. For $DD(n,q)$ at $n=400$ the spread over five
fits is around $150$ bits, larger than most of the differences being compared.
Where a sign is reported here it is the sign of an average, and the small numbers
should be read as saying that nothing was gained rather than as measurements of
how much was lost.

\section{The rule with no communities left}
\label{sec:gr-none}

\begin{figure}[t]
  \centering
  \includegraphics[width=\linewidth]{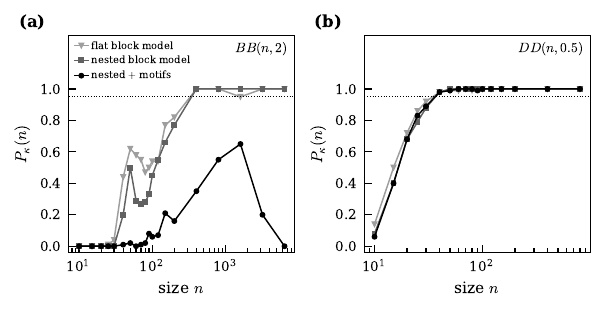}
  \caption{\textbf{Two rules built from squares, and only one of them says so.}
  Panels as in Figure~\ref{fig:gr-ramsey}: $P_\kappa$ from a hundred networks at
  each size up to $n=200$ and twenty above it. \textbf{(a)} $BB(n,2)$ closes one
  square per step on two new nodes, so its squares are nearly disjoint, the
  alphabet takes them up, and the communities the block model reports are largely
  dissolved. \textbf{(b)} $DD(n,q)$ at $q=\tfrac12$ closes squares in far
  greater numbers, but as overlapping $K_{2,d}$ bicliques rather than separable
  motifs; the alphabet plants none of them, and all three methods agree.}
  \label{fig:gr-squares}
\end{figure}

$BB(n,2)$ is the one rule in the table that the alphabet moves into the first
class. Both block models put its Ramsey community number at $400$; the motif
model never reaches the $95\%$ threshold at all. Its likelihood peaks near $0.65$
around $n=1600$ and then falls away --- $0.20$ at $3200$, and at $6400$ every one
of ten realisations returns a single community while the block model returns a
hundred and two (Figure~\ref{fig:gr-squares}a).

Pushed to larger sizes, that statement turns into something more awkward than a
suppression. The motif model's answer for $BB(n,2)$ is not small, it is
\emph{bimodal}. At $n=3200$, sixteen realisations in twenty return a single
community and the other four return between thirty-five and forty-one, which is
what the block model returns throughout. One description plants many squares and
needs no communities at all; the other finds the communities and plants fewer
squares; and which of them comes back is settled by where the search happens to
start.

Whether the criterion is \emph{indifferent} between them is a separate question,
and it has to be asked carefully, because the obvious way of asking it gives the
wrong answer. Comparing the mean $\Sigma$ of the realisations that landed on one
community with the mean of those that landed on many puts them eleven bits apart
in fifty-one thousand, which invites the conclusion that the two descriptions
cost the same. They do not. That comparison is between \emph{different networks},
sorted by which answer the search happened to find, so it measures which networks
are easy at least as much as which description is shorter.

The comparison to make is on one network at a time: fit it twice under the same
alphabet, once from a single block, so that motifs are promoted before any
partition exists, and once from the nested model's own partition, so that the
communities are there first and the motifs must pay to displace them. Both are
local minima --- neither search ever leaves the basin it started in --- so the
two descriptions can be priced against each other on identical data.

\begin{center}
\small\setlength{\tabcolsep}{4pt}
\begin{tabular}{lrr}
\hline
$BB(n,2)$, both fits on one network & $n=1600$ & $n=3200$\\
\hline
$\kappa$ from one block             & $1.0$ & $1.0$\\
$\kappa$ from the partition         & $22.6$ & $42.8$\\
$\phi$, links named, from one block & $0.57$ & $0.57$\\
$\phi$ from the partition           & $0.29$ & $0.27$\\
\hline
$\Delta\Sigma$ (bits)               & $-203$ & $-627$\\
$\Delta\Sigma/n$ (bits/node)        & $-0.13\pm0.05$ & $-0.20\pm0.08$\\
spread between networks (bits)      & $378$ & $903$\\
communities win                     & $40\%$ & $8\%$\\
\hline
\end{tabular}
\end{center}

\noindent Priced this way the motif description wins outright, and the margin per
node grows rather than shrinking. The criterion is not indifferent; it has
decided, and it decides more firmly as the network grows.

What is true is that the decision is smaller than the scatter. The
network-to-network spread is about half again the gap at both sizes, so an
individual realisation lands on either side, and that is what produces the
bimodality. It is also why the sorted comparison misleads: with a gap narrower
than the noise, sorting realisations by outcome and then averaging recovers
almost nothing of it.

So the squares and the communities are two ways of spending the same bits on the
same links, and for this rule they cost \emph{nearly} the same --- near enough
that the search's starting point decides any single case, not near enough that
the criterion is silent. That is why $BB(n,2)$ carries no motif-model exponent in
the master table: its mean $\kappa$ is the mixing ratio of two outcomes rather
than a law, and it turns over at $n=3200$ for that reason and no other.

There is a control for all this, and the rule that provides it differs from
$BB(n,2)$ in one place only. The bubble rule lays a chain of $L$ new nodes across
a random link and so closes a cycle of length $L+2$: squares at $L=2$, and
triangles at $L=1$. If what dissolved the communities was the square, or the mere
fact that a motif in the alphabet matches the rule, then $BB(n,1)$ should
dissolve the same way. It does not --- it sits in the previous section, its onset
merely postponed. Nor is the difference that triangles are harder to name than
squares: counting what the fits actually plant at $n=200$, $BB(n,1)$ plants $62$
triangles covering forty-seven per cent of its links and $BB(n,2)$ plants $40$
squares covering fifty-three per cent, and the master table's last column has
them at $54\%$ and $57\%$ at the largest sizes fitted. Comparable effort,
opposite outcome.

Something on the other side of the ledger must be doing it --- how much community
structure there was to destroy. $BB(n,2)$ is a marginal community-former to begin
with, its block-model Ramsey number of $400$ standing at four times $BB(n,1)$'s,
and a margin narrow enough for the starting partition to decide the outcome is
only reachable when the partition's own contribution is thin. A tenth of a bit
per node is what a thin one looks like.

That is an observation about $BB(n,2)$ rather than a rule, and it should not be
promoted into one: a thin partition is what makes the contest close, but it does
not by itself say which way the contest goes. Settling that needs the bubble
family swept across $L$ rather than sampled at two values of it --- a question
for the family, and not one this chapter can answer from two of its members. What
this section establishes is narrower and does not depend on the answer --- that an alphabet can abolish a rule's communities outright, and that
the rule it does it to is not the one a reading based on the motif's shape would
have picked.

\section{The lattices that describe themselves}
\label{sec:gr-square}

Two lattices are left, and they are the ones on which the alphabet does the most
and the search does the worst. Taking them together is what empties the fractal
class and puts both of them in the first.

The pseudofractal web hangs a new node on both ends of every link, closing a
triangle; the diamond replaces every link by two parallel paths of one node each,
closing a square. Both reach exactly two steps --- every new node is adjacent to
nodes that were already adjacent --- and both rebuild the whole lattice at once,
which is what keeps their cycles clean, since a rule that subdivides one link at a
time stretches the cycles it has already made. They differ in the cycle they
close and in almost nothing else.

Fitted, they look like a clean contrast. Both sit at the modular limit under the
nested block model, $1.02$ and $1.05$, and under the motif alphabet the
pseudofractal reads $0.99$ while the diamond falls to $0.31$; the diamond's fit
returns $3$, $5$ and $7$ groups at $n=684$, $2732$ and $10\,924$, a count
climbing by a fixed amount while the lattice quadruples. That contrast was the
original reading of this section, and it does not survive.

\subsection*{Both lattices tile themselves exactly}

Each of these rules gives every existing link exactly one new motif per
generation. So the motifs closed at the \emph{last} generation have distinct old
links and disjoint new chains, and between them they use every link in the
network exactly once. A perfect edge-disjoint tiling of the lattice by its own
motif exists at every size, in both lattices, and it can be written down from the
construction without fitting anything: $729$ triangles for the $2187$ links of
the pseudofractal at $n=1095$, $1024$ squares for the $4096$ links of the diamond
at $n=2732$.

Price that decomposition and compare it with what the search finds:

\begin{center}
\footnotesize\setlength{\tabcolsep}{5pt}
\begin{tabular}{lrrrrrr}
\hline
 & & \multicolumn{2}{c}{as fitted} & \multicolumn{2}{c}{the tiling} & \\
lattice & $n$ & $\Sigma$ & $\kappa$ & $\Sigma$ & $\kappa$ & cheaper by\\
\hline
pseudofractal & $366$    & $4\,474$   & $5$  & $3\,887$   & $1$ & $587$\\
pseudofractal & $1\,095$  & $15\,428$  & $14$ & $13\,728$  & $1$ & $1\,700$\\
pseudofractal & $3\,282$  & $53\,082$  & $44$ & $47\,673$  & $1$ & $5\,409$\\
diamond       & $684$    & $7\,812$   & $3$  & $7\,434$   & $1$ & $378$\\
diamond       & $2\,732$  & $37\,213$  & $5$  & $35\,666$  & $1$ & $1\,547$\\
diamond       & $10\,924$ & $171\,368$ & $7$  & $166\,876$ & $1$ & $4\,493$\\
\hline
\end{tabular}
\end{center}

\noindent The tiling is cheaper at every size, on both lattices, by a margin that
grows with the network. Handed to the optimiser as a starting state it does not
move off it. And it uses \emph{one} community: a single block, every link spoken
for by a motif, nothing left for a partition to explain.

So on the criterion this chapter has used throughout, the best available
description of the pseudofractal web has no communities --- not the fourteen the
fit reports at $n=1095$, and not the forty-four it reports at $n=3282$ --- and the
same is true of the diamond. The two rules are in the same class. Their
description lengths are not merely similar but equal to the bit where their sizes
coincide, $7434$ at $n=684$ and $35\,666$ at $n=2732$, because both are tiled by
their own motif on the same $n$ and $m$.

\subsection*{What that costs, and what it does not}

It costs the contrast. The pseudofractal was this section's foil --- the lattice
that keeps its law while the diamond loses it --- and there is no such foil.
Along with it goes the argument that reach cannot be the discriminator because
two rules of equal reach end up in different classes: they do not end up in
different classes. Chapter~\ref{ch:communities}'s ranking of a triangle rule
against a square rule may still be telling us something, but not this.

It does not cost the finding. The diamond's own result is strengthened rather
than weakened: an alphabet can take a network's communities away entirely, and
under the best description available the diamond has none at all, rather than the
vanishing density earlier arithmetic gave it. What was reported as $\kappa$
growing like $\log n$ was a partition the search was carrying and did not need.

Nor does it touch the comparison that made the finding safe. Reducing the
alphabet to the link alone recovers the ordinary degree-corrected model exactly
(Section~\ref{sec:gr-motifsbm}), and on the diamond it costs $8908$, $41\,929$
and $198\,987$ bits at the three sizes against the motif description's $7434$,
$35\,666$ and $166\,876$ --- margins of a sixth rather than the eighth the fitted
comparison gave. The motif description wins by more than was claimed, not less.

The awkward part is what it says about the instrument. These are the two networks
in the book whose optimal description can be written down by hand, and they are
the two where the greedy search is furthest from it --- by up to a tenth of the
whole description length. That is not a coincidence, and
Chapter~\ref{ch:bubbles} takes it up: a deterministic rule builds an exact
optimum, and an exact optimum is a thing a greedy search can exactly miss. Every
other rule in this book is stochastic, lays its motifs across links chosen at
random, and has no such optimum to be missed.

\section{What decides the class}
\label{sec:gr-decides}

Three classes, and two rules that changed class when the vocabulary changed. What
is it about a rule that decides which way it goes?

Not the motif's shape: triangles dissolve nothing on local search or $BB(n,1)$
and everything on the pseudofractal, and squares dissolve everything on
$BB(n,2)$ and the diamond while doing nothing whatever on $DD(n,q)$ or
duplication--split. What the master table's last column measures is the one
quantity that separates the cases at the ends: how many of a network's links an
edge-disjoint decomposition can take up.

At one end sit the two lattices, at a hundred per cent. Each gives every link one
new motif per generation, so the last generation's motifs tile the whole network,
and a partition is left with nothing to explain --- which is exactly what the fit
finds when it is started from the tiling, one community and no residue
(Section~\ref{sec:gr-square}). At the other sits duplication--divergence at
nought per cent, where every square is stacked on the same shared neighbourhood
and a decomposition that gives each link to one motif can name almost none of
them; there the alphabet never engages and the partition keeps all of its work.

That is the mechanism where it can be seen cleanly. Where the motif tiles the
network, the partition has nothing left to describe and the communities go. Where
the motifs overlap, they cannot absorb what the partition was explaining, and the
communities survive. The same overlap that made duplication's bicliques
unnameable is here doing something useful.

Both ends are extremes, and it is worth saying what makes them legible. A
hundred per cent against nought is not a marginal gap, and the lattices are
deterministic --- one network per size, no sampling, and a decomposition that can
be checked by construction rather than searched for.

Between the ends the account has nothing to say, and the last column says so
plainly.
$BB(n,1)$ names $54\%$ of its links and keeps its communities; $BB(n,2)$ names
$57\%$ and loses them; local search names $62\%$, more than either, and is in no
danger at all. Three rules within eight points of each other on the quantity that
is supposed to be deciding, and two different outcomes. Whatever separates
$BB(n,1)$ from $BB(n,2)$, it is not how much of the network their motifs can take
up, and no reading of that column will supply it.

What does separate them is a question about the bubble family rather than about
this chapter's classes, and the family repays being taken on its own terms:
$L$ is a dial, the rule is otherwise fixed, and the whole sequence can be fitted
with the alphabet extended to name each rule's own cycle. That is done
separately. What belongs here is the negative: the tiling account earns the two
lattices and does not earn the rest, and the difference between $BB(n,1)$ and
$BB(n,2)$ is left open by it.

What can be said from this side is what happens when nothing is taken up.
$BB(n,1)$ grows its communities as $n^{0.892}$ with the alphabet and $n^{0.886}$
without it, over $n=400$ to $3200$ --- indistinguishable. A motif the
decomposition cannot lay down over the network costs that network's density
nothing, whatever else is going on.

The caution generalises past any one rule. Wherever an alphabet can absorb the
same links that a partition would have explained, the two descriptions trade off
against each other, and the margin between them can fall below the
network-to-network scatter --- at which point the criterion still has a
preference but no single fit reliably reports it, and what comes back is decided
by where the search began. That is a limit on what any such method can be asked
to settle from one realisation, and not a defect of this implementation.
$BB(n,2)$ is that limit made visible: a real preference of a tenth of a bit per
node, carried on a scatter half again as large.

\begin{quote}
\emph{To follow up.} Chapter~\ref{ch:graphons} measured the community density of
three real growing networks with a vocabulary that could not name a triangle, and
the lattices show that such a vocabulary can flatter a network into a density it
does not have. Whether the Internet and the yeast interactome keep $\beta=1$
under the full alphabet is being measured and is not settled here; the
collaboration graph, at sixty thousand nodes, is beyond what the motif fit can
reach at present. The interesting outcome would be the empty class, which no rule
in this book occupies. A real network landing at $\tfrac12<\beta<1$ and staying
there under motif correction would be the first genuinely fractal community
structure the book has found --- Chapter~\ref{ch:communities}'s $0.61$ having
turned out to be the flat model's ceiling, and Chapter~\ref{ch:graphons}'s $0.80$
a sample that mixed grown networks with catalogued ones. The one caution the
lattices add is that such a claim is only as good as the fit behind it, and on a
real network there is no construction to check it against.
\end{quote}

\section{Which change moves which number}
\label{sec:gr-doubledissoc}

Three methods have been compared through these sections, and they differ in two
ways rather than one. Flat against nested is a difference of resolving power and
nothing else; nested against nested-with-motifs is a difference of vocabulary and
nothing else (Section~\ref{sec:gr-motifsbm}). Reading the master table of
Section~\ref{sec:gr-classes} across rather than down, the two differences turn
out to move two different quantities, and it is worth separating them because
one of them is a great deal better behaved than the other.

Where the first difference comes from was the subject of
Chapter~\ref{ch:graphons}, and was priced there exactly: the two models express
the same partitions and differ only in what naming one costs, the flat model
paying a charge quadratic in the number of groups and so meeting
Eq.~\eqref{eq:gr-flatceiling}'s ceiling of $\sqrt n$, the nested model
compressing the group-level structure instead and reaching
Eq.~\eqref{eq:gr-nestedceiling}'s $n/\log n$. Nothing in that arithmetic mentions
what the links are made of.

\emph{Resolution moves the exponent, and only the exponent.} Going from the flat
model to the nested one lifts $\beta$ by two to four tenths on every rule in the
table without exception, and moves the onset $r_\kappa$ by at most a seventh ---
$70$ to $80$, $90$ to $100$, $400$ to $400$, $40$ to $40$. It never changes a
rule's class, because it cannot: the flat model's fractal-looking exponents are
its ceiling, and lifting the ceiling reveals the density that was there all
along.

\emph{Vocabulary moves the onset, and sometimes the class.} Naming a rule's own
motif moves $r_\kappa$ by a factor of two to four --- $80$ to $200$ for local
search, $100$ to $400$ for $BB(n,1)$, and at least four and perhaps sixteen for
the diamond, whose generations are too far apart to pin down --- and leaves it
untouched where the alphabet has nothing to name, as with duplication--split and
duplication--divergence. On the five random rules where the alphabet cannot take
up the whole network it leaves $\beta$ alone to within a hundredth or two. On the
three where it can, it does something the resolution
knob cannot do at all: it moves the rule out of the constant-density class
entirely --- $BB(n,2)$, the diamond and the pseudofractal all into the first,
with no communities left at all.

So the tidy statement --- one knob for how finely a network can be cut, another
for whether there is anything to cut --- holds for most of the table and fails in
exactly the cases this chapter was written about. The failure is not a
complication to be regretted; it is the finding. A vocabulary that
\emph{cannot} take up the whole network leaves the exponent alone, and that is
the common case. A vocabulary that can take up the whole network takes over the
description, and then the exponent is not measuring the network's communities any
more but the leftovers of a tiling.

The asymmetry has a reason. The exponent is a statement about how finely a
network can be cut, and nothing about the vocabulary for describing links changes
how many groups a method is \emph{able} to resolve --- that ceiling is fixed by
what the description length charges for the block structure. But the vocabulary
does change how many groups are \emph{worth} resolving, by offering an
alternative account of the same links; and where that alternative can cover
everything, the number of groups worth resolving falls to almost none. The
ceiling and the demand are different things, and only the second is open to the
alphabet.

This also tidies up a conclusion that has been uncomfortable since
Chapter~\ref{ch:graphons} drew it: that Chapter~\ref{ch:communities}'s exponent
is in part a statement about the flat block model rather than about the networks.
It can now be said without reference to motifs at all. On the rules where the alphabet cannot tile, it is
irrelevant to $\beta$; the whole of the shift from $0.61$ is the nested model,
which is a standard tool and not one invented here. Whatever else is uncertain
about the machinery of these sections, that conclusion does not rest on it.

And it answers, with one qualification, the question Chapter~\ref{ch:graphons}
left at its door. What that chapter found was a density: with the resolution
lifted, the number of communities grows in proportion to the size, so $b=B/n$
becomes a number a network has rather than a rate at which something grows. The
obvious worry about such a quantity is that it might be an artefact of a poor
vocabulary --- that a model forced to describe every triangle as three separate
links would invent communities to absorb what it could not name, and that a
richer alphabet would dissolve them. On the rules whose motifs overlap it does
not: give the model triangles, squares and bicliques and the density stays where
the nested prior put it. The qualification is the three rules whose motifs do not
overlap --- $BB(n,2)$ and the two lattices --- where exactly that worry is
realised, and it is why Chapter~\ref{ch:graphons}'s conjecture has to be read as
a claim about the best available description rather than about the network
itself.

Two instrument effects had already been found and separated before this chapter
began, and it is worth noticing that all three were found the same way.
Chapter~\ref{ch:communities}'s $0.61$ became $0.80$ when the prior stopped
charging quadratically for groups, and $0.80$ became $1$ when the sample stopped
mixing grown networks with catalogued ones. Neither move was about networks. What
this chapter adds is a third of the same kind and a sharper one: change the
vocabulary and a network can change not merely its exponent but the class it
belongs to. A limit class, like an exponent, is a property of a network together
with the code used to describe it.

\chapter{Eureka: Motifs are blueprinted communities}
\label{ch:bubbles}

Chapter~\ref{ch:motifs} ended in a state that ought to be uncomfortable. The same
networks, counted by three methods differing only in what they are permitted to
say, gave community counts a hundred apart --- $BB(n,2)$ at $n=6400$ reads
$102$ communities under a nested block model and one under the same model with
squares in its alphabet. Three rules changed limit
class when nothing about them changed but the vocabulary. And the quantity
Chapter~\ref{ch:graphons} had offered as what a locally grown network converges
to --- the density of communities, constant in size and set by the rule --- came
out as one number under a block model and another under a block model that can
name a square. A quantity that moves that far when the describer's vocabulary
changes invites the question whether it was measuring the network at all.

This chapter is the answer, and the answer is not that the disagreement was a
mistake. It is that the two vocabularies were counting the same thing in
different units.

The hypothesis takes one sentence, and the rest of the chapter is what happens
when it is taken seriously. A motif and a community are the same object at two
scales. Both are the description saying \emph{these nodes belong together, and
here is how they are joined}. A motif says it from a blueprint --- a template
fixed in advance, identical in every copy, paid for once and then stamped out
wherever it fits. A community says it bespoke --- a size and an internal density
inferred for that group alone, paid for again for every group. Neither is more
real than the other. They are two ways of buying the same structure, and minimum
description length shops for the cheaper.

Follow that and a good deal of what looked like instability turns into
bookkeeping. A rule whose footprint can be stamped over the whole network needs
no bespoke groups and reports none. A rule whose footprint overlaps itself can be
only partly stamped, and whatever the blueprint cannot cover has to be bought at
the bespoke rate --- which is what a community is. The community count stops
being a property of the network and becomes the size of a remainder. What does
not move is the total.

That is a claim about descriptions rather than about any one network, so testing
it needs a family and not an example: rules that differ in their local structure
and in nothing else, so that the remainder can be made to grow and shrink while
everything around it is held still.

The bubble model is that family, and it is the reason this chapter exists.
$BB(n,L)$ of Chapter~\ref{ch:communities} lays a chain of $L$ new nodes across a
randomly chosen existing link, closing a fresh cycle of length $L+2$; it is the
only rule in this book with a knob that changes the local structure and nothing
else. The reach changes with $L$, and so does the density, but the construction,
the attachment rule and the growth are fixed. Sweeping $L$ from one to five, and
then building two variants of $L=2$ that break the pattern deliberately, puts
rules on both sides of Chapter~\ref{ch:motifs}'s divide and lets us ask what
survives the crossing. The answer, when it comes, is a single number that does
not care which side a rule is on --- but the route to it runs through an
alternation nobody predicted, a lattice that describes itself, and one experiment
whose result was called in advance and came out right.

\section{A dial for the shortest cycle}
\label{sec:bb-dial}

\begin{figure}[t]
  \centering
  \includegraphics[width=\linewidth]{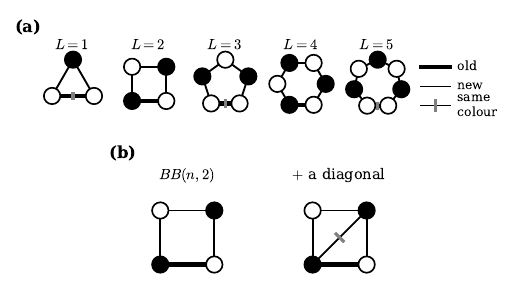}
  \caption{\textbf{Two-colouring the $L+2$ cycle.} It closes for even $L$. For
  odd $L$, or for any $L$ once a diagonal is added, one link is left with both
  ends the same colour.}
  \label{fig:bb-parity}
\end{figure}

What $L$ controls is the shortest cycle the rule ever closes, and through it the
whole cycle spectrum. Two faces sharing a link close a cycle of length $a+b-2$,
so the lengths present are generated by the motif and its gluings and everything
else is empty: $BB(n,1)$ carries every length, $BB(n,2)$ carries $4,6,8,10$,
$BB(n,3)$ carries $5$ and $8$, $BB(n,4)$ carries $6$ and $10$. The clustering
coefficient follows the same arithmetic --- $0.74$ at $L=1$ and exactly zero for
every $L$ above it, since no rule but the first closes a triangle.

Chapter~\ref{ch:communities} already measured what the dial does to the onset of
communities, and the answer was not a clean monotone. The Ramsey community number
climbs steeply with $L$, from $108$ at $L=1$ to fifty-six thousand at $L=9$, and
riding on the climb is an alternation: $246$ at $L=3$ against $326$ at $L=2$ and
$2201$ at $L=4$; $752$ at $L=5$ against $2201$ below it and $6595$ above.
\emph{Every} odd $L$ sits beneath both of its even neighbours. Communities come
on sooner when the cycle the rule closes is odd, and that chapter recorded the
effect without explaining it.

The measurements here reproduce it where the grid allows. Fitting the flat
degree-corrected block model on sizes $100,200,400,\ldots$, the onsets come out
at $90$, $400$, $400$, $1600$ and $800$ for $L=1$ to $5$, against the published
$108$, $326$, $246$, $2201$ and $752$. The two middle values collapse onto one
grid point --- $326$ and $246$ are not separable at a spacing of two --- but
$L=4$ against $L=5$ is resolved and comes out the right way round, $1600$ against
$800$, and the $L=5$ figure agrees with the published $752$ to within the grid.
That the odd rules form communities sooner is not an artefact of one
implementation.

\section{Extending the alphabet}
\label{sec:bb-alphabet}

Chapter~\ref{ch:motifs} fitted these networks with an alphabet of the link, the
triangle and the square. Run $BB(n,3)$ through it and the fit plants nothing:
there are no triangles in a pentagon-built network and no squares either, so the
enlarged model costs the price of announcing two totals that turn out to be zero
and buys nothing. Read the output naively and $BB(n,3)$ is a rule whose community
density is untouched by the motif alphabet --- filed alongside
duplication--divergence, in the class where the vocabulary makes no difference.

That reading would be worthless, and the reason is the lesson
Chapter~\ref{ch:motifs} spent itself establishing, turned on the instrument that
established it. A class is a property of a network together with the code used to
describe it. An alphabet with no symbol for a pentagon cannot be asked whether
naming pentagons dissolves communities; it can only report that it has no
pentagons to name. The whole point of the exercise is to give each rule a
vocabulary equal to its own footprint and then see what happens, and for $L>2$
that vocabulary did not exist.

So the alphabet is extended by the cycles above the square. An $L$-cycle has
$|\mathrm{Aut}|=2L$, the dihedral group of $L$ rotations and a reflection, which
is the number Eq.~\eqref{eq:gr-omega} divides by so that one planted copy is
charged once rather than once for every way of writing it down; the convention
checks out, since $k!/|\mathrm{Aut}|=(k-1)!/2$ returns twelve distinct
five-cycles on five labelled nodes, sixty six-cycles on six, and three hundred
and sixty seven-cycles on seven. The triangle and the square keep the
special-cased enumerators they always had, so nothing measured before this family
existed can have moved.

Every rule below is then fitted twice on the same networks: once with the
\emph{control} alphabet of link, triangle and square, which is
Chapter~\ref{ch:motifs}'s and which for $L>2$ has no symbol for the rule's own
cycle, and once with the \emph{test} alphabet, the same plus the $L+2$ cycle.

The control is not decoration. It is the check that the extended fit
reduces to the nested block model when it has nothing to name, and it does: on
$BB(n,3)$ at $n=3200$ the control returns $24.3$ communities against the nested
model's $24.1$. Whatever the test column then reports is the alphabet at work and
not a change of software.

\section{The family}
\label{sec:bb-family}

Seven rules, twenty realisations at each size, exponents fitted from $n=400$
upward so that every rule is past its own onset and none of them contributes a
stretch of $\kappa=1$ to a slope.

\begin{center}
\footnotesize\setlength{\tabcolsep}{3pt}
\begin{tabular}{lccrrrrr}
\hline
 & & & & & \multicolumn{2}{c}{$\beta$} & \\
rule & motif & bip. & $\langle k\rangle$ & $r_\kappa$ & nested & $+$mot. & named\\
\hline
$BB(n,1)$ & \ngon{3} & no & $4.0$ & $100$ & $0.89$ & $0.89$ & $54\%$\\
$BB(n,2)$ & \ngon{4} & yes & $3.0$ & $400$ & $0.97$ & --- & $57\%$\\
$BB(n,3)$ & \ngon{5} & no & $2.7$ & $400$ & $0.95$ & $0.96$ & $55\%$\\
$BB(n,4)$ & \ngon{6} & yes & $2.5$ & $1600$ & $1.07$ & --- & $56\%$\\
$BB(n,5)$ & \ngon{7} & no & $2.4$ & $800$ & $1.05$ & $1.08$ & $54\%$\\
\hline
$BB(n,2)+$ a link & \sqstub & yes & $4.0$ & $100$ & $0.92$ & $0.94$ & $5\%$\\
$BB(n,2)+$ a diagonal & \sqdiag & no & $4.0$ & $100$ & $0.97$ & $0.93$ & $64\%$\\
\hline
\end{tabular}
\end{center}

\noindent The exponents in the $\beta$ nested column all sit near one, which is
Chapter~\ref{ch:graphons}'s modular limit: under a nested block model every
member of this family has a constant density of communities. What the alphabet
does to that density is the last two columns, and it alternates.

$L=1$, $3$ and $5$ keep everything. Their exponents are unmoved to within two
hundredths --- $0.89$ against $0.89$, $0.95$ against $0.96$, $1.05$ against
$1.08$ --- while the alphabet works hard the whole way: $BB(n,3)$ plants pentagons
covering fifty-five per cent of its links and saves $2362$ bits at $n=3200$, and
$BB(n,5)$ plants heptagons covering fifty-four per cent and saves a thousand at
$n=1600$. These are not fits in which the enlarged model declined the offer. The
motifs are named, in quantity, at a large and growing saving, and the communities
do not notice.

$L=2$ and $L=4$ lose them. $BB(n,2)$ goes from $44.7$ communities under the
nested model at $n=3200$ to one; $BB(n,4)$ goes from $11.9$ to $2.65$, its
likelihood of finding any communities at all peaking at $0.55$ near $n=800$ and
falling to $0.25$ and $0.15$ at $1600$ and $3200$. Neither carries an exponent in
the table, and for the reason Section~\ref{sec:gr-none} gave for the first of
them: the answer is bimodal, most realisations returning a single community and a
minority returning what the block model returns, so the mean is a mixing ratio
rather than a law.

\begin{figure}[t]
  \centering
  \includegraphics[width=\linewidth]{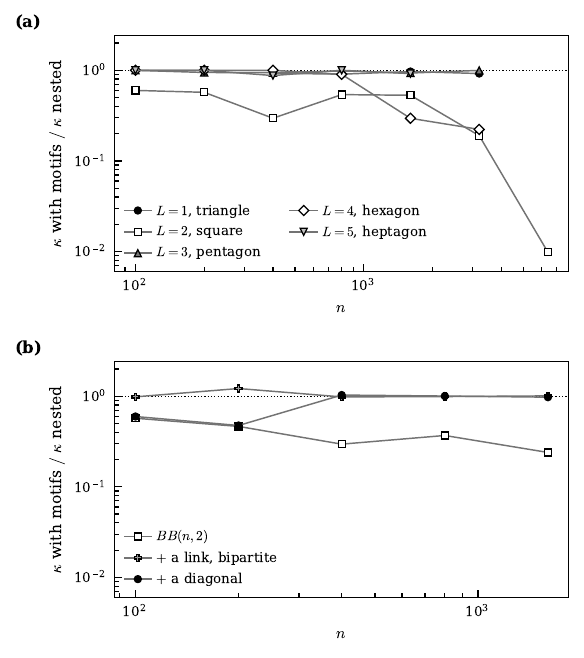}
  \caption{\textbf{The alternation, and the diagonal that breaks it.} Both panels
  plot the community count under the motif alphabet divided by the count the
  nested block model returns on the same networks: one where the communities
  survive, falling away where they are carried off. \textbf{(a)}~$L=1$ to $5$,
  odd $L$ filled and even $L$ open. \textbf{(b)}~$BB(n,2)$ against two variants
  that each add one link per step, of which only the diagonal destroys
  bipartiteness.}
  \label{fig:bb-alternation}
\end{figure}

The pattern is not subtle, and Figure~\ref{fig:bb-alternation}a is the whole of
it. Naming a rule's own motif abolishes its communities when $L$ is even and
leaves them exactly where they were when $L$ is odd.

The figure plots a ratio rather than the two community counts, and the reason is
worth a sentence, because it is the chapter's argument in miniature. These rules
differ several-fold in how many communities they have to begin with --- $BB(n,2)$
has three times $BB(n,4)$'s at the same size --- and that difference is exactly
what does \emph{not} decide the outcome. Dividing it out leaves one quantity per
rule, on a scale where surviving means one, and the five curves separate into two
families that nothing else about them separates.

\section{What does not decide}
\label{sec:bb-negative}

Two explanations were available before this sweep, and the table kills both.

\subsection*{Not how much of the network the motif covers}

Chapter~\ref{ch:motifs} found that on the two deterministic lattices the
discriminator is tiling: the diamond's squares cover ninety-two per cent of its
links and its density dies, the pseudofractal's triangles reach fifty-six and its
density survives. Extended to this family the account predicts that the rules
whose motifs cover more should be the ones that dissolve.

They are not distinguishable at all. The last column reads $54$, $57$, $55$, $56$
and $54$ per cent for $L=1$ to $5$ --- a spread of three points across a family
whose outcomes are opposite. The largest coverage in the whole family belongs to
$BB(n,2)$ at fifty-seven per cent and the smallest to $BB(n,1)$ and $BB(n,5)$ at
fifty-four, and one of those three dissolves while the other two do not. Whatever
tiling explains about the lattices, where the gap is ninety-two against
fifty-six, it explains nothing here.

\subsection*{Not how much there was to lose}

The other candidate was the other side of the ledger. A near-tie in description
length is easiest to reach when the partition was not contributing much in the
first place, and $BB(n,2)$ is a marginal community-former --- a Ramsey number of
$400$ against $BB(n,1)$'s $100$. A rule with thin community structure should be
the one whose communities the alphabet can carry off.

$BB(n,4)$ and $BB(n,5)$ settle that, and they settle it against. At $n=1600$ they
have almost exactly the same amount to lose, $6.10$ communities against $6.85$
under the nested model, and their alphabets name the same share of links, fifty-six
per cent against fifty-four. One collapses to a single community and the other
does not move. If thinness were deciding, the thinner of the two would go first;
$BB(n,5)$ is the thinner by every measure in the table --- fewer communities,
lower mean degree, a later onset than $L=3$ --- and it is the one that survives.

\section{Parity}
\label{sec:bb-parity}

What is left is the one property that has been sitting in plain view since
Section~\ref{sec:bb-dial}. $BB(n,L)$ closes cycles of length $L+2$ and glues
them at shared links, so for even $L$ every cycle in the network is even and the
network is bipartite; for odd $L$ it is not. Two-colouring a grown instance of
each confirms it directly rather than by inference from the cycle spectrum: $L=2$
and $L=4$ two-colour, $L=1$, $3$ and $5$ do not.

The two rules that dissolve are the two that are bipartite. That is the whole of
the regularity, and it holds across five rules whose coverage is constant, whose
community counts vary by a factor of three, and whose mean degrees run from $2.4$
to $4.0$. This is the geometry drawn in Figure~\ref{fig:bb-parity}.

It also connects to something Chapter~\ref{ch:communities} recorded and could not
explain. The alternation in the Ramsey number is the same alternation: every odd
$L$, which is to say every non-bipartite member of the family, forms communities
sooner than either of its even neighbours. So bipartiteness is doing two things
at once in this family --- making communities harder to form in the first place,
and making them easier to argue away once a vocabulary exists that can name the
rule's motif --- and the second was invisible until an alphabet was built that
could reach past the square.

\section{The diagonal}
\label{sec:bb-diagonal}

A regularity across five rules is not a mechanism, and parity in this family is
confounded with everything else $L$ controls. The way to test it is to break
bipartiteness while changing nothing else, and the bubble rule makes that easy,
because a square has diagonals.

Take $BB(n,2)$ and, having laid the chain $u$--$a$--$b$--$v$ across the link
$(u,v)$, add one diagonal. The chain is still two nodes, the reach is unchanged,
one fresh footprint is still laid down per step, and the rule is local in exactly
the sense it was before. But $a$ and $v$ sit on the same side of the bipartition
--- that is what a diagonal of a four-cycle is --- so the added link closes two
triangles and the network stops being bipartite. Parity predicts the communities
survive. Tiling predicts they dissolve, since the footprint still tiles.

The footprint is now $K_4$ minus an edge: four nodes, five links, two triangles
sharing the diagonal, with $|\mathrm{Aut}|=4$ because the two degree-two nodes may
swap and so may the two degree-three nodes, giving $k!/|\mathrm{Aut}|=6$, one for
each choice of the missing link. It has to go into the alphabet, for the reason
of Section~\ref{sec:bb-alphabet} in a subtler form: without it the fit does not
report nothing, it reports the \emph{pieces}. On twenty disjoint copies the
link-only description costs $1820$ bits, the triangle-and-square alphabet $1415$
--- naming each object as a square with one link left over --- and the alphabet
with the symbol $1028$, planting exactly twenty. A rule fitted without the symbol
would have been measured on a description of its footprint's fragments.

Three rules were grown, because the diagonal also adds a link and densification is
the alternative account of anything that moves:

\begin{center}
\footnotesize
\begin{tabular}{llcc}
\hline
rule & extra link per step & $\langle k\rangle$ & bipartite\\
\hline
$BB(n,2)$              & ---                                        & $3.00$ & yes\\
$+$ a link             & to a neighbour of $v$, the opposite colour & $3.97$ & yes\\
$+$ a diagonal         & to $v$, the same colour                    & $3.97$ & no\\
\hline
\end{tabular}
\end{center}

\noindent The result is in the master table, and it is unambiguous. The chorded
rule keeps its communities: from $n=400$ its count is indistinguishable from the
nested model's --- $11.95$ against $11.60$, $23.15$ against $23.05$, $43.25$
against $44.30$ --- with an exponent of $0.93$ against $0.97$. And it keeps them
while the alphabet takes up \emph{more} of it than any other network in this
book: two hundred and ninety-six chorded squares at $n=1600$, sixty-four per cent
of the links named, a saving of $1272$ bits.

That is the tiling account refuted on its own terms. A footprint that covers
sixty-four per cent of the network leaves the communities untouched, while one
covering fifty-seven per cent abolishes them, and the difference between the two
rules is one link per step and the parity it destroys.

Two cautions, because the experiment is not as clean as its result.

The bipartite control did not work. It was built to hold density fixed while
keeping the network two-colourable, and it does that, but the extra link turns
the rule's disjoint squares into overlapping ones and the alphabet is left with
almost nothing it can name --- five per cent of the links, against fifty-seven
without the extra link. Its communities survive, but they survive the way
duplication--divergence's do, because no motif was ever planted, and that says
nothing about parity. A rule that is bipartite, as dense as the chorded one, and
still has separable squares would close the argument; every link that can be
added to a bipartite rule creates overlapping even cycles, and no such rule has
been found.

So density is not controlled within the three rules of this experiment: the only
one that is both bipartite and has an alphabet that engages is $BB(n,2)$ itself,
at $\langle k\rangle=3$. What closes it is the family from outside. Mean degree
runs $4.00$, $3.00$, $2.67$, $2.50$, $2.40$ down the sweep, and the outcome
alternates against that ordering rather than following it: $L=1$ survives at the
highest density in the family and $L=5$ survives at the lowest, with the two
casualties in between. Nothing monotone in $\langle k\rangle$ can produce that.

\section{The same rule, made deterministic}
\label{sec:bb-deterministic}

Everything so far has been measured on a rule that chooses one link at random per
step. Make it deterministic --- lay a chain across \emph{every} link of the
current network, at every generation --- and the family becomes a sequence of
lattices, one network per size, in the way the pseudofractal web and the diamond
lattice are lattices. Call it $DBB(n,L)$.

Two of its members are already in this book under other names. $DBB(n,1)$ hangs
one node on both ends of every link and is exactly the pseudofractal web; it
reproduces $n$ and $m$ at every generation. $DBB(n,2)$ is a cousin of the diamond
lattice rather than the lattice itself --- both quadruple their links each
generation and agree on $n$ and $m$ at every size, but the diamond \emph{replaces}
each link by two parallel paths and leaves its endpoints unjoined, while $DBB$
keeps the link and runs the chain alongside it. The counts follow from the rule:
$m_t=(L+2)^t$, $n_t = 2 + L\big((L+2)^t-1\big)/(L+1)$, and
$\langle k\rangle\to 2(L+1)/L$, the same limiting density as the random family.

Removing the randomness was meant to remove a nuisance. There is one network per
size, nothing to average over, and the alternation --- if it is there --- has to
show up in a single object rather than in a mean. What it removed instead was the
cover the randomness had been providing, and the result is a caution that applies
to Chapter~\ref{ch:motifs} as much as to this one.

\subsection*{Every one of these lattices tiles perfectly}

A generation gives every existing link exactly one chain. So the motifs closed at
the \emph{last} generation have distinct old links and disjoint new chains, and
they use $(L+2)\,m_{t-1}=m_t$ links between them: a perfect edge-disjoint tiling
of the whole network by its own motif, at every size, for every $L$. Counted
directly, $729$ of $729$ triangles on the pseudofractal at $t=7$, $1024$ of
$1024$ squares on $DBB(n,2)$ at $t=6$, $625$ of $625$ pentagons on $DBB(n,3)$ at
$t=5$. The same holds for the diamond lattice.

That is a statement about the networks, proved by construction rather than
fitted. Set against it, what the fit reports is a statement about the search:

\begin{center}
\footnotesize\setlength{\tabcolsep}{4pt}
\begin{tabular}{lrrrr@{\hskip 10pt}rrr}
\hline
 & & \multicolumn{3}{c}{as fitted} & \multicolumn{3}{c}{seeded from the tiling}\\
lattice & $n$ & $\Sigma$ & $\kappa$ & named & $\Sigma$ & $\kappa$ & named\\
\hline
pseudofractal & $366$   & $4\,724$  & $3$  & $45\%$ & $\mathbf{3\,887}$  & $1$ & $100\%$\\
pseudofractal & $1\,095$ & $15\,487$ & $16$ & $56\%$ & $\mathbf{13\,728}$ & $1$ & $100\%$\\
$DBB(n,2)$    & $684$   & $8\,459$  & $1$  & $66\%$ & $\mathbf{7\,434}$  & $1$ & $100\%$\\
$DBB(n,2)$    & $2\,732$ & $42\,837$ & $26$ & $13\%$ & $\mathbf{35\,666}$ & $1$ & $100\%$\\
diamond       & $684$   & $7\,812$  & $3$  & $88\%$ & $\mathbf{7\,434}$  & $1$ & $100\%$\\
diamond       & $2\,732$ & $37\,213$ & $5$  & $88\%$ & $\mathbf{35\,666}$ & $1$ & $100\%$\\
\hline
\end{tabular}
\end{center}

\noindent The right-hand columns are the known tiling handed to the optimiser as
a starting state and allowed to move. It does not move: it stays on the tiling,
and the description is cheaper than the fitted one every time, by $378$ bits at
the smallest and $7172$ at the largest. So on the criterion this book uses
throughout, the best available description of the pseudofractal web has
\emph{one} community, not the sixteen the fit reports at $n=1095$ --- and the same
is true of the diamond.

Two of those rows have the same description length to the bit, $7434$ at $n=684$
and $35\,666$ at $n=2732$, because $DBB(n,2)$ and the diamond are both tiled by
squares on the same $n$ and $m$. Two lattices that Chapter~\ref{ch:motifs} placed
in different limit classes are, under their best descriptions, indistinguishable.

\subsection*{The same thing happens all the way up the family}

Section~\ref{sec:gr-square} makes this argument for the two lattices
Chapter~\ref{ch:motifs} fits, and finds that both of them have one community
under their best description rather than the fourteen and the seven the search
reports. What the sweep adds is that it is not a peculiarity of those two. It is
what every deterministic member of this family does, at every $L$, for a reason
that is written into the rule: give every link a motif at every generation and
the last generation's motifs are a tiling, whatever the motif is.

So the parity that organises the random family has nothing to organise here.
$DBB(n,L)$ is in the first class for every $L$ measured, odd and even alike, and
the alternation of Sections~\ref{sec:bb-family} to \ref{sec:bb-diagonal} is a
fact about rules that choose one link at a time.

\subsection*{Why the random family is not touched}

The randomness is what makes $BB(n,L)$ measurable. Its motifs are laid across
links chosen at random, so they overlap: a link that has already been built into
one motif can be chosen again, and no perfect tiling exists to be missed. The
fifty-four to fifty-seven per cent the alphabet names there is close to what is
available, not a sixth of it, and it is stable across sizes and realisations
rather than collapsing as the network grows. That is why the alternation of
Sections~\ref{sec:bb-family} to \ref{sec:bb-diagonal} is a measurement and this
section is a caution.

It is worth saying plainly which way that cuts. A deterministic rule looks like
the cleaner instrument --- no sampling, no averaging, one object per size --- and
it is the one on which the instrument fails. Stochasticity is not noise obscuring
the structure here. It is what keeps the structure within reach of a greedy
search, by ensuring there is no exact optimum to be exactly missed.

\section{Motifs are blueprinted communities}
\label{sec:bb-meso}

There is a way of reading this whole chapter that makes the alternation less
strange, and it is worth stating because it also says what the two vocabularies
have been arguing about.

A planted motif and a community are the same kind of object at two scales. Both
are the description saying \emph{these nodes belong together, and here is how
they are joined}. A motif is small, fixed and homogeneous: a handful of nodes, a
wiring given once by the symbol, every copy identical. A community is large,
variable and heterogeneous: as many nodes as the fit wants, an internal density
given by a row of the count matrix, no two alike. Community detection, put this
way, is the representation of a network by heterogeneous motifs of unbounded size
and arbitrary internal connectivity, and an alphabet of motifs is the same
representation with the size fixed and the wiring named in advance.

That is why the two compete. They are not a structure and a description of it;
they are two dialects for the same statement about the same links, and minimum
description length is choosing between dialects. Which is also why "the alphabet
abolished the communities" was always a slightly misleading way to put it. The
meso-structure did not go anywhere. It was renamed.

\begin{figure}[t]
  \centering
  \includegraphics[width=\linewidth]{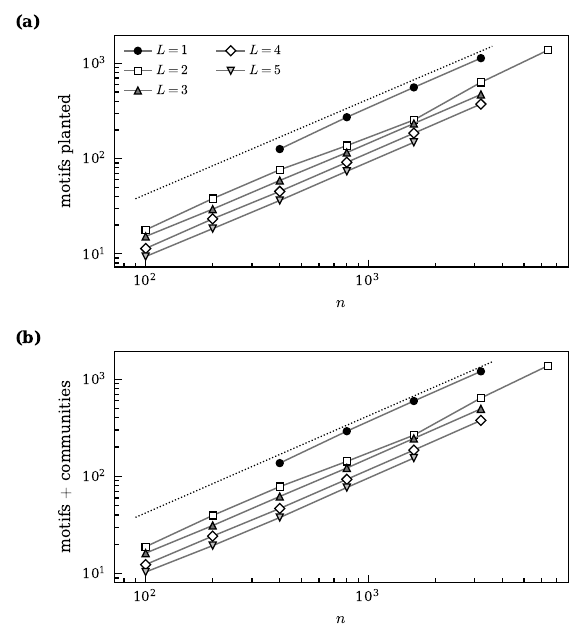}
  \caption{\textbf{Meso-structure, counted without regard to which vocabulary
  names it.} \textbf{(a)}~motifs planted, $\kappa_0$, and \textbf{(b)}~motifs plus
  communities, $\kappa_0+\kappa$, against size, with a line of slope one for the
  eye. Every rule is parallel to it in both panels, while $\kappa$ alone runs
  from $n^{0.25}$ to $n^{0.93}$ across the same five rules.}
  \label{fig:bb-meso}
\end{figure}

The counts say so, and they read more easily if the notation admits the point.
Write $\kappa_0$ for the number of motifs the fit plants, against $\kappa$ for the
number of communities it reports: a motif is the null case of a community, the
simplest form the thing can take, fixed in size and homogeneous in wiring, and
the index says so. Follow both along the family (Figure~\ref{fig:bb-meso}):

\begin{center}
\footnotesize
\begin{tabular}{lrrr}
\hline
rule & $\beta$ for $\kappa_0$ & $\beta$ for $\kappa$ & $\beta$ for $\kappa_0+\kappa$\\
\hline
$BB(n,1)$ & $1.06$ & $0.89$ & $1.04$\\
$BB(n,2)$ & $1.02$ & $\mathbf{0.25}$ & $1.01$\\
$BB(n,3)$ & $0.99$ & $0.93$ & $0.99$\\
$BB(n,4)$ & $1.01$ & $\mathbf{0.28}$ & $0.99$\\
$BB(n,5)$ & $1.00$ & $0.70$ & $0.98$\\
\hline
\end{tabular}
\end{center}

\noindent The middle column is the alternation this chapter has been about,
running from $n^{0.25}$ to $n^{0.93}$ with the parity of $L$. The outer two do
not alternate at all. Every rule in the family plants motifs in proportion to its
size, and every rule carries meso-structure in proportion to its size, with a
slope of one to within a few hundredths in all ten cases.

The two panels of Figure~\ref{fig:bb-meso} are nearly the same picture, and that
is the point rather than a defect of it: $\kappa$ is a few per cent of $\kappa_0+\kappa$
throughout, so the total is dominated by the motifs and the communities are a
small residue. What the parity of $L$ decides is the size of that residue --- how
much of the meso-structure is left over for the heterogeneous vocabulary once the
homogeneous one has taken what it can.

The density of that meso-structure is a constant of the rule, in exactly the
sense Chapter~\ref{ch:graphons} gave the density of communities: $0.38$ per node
for $L=1$, then $0.20$, $0.15$, $0.12$ and $0.09$ as the motif grows and fewer of
them fit. It falls with $L$ for a reason with no depth in it --- a longer cycle
uses more links per copy, and the mean degree is falling too --- but it is
constant in $n$ for each rule, which is the part that matters.

So Chapter~\ref{ch:graphons}'s conjecture can be restated in a form the diamond
does not break. What a locally grown network converges to is a constant density
of meso-structure. Whether a given description spends that density on motifs or
on communities is a question about the vocabulary, and the answer alternates with
the parity of the cycle; the density itself does not notice. Read that way,
$BB(n,2)$ and the diamond never lost anything. They were re-described, in a
dialect that happens to have a symbol for what they are made of.

\section{What four chapters add up to}
\label{sec:bb-conclusion}

This is the end of four chapters rather than of one, and the accounting is worth
doing in one place. Chapter~\ref{ch:communities} asked whether locally grown
networks have communities and found that they do, with a threshold size and an
exactly solvable case to show why. Chapter~\ref{ch:graphons} asked how many, and
found that the answer moved by four tenths in the exponent when the prior
changed. Chapter~\ref{ch:motifs} asked what the describer is allowed to
\emph{name}, and found that the answer could move by everything: a vocabulary
that can call a set of links a triangle sometimes reports no communities at all
where a vocabulary that cannot reports a constant density of them. This chapter
turned the one dial that changes a rule's local structure and nothing else.

Four chapters of numbers that disagree with each other invites the reading that
none of them means anything. The opposite is the case, and the last section says
why: they disagree because they are counting the same substance in two
currencies. Count it in one.

\begin{center}
\footnotesize\setlength{\tabcolsep}{5pt}
\begin{tabular}{llrr}
\hline
rule & motif & links named & $\beta$ for $\kappa_0+\kappa$\\
\hline
\multicolumn{4}{l}{\emph{no communities}}\\
\quad $BB(n,2)$ & \ngon{4} & $57\%$ & $\mathbf{1.01}$\\
\quad $BB(n,4)$ & \ngon{6} & $56\%$ & $\mathbf{0.99}$\\
\quad pseudofractal & \ngon{3} & $100\%$ & $\mathbf{1.00}$\\
\quad diamond & \ngon{4} & $100\%$ & $\mathbf{1.00}$\\
\hline
\multicolumn{4}{l}{\emph{constant density of communities}}\\
\quad $LS(n,1)$ & \ngon{3} & $62\%$ & $\mathbf{1.01}$\\
\quad $BB(n,1)$ & \ngon{3} & $54\%$ & $\mathbf{1.04}$\\
\quad $BB(n,3)$ & \ngon{5} & $55\%$ & $\mathbf{0.99}$\\
\quad $BB(n,5)$ & \ngon{7} & $54\%$ & $\mathbf{0.98}$\\
\quad $DS(n,1/3)$ & \ngon{4} & $26\%$ & $\mathbf{1.11}$\\
\quad $DS(n,0.5)$ & \ngon{4} & $12\%$ & $\mathbf{1.21}$\\
\quad $DD(n,0.5)$ & $K_{2,d}$ & $0\%$ & $\mathbf{1.11}$\\
\quad $BB(n,2)+$ a link & \sqstub & $5\%$ & $\mathbf{1.10}$\\
\quad $BB(n,2)+$ a diagonal & \sqdiag & $64\%$ & $\mathbf{0.94}$\\
\hline
\end{tabular}
\end{center}

\noindent Thirteen local rules, every one measured in this book, sorted into the
two limits they converge to under the best description available, and one
exponent. The exponent does not sort. It sits on one --- between $0.94$ and
$1.21$, and within a few hundredths of one wherever the alphabet engages at
all --- on both sides of a division that was supposed to separate a network with
communities from a network without any.

\subsection*{Two classes, and what actually separates them}

The classes are real. In the first, the best description uses a single group at
any size: the network is made of copies of one small object, laid down where the
rule put them, and once they are named there is nothing left for a partition to
say. In the second, naming the motifs leaves a residue that only heterogeneous
groups of unbounded size can describe, and the number of those groups grows in
proportion to the network.

What separates them is not how much meso-structure a rule builds. It is whether
the rule's own footprint can be laid down over the whole network without
overlapping itself. Where it can --- exactly, in the two lattices, which tile
themselves; nearly, in $BB(n,2)$ and $BB(n,4)$, whose fresh cycles land on new
nodes --- the blueprint suffices and the communities go. Where the footprints
pile onto each other, as duplication's bicliques do at nought per cent and local
search's triangles do at sixty-two, a decomposition that gives each link to one
motif cannot take up the network, and the leftover is what a community is.

That is the sense in which a motif is a blueprinted community. Both say
\emph{these nodes belong together, and here is how they are joined}. A motif says
it from a template fixed in advance, identical in every copy, and pays for the
template once. A community says it bespoke, with a size and an internal density
inferred for each one, and pays for every group separately. Minimum description
length is not choosing between structure and no structure. It is choosing how
much of the network can be described from a blueprint, and billing the remainder
at the bespoke rate.

\subsection*{The limit, restated}

Chapter~\ref{ch:graphons} conjectured that a locally grown network converges to
no kernel on the unit square, and that what characterises its limit is intensive:
the density of communities $b=B/n$, constant in size and set by the rule. The
diamond then broke it, and the repair is one substitution. The intensive quantity
is the density of \emph{meso-structure}, $(\kappa_0+\kappa)/n$, and it is
constant for every local rule in the table. The density of communities is
constant only for those in the second class, and zero for those in the first,
because $b$ measures how much of the meso-structure happened to need the
expensive vocabulary.

Read that way the two classes stop being a taxonomy of networks and become a
statement about compressibility. A rule in the first class builds a network that
is wholly blueprinted: one symbol, repeated, with no remainder. A rule in the
second builds one that is partly blueprinted, with a remainder that scales.
Nothing in either case converges to a shape one could draw on the unit square,
and the reason is the one Chapter~\ref{ch:graphons} gave --- the structure never
coarsens, because the rule keeps writing it at whatever scale the network has
reached --- but what fails to coarsen is now countable, and it is the same
quantity in both classes.

\subsection*{What the inference costs}

The other thing these chapters have between them established is how much of a
community count belongs to the counter. Three knobs, and each was found by
turning it and watching:

\emph{Resolution.} The flat block model cannot report more than about $\sqrt n$
groups. Chapter~\ref{ch:communities}'s $0.61$ was that ceiling as much as it was
the networks; the nested prior lifts it and the exponent goes to one.

\emph{Vocabulary.} What the model may name decides what it must partition. Give
it the rule's own motif and three of the eight rules in Chapter~\ref{ch:motifs}
leave the constant-density class outright.

\emph{Search.} The one this chapter had to learn the hard way. On the two
lattices whose optimal description can be written down by hand, the greedy fit
misses it by up to a tenth of the whole description length, and reports a
partition it does not need. Every published number here is a statement about a
network, a code, \emph{and} an optimiser, and only the first two are usually
declared.

The invariant survives all three. Whatever the resolution, whatever the alphabet,
whatever the search finds, the meso-structure comes out proportional to $n$.

\subsection*{Not settled}

Three things, and they are the obvious next measurements rather than
reservations.

Why parity decides the random family is not established. The story --- that a
square is what a bipartite network is made of, so on the even rules the alphabet
and the partition describe the same structure and either can stand in for the
other --- is plausible and untested. A chorded hexagon rule would test it: if
parity is the mechanism, $BB(n,4)$ with a diagonal should keep its communities
exactly as $BB(n,2)$ with one does.

Whether the meso-structure density is constant beyond this book's rules is
unmeasured. Every entry in the table is a growth rule whose construction is
known. The three real growing networks of Chapter~\ref{ch:graphons} are where to
take it, and the fit is within reach for two of them.

And whether the first class contains anything real is open. Its members here are
all lattices or near-lattices, built to tile; no real network is likely to be a
perfect blueprint of one motif. What a real network might plausibly be is a
member of the second class with an unusually large blueprinted fraction, and that
is a quantity this machinery can measure --- the last column of the table --- on
any network one can fit.

One last observation, which belongs to the book's argument rather than to this
family. Every result in this chapter required an alphabet that did not exist when
the chapter began. $BB(n,3)$ fitted with Chapter~\ref{ch:motifs}'s vocabulary
returns a clean, publishable, entirely misleading answer --- no motifs planted,
density untouched, filed with the rules the alphabet cannot reach. The finding
was not hiding in the data waiting to be noticed. It was unreachable until the
instrument was rebuilt, and the instrument was only rebuilt because a conjecture
about the family made a specific enough claim to be worth the trouble. A limit
class is a property of a network together with the code used to describe it, and
these four chapters are what that sentence costs when it is taken seriously.


\chapter{Boosting shortest path multiplicity}
\label{ch:distance}

\section{The redundancy of shortest paths}

The small-world property, met back in the introduction, promises that any two
nodes of a real network are joined by a \emph{short} path --- only a few hops
apart. It says nothing, though, about how \emph{many} short paths there are.
That count turns out to matter a great deal. If the only shortest route between
two nodes runs through a single chain of intermediaries, the connection is
fragile: cut one link and the shortest route is gone. If instead there are
several shortest routes side by side, the connection is robust, and whatever
flows between the two nodes --- a message, a signal, a delay --- can travel along
all of them at once. The quantity that captures this is the \emph{shortest path
multiplicity}: the number of distinct shortest routes between a pair of nodes. A
simple picture: if two friends can each be reached from you in two handshakes,
but by two completely different chains of acquaintances, the multiplicity of that
connection is two. In real networks this multiplicity is conspicuously high.

And it is not an isolated fact. \citet{deng2026} found that, across many real
networks, the abundance of multiple shortest paths is strongly correlated with
\emph{community structure} --- the very groupings of the previous chapter ---
more strongly than with any other standard measure. Two networks of the same size
and the same degree distribution can differ widely in their path multiplicity,
and the difference tracks how community-rich they are. A correlation like that
begs for an explanation. Does community structure somehow \emph{cause} paths to
multiply, or do both spring from a common deeper source? The previous chapter
hands us the obvious suspect: local growth rules, which we saw conjure
communities out of nothing but locality. This chapter, following
\citet{vazquezpaths2026}, completes the case by showing that those same local
rules also \emph{boost} the multiplicity of shortest paths --- so that the
striking link between paths and communities is not one causing the other, but the
shared shadow cast by the rule that grew the network.

The mechanism is easy to picture once named. Local rules build short
\emph{cycles} --- closed loops --- as a matter of course: a triangle whenever a
friend introduces a friend, a square whenever a gene is copied. And a cycle is
exactly a structure that offers more than one way around. In a network with no
loops at all --- a \emph{tree} --- there is precisely one route between any two
nodes; the more loops a network carries, the more alternative shortest routes it
affords. Path multiplicity is simply what local rules leave behind when they
close loops.

\section{Defining the multiplicity}

We need only one quantity. Write $\mu_{ij}(G)$ for the number of distinct
shortest paths between nodes $i$ and $j$ in a network $G$ --- counting every route
that ties for the minimum number of hops. Averaging this over all pairs of nodes,
and over many networks grown the same way, gives the \emph{average shortest path
multiplicity} $\langle\mu\rangle$, the single number we will follow as the
network grows to size $n$. A tree, with its unique routes, has $\langle\mu\rangle
= 1$ exactly; anything above one measures the redundancy that loops provide. The
whole chapter is about one question: how fast does $\langle\mu\rangle$ rise as the
network gets bigger, and does that rate betray whether the network was grown by a
local rule?

Four growth rules will do all the work, and we have met them all before. Their
definitions matter here more than they have so far, because we are about to
compute their exponents, so let us state them exactly.

\emph{Local search}, $LS(n,\ell)$, the search rule of Chapter~\ref{ch:search} in
undirected form. Start from two connected nodes. To add a node, pick an existing
node at random, take an $\ell$-step random walk from it, and link the newcomer to
every node the walk visited. It carries preferential attachment because the chance
of being visited, beyond the entry node, is proportional to degree; and it makes at
least one triangle at every step, between the entry point and the first neighbor
visited. For $\ell=1$ it \emph{is} the triadic-closure rule of
Chapter~\ref{ch:friends}.

\emph{Duplication--split}, $DS(n,q)$, the undirected cousin of the project rule of
Chapter~\ref{ch:projects}. Start from a cycle of four nodes, so that no node has
degree below two and no cycle is shorter than four. To add a node $i$, pick an
existing node $j$ at random. With probability $q$, make $i$ a duplicate of $j$,
linking $i$ to every neighbor of $j$. Otherwise pick a random neighbor $k$ of $j$
and \emph{split} the link: delete $(j,k)$ and create $(i,j)$ and $(i,k)$. It
carries preferential attachment because the chance that a node has a neighbor
duplicated is proportional to its degree. Duplication makes no triangles and
splitting destroys any that might form; its motif is the square.

The \emph{bubble} model, $BB(n,L)$. Start from a cycle of $L+2$ nodes. At each
step add a chain of $L$ new nodes and attach the two ends of the chain to the two
ends of a randomly chosen existing link, closing a fresh cycle of length $L+2$. It
carries preferential attachment because the chance of sitting at the end of the
chosen link is proportional to degree. $BB(n,1)$ is the Dorogovtsev--Mendes rule.

Against these we set the \emph{Barab\'asi--Albert} model $BA(n,m)$: start from a
complete graph of $m+1$ nodes, and let every newcomer attach $m$ links to nodes
chosen from the \emph{whole} network with probability proportional to degree. It
is our non-local control.

\section{The simplest case: a ring}

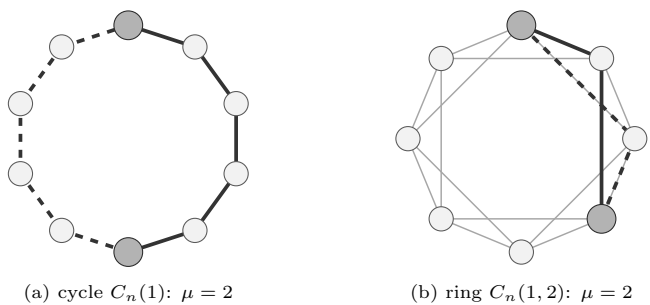
\begin{figure}[t]
\centering
\begin{adjustbox}{max width=\linewidth}\begin{tikzpicture}[
  nd/.style={circle,draw=black!65,fill=black!5,minimum size=3.2mm,inner sep=0pt},
  ep/.style={circle,draw=accent!80!black,fill=black!30,minimum size=3.8mm,inner sep=0pt},
  r1/.style={draw=black!80,line width=1.4pt},
  r2/.style={draw=black!80,line width=1.4pt,dashed},
  lnk/.style={draw=black!35,line width=0.55pt}]
  \begin{scope}
    \foreach \i in {1,...,10}{
      \pgfmathsetmacro\ang{90-(\i-1)*36}
      \coordinate (p\i) at (\ang:1.5);
    }
    \foreach \i in {1,2,3,4,5}{\pgfmathtruncatemacro\j{\i+1}\draw[r1] (p\i)--(p\j);}
    \foreach \i in {6,7,8,9}{\pgfmathtruncatemacro\j{\i+1}\draw[r2] (p\i)--(p\j);}
    \draw[r2] (p10)--(p1);
    \foreach \i in {2,3,4,5,7,8,9,10}{\node[nd] at (p\i){};}
    \node[ep] at (p1){}; \node[ep] at (p6){};
    \node[font=\scriptsize] at (0,-2.05) {(a) cycle $C_n(1)$: $\mu=2$};
  \end{scope}
  \begin{scope}[xshift=5.2cm]
    \foreach \i in {1,...,8}{
      \pgfmathsetmacro\ang{90-(\i-1)*45}
      \coordinate (q\i) at (\ang:1.5);
    }
    \foreach \a/\b in {1/2,2/3,3/4,4/5,5/6,6/7,7/8,8/1,1/3,2/4,3/5,4/6,5/7,6/8,7/1,8/2}
      {\draw[lnk] (q\a)--(q\b);}
    \draw[r1] (q1)--(q2)--(q4);
    \draw[r2] (q1)--(q3)--(q4);
    \foreach \i in {2,3,5,6,7,8}{\node[nd] at (q\i){};}
    \node[ep] at (q1){}; \node[ep] at (q4){};
    \node[font=\scriptsize] at (0,-2.05) {(b) ring $C_n(1,2)$: $\mu=2$};
  \end{scope}
\end{tikzpicture}\end{adjustbox}
\caption{\textbf{Multiplicity on a ring.} (a) On the plain cycle, the two arcs
between antipodal nodes tie, so both are shortest ($\mu=2$); every other pair has
one arc strictly shorter, hence a unique shortest path. (b) On the
next-nearest-neighbor ring, a step may cover one node or two, so for endpoints an
odd number of steps apart there are several ways to combine the steps into a
shortest path --- here the two two-hop routes from one highlighted node to the
other. Redundancy that was confined to the lone antipodal pair in (a) now appears
for pair after pair.}
\label{fig:mu-ring}
\end{figure}

Before any of that, it is worth feeling the measure on the plainest network that
has a loop at all --- the same ring we broke into communities in the previous
chapter. Place $n$ nodes on a circle, each joined to its two neighbors: the cycle
$C_n$. Between any two nodes there are two ways round, and the shortest path is
simply the shorter arc. Almost always one arc is strictly shorter than the other,
so the shortest path is \emph{unique} and $\mu_{ij}=1$. The single exception is a
pair of nodes exactly opposite each other on the circle: the two arcs then tie, and
both are shortest, so $\mu_{ij}=2$ (Figure~\ref{fig:mu-ring}).

Those antipodal pairs are rare --- there are only about $n/2$ of them among the
$\binom{n}{2}$ pairs in the ring --- so averaging gives
\begin{equation}
  \langle\mu\rangle = 1 + \frac{1}{n-1} \longrightarrow 1 .
  \label{eq:mu-ring}
\end{equation}
A plain ring, in other words, is barely more redundant than a tree. Its one loop
buys a second route for exactly one pair of nodes, the two farthest apart, and for
nobody else.

Give each node one more neighbor, though, and redundancy switches on
(Figure~\ref{fig:mu-ring}b). In the ring $C_n(1,2)$, where every node also links
to its second neighbors, a shortest path may step by one node or by two, and the
freedom to mix those steps creates ties. Between two nodes $j$ steps apart, when
$j$ is odd, there are exactly
\begin{equation}
  \mu = \frac{j+1}{2}
  \label{eq:mu-ring2}
\end{equation}
shortest paths --- one for each position of the lone single-step among the
double-steps --- while for even $j$ the all-doubles route is unique. The redundancy
now \emph{grows} with how far apart the endpoints are, instead of sitting at one,
and the average multiplicity climbs with the size of the ring rather than creeping
down to it.

That the two rings behave so differently comes down to a single number --- the same
one that decided their communities in the previous chapter. A connected network of
$n$ nodes and $m$ links carries $m-n+1$ independent cycles. The plain ring has
$m=n$: one loop, and we have just seen it lifts the multiplicity by a mere
$1/(n-1)$. The next-nearest ring has $m=2n$, so $m-n=n$ --- an \emph{extensive}
number of independent cycles --- and its multiplicity grows in step. This is
exactly the threshold at which Chapter~\ref{ch:communities} found the ring
beginning to break into communities: there, too, a network split precisely when it
carried more links than nodes. Both measures switch on together, at the same
lever. Redundant routes, like community structure, need loops in extensive number
--- supplied here by a ring of longer reach, and in the rest of this chapter by
growth rules that pile cycles up as the network grows.

\section{The baseline: random networks}

To judge what counts as ``boosted,'' we first need the expectation in the
\emph{absence} of any local structure. The natural reference is a random network,
and here we can do better than simulate --- we can calculate. The answer is
remarkably simple, and it turns on a single property of the degree distribution:
the \emph{average excess degree}
\begin{equation}
  c_2 = \frac{\langle k(k-1)\rangle}{\langle k\rangle},
  \label{eq:path-c2}
\end{equation}
which is the average number of \emph{further} links you find when you arrive at a
node by following one of its links --- its degree, less the link you came in on.
It is the natural branching factor of a network: from where you stand, $c_2$ is
how many ways the path onward can continue.

The cleanest way to see where Eq.~\eqref{eq:path-c2} comes from is to do the
simplest case first --- the classic random graph of Erd\H{o}s and R\'enyi, in which
every one of the $\binom{n}{2}$ possible links is present with the same probability
$p$ --- and then repair it for a general degree distribution.

\begin{calculation}{counting the shortest paths of a random graph}
Write $c=np$ for the mean degree and $D=\ln n/\ln c$ for the typical distance
between two nodes, and assume $c>1$ so the graph is connected. Fix two nodes. A
path of length $d$ between them uses $d-1$ intermediate nodes, in order, drawn from
the remaining $n-2$, so the number of \emph{candidate} paths is
\begin{equation}
  m_d = (n-2)(n-3)\cdots(n-d) \approx n^{\,d-1}
  \label{eq:path-md}
\end{equation}
for $d\ll n$. Each candidate is \emph{realized} only if all $d$ of its links happen
to be present, which has probability $p^d$; treating the candidates as independent,
the number realized is binomial, with mean
\begin{equation}
  \langle M_d\rangle = m_d\,p^{d} \approx \frac{(np)^{d}}{n} = c^{\,d-D}.
  \label{eq:path-Md}
\end{equation}
The shortest paths are those at the smallest $d$ for which any path exists at all,
so we must weight $\langle M_d\rangle$ by the probability $F_{d-1}$ that nothing
shorter exists:
\begin{equation}
  \langle\mu\rangle = \sum_{d}\langle M_d\rangle\,F_{d-1}.
  \label{eq:path-mu-sum}
\end{equation}
Since $\Pr(M_d=0)=(1-p^d)^{m_d}\approx e^{-\langle M_d\rangle}$ for $p\ll1$,
\begin{equation}
  F_d = \prod_{k=1}^{d}\Pr(M_k=0) = \exp\Big(-\sum_{k=1}^{d}\langle M_k\rangle\Big).
  \label{eq:path-Fsum}
\end{equation}
The sum in the exponent is a geometric series of ratio $c>1$, so it is dominated by
its last term, $\sum_{k\le d}\langle M_k\rangle\approx c^{\,d-D}$, and
\begin{equation}
  F_d \approx \exp\big(-c^{\,d-D}\big).
  \label{eq:path-Fd}
\end{equation}
Substituting Eqs.~\eqref{eq:path-Md} and \eqref{eq:path-Fd} into
Eq.~\eqref{eq:path-mu-sum},
\begin{equation}
  \langle\mu\rangle \approx \sum_{d} c^{\,d-D}\,\exp\big(-c^{\,d-D-1}\big).
  \label{eq:path-ersum}
\end{equation}
The summand is the function $x e^{-x/c}$ evaluated along a geometric ladder: it
climbs by a factor $c$ per hop while $d<D$, and is then cut down
super-exponentially. It therefore peaks sharply, one hop beyond the typical
distance, at $d=D+1$, and evaluating it there leaves
\begin{equation}
  \langle\mu\rangle \approx \frac{c}{e} + O(1),
  \label{eq:path-mfER}
\end{equation}
the $e$ being simply the value of $x e^{-x}$ at its maximum.
\end{calculation}

Equation~\eqref{eq:path-mfER} already tells us three things. If the graph is
\emph{sparse} --- $c$ fixed as $n$ grows, which holds for every real network
--- the multiplicity is a \emph{constant}. If it is \emph{minimally connected},
$c=b\ln n$, the multiplicity grows as $(b/e)\ln n$. And if it is \emph{dense},
$c\sim n$, it grows in proportion to $n$, recovering a result of
\citet{dong2025}. But a real network is sparse \emph{and} hub-dominated, and an
Erd\H{o}s--R\'enyi graph is neither. We need the same calculation for an arbitrary
degree distribution.

\begin{calculation}{the same count for any degree distribution}
Let the degrees follow $\pi_k$, with mean $c_1=\sum_k \pi_k k$ and excess degree
$c_2=\sum_k \pi_k k(k-1)/c_1$. (We write $\pi_k$ rather than $p_k$ here because $p$
is already the probability that a link exists.) The calculation runs exactly as
before; only the probability that the next link exists changes.

Build a candidate path node by node. The first step is from a node picked at
random to another node picked at random, which are linked with probability
$p_1=c_1/n$, with associated distance $D_1=\ln n/\ln c_1$. Every step after the
first is different, because we now stand at the \emph{end of a link}, and a node
reached by following a link is not a typical node: its degree is distributed as
$k\pi_k/c_1$, and one of those links is the one we arrived on, leaving $k-1$. The
next node is picked at random and has degree $k'$ with probability $\pi_{k'}$; the
chance the two are linked is $kk'/(nc_1)$, which on averaging over both gives
\begin{equation}
  p_2 = \frac{c_2 c_1}{n c_1} = \frac{c_2}{n},
  \qquad D_2 = \frac{\ln n}{\ln c_2}.
  \label{eq:path-p2}
\end{equation}
So a path of length $d$ costs one factor of $p_1$ and $d-1$ factors of $p_2$, and
Eqs.~\eqref{eq:path-Md} and \eqref{eq:path-Fd} become
\begin{equation}
  \langle M_d\rangle \approx c_1 c_2^{\,d-D_2-1},
  \qquad
  F_d \approx \exp\big(-c_1 c_2^{\,d-D_2-1}\big),
  \label{eq:path-Mdk}
\end{equation}
for $d>1$. Substituting into Eq.~\eqref{eq:path-mu-sum},
\begin{equation}
  \langle\mu\rangle \approx \sum_d c_1 c_2^{\,d-D_2-1}
  \exp\big(-c_1 c_2^{\,d-D_2-2}\big),
  \label{eq:path-ksum}
\end{equation}
whose summand again peaks sharply, now where $c_1 c_2^{\,d-D_2-2}=1$, leaving
\begin{equation}
  \langle\mu\rangle \approx \frac{c_2}{e} + O(1).
  \label{eq:path-mf}
\end{equation}
For an Erd\H{o}s--R\'enyi graph $c_2=c_1=c$ and we recover
Eq.~\eqref{eq:path-mfER}.
\end{calculation}

Equation~\eqref{eq:path-mf} is the baseline, and it is worth pausing on how much it
explains. The multiplicity of a random network is simply its branching factor,
divided by $e$. The mean degree does not appear; the whole distribution of degrees
enters through the single number $c_2$. And notice what happened between the two
boxes: $c_1$ was replaced by $c_2$ everywhere it mattered, because after the first
step a path is always standing at the end of a link, where the branching is the
\emph{excess} degree and not the degree.

Everything now hinges on how $c_2$ behaves as the network grows, and since $c_2$
involves $\langle k^2\rangle$, that is a question about hubs.

\begin{calculation}{how the excess degree grows with size}
Take a scale-free network, $p(k)\sim k^{-\gamma}$. For $\gamma>3$ the second moment
$\langle k^2\rangle$ converges, so $c_2$ tends to a constant and there is nothing
to discuss. For $\gamma\le3$ it diverges, and the divergence is cut off only by the
largest degree actually present. That largest degree follows from asking which $k$
is rare enough to appear about once in a network of $n$ nodes, $p(k_{\max})=1/n$:
\begin{equation}
  k_{\max}\sim n^{1/\gamma}.
  \label{eq:path-kmax}
\end{equation}
Now do the sum. For $\gamma=3$ the summand $k^2 p(k)\sim1/k$, so
\begin{equation}
  c_2 \sim \sum_{k=1}^{k_{\max}}\frac{1}{k} \sim \ln k_{\max} \sim \ln n,
  \label{eq:path-c2log}
\end{equation}
the logarithm of the size. For $2<\gamma<3$ the sum is dominated by its upper end,
$c_2\sim k_{\max}^{3-\gamma}$, and using Eq.~\eqref{eq:path-kmax},
\begin{equation}
  c_2 \sim n^{-1+3/\gamma}.
  \label{eq:path-c2pow}
\end{equation}
It is worth doing the $\gamma=3$ case explicitly, since three of our models share
it. There $p_k = A\,\Gamma(k)/\Gamma(3+k) = A/[(2+k)(1+k)k]$, so
\begin{equation}
  \langle k\rangle \sim A\sum_{k\ge2}\frac{1}{(2+k)(1+k)}
  = A\sum_{k\ge2}\Big(\frac{1}{k+1}-\frac{1}{k+2}\Big) \sim \frac{A}{3},
  \label{eq:path-k1exp}
\end{equation}
a telescoping sum that converges, while
\begin{equation}
  \langle k^2\rangle \sim A\sum_{k=1}^{k_{\max}}\frac{1}{k}
  \sim A\ln k_{\max} \sim \frac{A}{3}\ln n
  \label{eq:path-k2exp}
\end{equation}
does not. Their ratio is $c_2\sim\ln n$, confirming Eq.~\eqref{eq:path-c2log}.
\end{calculation}

Feeding the three cases into Eq.~\eqref{eq:path-mf} gives the baseline in full:
\begin{alignat}{2}
  \langle\mu\rangle &= a, &&\qquad \gamma>3,
  \label{eq:path-const}\\
  \langle\mu\rangle &= a + \tfrac{1}{e}\ln n, &&\qquad \gamma=3,
  \label{eq:path-log}\\
  \langle\mu\rangle &= a + b\,n^{\alpha}, \qquad \alpha = \tfrac{3}{\gamma}-1,
    &&\qquad 2<\gamma<3.
  \label{eq:path-power}
\end{alignat}
When $\gamma>3$ the multiplicity \emph{stops growing altogether}. At $\gamma=3$ it
creeps up like $\ln n$ --- about as slow as growth gets, since to add a fixed amount
you must \emph{multiply} the size --- and the predicted slope is not merely
logarithmic but exactly $1/e$. Below $\gamma=3$ the hubs are heavy enough to drive
a genuine power law. So there is no single random baseline: the bar that a local
rule must clear is set by the degree distribution it happens to produce, and we
must compare like with like.

\section{Which rule makes which exponent}

That last sentence obliges us to know the exponent of every model in this chapter.
Happily one calculation settles them all at once, because all four rules --- and
the shuffled versions we shall compare them against --- attach in the same way: a
node gains links at a rate that is affine in its degree.

\begin{calculation}{one rate equation for every model}
Let $n_k(n)$ be the expected number of nodes of degree $k$ once the network has
grown to $n$ nodes. Suppose a node of degree $k$ gains a link at rate
$(a+bk)/n$ --- a constant part $a$ and a preferential part $b$ --- and that the
newcomer arrives either with degree $2$, with probability $c$, or as a duplicate of
an existing node, at rate $d/n$. Then
\begin{equation}
\begin{split}
  n_k(n+1) = n_k(n) &+ \frac{a+b(k-1)}{n}\,n_{k-1} - \frac{a+bk}{n}\,n_k \\
  &+ c\,\delta_{k2} + \frac{d}{n}\,n_k .
\end{split}
\label{eq:path-rate}
\end{equation}
The first two terms are the traffic in and out of degree $k$; the last two account
for the newcomer's own degree. This is mean field: the expectation after the step
is computed from the expectations before it, ignoring the fluctuations of a single
addition. Look for a steady state in which the shape of the distribution is fixed
and only its size grows, $n_k(n)=n\,p_k$. Substituting and cancelling one power of
$n$ gives a recursion in $k$ alone,
\begin{equation}
  p_k = \frac{\eta+k-1}{\nu+k}\,p_{k-1}
  + \frac{c}{b}\,\frac{1}{\nu+k}\,\delta_{k2},
  \qquad
  \eta = \frac{a}{b},
  \quad
  \nu = \frac{1+a-d}{b},
  \label{eq:path-pkrec}
\end{equation}
which unrolls into a ratio of Gamma functions,
\begin{equation}
  p_k = \frac{c}{b}\,\frac{\Gamma(2+\nu)}{\Gamma(2+\eta)}\,
  \frac{\Gamma(k+\eta)}{\Gamma(k+1+\nu)} .
  \label{eq:path-pk}
\end{equation}
(Check it by induction: it holds at $k=2$, and substituting $p_{k-1}$ from
Eq.~\eqref{eq:path-pk} into Eq.~\eqref{eq:path-pkrec} returns $p_k$.) For
$k\gg\max(\eta,\nu)$ the ratio of Gammas behaves as $k^{\eta-\nu-1}$, a power law
$p_k\sim k^{-\gamma}$ with
\begin{equation}
  \gamma = 1 + \frac{1-d}{b} .
  \label{eq:path-gamma}
\end{equation}
Only two of the four constants survive: the preferential rate $b$, and the
duplication rate $d$.
\end{calculation}

Equation~\eqref{eq:path-gamma} is a small master key. Read off each rule's
constants and its exponent falls out.

For $BA(n,2)$ every newcomer brings two links and hands them out in proportion to
degree, so $a=0$, $b=1/2$, $c=1$, $d=0$, and
\begin{equation}
  \gamma_{BA} = 3 .
  \label{eq:path-gammaBA}
\end{equation}
For $LS(n,\ell)$ the walk lands on the entry node uniformly, which is the constant
part $a=1$, and thereafter in proportion to degree, giving
$b=\ell/\langle k\rangle=\ell/[2(\ell+1)]$ with $c=1$, $d=0$, so that
\begin{equation}
  \gamma_{LS} = \frac{3\ell+2}{\ell} .
  \label{eq:path-gammaLS}
\end{equation}
For $DS(n,q)$ duplication is the only source of preference, $b=q$, and it is also
the only way the newcomer copies a degree, $d=q$, with $a=0$ and $c=1-q$; then
$\eta=0$, $\nu=-1+1/q$ and
\begin{equation}
  \gamma_{DS} = \frac{1}{q} .
  \label{eq:path-gammaDS}
\end{equation}
For $BB(n,L)$ the chain's ends land on a random link, so $b=2/\langle
k\rangle=1/(L+1)$ with $a=0$, $c=1$, $d=0$, $\eta=0$, $\nu=L+1$, and
\begin{equation}
  \gamma_{BB} = L+2 .
  \label{eq:path-gammaBB}
\end{equation}

Two coincidences in this list are worth having. First, $DS(n,q)$ and $BB(n,L)$
produce the \emph{same} degree distribution whenever $q=(L+2)^{-1}$ --- so any
difference in their multiplicity cannot be blamed on their degrees. Second, and
this is what makes the comparisons of the next section fair, $DS(n,1/3)$,
$BB(n,1)$ and $BA(n,2)$ all sit at $\gamma=3$: three rules, one degree
distribution, and --- as we are about to see --- three quite different
multiplicities.

\begin{figure}[t]
\centering
\includegraphics[width=\linewidth]{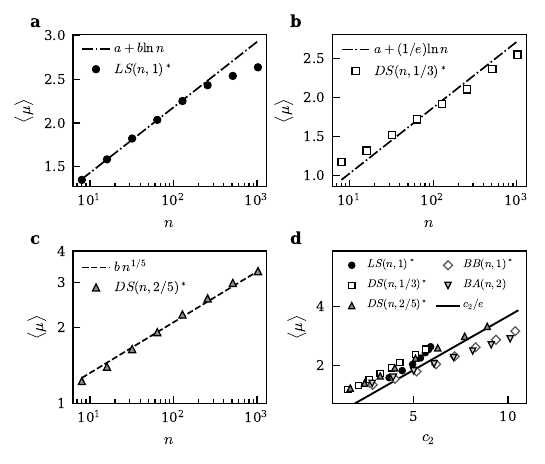}
\caption{\textbf{The random baseline.} Average shortest path multiplicity
$\langle\mu\rangle$ against network size $n$ for randomized networks (starred),
which keep each node's degree but destroy all local structure. (a) $LS(n,1)^*$,
degree exponent $\gamma=5$: growth stalls, the points falling away from the
fitted logarithm. (b) $DS(n,1/3)^*$, $\gamma=3$: logarithmic, near the predicted
slope $1/e$. (c) $DS(n,2/5)^*$, $\gamma=5/2$: the predicted power law $n^{1/5}$.
(d) All models, including the non-local $BA(n,2)$, plotted against the excess
degree $c_2$ collapse onto the mean-field line $c_2/e$ of
Eq.~\eqref{eq:path-mf}.}
\label{fig:mu-random}
\end{figure}

And now the baseline can be made concrete. Take a network grown by a local rule and
shuffle its links at random while leaving every node's degree untouched --- the same
configuration-model scramble used as a control in the previous chapter. The result
keeps the degree distribution and destroys everything else; we mark it with a star.
By Eq.~\eqref{eq:path-gammaLS} the randomized local-search network $LS(n,1)^*$ has
$\gamma=5$, and so by Eq.~\eqref{eq:path-const} its multiplicity must
\emph{stop growing}; it duly flattens out (Figure~\ref{fig:mu-random}a). By
Eq.~\eqref{eq:path-gammaDS} the randomized $DS(n,1/3)^*$ has $\gamma=3$, and creeps
up logarithmically, close to the predicted slope $1/e$
(Figure~\ref{fig:mu-random}b). The randomized $DS(n,2/5)^*$ has $\gamma=5/2$, for
which Eq.~\eqref{eq:path-power} predicts $\alpha=3/\gamma-1=1/5$, and it climbs as
$n^{1/5}$ (Figure~\ref{fig:mu-random}c). Plotted against $c_2$ rather than $n$, all
of them --- and the Barab\'asi--Albert model with them --- collapse onto the single
mean-field line $c_2/e$ of Eq.~\eqref{eq:path-mf} (Figure~\ref{fig:mu-random}d).
Three regimes, one formula, and no fitted parameter anywhere except the additive
constants.

The Barab\'asi--Albert model deserves its own note, because it is a
\emph{growing} network yet lands squarely on the random baseline. With $m=2$ it
has $\gamma=3$ by Eq.~\eqref{eq:path-gammaBA}, and its multiplicity creeps up
logarithmically, exactly as its reshuffled cousins do. It has no local structure to
lose: its newcomers attach across the whole graph and never consult a neighborhood.
This echoes the previous chapter, where the very same model failed to develop
communities. Lacking locality, it builds neither the modules nor the redundant
paths, and on both counts it behaves as though its links had been shuffled at
random.
\section{Local rules boost the multiplicity}

\begin{figure}[t]
\centering
\includegraphics[width=\linewidth]{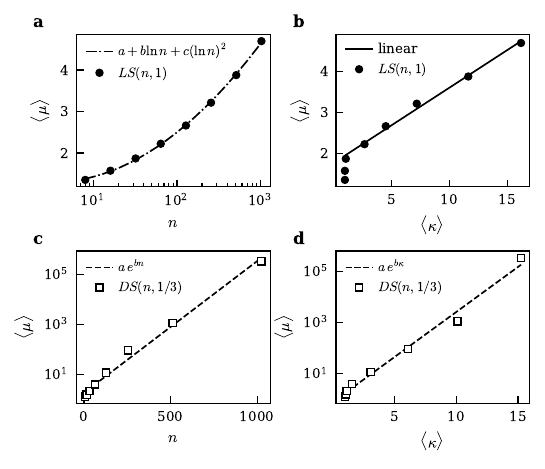}
\caption{\textbf{Local rules boost the multiplicity.} (a) Local search
$LS(n,1)$ grows log-quadratically, Eq.~\eqref{eq:path-typeI} --- while its
shuffled counterpart, with identical degrees, saturates
(Figure~\ref{fig:mu-random}a). (b) The same data against the inferred number of
communities $\langle\kappa\rangle$: linear. (c) Duplication--split $DS(n,1/3)$
grows exponentially, Eq.~\eqref{eq:path-typeII}; note the linear size axis and
the logarithmic multiplicity axis, and that the vertical scale spans five decades.
(d) Against $\langle\kappa\rangle$: exponential again. Communities are only
detectable beyond the Ramsey size $r_\kappa$, so the leftmost points in (b) and
(d) sit at $\langle\kappa\rangle=1$.}
\label{fig:mu-local}
\end{figure}

Now run the same models \emph{without} shuffling, and compare each against its own
starred baseline (Figure~\ref{fig:mu-local}). The multiplicity climbs decisively
faster. It helps to keep a ladder of growth rates in mind: constant, then
logarithmic, then \emph{log-quadratic}, then \emph{exponential} --- each rung
leaving the one below it far behind.

Most local rules --- local search and the bubble model among them --- sit on the
log-quadratic rung,
\begin{equation}
  \langle\mu\rangle = a + b\ln n + c(\ln n)^2,
  \label{eq:path-typeI}
\end{equation}
the extra squared term bending the curve clearly upward. For $LS(n,1)$ this is a
particularly sharp contrast: the unshuffled network grows without limit while the
shuffled one, with the very same degrees, stops growing altogether. Whatever is
doing the work is not in the degree distribution at all.

The duplication--split rule is different, and dramatically so: its multiplicity
grows \emph{exponentially} with size,
\begin{equation}
  \langle\mu\rangle = a\, e^{bn}, \qquad a,b>0.
  \label{eq:path-typeII}
\end{equation}
Exponential growth is a different world from logarithmic. Over the same range of
sizes in which its shuffled counterpart moves from $\langle\mu\rangle\approx1.2$
to $2.6$, the unshuffled $DS(n,1/3)$ runs from $1.3$ to more than $10^5$: by a
thousand nodes, a typical pair of nodes is joined by hundreds of thousands of
equally short routes. Copying, it turns out, is the great multiplier of shortest
paths, and that is what the title of this chapter names.

\section{Cycles and the boost}

Why should duplication outrun the rest? The answer is in the kind of loop each
rule makes. The triadic-closure rule of the friends chapter closes
\emph{triangles}, loops of length three --- but in a triangle the two nodes
flanking a third are themselves directly linked, so there is still only \emph{one}
shortest route between any pair. A triangle adds cohesion but no extra routes.
Duplication, by contrast, closes \emph{squares}, loops of length four, and a
square is the smallest loop that genuinely offers a choice: its two opposite
corners are two hops apart by \emph{two} different paths, one around each side.
That single observation is the seed of the boost; the box pushes it a little
further, and the takeaway follows.

\begin{figure}[t]
\centering
\begin{adjustbox}{max width=\linewidth}\begin{tikzpicture}[
  nd/.style={circle,draw=black!65,fill=black!4,minimum size=6mm,inner sep=0pt},
  ep/.style={circle,draw=accent!80!black,fill=black!30,minimum size=6mm,inner sep=0pt},
  lnk/.style={draw=black!45,line width=0.8pt},
  r1/.style={draw=black!80,line width=1.5pt},
  r2/.style={draw=black!80,line width=1.5pt,dashed}]
  \begin{scope}
    \node[ep] (A) at (0,1.0)    {};
    \node[nd] (B) at (-0.95,-0.5){};
    \node[ep] (C) at (0.95,-0.5){};
    \draw[lnk] (A)--(B);
    \draw[lnk] (B)--(C);
    \draw[r1]  (A)--(C);
    \node[font=\footnotesize] at (0,-1.4) {(a) triangle: one route};
  \end{scope}
  \begin{scope}[xshift=5.4cm]
    \node[ep] (A2) at (0,1.05)   {};
    \node[nd] (B2) at (1.15,0)   {};
    \node[ep] (C2) at (0,-1.05)  {};
    \node[nd] (D2) at (-1.15,0)  {};
    \draw[r1] (A2)--(B2); \draw[r1] (B2)--(C2);
    \draw[r2] (A2)--(D2); \draw[r2] (D2)--(C2);
    \node[font=\footnotesize] at (0,-1.55) {(b) square: two routes};
  \end{scope}
\end{tikzpicture}\end{adjustbox}
\caption{\textbf{Why squares multiply routes.} The two highlighted corners are the
endpoints. (a) In a triangle they are joined directly --- a single shortest route.
(b) In a square they are opposite corners, two steps apart by \emph{two} distinct
routes (drawn solid and dashed), one around each side. Local rules that close
four-cycles, such as gene
duplication, thread the network with these redundant shortest paths.}
\label{fig:path-square}
\end{figure}
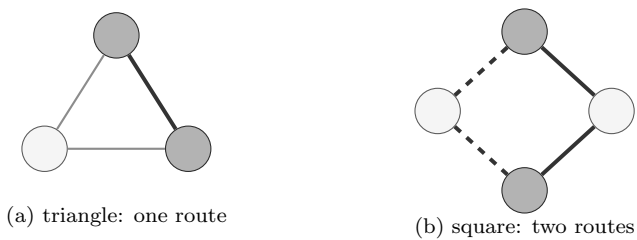

Figure~\ref{fig:path-square} makes the contrast concrete.

\begin{calculation}{why copying multiplies routes}
When a node is duplicated, the copy and the original share all their neighbors,
so any two of those neighbors sit at opposite corners of a four-cycle --- each
such pair instantly gains a second shortest route. A duplicated node of degree
$k$ creates $\tbinom{k}{2}=k(k-1)/2$ such squares in a single stroke. Moreover the
duplication rule carries its own preferential attachment, the chance of copying a
node being proportional to its degree, which (as in Chapter~\ref{ch:projects})
builds a heavy-tailed degree distribution with exponent $\gamma=1/q$. Squares are
thus laid down in proportion to $k(k-1)$ --- the very combination that
Eq.~\eqref{eq:path-c2} says controls the multiplicity --- on top of a distribution
with many high-degree nodes. Each round of copying compounds the squares laid down
by the last, and it is this compounding, rather than any one square, that is the
suspected origin of the exponential law, Eq.~\eqref{eq:path-typeII}, though a full
account of it remains open.
\end{calculation}

\begin{figure}[t]
\centering
\includegraphics[width=\linewidth]{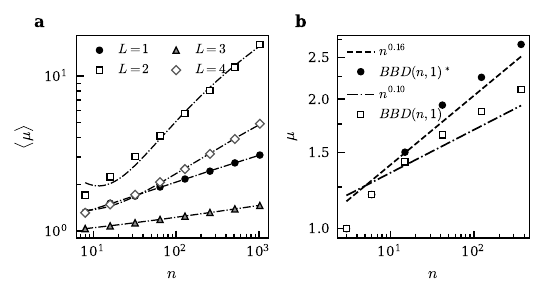}
\caption{\textbf{Cycle length and randomness.} (a) The bubble model $BB(n,L)$,
which closes loops of length $L+2$. Even $L$ (2 and 4) reaches a higher
multiplicity than odd $L$ (1 and 3), as the square-counting predicts --- yet every
$L$ stays log-quadratic, so cycle geometry alone does not buy the exponential.
(b) The deterministic bubble model $BBD(n,1)$ follows a power law $n^{0.10}$ ---
found exactly in Section~\ref{sec:mu-exact} --- falling \emph{below} its own
randomized version $BBD(n,1)^*$, which grows faster, as $n^{0.16}$
(Eq.~\eqref{eq:path-power}). Locality without randomness is not enough.}
\label{fig:mu-bubble}
\end{figure}

The lesson of the box is that squares, not triangles, are what multiply shortest
routes. The bubble model lets us test that directly, since it dials the loop
length by hand: a chain of $L$ added nodes makes a loop of length $L+2$. The
geometric intuition survives the test (Figure~\ref{fig:mu-bubble}a). Even-length
loops, whose opposite corners enjoy two equal routes, enrich the network more than
odd ones: $L=2$ and $L=4$ sit well above $L=1$ and $L=3$. But the test also sets a
limit on how much the geometry explains. Every bubble model, whatever its loop
length, stays on the log-quadratic rung. Even loops beat odd ones, and squares
beat triangles --- yet neither buys the exponential. The square-counting explains
why duplication \emph{outruns} triadic closure; the full ferocity of
duplication--split needs the compounding above.

One more clue shows that randomness itself plays a role
(Figure~\ref{fig:mu-bubble}b). The Dorogovtsev--Goltsev--Mendes ``pseudofractal''
network \citep{dorogovtsevgoltsev2002} is a \emph{deterministic} relative of the
bubble model, grown by a fixed clockwork rule rather than a random one: at every
step, new nodes are added to \emph{every} link at once. Its degree distribution is
scale-free with $\gamma = 1+\ln 3/\ln 2\approx 2.6$, so by
Eq.~\eqref{eq:path-power} its randomized version should show a power law with
$\alpha\approx0.16$ --- and it does. The deterministic original is a power law too,
but a \emph{slower} one. Being deterministic it can be solved exactly, as the next
section does, and the exponent comes out at $\alpha=0.100$, comfortably below the
$0.161$ of its own shuffle. So the deterministic construction falls \emph{below}
the random network built from its own degrees. The speed-up is not a property of
the loop geometry alone; a clockwork rule that closes loops everywhere at once does
worse than chance. It is the interplay of randomness and locality together, not
either by itself, that boosts the multiplicity.

\section{How fast could it possibly grow?}
\label{sec:mu-exact}

If randomness and locality together drive the multiplicity up, it is fair to ask
where the ceiling is. The stochastic rules can only be simulated, but two
\emph{deterministic} relatives of the bubble model can be solved exactly, and set
against each other they reveal what actually controls the growth. Both replace
every link at every step, so both are clockwork versions of the bubble rule; they
differ in a single detail. In the first, the subdivided link is \emph{kept}; in the
second, it is \emph{removed}. That one choice, it turns out, is the difference
between a power law and a stretched exponential.

Keep the link, and you have the deterministic bubble --- the pseudofractal web met
a moment ago. Start from a triangle and, at each step, add a new node across every
existing link while leaving the old links in place.

\begin{calculation}{the deterministic bubble grows as a power law}
The construction renormalizes: generation $G_{t+1}$ is three copies of $G_t$ glued
in pairs at the three corners of the original triangle. Because the two ends of
every subdivided link stay adjacent, geodesics inside a copy are undisturbed by the
rest of the graph, and the total number of geodesics $S_t=\sum_{i<j}\mu_{ij}$ obeys
a closed linear recursion. Its growth ratio converges to
\begin{equation}
  \frac{S_{t+1}}{S_t}\longrightarrow \lambda = 10.0494\ldots,
  \label{eq:bbd-lambda}
\end{equation}
the largest root of $\lambda^3-12\lambda^2+20\lambda-4=0$. Each generation triples
the node count and so multiplies the number of \emph{pairs} by nine; the average
multiplicity $\langle\mu\rangle=S_t/\binom{n_t}{2}$ therefore grows by $\lambda/9$
per generation. Since the size itself grows by a factor of three, this is a power
law in $n$,
\begin{equation}
  \langle\mu\rangle \sim n^{\alpha}, \qquad
  \alpha = \frac{\ln(\lambda/9)}{\ln 3} = 0.1004\ldots,
  \label{eq:bbd-alpha}
\end{equation}
matched to a relative error of $10^{-6}$ by the twelfth generation. So what looked
logarithmic over the reachable sizes (Figure~\ref{fig:mu-bubble}b) is in fact a
slow power law --- and slower than the $n^{0.161}$ of the network's own
degree-preserving shuffle, confirming from the exact side that the deterministic
rule lags its own randomization.

The average is not the whole story. The \emph{largest} multiplicity anywhere in
the graph grows faster than the mean: it is exactly the Fibonacci number
$\max_{ij}\mu_{ij}=F_{t+2}$, which in terms of the size is
$n^{\ln\varphi/\ln 3}=n^{0.438}$, with $\varphi=(1+\sqrt5)/2$ the golden ratio.
Even in this most sluggish of our constructions, the best-connected pair of nodes
already accumulates routes at a respectable clip.
\end{calculation}

\begin{figure}[t]
\centering
\includegraphics[width=\linewidth]{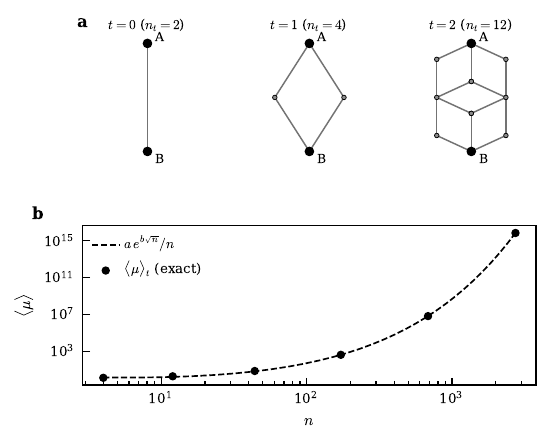}
\caption{\textbf{The diamond lattice, solved exactly.} (a) The diamond lattice:
every link is replaced, at every generation, by two parallel two-link paths. (b)
Its average multiplicity, computed exactly, follows the stretched exponential of
Eq.~\eqref{eq:diamond-n} across fifteen decades. With $2732$ nodes, a typical pair
of nodes is already joined by some $7\times10^{15}$ equally short routes --- the
fastest of the deterministic constructions, though still slower than the
exponential growth of duplication--split.}
\label{fig:mu-diamond}
\end{figure}

Now remove the link instead. Replacing rather than subdividing gives the
\emph{diamond lattice} \citep{migdal1975,kadanoff1976}. In its general $(b,s)$
form, start with a single link joining two poles $A$ and $B$ and, at each
generation, replace \emph{every} link by $b$ parallel paths of $s$ links each. The
standard diamond is $b=s=2$: every link becomes two parallel two-link paths
(Figure~\ref{fig:mu-diamond}a). It is the same lattice on which
Chapter~\ref{ch:communities} watched the Ramsey community number become a
renormalization-group crossing.

\begin{calculation}{the multiplicity between the poles}
The construction is exactly renormalizable, because distinct copies of the previous
generation meet only at their corners. Read it backwards --- deflation --- and
generation $G_t$ is $b$ parallel branches, each made of $s$ copies of $G_{t-1}$ in
series. Two rules suffice. Along a series, lengths add and multiplicities
\emph{multiply}, since a route must make each choice in turn. Across $b$ parallel
branches of \emph{equal} length, multiplicities \emph{add}, since each branch offers
its own routes and all tie. So the pole-to-pole distance $D_t$ and multiplicity
$P_t\equiv\mu_{AB}$ obey
\begin{equation}
  D_t = s\,D_{t-1} = s^{\,t},
  \qquad
  P_t = b\,P_{t-1}^{\,s},
  \label{eq:diamond-rec}
\end{equation}
with $P_0=1$. The second is nonlinear, but taking $\log_b$ linearizes it,
$\log_b P_t = 1 + s\log_b P_{t-1}$, a geometric recursion whose solution is
\begin{equation}
  P_t = b^{\,(s^{t}-1)/(s-1)}
  \;\xrightarrow{\ b=s=2\ }\;
  2^{\,2^{t}-1} = 2^{\,D_t-1},
  \label{eq:diamond-pole}
\end{equation}
a \emph{doubly} exponential count. (Equation~\eqref{eq:diamond-pole} has been
checked against direct construction of the lattice for $(b,s)$ equal to $(2,2)$,
$(3,2)$, $(2,3)$, $(4,2)$ and $(3,3)$.) The reason is transparent: a shortest route
must thread $(s^t-1)/(s-1)$ successive scales, and at each one it independently
picks one of $b$ parallel branches, so the choices multiply.
\end{calculation}

The poles are a special pair, so the average over \emph{all} pairs needs its own
argument --- and it is here that the diamond's rigidity pays off again.

\begin{calculation}{from the poles to the average}
The key geometric fact is that every node lies on some pole-to-pole geodesic,
\begin{equation}
  d(v,A) + d(v,B) = D_t ,
  \label{eq:diamond-onpath}
\end{equation}
proved by induction on $t$: within a replaced link the two poles of the sub-copy
inherit the property, and any detour around a parallel branch is strictly longer.
So a node is characterized by the single number $d(v,A)$ --- the lattice is, in
this sense, one-dimensional. Now split the pairs into those internal to one copy of
$G_{t-1}$ and those spanning two copies. The total geodesic count
$S_t=\sum_{u<v}\mu_{uv}$ then satisfies
\begin{equation}
  S_t = 4\,S_{t-1} + \mathrm{Inter}_t ,
  \label{eq:diamond-S}
\end{equation}
the four being the number of copies in the $b=s=2$ cell, and $\mathrm{Inter}_t$ a
convolution of the two copies' pole-path counts, weighted by the factor $P_{t-1}$
that any route pays to cross the shared corner. Carrying the recursion out exactly
gives the average $\langle\mu\rangle_t = S_t/\binom{n_t}{2}$ as
\begin{equation}
  \langle\mu\rangle_t \;\sim\; \kappa\,\frac{2^{\,2^{t}}}{4^{t}}
  \;\approx\; 2.1\,\frac{P_t}{n_t},
  \qquad \kappa \to 1.576 ,
  \label{eq:diamond-avg}
\end{equation}
where the prefactor $\kappa$ is the fixed point of the corner-transit convolution.
So the average merely tracks the pole value divided by the node count: the sum is
dominated by near-antipodal pairs, which inherit a geodesic count of order
$P_{t-1}^2\sim P_t$.
\end{calculation}

Equation~\eqref{eq:diamond-avg} is expressed in the generation $t$; the scaling with
size follows by inverting the node count.

\begin{calculation}{the stretched exponential}
Generation $t$ carries
\begin{equation}
  n_t = \tfrac{1}{3}\big(2\cdot4^{t}+4\big) \simeq \tfrac{2}{3}\,4^{t},
  \label{eq:diamond-nt}
\end{equation}
so $4^{t}\simeq\tfrac32 n$ and therefore $2^{t}\simeq\sqrt{3n/2}$. Substituting both
into Eq.~\eqref{eq:diamond-avg}, the doubly exponential in $t$ becomes a
\emph{stretched exponential} in $n$:
\begin{equation}
  \langle\mu\rangle \approx a\,e^{\,b\sqrt{n}},
  \qquad
  b = \sqrt{\tfrac32}\,\ln 2 \approx 0.849,
  \qquad
  a = \tfrac{2\kappa}{3} \approx 1.05 .
  \label{eq:diamond-n}
\end{equation}
The rate $b$ is exact; $a$ is the amplitude of the leading term, since
$\ln\langle\mu\rangle\simeq b\sqrt{n}+\ln a$ and the algebraic factor $1/n$ from
Eq.~\eqref{eq:diamond-avg} is subleading against $\sqrt{n}$. Retaining it, the
reduced form $\langle\mu\rangle\simeq a\,e^{\,b\sqrt{n}}/n$ reproduces the exact
values to better than $1\%$ from $t=4$ on.
\end{calculation}

The numbers are worth seeing, because they are exact rather than fitted
(Figure~\ref{fig:mu-diamond}b). By the sixth generation the diamond has only
$2732$ nodes, yet an average pair is joined by some $7\times10^{15}$ distinct
shortest paths. This is the fastest growth among the constructions we can solve by
hand --- but it is not, in fact, the fastest in the chapter. Duplication--split
grows \emph{exponentially} in the size, Eq.~\eqref{eq:path-typeII}, and an
exponential in $n$ outruns a stretched exponential in $\sqrt{n}$. Perfect,
maximally symmetric, deterministic branching is a formidable generator of
redundant routes, yet a stochastic rule that copies whole neighborhoods is more
formidable still. There is no ceiling here; there is a hierarchy, and copying sits
at the top of it.

The two exactly solvable constructions, placed side by side, say exactly what the
gap between them requires. Both are deterministic, both self-similar, both replace
every link at every step; they differ only in whether the subdivided link
survives. Keep it, and the two ends of every bubble stay one hop apart, so
distances grow only logarithmically and the multiplicity crawls up the power law
of Eq.~\eqref{eq:bbd-alpha}. Remove it, and the poles separate as $D_t=s^t$, so a
route must make an independent branch choice at every one of a growing number of
scales, and those choices compound into the stretched exponential of
Eq.~\eqref{eq:diamond-n}. It is not the closing of a cycle that decides how fast
routes multiply, but whether the cycle \emph{separates} the two nodes it joins.
A triangle joins adjacent nodes and adds nothing; a square holds its corners two
hops apart and doubles their routes; and a rule that drives its endpoints
geometrically apart, scale after scale, is what makes the count run away.

\section{Two sides of one coin}

Because local rules also create communities, and the number of communities grows
along with the network, we can ask how the multiplicity depends on the number of
communities directly. The relationship is a clean one
(Figure~\ref{fig:mu-local}b and d): linear for every local model tested, and ---
once again the exception --- exponential for duplication--split. This is the
empirical correlation of \citet{deng2026} restated from the inside: more
communities go hand in hand with more shortest paths.

The two exactly solvable constructions turn this from a plausible claim into a
demonstration. Both carry communities, and --- as Chapter~\ref{ch:communities}
found for each --- both carry the \emph{same} number of them, growing as
$\langle\kappa\rangle\sim\sqrt{n}$. Their community structure is therefore
identical in this respect, while their multiplicities could hardly differ more:
one a slow power law in size, the other a stretched exponential. Eliminating the
size between multiplicity and community count lays the two laws side by side,
\begin{equation}
  \langle\mu\rangle \sim \langle\kappa\rangle^{0.201}
  \quad\text{(bubble)},
  \qquad
  \langle\mu\rangle \sim a\,e^{\,b'\langle\kappa\rangle}
  \quad\text{(diamond)},
  \label{eq:path-mukappa}
\end{equation}
a weak power of the community count in the one case, an exponential in it in the
other. Were communities the \emph{cause} of redundant paths, two networks with the
same community growth would have to show the same $\langle\mu\rangle$ against
$\langle\kappa\rangle$. They flatly do not. The correlation is genuine in both, but
its very shape is inherited from the rule that grew the network, not from any
traffic between the paths and the communities.

But the point of the chapter is that this partnership is not a case of one
causing the other. ``Local rules boost shortest path multiplicity'' and
``communities accompany high shortest path multiplicity'' are two true statements
about the same networks, and as descriptions they say the same thing. Only one of
them, however, names a \emph{mechanism}. Communities are something an algorithm
reads off a finished network after the fact; the local rule is the process that
actually built that network, node by node, closing the very loops that both carve
it into communities and thread it with redundant routes. Community structure and
path multiplicity are two sides of a single coin --- and the coin is locality.

\section{Real networks}

\begin{figure}[t]
\centering
\includegraphics[width=\linewidth]{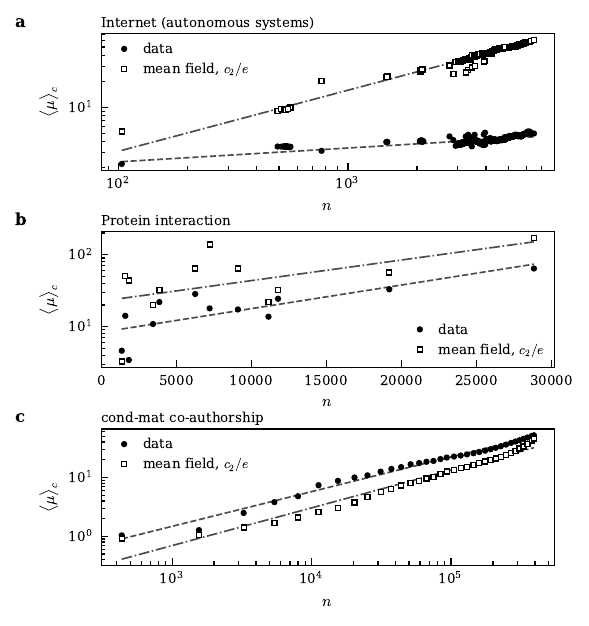}
\caption{\textbf{Real networks.} Average multiplicity over connected pairs,
$\langle\mu\rangle_c$ (Eq.~\eqref{eq:path-muc}), against size, with the mean-field
expectation $c_2/e$ for a random network of the same degrees.
(a) The Internet at the autonomous-system level: a power law, falling below the
random expectation --- the mark of disassortative, star-like wiring. (b) Protein
interaction networks of a dozen organisms: exponential growth, the duplication
signature, reaching $\langle\mu\rangle_c\approx64$ for the human network. (c)
\texttt{cond-mat} co-authorship: a power law lying \emph{above} the random
expectation, as assortative mixing among prolific authors provides many equally
short routes.}
\label{fig:mu-real}
\end{figure}

Models are one thing; the networks we started from are another. Three systems can
be caught at different stages of their growth, which is what we need if we are to
watch $\langle\mu\rangle$ move with $n$ (Figure~\ref{fig:mu-real}). The Internet
at the level of autonomous systems, mapped daily over several years
\citep{snapnets}. The protein interaction networks of a dozen organisms
\citep{stark2006}, which stand in for successive stages of evolution. And the
co-authorship network of the \texttt{cond-mat} preprint archive
\citep{cornellarxiv}, accumulated year by year. In each case we can also compute
the excess degree $c_2$ and so plot the mean-field expectation
$c_2/e$ alongside the measurement --- the random network with the same degrees.
The data are noisy and the systems are not stationary, so no strong claim is
available; but the trends are instructive.

One measurement choice needs stating, because these networks --- unlike the models
--- are not connected, and the connected fraction itself changes as they grow.
Averaging the multiplicity over \emph{all} pairs would then confound two things:
the redundancy of routes, which is what we are after, and the swelling of the
giant component, which is not. So we average only over pairs that are actually
joined, writing
\begin{equation}
  \langle\mu\rangle_c = \frac{\langle\mu\rangle}{\psi},
  \label{eq:path-muc}
\end{equation}
with $\psi$ the fraction of node pairs connected at all --- the convention of
\citet{deng2026}. The correction is nothing for the Internet, where $\psi=1$
throughout, but real for the rest: $\psi$ runs from $0.64$ to $1$ across the
protein networks and from $0.007$ to $0.89$ across the co-authorship snapshots,
the smallest values belonging to the earliest and sparsest. Reporting
$\langle\mu\rangle_c$ keeps the trends below about routes between reachable nodes,
not about how many nodes have become reachable.

In every system the multiplicity rises with size, and in every system it is
high: the human protein network reaches $\langle\mu\rangle_c\approx64$, meaning a
typical connected pair of proteins is joined by some sixty-four equally short
chains of interaction. The protein networks are also the ones that climb
\emph{exponentially} with size, which is the signature of duplication --- and gene
duplication is, of course, exactly how they are thought to grow
(Chapter~\ref{ch:duplication}). The Internet and the co-authorship network both
climb as power laws, more sedately.

The most interesting feature, though, is which side of the mean-field line each
system falls on. The Internet sits \emph{below} its random expectation, and it is
\emph{disassortative}, its hubs presiding over peripheries of poorly connected
nodes; such a star-like arrangement gives a route no alternatives, since
everything must pass through the hub. The co-authorship network sits \emph{above}
it, and it is \emph{assortative}, its well-connected authors writing with one
another, a clique of hubs being exactly what furnishes many equally short ways
across. The protein networks are disassortative and duly fall below the line. It
is tempting to read the deviation straight off the correlations --- disassortative
below, assortative above --- but that reading is a coincidence of these three
systems, not a law. The exactly solvable constructions settle it: the deterministic
bubble and the diamond are \emph{both} disassortative, yet the first falls far
below the mean-field line and the second towers enormously above it. The sign of
the correlations does not fix the sign of the deviation. What does is the
architecture the growth rule builds --- something the mean-field calculation, which
knows the network only through $c_2$, cannot yet capture. It is one of the clearer
invitations this book leaves open.

\section{Summary}

The shortest paths of real networks are richly redundant, and that redundancy
travels with their community structure. For a random network the redundancy is no
mystery: it is the branching factor, $\langle\mu\rangle\approx c_2/e$, and it
grows only as fast as the hubs allow --- not at all when $\gamma>3$,
logarithmically at $\gamma=3$, as a power law below it. Networks built by local
rules leave that baseline behind, log-quadratically for most rules and
exponentially for duplication, which closes the four-cycles that give opposite
corners two shortest routes apiece. Neither cycle geometry nor locality is
sufficient by itself: a deterministic rule that closes loops everywhere at once
does worse than chance. Two such rules can be solved exactly, and what separates
their power law from their stretched exponential is not the closing of a cycle but
whether the cycle drives its endpoints apart. Fastest of all, though, is none of
the clockwork constructions but the stochastic duplication--split rule, whose
exponential growth outruns even the diamond lattice: copying whole neighborhoods
is the most efficient generator of redundant routes we have found. Since the same
local rules also produce the communities, the observed tie between path
multiplicity and communities is best understood not as cause and effect but as two
consequences of one underlying local dynamics.


\chapter{Outlook}
\label{ch:outlook}

\section{One principle, many faces}

We have now followed a single idea through six concrete settings, and it is worth
gathering the threads before asking what lies beyond them. In every case the
network grew by a \emph{local} rule --- a newcomer acting only on a node it had
reached and that node's immediate neighbors --- and in every case the same family
of global regularities emerged unbidden.

Searchable networks, such as the citation graph and the World Wide Web, grow by
exploration: a newcomer surfs along links and connects to what it finds. Modeling
the surfer as the very random walk that defines Google's PageRank, we found that a
page's visiting probability rises in a straight line with its in-degree, so that
the local act of walking manufactures an effective ``rich get richer.'' From it
follow a power-law degree distribution whose exponent is fixed purely by how the
network is explored, a clustering hierarchy in which cliquishness falls with
degree, and generically disassortative correlations
(Chapter~\ref{ch:search}).

Social networks grow by introduction: a friend introduces a friend, closing a
triangle. Formalizing the friend-of-a-friend as a \emph{potential link}, born in
proportion to degree, we recovered the same rich-get-richer and the same
clustering hierarchy --- but now with \emph{assortative} correlations, the
well-connected keeping one another's company (Chapter~\ref{ch:friends}). The
contrast with the searchable networks is that chapter's quiet lesson: the sign of
the degree correlations is not a free parameter but a readout of whether the rule
reaches outward, to the periphery, or inward, within a neighborhood.

Biological networks grow by copying: a gene is duplicated, its protein inheriting
all of the ancestor's interactions, and the copies then diverge. Because a node
gains a link whenever one of its neighbors is duplicated, copying too is a
rich-get-richer engine; self-interactions seed the clustering hierarchy, and
coupling divergence to duplication adds a density transition and a multifractal
degree distribution born of inheritance, alongside the disassortative mixing
measured in real protein networks (Chapter~\ref{ch:duplication}). Here the three
mechanisms met in one sentence: a rule that attaches to a \emph{neighbor} of a
randomly chosen node attaches in proportion to degree, because a well-connected
node simply belongs to more neighborhoods. Preferential attachment, the
clustering hierarchy, and the degree correlations are all the generic harvest of
locality.

The same copying, carried into project schedules, refines each activity either by
duplicating it into parallel copies or by splitting it into specialized
sequential parts. The rule predicts identical power laws for an activity's
predecessors and successors, a critical path that vanishes as a fraction of large
projects, and a fractal rather than small-world geometry --- all borne out in
seventy-seven real construction projects --- and it relocates the risk of delay
from the celebrated critical path to a percolation threshold set by the whole
network's average degree (Chapter~\ref{ch:projects}).

Finally, locality proved to shape structure above the level of the individual
node. The same rules that set the degrees also segregate the network into
communities --- even when every node is identical --- a fact we made precise
through the Ramsey community number, the size beyond which communities are all but
certain (Chapter~\ref{ch:communities}). And not only whether communities appear,
but how many: across close to a hundred real networks the number of communities
was found to grow as a power law of size, $B\propto n^{\beta}$ with $\beta$ between
one half and two thirds --- super-fractal, and reproduced by the same local rules.
Locality also threads the network with redundant shortest paths, multiplying the
routes between nodes far faster than chance and in lockstep with those very
communities (Chapter~\ref{ch:distance}).

And with that, the book comes full circle. We set out asking where the global
regularities of real networks come from --- their power-law degrees, their
small-world distances, their clustering hierarchy, their degree correlations ---
and answered, one mechanism at a time, that none of them need be imposed from
above: each emerges from a way of growing in which no node ever uses more than
local knowledge. A node that surfs to a neighbor, befriends a friend, is copied
from an ancestor, or refines a task asks nothing of the network as a whole --- and
yet the networks these humble acts build carry, in the aggregate, the full and
unmistakable signature of real complex systems.

\section{What remains to be done}

If the thesis of this book is right, it opens more questions than it closes.

Several are concrete. The exponential growth of shortest path multiplicity under
the duplication--split rule --- so much faster than the log-quadratic growth of
every other rule we studied --- still wants a full explanation; the
square-counting argument shows why copying outruns triadic closure, but not why it
runs away exponentially. The exactly solvable constructions have sharpened the
question rather than closed it. They show that what governs the speed is not the
closing of a cycle but whether the cycle drives its endpoints apart: retaining the
subdivided link gives a power law, removing it a stretched exponential
(Chapter~\ref{ch:distance}). And they show that even the fastest of them, the
diamond lattice with its stretched exponential in $\sqrt{n}$, is beaten by the
stochastic copying rule, whose growth is exponential in $n$ outright. So the
open question is pointed: why does duplicating whole neighborhoods outrun even
maximally symmetric deterministic branching?

A second concrete gap sits beside it. The mean-field theory of the multiplicity
knows the network only through its excess degree $c_2$, and that is demonstrably
not enough: real networks fall on either side of the $c_2/e$ line, the co-authorship
graph above it and the Internet below. It is tempting to pin the split on the
degree correlations, but the solvable constructions forbid that shortcut --- the
deterministic bubble and the diamond are both disassortative, yet one sits far
below the line and the other far above. What sets the deviation is the architecture
the rule builds, not the sign of its correlations. A theory that predicts that
deviation from the rule itself would close the gap between the calculation and the
data, and would say something the models cannot yet say about which real systems
are route-rich and which are not.

The Ramsey community number has moved further still, though not in the direction
one might have hoped. That it depends on the method used to detect the communities
was always plain. Now that it can be computed exactly, we can say how far that
dependence goes: on one and the same web, under one and the same partition, a
plain block model puts $r_\kappa$ at $1095$ and a degree-corrected one at $42$, and
the two disagree about which cut is best (Chapter~\ref{ch:communities}). The dependence is not a rounding error to be
argued away; it is a factor of twenty-five, and it means a bare Ramsey number
quoted without its detection rule is not a property of the network at all. Yet the
same calculations hand us the beginnings of a repair. Underneath the
method-dependent threshold there is a method-independent quantity: the evidence
density flows to $\ln K$ per link, the entropy of resolving which of $K$
communities a link belongs to, and that value knows nothing of the lattice, the
degrees, or the priors. A definition of emergence built on the invariant rather
than on the crossing --- on the rate at which evidence accumulates rather than on
the size at which it clears an arbitrary bar --- would put the whole idea on firmer
ground.

One question spans both chapters at once, and it may be the most inviting of all.
The ring --- the barest case we can solve, a lattice with no growth and no
randomness in it --- turns out to be split into communities exactly when its links
outnumber its nodes, which is to say exactly when it carries an extensive number of
independent cycles. And cycles are also what multiply its shortest paths. Whether
that accounting is general, or an accident of the ring's simplicity --- whether the
surplus of links over nodes is the common currency of both communities and
redundant routes --- would fold the last two chapters into a single statement. The
book's thesis makes the question natural; it does not answer it.

Two smaller loose ends deserve naming, since both have the flavor of a clue rather
than a nuisance. The number of communities that best describes a self-similar
network grows as $\sqrt{n}$ --- proved exactly for the solvable lattices, where the
optimum sits at the geometric midpoint of the network's own hierarchy, and the
same $\sqrt{n}$ governs the shortest paths there. Real networks, though, divide
\emph{more} finely: their community count grows as $n^{\beta}$ with $\beta$
measured between one half and two thirds, above the self-similar value. What
carries a real network past the fractal $\sqrt{n}$ --- what extra structure, absent
from the self-similar models, its growth builds in --- is a concrete open question
the exponent now makes measurable. And the Ramsey number's climb with cycle length
carries a persistent even--odd alternation, odd cycles bringing communities on
sooner than the even cycles around them. It is too regular to be noise and has no
explanation.

Even the self-organized critical state of the exhaustive searcher, where a
scale-free network appears across a whole range of the control parameter rather
than at one fine-tuned value, deserves to be connected more tightly to the wider
physics of criticality.

Others are programmatic, and the largest of them is the price we paid for solving
anything exactly at all. Every closed form in the last two chapters --- the ring's
evidence, the diamond's renormalization group, the pseudofractal's counts --- lives
on a \emph{deterministic} lattice, a fixed construction rather than a grown
network, with the candidate partition handed to the calculation rather than found
by an algorithm. That is what made the algebra possible, and it is also the
catch: those results bound and illuminate the simulated ones, but they do not
reproduce them. Whether the renormalization-group map survives off the lattice ---
exactly, or on average, for networks grown at random by a local rule --- sets how
far this reading of emergence extends. It is a demanding question, because on
disordered networks detectability is itself a genuine phase transition, and the
threshold we have been computing would become the edge of a phase rather than the
crossing of a bar. The reward would be a theory of emergence for the networks we
actually care about, rather than for their crystalline cousins.

Throughout, we read each local rule \emph{forward}, from
mechanism to statistics; the empirical chapters hint at the reverse problem ---
reading the rule \emph{backward}, inferring from a single observed network which
local moves grew it, and with what parameters, as we did when fitting a
duplication index to project schedules. A general inverse theory, turning measured
correlations, clustering hierarchies and community structure into an estimate of
the underlying rule, would make these models genuinely diagnostic. And real
networks are surely not grown by one rule in isolation: a social graph is part
introduction and part search, a cellular network part duplication and part fresh
interaction. How competing local moves combine --- whether their signatures simply
add, or interfere to produce something neither would alone --- is largely open.

The setting, too, can be widened. The networks of this book are mostly
unweighted, undirected and static, whereas real ones carry weights, directions,
timestamps, and several kinds of link at once. Whether the locality principle
survives into weighted, temporal and multilayer networks --- and what new
regularities it might generate there --- is a natural frontier. So is the coupling
of growth to \emph{function}: the percolation of delays through a project schedule
was a first glimpse of a dynamics running on a network that local rules had built,
and the interplay between how a network grows and what it then does promises more
than either studied alone.

None of these questions threatens the central claim; each extends it. Local rules
are the driving mechanism; everything else follows.

\bibliographystyle{plainnat}
\bibliography{references}

\end{document}